\documentclass[a4paper,11pt]{article}
\pdfoutput=1 

\usepackage{jheppub} 

\usepackage[T1]{fontenc} 

\usepackage{amsmath,amssymb,amsthm,mathtools,dsfont,mathrsfs}
\usepackage[capitalize]{cleveref} 
\usepackage{upgreek}
\usepackage[usenames,dvipsnames]{xcolor}
\usepackage{float,subfig}
\usepackage{graphicx,epsfig}
\usepackage{makecell}
\usepackage{booktabs}

\newtheorem{theorem}{Theorem}[section] 

\newtheorem{proposition}[theorem]{Proposition} 
\newtheorem{claim}[theorem]{Claim}
\newtheorem{lemma}[theorem]{Lemma}
\newtheorem{conjecture}[theorem]{Conjecture}
\theoremstyle{definition}
\newtheorem{definition}[theorem]{Definition}

\newtheorem{deflem}[theorem]{Definition-Lemma}

\newtheorem{example}[theorem]{Example}

\theoremstyle{remark}

\numberwithin{equation}{section}
\renewcommand{\leq}{\leqslant}
\renewcommand{\geq}{\geqslant}
\renewcommand{\hat}{\widehat}

\newcommand{\xbf}{{\mathbf{x}}}
\newcommand{\ebf}{{\mathbf{e}}}
\newcommand{\fbf}{{\mathbf{f}}}
\newcommand{\kbf}{{\mathbf{k}}}
\newcommand{\sbf}{{\mathbf{s}}}
\newcommand{\tbf}{{\mathbf{t}}}
\newcommand{\ubf}{{\mathbf{u}}}
\newcommand{\pbf}{{\mathbf{p}}}

\renewcommand{\gg}{{\mathbb{G}}}
\newcommand{\hh}{{\mathbb{H}}}
\newcommand{\pp}{{\mathbb{P}}}
\newcommand{\rr}{{\mathbb{R}}}
\newcommand{\cc}{{\mathbb{C}}}
\newcommand{\uu}{{\mathbb{U}}}
\newcommand{\ww}{{\mathbb{W}}}
\newcommand{\vv}{{\mathbb{V}}}
\newcommand{\xx}{{\mathbb{X}}}
\newcommand{\zz}{{\mathbb{Z}}}

\newcommand{\cA}{\mathcal{A}}
\newcommand{\cB}{\mathcal{B}}
\newcommand{\cC}{\mathcal{C}}
\newcommand{\cG}{\mathcal{G}}
\newcommand{\cO}{\mathcal{O}}
\newcommand{\cP}{\mathcal{P}}
\newcommand{\cS}{\mathcal{S}}
\newcommand{\cD}{\mathcal{D}}
\newcommand{\cE}{\mathcal{E}}

\newcommand{\cH}{\mathcal{H}}
\newcommand{\cI}{\mathcal{I}}
\newcommand{\cJ}{\mathcal{J}}
\newcommand{\cK}{\mathcal{K}}
\newcommand{\cM}{\mathcal{M}}
\newcommand{\cN}{\mathcal{N}}

\newcommand{\cR}{\mathcal{R}}

\newcommand{\cT}{\mathcal{T}}
\newcommand{\cU}{\mathcal{U}}
\newcommand{\cV}{\mathcal{V}}
\providecommand{\sA}{\mathscr A}

\newcommand{\sE}{{\mathscr{E}}}

\newcommand{\zb}{{\bar{z}}}

\newcommand{\ph}{\varphi}
\newcommand{\botimes}{\bar{\otimes}}
\newcommand{\be}{\begin{equation}}
\newcommand{\ee}{\end{equation}}
\newcommand{\id}{\mathrm{id}}
\newcommand{\ba}{\begin{equation}\begin{aligned}}
\newcommand{\ea}{\end{aligned}\end{equation}}
\newcommand{\baa}{\begin{displaymath}\begin{aligned}}
\newcommand{\eaa}{\end{aligned}\end{displaymath}}

\newcommand{\wt}[1]{{\widetilde{#1}}}
\renewcommand{\sl}{\mathrm{SL}^+_b(2,\rr)}
\makeatletter
\newcommand{\thickhline}{%
    \noalign {\ifnum 0=`}\fi \hrule height 1pt
    \futurelet \reserved@a \@xhline
}
\makeatother

\title{Conformal field theory from\\ automorphic forms on a quantum group}

\author{Dalimil Maz\'{a}\v{c}}

\affiliation{Institut de Physique Th\'{e}orique, CEA, CNRS, Universit\'{e} Paris-Saclay, 91191 Gif-sur-Yvette, France}
\affiliation{Institut des Hautes \'{E}tudes Scientifiques, 91440 Bures-sur-Yvette, France}

\emailAdd{dalimil.mazac@ipht.fr}

\abstract{
We reformulate two-dimensional conformal field theory (2d CFT) in the language of noncommutative geometry. To this end, we introduce automorphic forms for the von Neumann algebraic quantum group $\mathrm{SL}^+_b(2,\mathbb{R})$. A quantum measure space is a von Neumann algebra equipped with a weight. We study such spaces carrying a weight-preserving action of $\mathrm{SL}^+_b(2,\mathbb{R})$. We argue that every unitary bosonic 2d CFT with central charge $c=1+6(b+b^{-1})^2>1$ is equivalent to a quantum measure space with an action of $\mathrm{SL}^+_b(2,\mathbb{R})\times\mathrm{SL}^+_b(2,\mathbb{R})$. In this dictionary, Virasoro primary operators are quantum automorphic forms and their scaling dimensions correspond to the spectrum of a quantum Laplace operator. The Virasoro crossing equations express associativity of multiplication on the quantum measure space. For compact theories, the algebra is a type III$_1$ factor. Liouville CFT is realized as $\mathrm{SL}^+_b(2,\mathbb{R})$ acting on itself by left and right translations. Virasoro boundary conditions correspond to quantum measure spaces with an action of a single copy of $\mathrm{SL}^+_b(2,\mathbb{R})$. When the boundary spectrum is discrete, these are noncommutative counterparts of compact hyperbolic surfaces. The formulation accommodates models both with and without extended chiral algebras. It bridges the Virasoro conformal bootstrap with categorical and von Neumann algebraic approaches to 2d CFT.
}

\begin{document} 
\maketitle
\flushbottom

\newpage
\section{Introduction}\label{sec:intro}
\subsection{The landscape of two-dimensional CFTs}\label{ssec:introLandscape}
Unitary two-dimensional conformal field theories (2d CFTs) with central charge $c>1$ represent a natural frontier in our understanding of quantum field theory. If we order CFTs by spacetime dimension and number of degrees of freedom, these are the first theories in which the infinite-dimensional nature of QFT confronts us head-on. The goal of this paper is to propose a new formulation of these theories, designed specifically with the irrational ones in mind. Before describing it, let us explain why such a reformulation is in order.

We consider bosonic unitary 2d CFTs with equal left- and right-moving central charges $c$. A unitary 2d CFT is completely characterized by its spectral data: the scaling dimensions and spins of its Virasoro primary operators, together with their structure constants. There is a sharp dichotomy between $c<1$ and $c\geq 1$. For $c<1$, unitarity forces the theory to be a minimal model~\cite{BPZ1984}, containing only finitely many Virasoro primaries. The consistent models are classified by pairs of ADE Dynkin diagrams~\cite{FriedanQiuShenker1984,FriedanQiuShenker1986,GoddardKentOlive1986,CappelliItzyksonZuber1987,KawahigashiLongo2004}, and the classification problem is finite-dimensional and purely algebraic. For $c\geq 1$, by contrast, the number of Virasoro primaries in a unitary theory is necessarily infinite~\cite{Cardy1986}. In the borderline case $c=1$, the classification of theories with a discrete spectrum has recently been completed~\cite{Eberhardt:2026qqm}, building on~\cite{Ginsparg1988}. For $c>1$, a classification appears to be out of reach. This is the regime we focus on here.

In special cases, a theory with $c>1$ has enough conserved currents that its spectrum organizes into finitely many representations of an extended chiral algebra. Such theories are rational with respect to that algebra: their structure is captured by finite algebraic data~\cite{MooreSeiberg:1988Polynomial,MooreSeiberg:1989Classical,Fuchs:2002cm}, and constructing them reduces to solving a finite system of algebraic equations. Rationality, however, depends on the choice of chiral algebra. Throughout this paper, primaries are Virasoro primaries and rationality is understood with respect to the Virasoro algebra unless stated otherwise. In this sense every theory with $c>1$ is irrational. Roughly speaking, irrational CFTs are to rational ones what manifolds are to finite sets. The language of noncommutative geometry makes this analogy concrete, as we discuss in~\cref{ssec:spacesAlgebras}.

There is no shortage of unitary CFTs with $c>1$. A special role is played by Liouville theory, a family of CFTs parametrized by $c$ with a continuous spectrum of Virasoro primaries. It is unitary for all $c>1$, admits an exact solution~\cite{DornOtto1994,ZamolodchikovZamolodchikov1996,Teschner1995ThreePoint,PonsotTeschner1999,PonsotTeschner2001,Teschner2001Liouville,Teschner2004Vertex}, and, for $c>25$, has a rigorous probabilistic construction~\cite{KupiainenRhodesVargas2020,GuillarmouKupiainenRhodesVargas2024}. Our main interest, however, lies in \emph{compact} CFTs, whose Hamiltonian on the circle has a discrete spectrum containing the vacuum. Exactly solvable examples abound: theories of several free bosons with compact target space, WZW models of compact groups, tensor products of minimal models, and orbifolds of all of these. Rationality and solvability should not be conflated. For example, free bosons at a generic point of their moduli space are exactly solvable, but they are not rational even with respect to their maximal chiral algebra.

The solvable examples are not expected to capture the spectral complexity of a general compact unitary CFT with $c>1$. That complexity is visible, for example, in superconformal sigma models with compact Calabi--Yau targets, such as K3 surfaces or Calabi--Yau threefolds. In the large-volume regime, the dimensions of low-lying unprotected scalar operators are proportional to eigenvalues of the target-space Laplacian, so their spectra are not expected to admit an explicit solution at generic points of moduli space. Indeed, numerical studies of these spectra exhibit random-matrix statistics~\cite{AfkhamiJeddiAshmoreCordova2022}. Gauging fermion parity then produces consistent bosonic theories that retain these operators. More generally, we expect a generic compact unitary CFT with $c>1$ to have no Virasoro primary conserved currents at all. We will refer to such theories as \emph{generic Virasoro CFTs}. They should exhibit a highly irregular spectrum of transcendental scaling dimensions, which makes them a natural stepping stone toward interacting theories in higher dimensions, such as the 3d Ising CFT~\cite{ElShowk2012Ising,Kos2016Islands,PolandRychkovVichi2018}. Candidate examples can be reached by renormalization group flows from solvable theories~\cite{AntunesBehan2022,AntunesBehan2024}, but no generic Virasoro CFT has been conclusively constructed, and we currently lack the tools to do so systematically. See~\cite{BenjaminFitzpatrickLiThaler2026} for a recent numerical search for solutions of the modular bootstrap.

Unitary 2d CFTs with $c>1$ already admit good axiomatizations. The most general is due to Segal~\cite{Segal:2004}, who described a 2d CFT as a functor from the category of one-manifolds and conformal cobordisms to the category of Hilbert spaces. A more concrete one is the Virasoro conformal bootstrap~\cite{BPZ1984}, which morally follows from Segal's axioms. Its starting point is the Hilbert space on the circle, which carries a unitary action of two commuting copies of the Virasoro algebra and decomposes into irreducible representations labeled by Virasoro primaries. Functoriality then determines every correlation function on every Riemann surface in terms of the spectrum and structure constants, and the consistency of this sewing procedure becomes the Virasoro bootstrap equations. Crossing symmetry of four-point functions on the sphere and modular covariance of one-point functions on the torus suffice to guarantee consistency at any genus~\cite{Sonoda:1988I,Sonoda:1988SewingII,MooreSeiberg:1988Polynomial,MooreSeiberg:1989Classical}. The Virasoro bootstrap thus provides a precise set of axioms: an infinite system of constraints on an infinite set of spectral data. Other approaches include the operator-algebraic framework of conformal nets~\cite{BrunettiGuidoLongo:1993Modular,LongoRehren:1995Nets,KawahigashiLongoMuger:2001Modularity,LongoRehren:2004Boundary,CarpiKawahigashiLongo:2013Boundary,BischoffKawahigashiLongo:2015Boundary,Kawahigashi:2015Conformal}, the categorical construction of rational CFTs~\cite{Fuchs:2002cm,FjelstadEtAl:2006Sewing,Fjelstad:2006aw,Kong:2009inh}, and the probabilistic construction of CFTs from the path integral, which has led to a verification of Segal's axioms in Liouville theory~\cite{DKRV2016,KupiainenRhodesVargas2020,GuillarmouKupiainenRhodesVargas2024,GKRV:Segal}.

The bootstrap equations are expected to encode the answers to many outstanding questions about 2d CFT. Do generic Virasoro CFTs exist, and what are their spectral statistics at high energy? What is the maximal spectral gap at fixed $c$ and as $c\to\infty$~\cite{Hellerman2009Universal,FriedanKeller2013,CollierLinYin2016,AfkhamiJeddiHartmanTajdini2019,HartmanMazacRastelli2019,Perlmutter:2026ujr}? What is the infimum of central charges $c>1$ realized by compact theories? Are generic theories isolated, like the 3d Ising CFT, or do they come in continuous families, like many of the solvable examples? Extracting such answers from the bootstrap equations has proved difficult. In practice, one can impose only a small subset of the equations at a time, and the solutions of a subset behave qualitatively differently from the solutions of the full system.

Beyond these practical difficulties, the bootstrap equations have a conceptual shortcoming. They clearly axiomatize a beautiful class of mathematical objects, but they describe each object in a particular basis, namely the basis of Virasoro primary operators. To see why this matters, consider an analogous problem in geometry. One can describe a compact Riemannian manifold $M$ purely through the spectrum of its Laplacian, $-\nabla^2_M \phi_i=\lambda_i \phi_i$, together with the triple overlaps $\int_M \phi_i \phi_j \phi_k$ of its eigenfunctions. The spectrum alone does not determine $M$~\cite{Kac1966,Milnor1964,Vigneras1980,Sunada1985}, but together with the triple overlaps it does~\cite{GelfandNaimark1943}. Associativity of pointwise multiplication imposes an infinite set of consistency conditions on these data, closely analogous to crossing symmetry~\cite{BonifacioHinterbichler:2019Geometry,BonifacioHinterbichler:2020Einstein,Bonifacio:2021msa,Kravchuk:2021akc,Bonifacio:2021aqf,BonifacioMazacPal:2023}. In fact, these conditions provide a complete axiomatization of compact hyperbolic surfaces or orbifolds in terms of their spectral data~\cite{Adve:2025sld}. Yet imagine someone who has never encountered the concept of a manifold being presented with the eigenvalues, the triple overlaps, and the equations they satisfy. Would they be able to visualize what a manifold is? In conformal field theory, we are still, to a large extent, in the position of this person: we can work with the spectral data and equations they satisfy, but not directly with the underlying object that unites them. This is the question we address in this work:
\begin{center}
\emph{If Virasoro primaries are the analogue of Laplace eigenfunctions,\\ what is the analogue of the manifold?}
\end{center}
Equivalently, what is the basis-independent mathematical object that the Virasoro bootstrap axiomatizes?

In brief, our answer is a von Neumann algebra equipped with an action of the locally compact quantum group $\sl$ and an invariant notion of integration. In other words: a \emph{quantum measure space} with a specific quantum symmetry. We think of the von Neumann algebra as the algebra of bounded functions on a noncommutative space~\cite{Connes1994}, and it takes over the role of the algebra of observables. Conformal boundary conditions correspond to quantum measure spaces with a single action of $\sl$, and bulk CFTs to quantum measure spaces with commuting left and right actions. The description applies whether or not the theory possesses an extended chiral algebra. The rest of this introduction explains the proposal in more detail.

\subsection{Hyperbolic surfaces and the automorphic bootstrap}\label{ssec:introClassical}
Our construction is modeled on the spectral theory of compact hyperbolic surfaces, whose relation to the conformal bootstrap was developed in~\cite{Kravchuk:2021akc,Bonifacio:2021aqf} and is reviewed in~\cref{sec:bootstrapClassical}. \Cref{sec:bootstrapQuantum} develops its quantum counterpart in parallel. Let $G = \mathrm{PSL}(2,\rr)$ and let $\Gamma<G$ be a discrete cocompact torsion-free subgroup, so that the quotient $\Sigma$ of the hyperbolic plane $\hh$ by $\Gamma$ is a compact hyperbolic surface. Instead of $\Sigma$ itself, it is convenient to consider the coset space $X = G/\Gamma$, which encodes the same geometric object. The group $G$ acts on $X$ by left translations, and the Haar measure of $G$ descends to an invariant measure on $X$. Therefore $L^2(X)$ is a unitary representation of $G$, and since $X$ is compact, it decomposes as a discrete direct sum of irreducibles. The trivial representation appears exactly once, accounting for the constant functions. The remaining summands belong to the complementary, principal, and discrete series. The quadratic Casimir of $\mathfrak{sl}(2,\rr)$ acts on functions on $X$ invariant under the maximal compact subgroup as the Laplacian of $\Sigma$, and the Laplace eigenvalue $\lambda = \frac{1}{4}+\nu^2$ labels the corresponding representation. In particular, the trivial representation corresponds to $\nu = \frac{i}{2}$, the complementary series to $\nu\in i(0,\frac{1}{2})$, and the principal series to $\nu\geq 0$. The spectrum
\be
0 = \lambda_0 < \lambda_1 \leq \lambda_2 \leq \ldots \rightarrow \infty
\label{eq:LaplaceSpectrum}
\ee
is typically a sequence of transcendental numbers with irregular spacings. On a generic $\Sigma$, its high-energy behavior is expected to obey random-matrix statistics~\cite{BohigasGiannoniSchmit1984,Bogomolny1997Arithmetical}, and its asymptotic density is governed by the Weyl law~\cite{Weyl1912}.

$K$-finite vectors transforming irreducibly under $G$ give rise to automorphic forms: smooth, hence bounded, functions on $X$. Their products expand back in automorphic forms, with coefficients given by integrals of triple products, the analogues of CFT structure constants. In this language, the consistency conditions discussed above express the associativity of this product expansion~\eqref{eq:bootstrapc}. They are almost identical to four-point crossing equations of a one-dimensional CFT with $\mathrm{SL}(2,\rr)$ symmetry. Besides being complete~\cite{Adve:2025sld} (\cref{ssec:classicalConverse}), these hyperbolic bootstrap equations are effective tools, leading to bounds on spectral gaps~\cite{Kravchuk:2021akc,Bonifacio:2021aqf}, triple products, and hence $L$-functions~\cite{Adve:2025rvf}.

What the spectral description of hyperbolic surfaces lacks is supplied by geometry. Hyperbolic surfaces have many complementary realizations: as quotients of $\hh$, as Riemannian or complex manifolds, as algebraic curves, and, in the arithmetic case, as adelic quotients. The interplay of these points of view is what makes the theory so rich. Geometry also tells us how to produce solutions of the hyperbolic bootstrap systematically: we simply choose a torsion-free cocompact lattice $\Gamma<G$. No such procedure is known for the Virasoro bootstrap. Our aim is to provide 2d CFT with an analogous geometric realization.

The most obvious difference between the two settings is commutativity. Functions on $X$ multiply commutatively, and there is little distinction between states, which are elements of $L^2(X)$, and operators, which are bounded functions acting on $L^2(X)$ by multiplication. The operator algebra of a quantum field theory, on the other hand, is noncommutative. Its noncommutativity is, however, of a mild and tightly controlled kind: spacelike-separated operators commute, and the failure of commutativity at timelike separation is governed by the monodromies of the conformal blocks. This suggests generalizing hyperbolic surfaces to the noncommutative setting. Remarkably, perhaps the simplest such generalization reproduces precisely the notion of a conformal boundary condition with $c>1$, and an analogous construction produces bulk 2d CFTs.

\subsection{Noncommutative spaces and the quantum group \texorpdfstring{$\sl$}{SL+b(2,R)}}\label{ssec:introQG}
In ordinary geometry, a space can be studied through the commutative algebra of functions on it. In the measure-theoretic setting, which is the relevant one for us, measure spaces correspond to commutative von Neumann algebras equipped with a faithful normal semifinite weight, the latter playing the role of integration~\cite{BlecherGoldsteinLabuschagne2022}. Following the philosophy of noncommutative geometry~\cite{Connes1994}, we drop the commutativity assumption and define a \emph{quantum measure space} $\xx$ to be a von Neumann algebra $N$, which we write suggestively as $N = L^{\infty}(\xx)$, equipped with such a weight $\theta$, schematically $\theta(f) = \int_{\xx} f\,d\mu$. When $\theta(1)=1$, the weight is a state, the analogue of a probability measure. A von Neumann algebra by itself retains very little geometric information; for instance, $L^\infty$ of the circle and of an interval are isomorphic. The geometry is restored by additional structure, which in our case is supplied by an ergodic action of a quantum group. We review this point of view, starting from finite sets, in~\cref{ssec:spacesAlgebras}.

To generalize hyperbolic surfaces, we must also generalize the group $\mathrm{PSL}(2,\rr)$. A robust framework that fits the bill already exists: the theory of locally compact quantum groups~\cite{Kac1963,VainermanKac1974,EnockSchwartz1992,BaajSkandalis1993,MasudaNakagami1994,Woronowicz1987,KustermansVaes2000,KustermansVaes2003}. A locally compact quantum group $\gg$ is described by a von Neumann algebra $L^{\infty}(\gg)$ of ``functions on $\gg$'', together with a coproduct encoding the group law and invariant weights playing the role of Haar measures. Classical locally compact groups and their Pontryagin duals are special cases The theory of unitary representations, actions on quantum measure spaces, and harmonic analysis all carry over. We review this framework in~\cref{sec:LCQG}.

The relevant quantum group for 2d CFT is $\sl$, which can be thought of as a quantization of the totally positive part of $\mathrm{SL}(2,\rr)$~\cite{PonsotTeschner1999,Ip2013QuantumPlane,DeCommerSchraderShapiroVoigt2025}. It depends on a parameter $b$, with $\sl$ and $\mathrm{SL}^{+}_{1/b}(2,\rr)$ being canonically isomorphic. The relevant quantum groups are thus labeled by
\be
Q = b + b^{-1}\in(0,\infty)\,,\qquad c = 1+6Q^2\,,
\label{eq:introQc}
\ee
where $b>0$ corresponds to $c\geq 25$ and $|b|=1$ to $1<c\leq 25$. For $b>0$, the structure of a locally compact quantum group follows from~\cite{DeCommerSchraderShapiroVoigt2025}; for $|b|=1$ we provide evidence that it persists (\cref{clm:GbLCQG}). It should be stressed that $\sl$ is not a deformation of $\mathrm{SL}(2,\rr)$: there is no value of $b$ at which it reduces to the classical group. It is an intrinsically quantum object. Its dual is generated by Faddeev's modular double of $\cU_q(\mathfrak{sl}(2,\rr))$~\cite{Faddeev1999ModularDouble,BytskoTeschner2003}, so the modular double plays the role of the Lie algebra, and $\sl$ is its global version, much as $\mathrm{SL}(2,\rr)$ is the global version of $\mathfrak{sl}(2,\rr)$. We review the explicit construction of $\sl$ and its relation to the modular double in~\cref{sec:slb2r}.

The basic reason why $\sl$ is related to 2d CFT is a close parallel between its representation theory and that of the Virasoro algebra at central charge~\eqref{eq:introQc}. This parallel was uncovered by Ponsot and Teschner~\cite{PonsotTeschner1999,PonsotTeschner2001}, and developed further in~\cite{BytskoTeschner2003,Teschner2001Liouville,Teschner2004Vertex,NidaievTeschner2013,TeschnerVartanov2014}. They studied a family of positive principal series representations of the modular double, labeled by $\nu\geq 0$. These correspond to unitary lowest-weight Virasoro modules with conformal weight
\be
h = \frac{Q^2}{4}+\nu^2\,,
\label{eq:introHnu}
\ee
thus covering the range $h\geq\frac{c-1}{24}$. Crucially, the $6j$ symbols of these representations coincide with the crossing kernels of Virasoro four-point conformal blocks~\cite{PonsotTeschner1999,PonsotTeschner2001,TeschnerVartanov2014}. Heuristically, there is a Kazhdan--Lusztig-type correspondence between the modular double with its ordinary tensor product and the Virasoro algebra with the monoidal structure of chiral vertex operators; see~\cite{Teschner2001Liouville,Teschner2004Teichmuller,NidaievTeschner2013} for related discussions. Ponsot and Teschner used it to prove that the DOZZ structure constants of Liouville theory lead to crossing-symmetric four-point functions~\cite{PonsotTeschner1999,PonsotTeschner2001,Teschner2001Liouville}. More broadly, the idea of formulating crossing symmetry in terms of harmonic analysis of the symmetry group was advocated in~\cite{Gadde2017,HogervorstVanRees2017}. The Liouville spectrum consists only of scalar primaries with $h\geq\frac{c-1}{24}$, so the principal series suffices there. A general CFT, however, also contains primaries with $0\leq h<\frac{c-1}{24}$, including the vacuum.

Working with $\sl$ leads to a clean representation theory, in which the principal series is naturally supplemented by the complementary series and the trivial representation (\cref{sec:reps}). The positive principal series gives rise to unitary irreducible representations $\pp_{\nu}$ of $\sl$, $\nu\geq 0$, acting on Hilbert spaces $\cH_\nu$, which suffice for the Plancherel decomposition of $L^2(\sl)$ (\cref{clm:plancherel}). A modified inner product makes the analytic continuation of $\pp_{\nu}$ to $\nu\in i(0,\frac{Q}{2})$ unitary as well (\cref{ssec:unitaryIrreps}). These complementary series representations cover $0<h<\frac{c-1}{24}$, and for $b>0$ they also give new positive representations of the modular double (\cref{ssec:posRepMD}). The trivial representation, describing the vacuum, arises as the limit $\nu\rightarrow\frac{iQ}{2}$ of the complementary series, in a way that mirrors the decoupling of the null vector $L_{-1}|h\rangle$ as $h\rightarrow 0$ (\cref{ssec:CfromP}). We conjecture that the trivial, complementary, and principal series exhaust the unitary dual of $\sl$ (\cref{conj:unitaryDual}). If so, unitary irreducible representations of $\sl$ are in one-to-one correspondence with unitary lowest-weight modules of the Virasoro algebra via~\eqref{eq:introHnu}, as summarized in~\cref{tab:RepCor}. The parallel with $\mathrm{PSL}(2,\rr)$ is close. The trivial representation sits at $\nu = \frac{iQ}{2}$ instead of $\frac{i}{2}$. One notable difference is that $\sl$ has no analogue of the discrete series. The role of the Laplacian is played by the quadratic Casimir $C = FE-qK^2-q^{-1}K^{-2}$ of the modular double~\cite{PonsotTeschner2001}, together with its counterpart $\check{C}$ obtained by $b\rightarrow b^{-1}$~\eqref{eq:Casimir}.

\subsection{Conformal boundary conditions as quantum measure spaces}\label{ssec:introBoundary}
Consider now a unitary Virasoro boundary condition $\mathbb{S}$ with $c>1$ and a discrete spectrum of boundary primaries. Our formulation recasts $\mathbb{S}$ as a quantum measure space $\xx$, a noncommutative version of the coset space $X = G/\Gamma$. The von Neumann algebra $N = L^{\infty}(\xx)$ plays the role of boundary observables. $\theta$ is a state. Furthermore, $\xx$ carries an action $\alpha$ of $\gg_b=\sl$ which preserves $\theta$, the noncommutative counterpart of the action of $G$ on $X$ by left translations. We call such a triple $(N,\theta,\alpha)$, with $\theta$ more generally allowed to be a weight, a \emph{quantum $\gg_b$-space} (\cref{def:quantumGspace}).

The GNS construction for $(N,\theta)$ provides a Hilbert space $L^2(\xx)$, the direct analogue of $L^2(X)$, and the action of $\gg_b$ makes it into a unitary representation of $\gg_b$~\cite{Vaes2001UnitaryImplementation}. For the spaces describing compact boundary conditions, it decomposes as a discrete direct sum of the irreducibles discussed above, with finite multiplicities
\be
L^2(\xx) \cong \bigoplus_{i=0}^{\infty}\cH_{\nu_i}\,.
\label{eq:introBoundarySpec}
\ee
The summands are in one-to-one correspondence with the Virasoro primary operators living on $\mathbb{S}$, and~\eqref{eq:introHnu} gives their conformal weights $h_i$. The trivial representation $\nu_0=\frac{iQ}{2}$ is the vacuum. When the action is ergodic, the vacuum appears exactly once, as befits an elementary boundary condition. The resulting spectrum
\be
0 = h_0 < h_1 \leq h_2 \leq \ldots \rightarrow \infty
\label{eq:PrimarySpectrum}
\ee
is the quantum analogue of the Laplace spectrum~\eqref{eq:LaplaceSpectrum}. It is encoded in the spectrum of quantum Laplace operators on $\xx$, which arise from the Casimirs $C$ and $\check{C}$. By analogy with hyperbolic surfaces, we expect that for a generic $\xx$, the spectrum~\eqref{eq:PrimarySpectrum} is transcendental with irregular spacing and exhibits random-matrix statistics at high energy.\footnote{See ~\cite{HaehlMarteauReevesRozali2023,DiUbaldoPerlmutter2023} for related statements.} However, it is qualitatively denser than its classical counterpart, exhibiting Cardy growth~\cite{Cardy1986} rather than the polynomial Weyl law.

The von Neumann algebra $L^{\infty}(\xx)$ is far from classical. While $L^{\infty}(X)$ is abelian, we prove that $L^{\infty}(\xx)$ is a type III$_1$ factor whenever $\xx$ is ergodic, has a discrete spectrum, and is not a point (\cref{thm:typeIII}). The proof builds on a result of Marrakchi and Vaes~\cite{MarrakchiVaes:2024ErgodicStates}, and ultimately reflects the fact that the modular flow of $\theta$ is implemented by the scaling group of $\gg_b$, which has no classical counterpart. Every elementary compact unitary conformal boundary condition with $c>1$ and at least one nonidentity boundary primary is thus expected to give rise to an ergodic action of a noncompact quantum group on a type III$_1$ factor with an invariant state.

Virasoro primary operators acquire an analytic interpretation. They correspond to eigenfunctions of the quantum Laplace operators on $\xx$, the quantum analogues of automorphic forms. One can thus say that the spectral theory of 2d CFTs is the quantum version of the spectral theory of automorphic forms. Concretely, a quantum automorphic form $\Psi_i(f)\in L^{\infty}(\xx)$ is obtained by lifting an analytic vector $f$ in the $i$-th summand of~\eqref{eq:introBoundarySpec} from $L^2(\xx)$ to $L^{\infty}(\xx)$ via the inverse of the GNS map (\cref{def:quantumAutomorphicForms}). Since quantum automorphic forms are bounded operators, they can be multiplied, and their products can be integrated against $\theta$:
\be
\theta\big(\Psi_{i_1}(f_1)\cdots\Psi_{i_n}(f_n)\big)\,.
\label{eq:introCor}
\ee
We argue that these noncommutative integrals reproduce the correlation functions of boundary primary operators on the disk. In particular, invariance of $\theta$ fixes the two- and three-point functions up to constants, and the boundary structure constants $c_{ijk}$ are integrals of triple products of quantum automorphic forms (\cref{clm:threePoint}). Quantum automorphic forms thus realize Virasoro conformal families directly as bounded operators transforming irreducibly under the symmetry. This is an attractive feature of the framework. In conformal nets~\cite{BrunettiGuidoLongo:1993Modular,LongoRehren:1995Nets,KawahigashiLongoMuger:2001Modularity}, by contrast, a smeared primary field is typically unbounded, and bounded functions of it, such as its exponential, mix different conformal families.

Ergodic quantum $\gg_b$-spaces with an invariant state and a discrete spectrum, subject to two mild technical conditions explained in~\cref{sec:bootstrapQuantum}, will be called \emph{quantum boundary spaces} (\cref{def:quantumBoundarySpace}). The first condition is the existence of quantum automorphic forms, which we conjecture to be automatic. The second is a compatibility between the modular flow of $\theta$ and the decomposition~\eqref{eq:introBoundarySpec}.

The main evidence that this reformulation is correct comes from four-point correlations, i.e.~the $n=4$ case of~\eqref{eq:introCor}. Quantum automorphic forms admit a product expansion convergent in the GNS Hilbert-space norm (\cref{prop:productExpansion}). Together with the KMS condition for $\theta$, it allows us to expand the four-point correlation in two different ways, the s-channel and the t-channel. Equating the two gives the \emph{quantum associativity equations} (\cref{thm:boundaryAssociativity})
\be
\sum_{m}c_{ijm}c_{k\ell m}\,\cB^s_{\nu_i,\nu_j,\nu_k,\nu_\ell;\nu_m}
=
\sum_{m}c_{\ell i m}c_{jkm}\,\cB^t_{\nu_i,\nu_j,\nu_k,\nu_\ell;\nu_m}\,.
\label{eq:introAssoc}
\ee
Here $\cB^s$ and $\cB^t$ are invariant quadrilinear forms on analytic test vectors, the quantum group analogues of s- and t-channel conformal blocks (\cref{sec:invariantForms}). Equations~\eqref{eq:introAssoc} are the noncommutative counterpart of the hyperbolic bootstrap equations~\eqref{eq:bootstrapc}, with $\mathrm{PSL}(2,\rr)$ replaced by $\gg_b$. We argue that~\eqref{eq:introAssoc} are fully equivalent to the Virasoro crossing equations for four points on the boundary of a disk
\be
\sum_{m}c_{ijm}c_{k\ell m}
\cG_{h_i,h_j,h_k,h_\ell;h_m}(\chi)=\sum_{m}c_{\ell i m}c_{jkm}
\cG_{h_i,h_\ell,h_k,h_j;h_m}(1-\chi)\,.
\label{eq:introBVirBootBdry}
\ee
We give a formal argument for the equivalence based on the identification of the $6j$ symbols of $\gg_b$ with the Virasoro fusion kernel. Indeed, $\cB^s$ and $\cB^t$ are related by the same kernel as Virasoro blocks~\eqref{eq:recouplingB}. More convincingly, we also test the equivalence numerically: exact BCFT data of compact free bosons with $c=19$ and $c=31$ solves finite-$m$ truncations of~\eqref{eq:introAssoc} to excellent accuracy, see~\cref{app:freeBosonNumerics}. This test demonstrates that the equivalence works seamlessly also for the identity exchange, and for external and internal complementary series states. Finally, we give numerical evidence that~\eqref{eq:introAssoc} can be obtained by applying continuous linear functionals to~\eqref{eq:introBVirBootBdry}, see~\cref{app:functionalNumerics}.

The above arguments lead us to conjecture that quantum boundary spaces are in one-to-one correspondence with elementary, compact, unitary, Virasoro-preserving conformal boundary conditions with $c>1$ (\cref{conj:equivalence1}).

\subsection{Bulk CFT and locality}\label{ssec:introBulk}
Compact unitary bosonic 2d CFTs with $c>1$ are described in a similar way (\cref{sec:bulkObjects}). They correspond to quantum measure spaces carrying a pair of commuting left and right actions of $\gg_b$, both preserving the state $\theta$, which we call \emph{quantum $\gg_b$-bispaces} (\cref{def:BulkO}). This makes $L^2(\xx)$ into a unitary representation of $\gg_b\times\gg_b$. Its decomposition into irreducibles,
\be
L^2(\xx) \cong \bigoplus_{i=0}^{\infty}\cH_{\nu^L_i}\otimes\cH_{\nu^R_i}\,,
\label{eq:introBulkSpec}
\ee
gives a discrete spectrum of left- and right-moving conformal weights $((h^L_i,h^R_i))_{i=0}^{\infty}$, where $(h^L_0,h^R_0)=(0,0)$ is the Virasoro identity module. Up to exceptional coincidences, it is encoded in the joint spectrum of the commuting quantum Laplace operators of the two actions. Just as for boundary conditions, $L^{\infty}(\xx)$ is then a type III$_1$ factor (\cref{thm:typeIIIBulk}). Summands that are trivial for one of the two actions correspond to holomorphic or antiholomorphic currents, so theories with extended chiral algebras fit naturally into the framework. In particular, generic Virasoro CFTs, which have no Virasoro primary conserved currents, correspond precisely to bispaces that are ergodic separately under the left and under the right action (\cref{prop:separateErgodicity}).

The new ingredient in the bulk is locality, which restricts the degree of noncommutativity of the operator algebra. In bosonic 2d CFTs, it implies that the spins are integers and that the structure constants are symmetric or antisymmetric, depending on the parity of the total spin:
\be
s_i = h^L_i - h^R_i \in \zz\,,\qquad C_{jik} = (-1)^{s_i+s_j+s_k}C_{ijk}\,.
\label{eq:introLocality}
\ee
We use the unitary R-matrix of $\gg_b$~\cite{Faddeev1999ModularDouble,BytskoTeschner2003} (\cref{clm:RmatrixGb}) to formulate locality directly at the level of $\xx$. Intuitively, the requirement is that the product in $L^{\infty}(\xx)$ be commutative up to a braiding determined by the R-matrix, with the left and right actions braiding in opposite directions. We call this property \emph{braided commutativity} (\cref{def:bulkBraidedCommutativity}), and prove that it is equivalent to the locality constraints~\eqref{eq:introLocality} (\cref{prop:bulkBraidingCriterion}). Braided commutative bispaces with a discrete spectrum, subject to the same technical conditions as in the boundary case, will be called \emph{quantum bulk spaces} (\cref{def:quantumBulkSpace}). Their spectral data satisfy bulk quantum associativity equations
\begin{align}
&\sum_{m\in I}C_{ijm}C_{k\ell m}
\bigl(\cB^s_{\nu_i^L,\nu_j^L,\nu_k^L,\nu_\ell^L;\nu_m^L}
\boxtimes\wt{\cB}^s_{\nu_i^R,\nu_j^R,\nu_k^R,\nu_\ell^R;\nu_m^R}\bigr)(F)
\notag\\
&\qquad=\sum_{m\in I}C_{\ell im}C_{jkm}
\bigl(\cB^t_{\nu_i^L,\nu_j^L,\nu_k^L,\nu_\ell^L;\nu_m^L}
\boxtimes\wt{\cB}^t_{\nu_i^R,\nu_j^R,\nu_k^R,\nu_\ell^R;\nu_m^R}\bigr)(F)\,.
\label{eq:assocBulkIntro}
\end{align}
The same arguments as in the boundary case then provide strong evidence that these equations are fully equivalent to the bulk Virasoro crossing equations. We conjecture that quantum bulk spaces are in one-to-one correspondence with compact unitary bosonic 2d CFTs \emph{at genus zero} (\cref{conj:bulkEquivalence}). We will return to the role of modular invariance below.

\subsection{Liouville theory}\label{ssec:introLiouville}
Liouville theory provides an exactly solvable test of the framework (\cref{sec:Liouville}). Already in~\cite[Section~5]{PonsotTeschner1999}, Ponsot and Teschner argued that Liouville theory arises from the regular representation of $\gg_b$. This is a special case of our proposal, where $\xx=\gg_b$ itself, left and right translations provide the commuting pair of actions, and the Haar weight of $\gg_b$ takes the place of the state $\theta$ (\cref{ex:canonicalBispace}). Since the Haar weight is not normalizable, there is no normalizable vacuum. The decomposition of $L^2(\gg_b)$ given by the Plancherel theorem has a continuous spectrum supported on the diagonal principal series $h^L=h^R\geq\frac{c-1}{24}$. This is precisely the spectrum of Liouville theory~\eqref{eq:LiouvilleSpec}. The quantum automorphic forms of $\gg_b$ can be written down explicitly, and applying the Haar weight to their triple products recovers the DOZZ formula for the Liouville structure constants~\cite{DornOtto1994,ZamolodchikovZamolodchikov1996,Teschner1995ThreePoint,KupiainenRhodesVargas2020}. Associativity becomes the orthogonality of the recoupling transformation~\eqref{eq:LiouvilleAssociativity}, as in~\cite{PonsotTeschner1999,PonsotTeschner2001}.

Boundary conditions of Liouville theory also admit simple descriptions (\cref{ssec:LiouvilleBoundaries}). The ZZ brane~\cite{ZZ2001} corresponds to the trivial quantum measure space, a single point, with $L^{\infty}(\xx)=\cc$ (\cref{ex:trivialGSpace}). A stack of $r$ coincident ZZ branes generalizes this to the algebra $M_r(\cc)$ of $r\times r$ matrices, with $\theta$ given by the normalized trace and the trivial action of $\gg_b$ (\cref{ex:ZZstack}). The continuous family of FZZT branes~\cite{FZZ2000,Teschner2000Boundary} arises by taking $L^{\infty}(\xx)$ to be the algebra of all bounded operators on the Hilbert space $\cH_\sigma$ of a principal series representation (\cref{ex:innerGSpace}). The FZZT boundary structure constants~\cite{PonsotTeschner2002Boundary} then follow by applying the invariant weight to a product of three bounded operators transforming irreducibly under $\gg_b$ (\cref{clm:FZZTStructureConstants}). All of these boundary conditions share Liouville theory as their bulk, which in our language is explained by a notion of equivariant Morita equivalence (\cref{ssec:Morita}). Liouville theory can thus be seen as a quantum analogue of the hyperbolic plane, or more precisely of its frame bundle $\mathrm{PSL}(2,\rr)$. Its exact solvability is a direct consequence of harmonic analysis on $\gg_b$.

This suggests a precise version of a rough heuristic: compact unitary 2d CFTs are to Liouville theory what compact hyperbolic surfaces are to the hyperbolic plane. Every compact hyperbolic surface locally looks like the hyperbolic plane, and correspondingly, average properties of high-energy Laplace eigenfunctions, such as their squared triple overlaps, agree with predictions on the hyperbolic plane~\cite{Wyman:2022TripleProducts}. Similarly, the universal high-energy behavior of compact 2d CFTs is controlled by Liouville theory~\cite{Collier:2019weq,NumasawaTsiares:2022Boundary}. In our language, this reflects the expectation that a quantum measure space describing a compact theory is locally indistinguishable from $\gg_b$ itself.

Our formulation is thus the natural common generalization of the automorphic bootstrap of~\cite{Kravchuk:2021akc} on the one hand, and of the work of Ponsot and Teschner on Liouville theory~\cite{PonsotTeschner1999,PonsotTeschner2001,Teschner2001Liouville} on the other. It generalizes the former from $\mathrm{PSL}(2,\rr)$ to $\gg_b$, and the latter from Liouville theory to arbitrary unitary 2d CFTs with $c>1$.

\subsection{Quantum boundary spaces as fundamental objects}\label{ssec:introPicture}
To summarize, there are two natural classes of quantum measure spaces with $\gg_b$ symmetry: quantum boundary spaces, which carry a single action of $\gg_b$, and quantum bulk spaces, which carry a commuting left-right pair of such actions. Diagonalizing the action on a quantum boundary space produces a boundary spectrum $h_i$ and boundary structure constants $c_{ijk}$, which solve the quantum associativity equations, and hence the four-point Virasoro crossing equations on the disk for all combinations of external primaries. Similarly, diagonalizing the actions on a quantum bulk space produces a bulk spectrum $(h^L_i,h^R_i)$ and bulk structure constants $C_{ijk}$, which are consistent with locality and with all Virasoro bootstrap constraints at genus zero.

It is natural to ask whether these constraints fully characterize the underlying quantum measure spaces. An important positive result of this kind was obtained by Adve in the commutative setting~\cite{Adve:2025sld}: every solution of the hyperbolic bootstrap equations of~\cite{Kravchuk:2021akc,Bonifacio:2021aqf} with a discrete spectrum arises from a unique compact hyperbolic surface or orbifold. We conjecture that the analogue of Adve's theorem holds also in the noncommutative setting, both for quantum boundary and bulk spaces: any candidate spectral data with a discrete spectrum that solve the quantum associativity equations arise from a unique quantum measure space (Conjectures~\ref{conj:boundaryReconstruction} and~\ref{conj:converseTheorem}). The uniqueness part is not surprising, since the spectral data determine all matrix elements of the quantum automorphic forms. The nontrivial part is existence, in particular showing that the associativity equations guarantee that the reconstructed operators are bounded. Together with the equivalence of Virasoro crossing and quantum associativity, this conjecture implies that every elementary compact unitary conformal boundary condition with $c>1$ defines a quantum boundary space, and every compact unitary 2d CFT with $c>1$ a quantum bulk space. Our proposal then provides a concrete algorithm for building type III$_1$ factors (or $\cc$) with an action of $\gg_b$ and an invariant state directly from CFT data (\cref{sec:discussion}).

Conversely, does every quantum bulk space define a 2d CFT, and every quantum boundary space a conformal boundary condition? For bulk spaces, the answer is no. Quantum bulk spaces satisfy all sewing constraints at genus zero, but they are not required to be modular invariant. For example, the trivial algebra $L^{\infty}(\xx)=\cc$ is a perfectly consistent quantum bulk space. It describes the genus-zero theory containing only the identity and its Virasoro descendants, which is inconsistent on the torus. Modular invariance is therefore an additional constraint on quantum bulk spaces, which intuitively expresses a kind of maximality or completeness of $L^{\infty}(\xx)$ (\cref{ssec:QGmodularInvariance}).

For boundary spaces, on the other hand, the answer is plausibly yes, as stated in~\cref{conj:equivalence1}. Four-point boundary crossing is the only sewing constraint of a boundary conformal field theory that involves no bulk data~\cite{CardyLewellen1991,Lewellen:1992Sewing}. At the same time, all bulk data can be reconstructed from the spectrum and structure constants of any single boundary condition~\cite{Fjelstad:2006aw,Kong:2009inh,GaberdielRunkel:2008Bulk,FuchsSchweigert:2021Bulk}, by considering the bulk and boundary channels of the annulus with boundary insertions and of the sphere with three disks removed; see~\cite{BrehmRunkel:2022Holes,ChengEtAl:2025Reconstruction,ChenEtAl:2025Liouville} for related constructions. It is plausible that four-point crossing of the boundary data suffices to guarantee that the reconstructed bulk data are consistent on surfaces of all genera, and are in particular modular invariant (\cref{conj:bulkFromBoundary}). This possibility was also raised in~\cite{Jafferis:2025yxt}. For rational CFTs, the analogous statement follows from the construction and uniqueness theorems for RCFT correlators of Fjelstad, Fuchs, Runkel and Schweigert (FRS)~\cite{Fuchs:2002cm,FjelstadEtAl:2006Sewing,Fjelstad:2006aw}.

In fact, our framework can be viewed as an adaptation of the FRS approach from rational to irrational CFTs. In a rational CFT whose chiral algebra has representation category $\mathscr{C}$, the boundary fields on an elementary boundary condition form a Frobenius algebra $A$ in $\mathscr{C}$ (more precisely, a haploid special symmetric one), and every such algebra arises in this way. The bulk CFT can be fully reconstructed from $A$ as the full center $Z(A)$, a commutative Frobenius algebra in $\mathscr{C}\boxtimes\mathscr{C}^{\mathrm{rev}}$ which is modular invariant by construction. Two boundary conditions have the same bulk precisely when they are Morita equivalent~\cite{Kong:2007yv}.

Quantum boundary spaces are irrational analogues of the algebras $A$, with $\mathscr{C}$ replaced by the representation category of $\gg_b$, or equivalently of $\mathrm{Vir}_c$. Quantum bulk spaces are analogues of the commutative algebras in $\mathscr{C}\boxtimes\mathscr{C}^{\mathrm{rev}}$, with braided commutativity playing the role of commutativity (\cref{ssec:QGmodularInvariance,ssec:BulkFromBoundary}). An irrational version of the full center construction, producing a quantum $\gg_b$-bispace from a quantum boundary space, would therefore be a natural route to~\cref{conj:bulkFromBoundary}, with bulk modular invariance guaranteed by construction. In the simplest example, the full center of the point $L^{\infty}(\xx)=\cc$ should be the canonical bispace, i.e.~$\gg_b$ acting on itself by left and right translations, which describes Liouville theory, the common bulk of the ZZ and FZZT branes.

Combining these conjectures leads to an appealing picture of unitary two-dimensional conformal field theory with $c>1$. Its fundamental objects are the quantum boundary spaces: ergodic quantum measure spaces with an action of $\sl$, an invariant state, and a discrete spectrum, subject to no further constraints. They are the noncommutative analogues of compact hyperbolic surfaces. Bulk CFTs are a derived notion, obtained from quantum boundary spaces by an analytic analogue of the full center, which guarantees their modular invariance. With a bit of optimism, one can say that two-dimensional conformal field theory is the theory of quantum hyperbolic surfaces.

\begin{table}[t]
\centering
\small
\renewcommand{\arraystretch}{1.45}
\begin{tabular}{>{\raggedright\arraybackslash}p{4.3cm} >{\raggedright\arraybackslash}p{4.6cm} >{\raggedright\arraybackslash}p{4.6cm}}
\toprule
\emph{classical analogue} & \multicolumn{2}{c}{\emph{conjecturally equivalent}}\\
\cmidrule(r){1-1}\cmidrule(l){2-3}
compact hyperbolic surface & quantum boundary space & Virasoro boundary condition (elementary compact unitary)\\
\midrule
$G=\mathrm{PSL}(2,\rr)$ & $\gg_b = \sl$ & $\mathrm{Vir}_c$, $c = 1+6(b+b^{-1})^2$\\
$X = G/\Gamma$ & $\xx = (N,\theta,\alpha)$ & boundary condition\\
Haar measure & state $\theta$ & disk expectation value\\
left $G$-translations & $\gg_b$-action $\alpha$ & Virasoro action\\
abelian von Neumann algebra $L^{\infty}(X)$ & type III$_1$ factor $N$, or $\cc$ &  boundary observables\\
Hilbert space $L^2(X)$ & GNS Hilbert space & states on the interval\\
Laplace eigenvalue $\lambda = \frac14+\nu^2$ & irreducible representation label $h = \frac{Q^2}{4}+\nu^2$ & conformal weight $h$\\
automorphic form $\phi_i$ & quantum automor.~form $\Psi_i$ & boundary primary operators\\
$\int_X \phi_i\phi_j\phi_k$ &  $\theta(\Psi_i\Psi_j\Psi_k)$ & boundary OPE coef.~$c_{ijk}$\\
associativity~\eqref{eq:bootstrapc} & associativity~\eqref{eq:BAssoc2} & crossing~\eqref{eq:VirBoot}\\
converse theorem~\cite{Adve:2025sld} & \cref{conj:boundaryReconstruction} & \cref{conj:equivalence1}\\
a point & $\cc$ & ZZ brane\\
$G$ with left and right translations & $\gg_b$ with left and right translations & Liouville CFT\\
\bottomrule
\end{tabular}
\caption{The dictionary between compact hyperbolic surfaces, quantum boundary spaces, and conformal boundary conditions at $c>1$. The first column is related to the other two only by analogy, while the second and third columns are conjecturally equivalent (\cref{conj:equivalence1}). The last row concerns bulk objects.}
\label{tab:introDictionary}
\end{table}

\subsection{Summary of results and organization}\label{ssec:introResults}
The dictionary underlying our proposal is summarized in~\cref{tab:introDictionary}. The main results of this paper are the following.
\begin{itemize}
\item \textbf{Representations of $\sl$.} We construct the complementary series of unitary irreducible representations of $\sl$ (\cref{ssec:unitaryIrreps}) and realize the trivial representation as its limit (\cref{ssec:CfromP}). For $b>0$, this yields new positive representations of the modular double (\cref{ssec:posRepMD}). We conjecture that these, together with the principal series, exhaust the unitary dual (\cref{conj:unitaryDual}), in one-to-one correspondence with unitary Virasoro modules.
\item \textbf{Quantum automorphic forms.} For ergodic quantum $\gg_b$-spaces with an invariant state and a discrete spectrum, we define quantum automorphic forms $\Psi_i(f)\in N$ by lifting analytic test vectors $f$ in the summands of $L^2(\xx)$ to $N$ via the inverse of the GNS map (\cref{def:quantumAutomorphicForms}). We conjecture that all such vectors lift (\cref{conj:regularity}), in which case $\Psi_i$ is continuous from analytic test vectors to $N$ with the operator norm (\cref{lem:PsiCts}). Their correlations~\eqref{eq:introCor} obey a twisted cyclic symmetry expressing the KMS condition, and invariance of $\theta$ fixes the two- and three-point correlations in terms of invariant bilinear and trilinear forms, which defines the structure constants $c_{ijk}$ (\cref{lem:2ptcor,clm:threePoint}). The same holds for quantum $\gg_b$-bispaces (\cref{sec:bulkObjects}).
\item \textbf{Structure of quantum measure spaces.} Every nontrivial ergodic spectrally compact quantum $\gg_b$-space or $\gg_b$-bispace is a type III$_1$ factor (\cref{thm:typeIII,thm:typeIIIBulk}).
\item \textbf{Associativity.} Quantum automorphic forms admit a product expansion convergent in the GNS Hilbert-space norm. The spectral data of quantum boundary spaces and quantum bulk spaces satisfy the quantum associativity equations~\eqref{eq:BAssoc2} and~\eqref{eq:assocBulk} (\cref{thm:boundaryAssociativity,prop:bulkFourPoint}).
\item \textbf{Locality.} Braided commutativity of a quantum $\gg_b$-bispace is equivalent to spin integrality and the (anti)symmetry of the bulk structure constants (\cref{prop:bulkBraidingCriterion}).
\item \textbf{Conjectures and evidence.} We conjecture that quantum boundary and bulk spaces are equivalent to compact unitary conformal boundary conditions and genus-zero bulk CFTs with $c>1$ (Conjectures~\ref{conj:equivalence1} and~\ref{conj:bulkEquivalence}). We also conjecture that they can be reconstructed from their spectral data (Conjectures~\ref{conj:boundaryReconstruction} and~\ref{conj:converseTheorem}), and that boundary data determine the bulk CFT (\cref{conj:bulkFromBoundary}). The boundary equivalence is supported by a formal argument and from numerical tests (\cref{app:freeBosonNumerics,app:functionalNumerics}).
\end{itemize}

The paper is organized as follows. \Cref{sec:bootstrapClassical,sec:bootstrapQuantum} form a classical-quantum pair. \Cref{sec:bootstrapClassical} reviews the automorphic bootstrap on $\mathrm{PSL}(2,\rr)$. \Cref{sec:bootstrapQuantum} develops its quantum counterpart in parallel: it states our main constructions and results in a streamlined but mathematically precise form, and discusses the relation to the FRS construction and the role of boundary conditions. \Cref{tab:introDictionary} serves as a bridge between the two. \Cref{sec:LCQG} is a self-contained review of locally compact quantum groups, including representations and actions on quantum measure spaces, starting from finite and locally compact groups. \Cref{sec:slb2r} constructs $\sl$ explicitly and explains its relation to the modular double. \Cref{sec:reps} constructs the unitary representations of $\sl$, including the complementary series, and discusses the Plancherel theorem. \Cref{sec:invariantForms} develops invariant forms on analytic test vectors and their relation to $6j$ symbols. \Cref{sec:boundary,sec:bulkObjects} contain the precise definitions, statements, and proofs for quantum $\gg_b$-spaces and bispaces, including the R-matrix and braided commutativity, and formulate the correspondence with boundary and bulk CFT. \Cref{sec:Liouville} treats Liouville theory and its boundary conditions. We conclude in~\cref{sec:discussion} with a summary and an outlook. The appendices contain the numerical tests. Readers who want the results first may read~\cref{sec:bootstrapQuantum,sec:boundary,sec:bulkObjects}, and return to the quantum group machinery afterwards.

We distinguish between technical claims and conjectures. Statements labeled \emph{Claims} are expected to follow from existing methods, but we do not provide complete proofs, and some analytic details remain to be verified. Statements labeled \emph{Conjectures} concern more substantial questions whose resolution may require new ideas. Theorems and propositions are proved assuming the Claims and explicitly stated conjectural inputs they rely on. Throughout the paper, all Hilbert spaces are assumed to be separable, and all von Neumann algebras are assumed to have a separable predual.

\section{Classical automorphic forms and the conformal bootstrap}\label{sec:bootstrapClassical}
In this section, we will review the interplay between automorphic forms on $\mathrm{PSL}(2,\rr)$ and representation theory, as well as their close connection to the conformal bootstrap. In the rest of the article, we will generalize the present section to noncommutative geometry by replacing $\mathrm{PSL}(2,\rr)$ with the locally compact quantum group $\sl$. This replacement directly leads to a reformulation of two-dimensional conformal field theory.

The ideas reviewed in this section originate in the classic work of Bernstein and Reznikov~\cite{BernsteinReznikov2004,BernsteinReznikov2005,BernsteinReznikov2010}, who applied associativity to bound trilinear periods and thus triple product $L$-functions. The close parallel with the modern conformal bootstrap was pointed out in~\cite{Kravchuk:2021akc}, where associativity was applied to bound spectral gaps. Our exposition combines elements of these references.

\subsection{Automorphic forms on $\mathrm{PSL}(2,\rr)$}
Let $\Sigma$ be a closed connected oriented hyperbolic surface. Such a surface can be equivalently described as $\Sigma =  \hh/\Gamma$, where $\hh$ is the upper half-plane with the hyperbolic metric, and $\Gamma$ is a discrete cocompact torsion-free subgroup of $G = \mathrm{PSL}(2,\rr)$. $\hh$ itself is a quotient of $G$ by a maximal compact subgroup $K = \mathrm{PSO}(2,\rr)$. We thus have
\be
\Sigma =  K \backslash G / \Gamma\,.
\ee
There is then a one-to-one correspondence between closed connected oriented hyperbolic surfaces, considered up to orientation-preserving isometry, and discrete cocompact torsion-free subgroups of $G$, considered up to conjugation in $G$. From now on, we will call these objects compact hyperbolic surfaces for simplicity.

Many fruitful ideas in mathematics revolve around \emph{automorphic forms} on $\Sigma$. Two classes of automorphic forms play a distinguished role:
\begin{enumerate}
\item Maass forms, defined as Laplace eigenfunctions on $\Sigma$, i.e.~smooth functions \mbox{$h_i:\hh\rightarrow \cc$}, such that
\be
-\nabla_{\hh}^2 \,h_i(z) = \lambda_i \,h_i(z)\,,\qquad h_i\left(\frac{az+b}{cz+d}\right) = h_i(z)\,,\quad\text{for all}\quad \small\begin{pmatrix}a & b\\ c& d\end{pmatrix}\in \Gamma\,.
\ee
\item Holomorphic forms, defined as holomorphic functions $f_j:\hh\rightarrow \cc$, such that
\be
f_j\left(\frac{az+b}{cz+d}\right) = (cz+d)^{k_j} f_j(z)\,,\quad\text{for all}\quad \small\begin{pmatrix}a & b\\ c& d\end{pmatrix}\in \Gamma\,.
\ee
\end{enumerate}
Here $\lambda_i\in\rr_{\geq 0}$ is the Laplace eigenvalue of the Maass form $h_i$, and $k_j\in 2\zz_{>0}$ is the weight of the holomorphic form $f_j$. Every compact hyperbolic surface has a discrete spectrum of Laplace eigenvalues, and a spectrum of holomorphic forms, each having finite multiplicities
\be
0=\lambda_0 < \lambda_1 \leq \lambda_2\leq \ldots \rightarrow \infty\,,\qquad
2\leq k_1\leq k_2 \leq \ldots \rightarrow \infty\,.
\ee
On a generic compact hyperbolic surface, the Laplace spectrum is highly erratic, exhibiting symptoms of quantum chaos. We will see that the Laplace eigenfunctions have a close analogue in the noncommutative setting. They correspond to Virasoro primary operators. The spectrum of Laplace eigenvalues corresponds to the spectrum of conformal scaling dimensions. Holomorphic forms do not seem to have a direct analogue for $\sl$.

Automorphic forms on $\Sigma$ can be defined purely representation theoretically as follows. $G$ is a unimodular locally compact group, i.e.~it comes equipped with a Haar measure $\mu_G$, simultaneously invariant under left and right translations. Since $\Gamma < G$ is unimodular, $\mu_G$ gives rise to a measure $\mu$ on $G / \Gamma$, invariant under left translations. We normalize $\mu$ to have unit total mass. We then consider the Hilbert space
\be
\cK = L^2(G/\Gamma, \mu)\,.
\ee
It is analogous to the Hilbert space of an irrational 2d CFT on the unit interval with identical boundary conditions at the endpoints. Through the left action of $G$ on $G/\Gamma$, $\cK$ becomes a strongly continuous unitary representation of $G$. Since $G/\Gamma$ is compact, $\cK$ admits a decomposition as an infinite direct sum of unitary irreducible representations of $G$
\be
\cK \cong \cc \oplus \bigoplus\limits_{i=1}^{\infty} \cP_{\lambda_i} \oplus \bigoplus\limits_{j=1}^{\infty}(\cD_{k_j}\oplus \cD^c_{k_j})\,.
\label{eq:L2GGamma}
\ee
With the exception of the trivial representation $\cc$, each of the direct summands is infinite-dimensional. The irreducible terms correspond to the specific automorphic forms appearing in the spectrum of $\hh/\Gamma$. The trivial representation, always present for surfaces of finite volume, corresponds to the constant functions, i.e.~Laplace eigenfunctions with $\lambda=0$. The representations $\cP_{\lambda_i}$ denote the principal and complementary series, and correspond to Laplace eigenfunctions on $\Sigma$ of eigenvalue $\lambda_i>0$ with
\ba
\,&\text{complementary series:} \quad && \lambda \in (0,\tfrac{1}{4})\,,\\
&\text{principal series:} \quad &&\lambda \in [\tfrac{1}{4},\infty)\,.
\ea
The remaining terms are the discrete series, corresponding to the holomorphic forms on $\Sigma$ together with their antiholomorphic conjugates. Automorphic forms are then defined as $K$-finite vectors inside the irreducible terms in~\eqref{eq:L2GGamma}. In particular, the Maass form $h_i$ lives in the one-dimensional $K$-invariant subspace of $\cP_{\lambda_i}\subset \cK$. We see that finding the often erratic spectrum of the Laplacian on $\Sigma$ is a part of the task of decomposing $\cK$ into irreducible representations. This is analogous to decomposing the Hilbert space of a conformal field theory into irreducible lowest-weight modules of the Virasoro algebra.

\subsection{Multiplication}
The possible spectra that can appear on the right-hand side of~\eqref{eq:L2GGamma} are highly constrained by the compatibility of $G$-equivariance with associativity of pointwise multiplication of functions on $G/\Gamma$. However, some analytic care is needed to formulate this idea correctly. For comparison, imagine replacing $G/\Gamma$ by a finite set $X$ with counting measure. Pointwise multiplication then gives rise to a bounded bilinear map \mbox{$L^2(X)\times L^2(X)\rightarrow L^2(X)$}. This map extends to a bounded linear map of Hilbert spaces
\be
L^2(X)\otimes L^2(X)\rightarrow L^2(X)\,,
\ee
where $\otimes$ denotes the Hilbert-completed tensor product. The situation is very different for $G/\Gamma$. Here, the pointwise product of two elements of $L^2(G/\Gamma)$ may fail to belong to $L^2(G/\Gamma)$. The same problem already occurs on $[0,1]$: the function $f(x) = x^{-\frac{1}{4}}$ belongs to $L^2([0,1])$ but $f^2$ does not. Thus, multiplication defines neither a bilinear map $\cK \times \cK \rightarrow \cK$, nor a bounded linear map $\cK \otimes \cK \rightarrow \cK$.

A simple resolution of this problem is to restrict multiplication to smooth functions, denoted $C^{\infty}(G/\Gamma)$. Since the product of smooth functions is itself smooth, multiplication defines a bilinear map
\be
m: C^{\infty}(G/\Gamma)\times C^{\infty}(G/\Gamma)\rightarrow C^{\infty}(G/\Gamma)\,.
\ee
$m$ is jointly continuous with respect to the Fr\'{e}chet topology on $C^{\infty}(G/\Gamma)$. Since $G/\Gamma$ is compact, we have $C^{\infty}(G/\Gamma)\subset \cK$, with $C^{\infty}(G/\Gamma)$ being dense in $\cK$. Furthermore, automorphic forms, i.e.~$K$-finite vectors in the irreducible summands of~\eqref{eq:L2GGamma}, are automatically smooth. The notion of smoothness is incorporated into the representation-theoretic setting as follows. One defines the abstract space of smooth vectors $\cP_{\lambda}^{\infty}$ in the irreducible representations as
\be
\cP_{\lambda}^{\infty} := \{v \in \cP_{\lambda}\,:\, G\ni g\mapsto g\cdot v\in \cP_{\lambda}\text{ is smooth}\}\,.
\ee
$\cP_{\lambda}^{\infty}$ comes naturally equipped with a Fr\'{e}chet topology. It is a dense subspace of $\cP_{\lambda}$ containing all of its $K$-finite vectors. However, not all elements of $\cP_{\lambda}^{\infty}$ are $K$-finite. Note that the action of $G$ preserves $\cP_{\lambda}^{\infty}$. Let us also define
\be
\psi_i : \cP_{\lambda_i} \rightarrow \cK
\label{eq:psiClassical}
\ee
to be the isometric intertwining embeddings provided by~\eqref{eq:L2GGamma}. The corresponding adjoint maps \mbox{$\psi^*_i : \cK \rightarrow \cP_{\lambda_i}$} satisfy $\psi^*_i \psi_{i} =\id_{\cP_{\lambda_i}}$ and $\psi^*_i \psi_{j}=0$ for $i\neq j$. Crucially, we have
\be
\psi_i(\cP_{\lambda_i}^{\infty}) = \psi_i(\cP_{\lambda_i}) \cap C^{\infty}(G/\Gamma)\,,
\ee
i.e.~abstract smooth vectors in $\cP_{\lambda_i}$ correspond precisely to smooth functions on $G/\Gamma$ in the image of $\psi_i$. This means pointwise multiplication restricts to the irreducible components through the bilinear maps
\ba
\,&m^{k}_{ij}:\; \cP_{\lambda_i}^{\infty}\times \cP_{\lambda_j}^{\infty} \rightarrow \cP_{\lambda_k}^{\infty}\,,\\
&m^{k}_{ij}(v_1,v_2):= \psi^*_k(m(\psi_i(v_1),\psi_j(v_2)))\,.
\label{eq:mijk}
\ea
Furthermore, $m^{k}_{ij}$ are $G$-equivariant and jointly continuous with respect to the respective Fr\'{e}chet topologies. However, $m^{k}_{ij}$ may not extend to bounded linear maps of Hilbert spaces $\cP_{\lambda_i}\otimes \cP_{\lambda_j}\rightarrow \cP_{\lambda_k}$. To complete the picture, write the left-invariant integration on $G/\Gamma$ as
\be
\theta: L^{1}(G/\Gamma)\rightarrow \cc\,,\qquad \theta(F) = \int\limits_{G/\Gamma} F(g) \,d\mu(g)\,.
\ee
We can then define the multilinear, $G$-invariant, and jointly continuous maps
\ba
\,&\cC^{(n)}_{i_1,\ldots, i_n} : \cP_{\lambda_{i_1}}^{\infty}\times\ldots \times \cP_{\lambda_{i_n}}^{\infty} \rightarrow \cc\,,\\
& \cC^{(n)}_{i_1,\ldots, i_n}(v_1,\ldots,v_n):= \theta(m(\psi_{i_1}(v_1),\ldots,\psi_{i_n}(v_n)))\,,
\ea
where $m:C^{\infty}(G/\Gamma)^n\rightarrow C^{\infty}(G/\Gamma)$ is the pointwise product of $n$ functions. $\cC^{(n)}_{i_1,\ldots, i_n}$ are analogous to $n$-point correlation functions in a conformal field theory. In general, they do not extend to bounded linear maps of Hilbert spaces $\cP_{\lambda_{i_1}}\otimes\ldots \otimes \cP_{\lambda_{i_n}}\rightarrow \cc$.

\subsection{Trilinear forms and the product expansion}
Analysis of correlations $\cC^{(n)}_{i_1,\ldots, i_n}$ is simplified by noting that $G$-invariance strongly constrains continuous multilinear forms on $\cP_{\lambda}^{\infty}$. In particular, any such map $\cP_{\lambda_1}^{\infty} \times \cP_{\lambda_2}^{\infty}\rightarrow \cc$ vanishes unless $\lambda_1=\lambda_2$. In that case, the space is one-dimensional, generated by the symmetric bilinear form
\be
\cD_{\lambda}: \cP_{\lambda}^{\infty} \times \cP_{\lambda}^{\infty}\rightarrow \cc\,,\qquad
\cD_{\lambda}(v_1,v_2) := \langle J_{\lambda} v_1, v_2\rangle_{\lambda}\,.
\ee
Here $J_{\lambda}:\cP_{\lambda}\rightarrow \cP_{\lambda}$ and $\langle\cdot,\cdot \rangle_{\lambda}$ are respectively the $G$-invariant real structure, and inner product on $\cP_{\lambda}$. $\cD_{\lambda}$ does not extend to a bounded linear intertwiner $\cP_{\lambda}\otimes \cP_{\lambda}\rightarrow \cc$ at the Hilbert space level. Indeed, there does not exist any (nonvanishing) such intertwiner. We can always ensure that all $\psi_i$ intertwine $J_{\lambda_i}$ with the canonical real structure on $\cK$, provided by pointwise complex conjugation. In that case
\be
\cC^{(2)}_{i,i} = \cD_{\lambda_{i}} \quad\text{and}\quad \cC^{(2)}_{i,j} = 0 \quad\text{for}\quad i\neq j\,.
\ee
Similarly, the space of continuous, trilinear, $G$-invariant maps $\cP_{\lambda_1}^{\infty} \times \cP_{\lambda_2}^{\infty}\times \cP_{\lambda_3}^{\infty} \rightarrow \cc$ is at most two-dimensional, generated by
\be
\cT^{+}_{\lambda_1,\lambda_2,\lambda_3},\, \cT^{-}_{\lambda_1,\lambda_2,\lambda_3}: \,\cP_{\lambda_1}^{\infty} \times \cP_{\lambda_2}^{\infty}\times \cP_{\lambda_3}^{\infty} \rightarrow \cc\,,
\ee
where the sign in the superscript refers to their parity under an orientation-reversing involution in $\mathrm{PGL}(2,\rr)$. We can normalize $\cT^{\pm}_{\lambda_1,\lambda_2,\lambda_3}$ to be compatible with the real structure
\be
\cT^{\pm}_{\lambda_1,\lambda_2,\lambda_3}(J_{\lambda_1} v_1,J_{\lambda_2} v_2,J_{\lambda_3} v_3) =
\overline{\cT^{\pm}_{\lambda_1,\lambda_2,\lambda_3}(v_1,v_2,v_3)}\,,
\ee
which fixes each form up to a nonzero real factor. Furthermore, we can choose this factor so that $\cT^{+}_{\lambda_1,\lambda_2,\lambda_3}(v_1,v_2,v_3)$ and $\cT^{-}_{\lambda_1,\lambda_2,\lambda_3}(v_1,v_2,v_3)$ are respectively fully symmetric and antisymmetric under simultaneous permutations of $\lambda_i$ and $v_i$.

It follows that we must have, for any three nonconstant Laplace eigenfunctions in the spectrum,
\be
\cC^{(3)}_{i,j,k} = c^{+}_{ijk} \cT^{+}_{\lambda_i,\lambda_j,\lambda_k} + c^{-}_{ijk} \cT^{-}_{\lambda_i,\lambda_j,\lambda_k}
\ee
for some $c^{+}_{ijk}, c^{-}_{ijk} \in \rr$. Commutativity of pointwise multiplication makes $c^{+}_{ijk}$ and $c^{-}_{ijk}$ respectively fully symmetric and antisymmetric in $i,j,k$. $c^{\pm}_{ijk}$ are analogous to the OPE coefficients in conformal field theory.

Finally, the space of continuous, bilinear, $G$-equivariant maps $\cP_{\lambda_2}^{\infty}\times \cP_{\lambda_3}^{\infty} \rightarrow \cP_{\lambda_1}^{\infty}$ is generated by $\cT^{\lambda_1,\pm}_{\lambda_2,\lambda_3}$, which are uniquely fixed by demanding
\be
\cT^{\pm}_{\lambda_1,\lambda_2,\lambda_3}(v_1,v_2,v_3) = \cD_{\lambda_1}(v_1,\cT^{\lambda_1,\pm}_{\lambda_2,\lambda_3}(v_2,v_3))
\ee
for all $v_1\in \cP_{\lambda_1}^{\infty}$, $v_2\in \cP_{\lambda_2}^{\infty}$, $v_3\in \cP_{\lambda_3}^{\infty}$.

$\cT^{\pm}_{\lambda_1,\lambda_2,\lambda_3}$ never extends to a bounded linear intertwiner $\cP_{\lambda_1}\otimes \cP_{\lambda_2}\otimes \cP_{\lambda_3}\rightarrow \cc$ at the Hilbert space level. $\cT^{\lambda_1,\pm}_{\lambda_2,\lambda_3}$ only extends to $\cP_{\lambda_2}\otimes \cP_{\lambda_3}\rightarrow \cP_{\lambda_1}$ in exceptional cases.\footnote{It extends if and only if all three parameters lie in the complementary series and satisfy $\sigma_1 = \sigma_2+\sigma_3$ with $\lambda_j = \sigma_j(1-\sigma_j)$, $0<\sigma_j<1/2$. If we include the discrete series, there are additional cases.} Indeed, the (non)existence of such intertwiners is irrelevant to our discussion, which instead requires continuity in the Fr\'{e}chet topologies.

Note that
\ba
\cC^{(3)}_{i,j,k}(v_1,v_2,v_3) &= \langle \overline{\psi_i(v_1)},m(\psi_j(v_2),\psi_k(v_3))\rangle_{\cK}\\
&= \langle J_{\lambda_i}v_1 ,m^{i}_{jk}(v_2,v_3)\rangle_{\lambda_i}\\
&= \cD_{\lambda_i}(v_1 ,m^{i}_{jk}(v_2,v_3))\,,
\ea
where $m^{i}_{jk}$ are the multiplication maps of~\eqref{eq:mijk}. It follows that
\be
m^i_{jk} = c^{+}_{ijk} \cT^{\lambda_i,+}_{\lambda_j,\lambda_k} + c^{-}_{ijk} \cT^{\lambda_i,-}_{\lambda_j,\lambda_k}\,,
\ee
i.e.~$c^{\pm}_{ijk}$ play the role of structure constants in the basis of automorphic forms. The analogous uniqueness results apply also to the discrete series. We refer the reader to~\cite{Kravchuk:2021akc,Bonifacio:2021aqf} for details.

Let us then pick two vectors $v_1\in \cP_{\lambda_j}^{\infty}$, $v_2\in \cP_{\lambda_k}^{\infty}$ and consider the pointwise product $m(\psi_j(v_1),\psi_k(v_2)) \in C^{\infty}(G/\Gamma)$. Completeness of the direct sum decomposition~\eqref{eq:L2GGamma} implies the convergent expansion
\be
m(\psi_j(v_1),\psi_k(v_2)) =  \delta_{jk} \, \cD_{\lambda_j}(v_1,v_2) + \sum\limits_{i=1}^{\infty}\sum\limits_{\epsilon=\pm} c^{\epsilon}_{ijk}  \psi_i(\cT^{\lambda_i,\epsilon }_{\lambda_j,\lambda_k}(v_1,v_2))+\text{discrete series}\,.
\label{eq:OPEc}
\ee
The full expansion, including the discrete series contributions, converges both in the norm of $\cK$ and the Fr\'{e}chet topology of $C^{\infty}(G/\Gamma)$. This formula is analogous to the operator product expansion in conformal field theory.

\subsection{Conformal bootstrap and the converse theorem}\label{ssec:classicalConverse}
Consider the four-point correlation $\cC^{(4)}_{i,j,k,\ell}(v_1,v_2,v_3,v_4)$ for $v_1\in \cP_{\lambda_i}^{\infty}$, $v_2\in \cP_{\lambda_j}^{\infty}$, $v_3\in \cP_{\lambda_k}^{\infty}$, $v_4\in \cP_{\lambda_\ell}^{\infty}$. We can apply the expansion~\eqref{eq:OPEc} to its last two entries $m(\psi_k(v_3),\psi_\ell(v_4))$. $L^2$-convergence guarantees the sum can be exchanged with $\theta$, which leads to the formula
\ba
\cC^{(4)}_{i,j,k,\ell}(v_1,v_2,v_3,v_4) =  &\delta_{ij}\delta_{k\ell} \, \cD_{\lambda_i}(v_1,v_2)\cD_{\lambda_k}(v_3,v_4)\\
&\qquad+\sum\limits_{m=1}^{\infty}\sum\limits_{\epsilon,\epsilon'=\pm} c^{\epsilon}_{ijm}c^{\epsilon'}_{k\ell m} \cB^{\epsilon,\epsilon'}_{\lambda_i,\lambda_j,\lambda_k,\lambda_\ell;\lambda_m}(v_1,v_2,v_3,v_4)\\
&\qquad\qquad+\sum \text{discrete series}\,,
\label{eq:C4}
\ea
where we define
\ba
\,&\cB^{\epsilon,\epsilon'}_{\lambda_1,\lambda_2,\lambda_3,\lambda_4;\lambda_5}:\;
 \cP_{\lambda_1}^{\infty} \times \cP_{\lambda_2}^{\infty}\times \cP_{\lambda_3}^{\infty} \times \cP_{\lambda_4}^{\infty} \rightarrow \cc\\
&\cB^{\epsilon,\epsilon'}_{\lambda_1,\lambda_2,\lambda_3,\lambda_4;\lambda_5}(v_1,v_2,v_3,v_4):= 
\cT^{\epsilon}_{\lambda_1,\lambda_2,\lambda_5}(v_1,v_2,\cT^{\lambda_5,\epsilon'}_{\lambda_3,\lambda_4}(v_3,v_4))\\
&\qquad\qquad\qquad\qquad\qquad\quad\quad\!=\cD_{\lambda_5}(\cT^{\lambda_5,\epsilon}_{\lambda_1,\lambda_2}(v_1,v_2),\cT^{\lambda_5,\epsilon'}_{\lambda_3,\lambda_4}(v_3,v_4))
\,.
\ea
$\cB^{\epsilon,\epsilon'}_{\lambda_1,\lambda_2,\lambda_3,\lambda_4;\lambda_5}$ is the analogue of the conformal block. It is a continuous, quadrilinear, $G$-invariant map, determined by representation theory of $\mathrm{PSL}(2,\rr)$. \eqref{eq:C4}~is in turn the analogue of the conformal block expansion. 

Since pointwise multiplication is commutative, we must have
\be
\cC^{(4)}_{i,j,k,\ell}(v_1,v_2,v_3,v_4) = \cC^{(4)}_{j,k,\ell,i}(v_2,v_3,v_4,v_1)\,.
\label{eq:permutation}
\ee
This symmetry is not manifest in the expansion~\eqref{eq:C4}. Applying~\eqref{eq:C4} to both sides of~\eqref{eq:permutation} leads to
\ba
\,&\delta_{ij}\delta_{k\ell} \, \cD_{\lambda_i}(v_1,v_2)\cD_{\lambda_k}(v_3,v_4)+\sum\limits_{m=1}^{\infty}\sum\limits_{\epsilon,\epsilon'=\pm} c^{\epsilon}_{ijm}c^{\epsilon'}_{k\ell m} \cB^{\epsilon,\epsilon'}_{\lambda_i,\lambda_j,\lambda_k,\lambda_\ell;\lambda_m}(v_1,v_2,v_3,v_4) \\
&\qquad+ \text{discrete series}\\
&=\delta_{jk}\delta_{\ell i} \, \cD_{\lambda_j}(v_2,v_3)\cD_{\lambda_\ell}(v_4,v_1)+\sum\limits_{m=1}^{\infty}\sum\limits_{\epsilon,\epsilon'=\pm} c^{\epsilon}_{jkm}c^{\epsilon'}_{\ell i m} \cB^{\epsilon,\epsilon'}_{\lambda_j,\lambda_k,\lambda_\ell,\lambda_i;\lambda_m}(v_2,v_3,v_4,v_1) \\
&\qquad+ \text{discrete series}\,.
\label{eq:bootstrapc}
\ea
To every compact hyperbolic surface $\Sigma$, we can associate its spectral data, consisting of the countably infinite set of real numbers $\lambda_i(\Sigma)$ with $i\in\zz_{>0}$, and $c^{\pm}_{ijk}(\Sigma)$ with $i,j,k\in\zz_{>0}$, together with their analogues involving holomorphic forms.~\eqref{eq:bootstrapc} represents an infinite set of identities that the spectral data of any $\Sigma$ must satisfy.\footnote{There are also additional identities arising from four-point correlations including the holomorphic forms.} The identities implement associativity of pointwise multiplication in the basis of automorphic forms, and we will call them associativity equations. They are universal in the sense of being independent of $\Sigma$. They are analogous to the conformal bootstrap equations, which constrain the scaling dimensions and structure constants of conformal field theories.

The associativity equations imply strong constraints on both the Laplace spectrum~\cite{Bonifacio:2021aqf,Kravchuk:2021akc}, and the structure constants~\cite{BernsteinReznikov2010}. When $\Sigma$ is an arithmetic hyperbolic surface, the Watson--Ichino formula relates structure constants to central values of certain triple product $L$-functions. The associativity equations with external discrete series then imply Weyl-strength subconvex bounds on the corresponding $L$-functions~\cite{Adve:2025rvf}.

Remarkably, the full set of associativity equations completely characterizes compact hyperbolic surfaces or orbifolds~\cite{Adve:2025sld}. More precisely, a candidate set of spectral data solves the full set of associativity equations if and only if there exists a compact hyperbolic surface or orbifold which realizes this data, assuming the candidate Laplace spectrum is discrete. In that light, one can say that the associativity equations \emph{define} compact hyperbolic surfaces or orbifolds.

\section{Automorphic forms on $\sl$ and Virasoro CFTs}\label{sec:bootstrapQuantum}
The purpose of this section is to overview our main results in a mathematically precise language. The idea is to generalize the reasoning of Section~\ref{sec:bootstrapClassical} from $\mathrm{PSL}(2,\rr)$ to the locally compact quantum group $\sl$. This will unavoidably lead us to rediscover two-dimensional conformal field theory with $c>1$ as well as the notion of conformal boundary conditions.
\subsection{$\gg_b=\sl$ and quantum $\gg_b$-spaces}\label{ssec:QGsl2}
The first step is to replace the locally compact group $\mathrm{PSL}(2,\rr)$ by the locally compact quantum group $\gg_b:=\sl$. This is not a group. Instead, it is realized through a von Neumann algebra $M$. $M$ represents the noncommutative algebra of essentially bounded complex functions on $\sl$, i.e.~$M = L^{\infty}(\sl)$. The group multiplication law is realized through the coproduct $\Delta$, which is a normal unital $*$-homomorphism of von Neumann algebras $\Delta: M\rightarrow M\botimes M$. The notion of invariant integration on $\sl$ is realized as a normal semifinite faithful weight $\ph$ on $M$. $\ph$ is simultaneously left- and right-invariant with respect to $\Delta$. In other words, $\sl$ is unimodular. We have $\ph(1_M) = \infty$, meaning that $\sl$ is noncompact. $M$, $\Delta$ and $\ph$ are described in detail in Section~\ref{sec:slb2r}.

The next natural step is to generalize homogeneous spaces $\mathrm{PSL}(2,\rr)/\Gamma$ to the noncommutative setting. This is considerably more subtle than in the classical case. There exists an elegant definition of a closed quantum subgroup of a locally compact quantum group~\cite{Vaes2005Induction}. However, the natural noncommutative analogues of homogeneous spaces~\cite{Kasprzak2012QuantumHomogeneousSpaces,KasprzakSoltan2014Embeddable} are strictly more general than quotients of a locally compact quantum group by a closed quantum subgroup.

We do not commit to any of the existing definitions. Instead, we proceed as follows. Initially, we replace $X = \mathrm{PSL}(2,\rr)/\Gamma$ by a von Neumann algebra $N$. $N$ represents the noncommutative algebra of essentially bounded functions on a quantum measurable space $\xx$, i.e.~$N = L^{\infty}(\xx)$. Since $\xx$ is not a quantum group, $N$ is not equipped with a coproduct. On the other hand, we do require that $\sl$ acts on $\xx$. The action is realized as a (left) coaction of $(M,\Delta)$ on $N$, i.e.~a normal, injective, unital $*$-homomorphism $\alpha: N \rightarrow M\botimes N$, compatible with $\Delta$ in the sense of~\eqref{eq:actionDef}. We also impose that the action is ergodic, i.e.~that the fixed-point algebra $N^{\alpha}:=\{a\in N\,:\,\alpha(a) = 1_M\otimes a\}$ is trivial, $N^{\alpha}=\cc 1_N$. Finally, we equip $\xx$ with invariant integration. This is realized as a faithful normal state $\theta: N\rightarrow \cc$. $\theta$ being a state means $\xx$ has finite volume. Invariance of $\theta$ under the action is expressed as
\be
(\id\otimes \theta)\alpha(a) = \theta(a) 1_M\,,\qquad a\in N\,.
\label{eq:thetaInvariance}
\ee
In summary, we realize the noncommutative analogue of $\mathrm{PSL}(2,\rr)/\Gamma$ as the triple $(N,\theta,\alpha)$. More generally, a pair $(N,\theta)$, with $\theta$ a normal semifinite faithful weight on $N$, is called a quantum measure space, and a triple $(N,\theta,\alpha)$ with $\alpha$ a $\theta$-preserving action of $\gg_b$ is called a \emph{quantum $\gg_b$-space} (\cref{def:quantumGspace}). We are thus considering ergodic quantum $\gg_b$-spaces whose weight $\theta$ is a state. Note that when a classical noncompact group $G$ acts on a classical measure space $X$, ergodicity and finite volume are not sufficient for transitivity.\footnote{An example is provided by the action of $\rr$ on $\rr^2/\zz^2$ given by $t\cdot (x,y) = (x+t,y+\gamma \,t)$ with $\gamma\notin\mathbb{Q}$. This is ergodic but not transitive.} We will soon supplement our requirements on $(N,\theta,\alpha)$ by a spectral condition, spectral compactness, whose classical counterpart does ensure transitivity.

The data $(N,\theta,\alpha)$ is sufficient to set up harmonic analysis on the quantum space $\xx$. To proceed, we need a generalization of the Hilbert space $L^2(\mathrm{PSL}(2,\rr)/\Gamma)$. This is provided by the GNS construction for the pair $(N,\theta)$. We will denote the GNS Hilbert space interchangeably as $\cK_{\theta}$ or $L^2(\xx,\theta)$. The GNS map $\Lambda_\theta: N \rightarrow \cK_{\theta}$ generalizes the inclusion $L^{\infty}(X) \hookrightarrow L^2(X)$. In perfect analogy with the classical situation, the coaction $\alpha$ gives $\cK_{\theta}$ the structure of a unitary representation $\uu$ of $\sl$~\cite{Vaes2001UnitaryImplementation} (\cref{thm:UImplementation}).

We would like to decompose $\cK_{\theta}$ into unitary irreducible representations of $\sl$. In Section~\ref{sec:reps}, we construct a family of unitary irreducible representations $\pp_{\nu}$ of $\sl$ on Hilbert spaces $\cH_{\nu}$. Here $\nu \in \rr \cup i (-\frac{Q}{2},\frac{Q}{2})$ is a parameter. Representations $\pp_{\nu}$ and $\pp_{-\nu}$ are equivalent. $\pp_{\nu}$ with $\nu\in \rr$ is analogous to the principal series of $\mathrm{PSL}(2,\rr)$, while $\pp_{\nu}$ with $\nu \in i (-\frac{Q}{2},\frac{Q}{2})\setminus\{0\}$ is analogous to the complementary series, and we adopt these names for $\pp_\nu$ as well. The underlying Hilbert spaces $\cH_{\nu}$ are completions of spaces of functions on the real line. In particular, $\cH_{\nu}\cong L^2(\rr)$ for the principal series. We conjecture that together with the trivial representation, the complementary and principal series furnish the complete set of unitary irreducible representations of $\sl$ (\cref{conj:unitaryDual}).

$\theta$ being a state and ergodicity of $\alpha$ together ensure that the trivial representation appears in $\cK_{\theta}$ with unit multiplicity. Irrespective of whether \cref{conj:unitaryDual} holds, we will assume that besides the trivial representation, $\cK_{\theta}$ decomposes as a direct sum of complementary and principal series:
\be
\cK_{\theta} \cong \cc \oplus \bigoplus_{j\in I} \cH_{\nu_j}\,.
\label{eq:QGKthetaD}
\ee
Here $I$ is a countable indexing set and $\nu_j \in \rr_{\geq 0}\cup i(0,\frac{Q}{2})$. Unlike in Sections~\ref{sec:boundary} and~\ref{sec:bulkObjects}, where the trivial summand is labeled by $0\in I$, in this section $I$ labels only the nontrivial summands. We call $(\nu_j)_{j\in I}$ the spectrum of $\xx$ and assume it is discrete with finite multiplicities. Since $\theta$ is a state, this means that $(N,\theta,\alpha)$ is \emph{spectrally compact} (\cref{def:spectralCompactness}). The relation between ergodic spectrally compact quantum $\gg_b$-spaces and quantum homogeneous spaces in the sense of~\cite{Kasprzak2012QuantumHomogeneousSpaces,KasprzakSoltan2014Embeddable} is unclear. In particular, it would be interesting to understand whether every ergodic spectrally compact quantum $\gg_b$-space is homogeneous in that sense. The analogous implication does hold in the classical setting, following from~\cite{Adve:2025sld}.

\subsection{Quantum automorphic forms and multiplication}\label{ssec:QGauto}
The decomposition~\eqref{eq:QGKthetaD} provides a complete set of isometric equivariant embeddings
\be
\psi_i: \cH_{\nu_i}\rightarrow \cK_{\theta}\,.
\ee
All of the complementary and principal series representations of $\sl$ admit a real structure $J_{\nu}$ (\cref{clm:Jnu}). $\cK_{\theta}$ carries a canonical real structure given by the modular conjugation $J_{\theta}$ for the pair $(N,\theta)$. $J_{\theta}$ is the quantum analogue of pointwise complex conjugation of elements of $L^2(G/\Gamma)$. It is provided by the Tomita--Takesaki theory. We can (and will) choose each $\psi_i$ so that it intertwines the respective real structures, see~\eqref{eq:psiReal}.

A novel feature of the noncommutative setting is the modular flow of $\theta$ on $N$. It is given by $a \mapsto \nabla^{it}_{\theta}a \nabla^{-it}_{\theta}$, where $\nabla_{\theta}$ is the modular operator. We prove that in the absence of multiplicities in the spectrum, the modular flow preserves the image of each $\psi_i$ (\cref{prop:modularFlowPsi}). More generally, it can mix representations with the same $\nu_i$. We specialize to quantum $\gg_b$-spaces such that no mixing occurs, and call them \emph{symmetric} (\cref{def:symmetry}). This is analogous to the symmetry requirement for symmetric Frobenius algebras in the FRS construction~\cite{Fuchs:2002cm}. Even without assuming symmetry, we deduce that the only elements of $N$ fixed by the modular flow are multiples of $1_N$. It follows that if $N$ is nontrivial, it is a type III$_1$ factor (\cref{thm:typeIII}), using a result of~\cite{MarrakchiVaes:2024ErgodicStates}.

In analogy with the classical case, we expect that compatibility of algebra multiplication in $N$ with the $\sl$ action implies strong constraints on the possible spectra that can appear on the right-hand side of~\eqref{eq:QGKthetaD}. In implementing this idea, we encounter the same problem as in the classical case: multiplication does \emph{not} give rise to a bilinear map
\be
\cK_{\theta}\times \cK_{\theta} \rightarrow \cK_{\theta}\,.
\ee
Rather, multiplication is only defined on the image of the GNS map $\Lambda_{\theta}(N)$, where it acts as
\be
(\Lambda_{\theta}(a),\Lambda_{\theta}(b))\mapsto \Lambda_{\theta}(ab)\,,\qquad a,b\in N\,.
\ee
To deduce the quantum analogue of the associativity equations, the challenge is therefore to identify dense subspaces $\cH^{\infty}_{\nu_i}$ of $\cH_{\nu_i}$, generalizing the spaces of smooth vectors, and having the crucial property
\be
\psi_i(\cH^{\infty}_{\nu_i}) \subset \Lambda_{\theta}(N)\,.
\ee
To the best of our knowledge, there is currently no general definition of smooth vectors in unitary representations of locally compact quantum groups. To avoid this problem, we replace $\cH^{\infty}_{\nu_i}$ by the Fr\'{e}chet space $\cE_{\nu_i} \subset \cH_{\nu_i}$ of analytic test vectors, i.e.~entire functions with rapid decay in horizontal strips (\cref{def:ESpace}). The choice of $\cE_{\nu_i}$ is ad hoc but has the important property that $\psi_i(\cH_{\nu_i})\cap \Lambda_{\theta}(N) \neq \{0\}$ implies $\psi_i(\cE_{\nu_i})\subset \Lambda_{\theta}(N)$ (\cref{clm:EautomaticInclusion}). Furthermore, $\cE_{\nu_i}$ is dense in $\cH_{\nu_i}$. By analogy with the classical case, we expect that the image of each $\psi_i$ contains a nonzero bounded element, and thus $\psi_i(\cE_{\nu_i})\subset \Lambda_{\theta}(N)$ for all $i \in I$ (\cref{conj:regularity}). Ergodic spectrally compact quantum $\gg_b$-spaces with this property are called \emph{regular} (\cref{def:regularity}). We assume regularity in the rest of this section.

It follows that upon restricting $\cH_{\nu_i}$ to $\cE_{\nu_i}$, we can lift the image of $\psi_i$ to $N$ via the inverse GNS map. Thus, define, for each $i \in I$
\be
\Psi_i: \cE_{\nu_i}\rightarrow N\,,\qquad \Psi_i := \Lambda_{\theta}^{-1}\psi_i\,.
\ee
This is well-defined because $\theta$ is faithful and thus $\Lambda_{\theta}$ is injective. $\Psi_i$ is continuous with respect to the Fr\'{e}chet topology on $\cE_{\nu_i}$ and the operator norm on $N$ (\cref{lem:PsiCts}). A \emph{quantum automorphic form} for $\gg_b$ is an element of $\Psi_i(\cE_{\nu_i})$ for some $i\in I$ (\cref{def:quantumAutomorphicForms}).

\subsection{Trilinear forms and the product expansion}\label{ssec:QGtrilinear}
We can now define $n$-point correlations by applying $\theta$ to a product of quantum automorphic forms:
\be
\cC^{(n)}_{i_1,\ldots,i_n}(f_1,\ldots,f_n) := \theta(\Psi_{i_1}(f_1)\ldots \Psi_{i_n}(f_n))\,,\qquad f_1\in\cE_{\nu_{i_1}}\,,\,\ldots\,,\,f_n\in\cE_{\nu_{i_n}}\,.
\label{eq:QGCorDef}
\ee
$\cC^{(n)}_{i_1,\ldots,i_n}$ are multilinear forms on $\cE_{\nu_{i_1}}\times\cdots\times\cE_{\nu_{i_n}}$, jointly continuous with respect to the Fr\'{e}chet topologies. They are constrained by invariance of $\theta$ under the coaction $\alpha$, i.e.~\eqref{eq:thetaInvariance}. In particular, $\cC^{(1)}_{i} = 0$ for all $i \in I$, and, using symmetry (\cref{lem:2ptcor}),
\be
\cC^{(2)}_{i,i} = \cD_{\nu_i}\quad\text{for}\quad i\in I\,,\qquad \cC^{(2)}_{i,j}=0\quad\text{for}\quad i\neq j\,.
\ee
Here $\cD_{\nu}:=\cD_{\nu,\nu}$ is the invariant bilinear form of \cref{def:bilinearForm}, which is the unique (up to a rescaling) continuous $\sl$-invariant form on $\cE_{\nu}\times \cE_{\nu}$ (\cref{clm:bilinearUniqueness}).\footnote{We use the same notation $\cD_{\nu}$ and $\cT_{\nu_1,\nu_2,\nu_3}$ for the bilinear and trilinear invariant forms of $\sl$ as we did for $\mathrm{PSL}(2,\rr)$, although the forms are not the same.} Similarly,
\be
\cC^{(3)}_{i,j,k} = c_{ij k}\, \cT_{\nu_i,\nu_j,\nu_k}\,,\qquad i,j,k \in I\,,
\ee
where $\cT_{\nu_1,\nu_2,\nu_3}$ is the continuous invariant trilinear form on $\cE_{\nu_1}\times \cE_{\nu_2}\times \cE_{\nu_3}$ of \cref{def:trilinearForm}, conjecturally unique up to a rescaling (\cref{conj:trilinearUniqueness}); the existence of $c_{ijk}$ relies on this uniqueness (\cref{clm:threePoint}). The constants $c_{ijk}\in \cc$ generalize trilinear periods of automorphic forms to the noncommutative setting.

The KMS condition for $\theta$ and the KMS cyclic symmetry~\eqref{eq:TKMS} of $\cT_{\nu_1,\nu_2,\nu_3}$ imply that $c_{ijk}$ is cyclically symmetric
\be
c_{ijk} = c_{jki} = c_{kij}\,.
\label{eq:QGcyclicity}
\ee
The identity $\overline{\theta(a)}=\theta(a^*)$ and the reality property~\eqref{eq:TReality} of $\cT_{\nu_1,\nu_2,\nu_3}$ imply
\be
\overline{c_{ijk}} = c_{kji}\,.
\label{eq:QGconjugation}
\ee
We deduce from~\eqref{eq:QGKthetaD} that the same constants $c_{ijk}$ encode the action of the operator $\Psi_j(f_1) \in N$ on the vector $\psi_k(f_2) \in \cK_{\theta}$, where $f_1\in \cE_{\nu_j}$, $f_2\in \cE_{\nu_k}$ (\cref{prop:productExpansion}):
\be
\Psi_j(f_1)\psi_k(f_2) = \delta_{jk} \cD_{\nu_j}(f_1,f_2)\Lambda_{\theta}(1_N) + \sum\limits_{i\in I} c_{ijk} \, \psi_i(\cT^{\nu_i}_{\nu_j,\nu_k}(f_1,f_2))\,.
\label{eq:quantumOPE}
\ee
Here $\cT^{\nu_1}_{\nu_2,\nu_3}: \cE_{\nu_2}\times \cE_{\nu_3}\rightarrow \cE_{\nu_1}$ is the unique continuous bilinear map such that
\be
\cT_{\nu_1,\nu_2,\nu_3}(f_1,f_2,f_3) = \cD_{\nu_1}(f_1 , \cT^{\nu_1}_{\nu_2,\nu_3}(f_2,f_3))\,, \qquad f_1\in \cE_{\nu_1}\,,\,f_2\in \cE_{\nu_2}\,,\,f_3\in \cE_{\nu_3}\,,
\ee
see~\eqref{eq:2To1Def}. $\cT^{\nu_1}_{\nu_2,\nu_3}$ is equivariant when its domain and range are considered as subsets of $\cH_{\nu_2}\otimes\cH_{\nu_3}$ and $\cH_{\nu_1}$. The sum in~\eqref{eq:quantumOPE} converges in the norm of $\cK_{\theta}$.

\subsection{Quantum associativity equations and boundary CFT}\label{ssec:QGassoc}
Using~\eqref{eq:quantumOPE}, we prove that four-point correlations admit a pair of absolutely convergent expansions (\cref{prop:fourPoint}), for all $i,j,k,\ell\in I$ and all $f_{1}\in \cE_{\nu_i}$, $f_{2}\in \cE_{\nu_j}$, $f_{3}\in \cE_{\nu_k}$, $f_{4}\in \cE_{\nu_\ell}$:
\begin{align}
\,&\cC^{(4)}_{i,j,k,\ell}(f_1,f_2,f_3,f_4)\nonumber \\
&= \delta_{ij}\delta_{k\ell} \,\cD_{\nu_i}(f_1,f_2)\,\cD_{\nu_k}(f_3,f_4) + \sum\limits_{m\in I} c_{ijm}c_{k\ell m}\, \cB^{s}_{\nu_i,\nu_j,\nu_k,\nu_\ell;\nu_m}(f_1,f_2,f_3,f_4) \label{eq:QGcrossingS} \\
&= \delta_{i\ell}\delta_{j k} \,\cD_{\nu_i}(f_1,f_4)\,\cD_{\nu_j}(f_2,f_3) + \sum\limits_{m\in I} c_{\ell i m}c_{j k m}\, \cB^{t}_{\nu_i,\nu_j,\nu_k,\nu_\ell;\nu_m}(f_1,f_2,f_3,f_4)\,.
\label{eq:QGcrossingT}
\end{align}
Here $\cB^{s}_{\nu_1,\nu_2,\nu_3,\nu_4;\nu_5}$, $\cB^{t}_{\nu_1,\nu_2,\nu_3,\nu_4;\nu_5}$ are universal continuous quadrilinear forms on $\cE_{\nu_1}\times\cE_{\nu_2}\times\cE_{\nu_3}\times\cE_{\nu_4}$, expected to be invariant (\cref{def:quadrilinarForms}), obtained by composing the equivariant maps as follows:
\ba
\cB^{s}_{\nu_1,\nu_2,\nu_3,\nu_4;\nu_5}(f_1,f_2,f_3,f_4) &= \cT_{\nu_1,\nu_2,\nu_5}(f_1,f_2,\cT^{\nu_5}_{\nu_3,\nu_4}(f_3,f_4))\,,\\
\cB^{t}_{\nu_1,\nu_2,\nu_3,\nu_4;\nu_5}(f_1,f_2,f_3,f_4) &= \cT_{\nu_1,\nu_5,\nu_4}(f_1,\cT^{\nu_5}_{\nu_2,\nu_3}(f_2,f_3),f_4)\,.
\ea
A regular, symmetric, ergodic, spectrally compact quantum $\gg_b$-space is called a \emph{quantum boundary space} (\cref{def:quantumBoundarySpace}). The equality of~\eqref{eq:QGcrossingS} and~\eqref{eq:QGcrossingT} constitutes an infinite set of universal constraints satisfied by the spectral data $(\nu_i)_{i\in I}$, $(c_{ijk})_{i,j,k\in I}$ of any quantum boundary space (\cref{thm:boundaryAssociativity}). The identities express compatibility of algebra multiplication in $L^{\infty}(\xx)$ and the action of $\sl$ on $\xx$ in the basis of quantum automorphic forms. We will call them quantum associativity equations.

In direct analogy with Adve's converse theorem~\cite{Adve:2025sld}, it is natural to conjecture that any candidate spectral data $(\nu_i)_{i\in I}$, $(c_{ijk})_{i,j,k\in I}$ satisfying all quantum associativity equations necessarily arises from a quantum boundary space, i.e.~from a triple $(N,\theta,\alpha)$ satisfying the properties spelled out above, unique up to equivariant state-preserving isomorphism (\cref{conj:boundaryReconstruction}).

We further argue that the objects satisfying quantum associativity equations are already familiar in the mathematical physics literature. We conjecture that they are precisely the same as elementary compact unitary conformal boundary conditions in 2d conformal field theories with central charge $c = 1+ 6Q^2 > 1$ (\cref{conj:equivalence1}). Under this correspondence, the spectrum $(\nu_i)_{i\in I}$ of quantum automorphic forms on $\xx$ maps to the spectrum of scaling dimensions $(h_i)_{i\in I}$ of Virasoro primary operators on the boundary condition via
\be
h_i = \frac{Q^2}{4} + \nu^2_i\,.
\ee
The trivial representation $\cc \subset \cK_{\theta}$ maps to the vacuum module of the boundary Virasoro algebra. The correspondence between unitary irreducible representations of $\sl$ and unitary irreducible lowest-weight modules of the Virasoro algebra is summarized in Table~\ref{tab:RepCor}. The trilinear periods $c_{ijk}$ of quantum automorphic forms on $\xx$ map to the structure constants of the boundary operator product expansion. The latter are known to satisfy the cyclicity and conjugation properties~\eqref{eq:QGcyclicity},~\eqref{eq:QGconjugation}. 

\begin{table}[h!]
\centering
\renewcommand{\arraystretch}{1.4}
\begin{tabular}{ll}
\hline
$\mathrm{SL}_b^+(2,\mathbb R)$ & Virasoro$_c$ \\
\hline
Trivial representation $\mathbf{1}$
& Vacuum module $h=0$ \\

Complementary series: $\pp_{i\sigma}$,
$0<\sigma<\frac{Q}{2}\qquad$
& $0<h<\frac{c-1}{24}$ \\

Principal series: $\pp_{\nu}$,
$\nu\in\mathbb R_{\geq 0}$
& $h\geq\frac{c-1}{24}$ \\
\hline
\end{tabular}
\caption{The correspondence between unitary representations of $\sl$ and unitary irreducible lowest-weight modules of the Virasoro algebra at central charge $c=1+6Q^2$.}
\label{tab:RepCor}
\end{table}

Finally, we argue that the quantum associativity equations~\eqref{eq:QGcrossingS}=\eqref{eq:QGcrossingT} are precisely equivalent to the crossing equations for four Virasoro primaries on the boundary of the disk (Conjectures~\ref{conj:equivalence2} and~\ref{conj:equivalence3}). This can be seen heuristically from the equality between the 6j symbol of $\sl$ and the crossing kernel of Virasoro conformal blocks. We provide additional evidence for the equivalence, incorporating also the complementary series and the trivial representation.

By combining this conjectured equivalence with the conjectured quantum analogue of Adve's converse theorem, we are led to conjecture that any elementary compact unitary conformal boundary condition of a 2d CFT with $c = 1 + 6Q^2$ gives rise to a quantum boundary space (\cref{conj:equivalence1}). Nonelementary boundary conditions also arise naturally in this language: they correspond to dropping the requirement that $\alpha$ is ergodic.

The present work provides a recipe for constructing $(N,\theta,\alpha)$ from the spectral data of the boundary condition, whose validity rests on \cref{conj:boundaryReconstruction}. There exist many exactly solvable conformal boundary conditions with $c>1$, for example in theories of free bosons. Our construction thus (conjecturally) gives rise to many examples of type III$_1$ von Neumann factors carrying the extra structure provided by $\theta$ and $\alpha$.

It would be very interesting to understand the precise relationship between $(N,\theta,\alpha)$ that we associate to a given boundary condition, and von Neumann algebras of local observables associated to spacetime regions. In the context of rational 2d CFTs, the relevant description is known under the name conformal nets~\cite{Kawahigashi:2015Conformal,BrunettiGuidoLongo:1993Modular,LongoRehren:1995Nets,KawahigashiLongoMuger:2001Modularity}. References which treat boundary conditions of rational CFTs in this language include~\cite{LongoRehren:2004Boundary,CarpiKawahigashiLongo:2013Boundary,BischoffKawahigashiLongo:2015Boundary}.

\subsection{Quantum $\gg_b$-bispaces}\label{ssec:QGbispaces}
Given the conjectured equivalence between conformal boundary conditions and quantum boundary spaces, it is natural to look for an interpretation of full bulk 2d CFTs in the same language. We conjecture that they correspond to a specific kind of quantum $\gg_b$-bispaces, i.e.~quantum measure spaces carrying commuting left and right actions of $\gg_b$, or equivalently an action of the locally compact quantum group $\sl\times\sl$.

We will adopt the same notation as in the preceding discussion. We model the quantum space $\xx$ by a von Neumann algebra $N = L^{\infty}(\xx)$. We furnish $N$ with a faithful normal state $\theta$. We further equip $\xx$ with a commuting pair of left and right actions of $\sl$. These are modelled by a left coaction $\alpha_L : N \rightarrow M\botimes N$ and a commuting right coaction $\alpha_R : N \rightarrow N\botimes M$, each preserving $\theta$. The quadruple $(N,\theta,\alpha_L,\alpha_R)$ is called a \emph{quantum $\gg_b$-bispace} (\cref{def:BulkO}). We assume that the bispace is ergodic, i.e.~that $N^{\alpha_L}\cap N^{\alpha_R}=\cc 1_N$: $\alpha_L(a)=1_M\otimes a$ and $\alpha_R(a) = a\otimes 1_M$ simultaneously hold only if $a \in \cc 1_N$.

The GNS construction for $(N,\theta)$ provides a realization of $N$ on a Hilbert space $\cK_{\theta} = L^2(\xx,\theta)$, as well as the GNS map $\Lambda_{\theta}: N\rightarrow \cK_{\theta}$. The commuting pair of coactions gives $\cK_{\theta}$ the structure of a unitary representation $\uu_{LR}$ of $\sl\times \sl$. Ergodicity implies the trivial representation appears in $\cK_{\theta}$ with unit multiplicity. We assume that $\cK_{\theta}$ decomposes as a direct sum of the external tensor product representations $\pp_{\nu^L}\boxtimes \pp_{\nu^R}$ on the tensor product Hilbert spaces $\cH_{\nu^L}\otimes \cH_{\nu^R}$
\be
\cK_{\theta} \cong \cc \oplus \bigoplus\limits_{i\in I }\cH_{\nu^L_i}\otimes \cH_{\nu^R_i}\,.
\label{eq:QGBulkDecomposition}
\ee
The left and right factors of $\sl\times \sl$ act respectively on $\cH_{\nu^L_i}$ and $\cH_{\nu^R_i}$. We allow for nontrivial direct summands $j \in I$ such that either $\pp_{\nu^L_j}$ or $\pp_{\nu^R_j}$ is the trivial representation (but not both simultaneously). We write this as~$\nu^L_j = iQ/2$ or $\nu^R_j = i Q/2$ and set $\cH_{i\frac{Q}{2}}:=\cc$. These correspond to Virasoro primary holomorphic and antiholomorphic conserved currents in the 2d CFT. We assume the spectrum $((\nu^L_i,\nu^R_i))_{i \in I}$ is discrete and has finite multiplicities, i.e.~that the bispace is spectrally compact (\cref{def:bulkSpectralCompactness}).

We consider the equivariant isometric embeddings
\be
\psi_i: \cH_{\nu^L_i}\otimes \cH_{\nu^R_i}\rightarrow \cK_{\theta}\,,\qquad i \in I\,,
\ee
chosen to intertwine the respective real structures (\cref{claim:realityBulk}). The modular flow of $\theta$ acts within each isotypic subspace. It follows that if $N$ is nontrivial, it is a type III$_1$ factor (\cref{thm:typeIIIBulk}). We assume again that the modular flow does not mix degenerate $i$, i.e.~that $(N,\theta,\alpha_L,\alpha_R)$ is symmetric (\cref{def:symmetricBulkObj}), an analogue of a symmetric Frobenius algebra object in the language of~\cite{Fuchs:2002cm}.

For each $i\in I$, we define the space of analytic test vectors $\sE_i := \cE_{\nu^L_i}\hat{\otimes}_{\pi}\cE_{\nu^R_i}\subset \cH_{\nu^L_i}\otimes \cH_{\nu^R_i}$, with the convention $\cE_{i\frac{Q}{2}}=\cc$; thus $\sE_i \cong \cE_{\nu^R_i}$ or $\sE_i \cong \cE_{\nu^L_i}$ if $\pp_{\nu^L_i}$ or $\pp_{\nu^R_i}$ is trivial, respectively. We assume that the bispace is regular, i.e.~that $\psi_i(\sE_i) \subset \Lambda_{\theta}(N)$ for all $i\in I$ (\cref{def:bulkRegularity}); conjecturally, this is automatic (\cref{conj:bulkRegularity}). We define the GNS lifts
\be
\,\Psi_i : \sE_i \rightarrow N\,,\qquad \Psi_i := \Lambda^{-1}_{\theta}\psi_i\,.
\ee
Elements of $\Psi_i(\sE_i)$ are again called quantum automorphic forms; they are the working analogues of automorphic forms on $\sl\times \sl$. Correlations are again defined by applying $\theta$ to products of quantum automorphic forms
\be
\cC^{(n)}_{i_1,\ldots,i_n}(F_1,\ldots,F_n) := \theta(\Psi_{i_1}(F_1)\ldots \Psi_{i_n}(F_n))\,,\qquad F_1\in\sE_{i_1}\,,\,\ldots \,,\,F_n\in\sE_{i_n}\,.
\ee
We have $\cC^{(1)}_i = 0$ for all $i\in I$ and (see~\eqref{eq:twoPointBulk})
\be
\cC^{(2)}_{ij} = \delta_{ij}\; \cD_{\nu^L_i}\boxtimes \wt{\cD}_{\nu^R_i}\,.
\ee
The tilde signifies mirror reversal of the arguments. The product $\boxtimes$ recombines matching left- and right-moving entries. The three-point correlations take the form
\be
\cC^{(3)}_{ijk} = C_{ijk}\; \cT_{\nu^L_i,\nu^L_j,\nu^L_k}\boxtimes \wt{\cT}_{\nu^R_i,\nu^R_j,\nu^R_k}
\ee
with $C_{ijk}\in\cc$ a priori (\cref{prop:cBulkSym}, again relying on \cref{conj:trilinearUniqueness}). It is capitalized to distinguish it from the boundary OPE coefficients $c_{ijk}$. $C_{ijk}$ also automatically exhibits cyclicity and conjugation symmetry
\be
C_{ijk} = C_{jki}=C_{kij}\,,\qquad \overline{C_{ijk}} = C_{kji}\,.
\ee
Four-point correlations admit a pair of absolutely convergent expansions (\cref{prop:bulkFourPoint})
\ba
\cC^{(4)}_{i,j,k,\ell}&= \sum\limits_{m\in I_0} C_{ijm}C_{k\ell m}\, \cB^{s}_{\nu^L_i,\nu^L_j,\nu^L_k,\nu^L_\ell;\nu^L_m}\boxtimes \wt{\cB}^{s}_{\nu^R_i,\nu^R_j,\nu^R_k,\nu^R_\ell;\nu^R_m} \\
&=  \sum\limits_{m\in I_0} C_{\ell i m}C_{j k m}\, \cB^{t}_{\nu^L_i,\nu^L_j,\nu^L_k,\nu^L_\ell;\nu^L_m}\boxtimes \wt{\cB}^{t}_{\nu^R_i,\nu^R_j,\nu^R_k,\nu^R_\ell;\nu^R_m}\,,
\label{eq:QGAssocBulk}
\ea
where we included the trivial representation in the sum by setting $I_0 = \{0\} \sqcup I$ and
\be
C_{ij0} = C_{i0j}=C_{0ij}=\delta_{ij}\,.
\ee

\subsection{Braided commutativity and bulk CFT}
So far, the discussion closely parallels that of quantum $\gg_b$-spaces. The main new feature of quantum $\gg_b$-bispaces is the notion of braided commutativity. $\sl$ admits a quasitriangular structure, provided by the universal R-matrix $\cR$~\cite{Faddeev1999ModularDouble,BytskoTeschner2003} (\cref{def:RMatrix,clm:RmatrixGb}). $\cR$ is a unitary operator in $\hat{M}\botimes \hat{M}$, where $\hat{M} = L^{\infty}(\hat{\gg}_b)$. Its main role is to furnish a natural family of equivariant isomorphisms $\cH_1\otimes \cH_2\rightarrow \cH_2\otimes \cH_1$, for an arbitrary pair $\cH_1,\cH_2$ of unitary representations of $\sl$. Given $(N,\theta,\alpha_L,\alpha_R)$ as above, we use the R-matrix to define the braiding operator $\beta_N$, which is a certain distinguished unitary on $\cK_{\theta}\otimes\cK_{\theta}$. We impose that $(N,\theta,\alpha_L,\alpha_R)$ is \emph{braided commutative} (\cref{def:bulkBraidedCommutativity}), in the sense
\be
\mu_N \beta_N = \mu_N\,,
\label{eq:QGbraidedCom}
\ee
where $\mu_N: \Lambda_{\theta}(a)\otimes\Lambda_{\theta}(b)\mapsto \Lambda_{\theta}(a b)$ is the multiplication map.~\eqref{eq:QGbraidedCom} is given a precise meaning on the dense subspace of $\cK_{\theta}\otimes\cK_{\theta}$ spanned by $\psi_i(\sE_i)\otimes\psi_j(\sE_j)$, with $i,j\in I_0$. We show that~\eqref{eq:QGbraidedCom} is equivalent to the constraints (\cref{prop:bulkBraidingCriterion})
\be
s_i:=(\nu^L_i)^2-(\nu^R_i)^2 \in \zz\,,\qquad
C_{jik} = (-1)^{s_i+s_j+s_k}C_{ijk}\,.
\ee
It follows that $C_{ijk}$ is real and fully symmetric when $s_i+s_j+s_k$ is even. It is imaginary and fully antisymmetric when $s_i+s_j+s_k$ is odd.

A regular, symmetric, ergodic, spectrally compact, braided commutative quantum $\gg_b$-bispace is called a \emph{quantum bulk space} (\cref{def:quantumBulkSpace}). We conjecture that a quantum bulk space $(N,\theta,\alpha_L,\alpha_R)$ is equivalent to a bosonic compact unitary 2d CFT with central charge $c=1+6Q^2$, consistent at genus zero (\cref{conj:bulkEquivalence}). Indeed, the spectrum $((\nu^L_i,\nu^R_i))_{i\in I}$ maps to the spectrum of left- and right-moving scaling dimensions $((h^L_i,h^R_i))_{i\in I}$ of bulk Virasoro primary operators via
\be
h^L_i = \frac{Q^2}{4}+ (\nu^L_i)^2\,,\qquad h^R_i = \frac{Q^2}{4}+ (\nu^R_i)^2\,,\qquad i\in I\,.
\ee
The total scaling dimension is $\Delta_i = h^L_i+h^R_i$, and the conformal spin is $s_i = h^L_i-h^R_i$. Braided commutativity implies $s_i\in\zz$ as required by bosonic locality. The trilinear periods $C_{ijk}$ are identified with the structure constants of the bulk operator product expansion. They have the correct symmetry and reality properties.

If, as conjectured, the quantum associativity equations \eqref{eq:QGcrossingS}=\eqref{eq:QGcrossingT} are equivalent to Virasoro four-point crossing on the boundary of a disk, then~\eqref{eq:QGAssocBulk} are expected to be equivalent to Virasoro four-point crossing on a sphere (\cref{ssec:bulkEquivalence}). It would follow that the data $((h^L_i,h^R_i))_{i\in I}$, $(C_{ijk})_{i,j,k\in I}$ arising from the quantum bulk space $(N,\theta,\alpha_L,\alpha_R)$ satisfy all sewing constraints at genus zero.

Conversely, the above reasoning is expected to produce a type III$_1$ factor $N$ from any bosonic compact unitary 2d CFT of central charge $c=1+6Q^2>1$ (see Conjectures~\ref{conj:converseTheorem} and~\ref{conj:bulkEquivalence}). Our construction gives a concrete recipe for building $(N,\theta,\alpha_L,\alpha_R)$ from the CFT data.

\subsection{Modular invariance}\label{ssec:QGmodularInvariance}
Consistency at genus $>0$ implies additional constraints on the spectral data $((h^L_i,h^R_i))_{i\in I}$, $(C_{ijk})_{i,j,k\in I}$. For consistency at any genus and any number of insertions, it is sufficient to additionally require consistency of all torus one-point functions, a.k.a.~modular invariance~\cite{Sonoda:1988SewingII,MooreSeiberg:1988Polynomial}. Intuitively, modular invariance demands that the algebra $N$ be ``as large as possible'' subject to the previous (genus-zero) constraints. For example, the trivial quantum $\gg_b$-bispace
\be
N = \cc\,,\qquad \theta(z) = z\,,\qquad \alpha_L(z)=\alpha_R(z) = z 1_M
\ee
satisfies all of the previous constraints, but is not modular invariant. It corresponds to the theory containing only the Virasoro vacuum module $\cP(0,c)\otimes \cP(0,c)$. This theory is only consistent at genus zero. It would be desirable to impose modular invariance as an additional structural property of the quantum bulk space $(N,\theta,\alpha_L,\alpha_R)$ similarly to how we treated braided commutativity, i.e.~without invoking its spectral data.

To clarify the meaning of modular invariance, it is useful to compare our construction with the situation in RCFT, following the classic work of Fuchs, Runkel, and Schweigert~\cite{Fuchs:2002cm}. In RCFT, representation theory of the chiral algebra is distilled into a modular tensor category $\mathscr{C}$. Objects of $\mathscr{C}$ are the chiral algebra modules, with module homomorphisms serving as morphisms. Chiral vertex operators endow $\mathscr{C}$ with a monoidal structure. In the language of~\cite{Fuchs:2002cm}, a bulk 2d RCFT consistent at genus zero is equivalent to
\be
\text{a haploid commutative symmetric Frobenius algebra object }B\text{ in }\mathscr{C}\boxtimes \mathscr{C}^{\mathrm{rev}}\,.
\label{eq:FRSBulk1}
\ee
Our definition of a quantum bulk space $(N,\theta,\alpha_L,\alpha_R)$ generalizes~\eqref{eq:FRSBulk1} to the irrational, analytic setting. Indeed, the von Neumann algebra $N$ generalizes the algebra object $B$. The decomposition~\eqref{eq:QGBulkDecomposition} of the GNS Hilbert space generalizes the decomposition of $B$ into simple modules. The state $\theta$ provides the analogue of the Frobenius structure. The symmetry condition, i.e.~the requirement that the modular flow of $\theta$ does not mix images of degenerate $\psi_i$, generalizes symmetry of $B$. Braided commutativity~\eqref{eq:QGbraidedCom} implements commutativity of $B$. Finally, joint ergodicity of $\alpha_L$, $\alpha_R$ implements haploidity of $B$, i.e.~uniqueness of the vacuum.\footnote{In fact, $(N,\theta,\alpha_L,\alpha_R)$ also implements unitarity, so it would be more precise to say that it is an analytic version of a haploid commutative $Q$-system in $\mathscr{C}\boxtimes\mathscr{C}^{\mathrm{rev}}$, where $\mathscr{C}$ is a unitary modular tensor category.}

The quantum dimension of any algebra object $B$ of the type~\eqref{eq:FRSBulk1} necessarily satisfies the upper bound
\be
\mathrm{dim}(B) \leq \sum\limits_{i \in \mathrm{Irr}(\mathscr{C})} d_i^2\,,
\label{eq:dimBbound}
\ee
where $d_i$ is the quantum dimension of the $i$th simple module of $\mathscr{C}$. To obtain an RCFT consistent also at all higher genera, it is sufficient to supplement~\eqref{eq:FRSBulk1} with the requirement that the upper bound~\eqref{eq:dimBbound} is saturated. This is a precise sense in which modular invariance is equivalent to maximality of the algebra.

We do not know how to generalize saturation of~\eqref{eq:dimBbound} to the setting of quantum bulk spaces. Instead, we propose the following route to modular invariance, also inspired by~\cite{Fuchs:2002cm}.

\subsection{The central role of boundary conditions}\label{ssec:BulkFromBoundary}
The central idea of~\cite{Fuchs:2002cm} is that the fundamental objects in conformal field theory are not bulk CFTs but rather elementary conformal boundary conditions.\footnote{The author thanks Yichul Choi and Zhuo Yang for discussions on this topic.} Assuming $\mathscr{C}$ is unitary, an elementary boundary condition preserving the underlying chiral algebra is equivalent to
\be
\text{a haploid symmetric Frobenius algebra object }A\text{ in }\mathscr{C}\,.
\label{eq:FRSBoundary}
\ee
$A$ gives an intrinsic description of the boundary condition without any reference to the bulk theory. The decomposition of $A$ into simple objects gives the spectrum of boundary primary operators. The algebra product of $A$ encodes the structure constants of the boundary OPE. In contrast with the definition of a bulk CFT,~\eqref{eq:FRSBoundary} does not require commutativity or modular invariance.

The notion of a quantum boundary space $(N,\theta,\alpha)$ described in Sections~\ref{ssec:QGsl2}--\ref{ssec:QGassoc} generalizes precisely~\eqref{eq:FRSBoundary} to the irrational setting. The justification is the same as in the bulk case, except for the absence of braided commutativity.

The remarkable findings of~\cite{Fuchs:2002cm,Fjelstad:2006aw,Kong:2009inh,Kong:2007yv} are:
\begin{enumerate}
\item Every boundary object $A$ of the type~\eqref{eq:FRSBoundary} is realized as a boundary condition for a unique modular invariant bulk object $B$ of the type~\eqref{eq:FRSBulk1}.
\item $B$ can be recovered from $A$ through the full center construction $B \cong Z(A)$.
\end{enumerate}
These two theorems justify the point of view that boundary conditions are the fundamental objects in conformal field theory. Modular invariant bulk CFTs are in one-to-one correspondence with equivalence classes of conformal boundary conditions, under the equivalence relation $A_1\sim A_2$ iff $Z(A_1) \cong Z(A_2)$, also known as Morita equivalence.

It is natural to wonder to what extent these statements generalize to the wild world of unitary 2d CFTs with $c>1$. Their natural extrapolation can be phrased as follows:
\begin{conjecture}\label{conj:bulkFromBoundary}
Fix a central charge $c>1$. Let $(h_i)_{i = 0}^{\infty}$ with $h_i \geq 0$ be a discrete candidate spectrum of boundary Virasoro primaries with a unique vacuum $h_0 = 0$. Let $c_{ijk}\in \cc$ be candidate boundary structure constants satisfying $c_{ijk}=c_{jki}=c_{kij}$, $\overline{c_{ijk}}=c_{kji}$, $c_{0ij} = \delta_{ij}$. Suppose this data solves all boundary Virasoro four-point crossing equations.

There then necessarily exists a unique bulk CFT $B$, consistent at any genus, and a unique elementary conformal boundary condition $A$ for $B$ which realizes $(h_i)_{i = 0}^{\infty}$, $(c_{ijk})_{i,j,k =0}^{\infty}$. Furthermore, the bulk spectrum, bulk structure constants, and bulk-to-boundary OPE coefficients can all be reconstructed from $(h_i)_{i = 0}^{\infty}$, $(c_{ijk})_{i,j,k =0}^{\infty}$.
\end{conjecture}
At first sight, this conjecture sounds too good to be true. Upon closer inspection, there does not seem to be any reason to doubt it. The reconstruction part is in principle straightforward. Bulk spectrum and bulk-to-boundary OPE coefficients can be extracted from the bulk channel of the annulus with boundary insertions. The latter is in turn fully determined by $(h_i)_{i = 0}^{\infty}$, $(c_{ijk})_{i,j,k =0}^{\infty}$ through the boundary channel expansion. Similarly, one can use the sphere with three disks removed to reconstruct bulk structure constants from the boundary data. It is plausible that this procedure defines a 2d CFT which is consistent at any genus, assuming nothing more than full boundary crossing symmetry of the seed boundary data. It would be nice to prove this. See~\cite{GaberdielRunkel:2008Bulk,FuchsSchweigert:2021Bulk,BrehmRunkel:2022Holes,ChengEtAl:2025Reconstruction,ChenEtAl:2025Liouville,Jafferis:2025yxt} for results relevant to this discussion.

The above point of view resonates with the formulation of 2d CFT in terms of quantum $\gg_b$-spaces. It indicates that the fundamental objects in irrational 2d CFT are quantum boundary spaces $(N,\theta,\alpha)$, as described in Sections~\ref{ssec:QGsl2}--\ref{ssec:QGassoc}. Conjecturally, they are in one-to-one correspondence with elementary conformal boundary conditions (\cref{conj:equivalence1}) and are subject to no further constraints.

One should be able to develop an analytic analogue of the full center construction $A\mapsto Z(A)$. Starting from an arbitrary quantum boundary space $(N,\theta,\alpha)$, the construction would produce a quantum $\gg_b$-bispace $(N',\theta',\alpha_L,\alpha_R) := Z(N,\theta,\alpha)$, representing the bulk 2d CFT for which $(N,\theta,\alpha)$ represents a boundary condition. $Z(N,\theta,\alpha)$ would be braided commutative and modular invariant by construction.

Once the analytic analogue of $Z$ is available, one will be justified in saying that irrational 2d CFT is the theory of quantum hyperbolic surfaces.

\section{Locally compact quantum groups}\label{sec:LCQG}
The goal of this section is to provide an accessible review of locally compact quantum groups in the von Neumann algebraic setting. The modern formulation used here was developed by Kustermans and Vaes in~\cite{KustermansVaes2000,KustermansVaes2003}, building on a number of previous foundational works, particularly~\cite{Kac1963,VainermanKac1974,Woronowicz1987,Woronowicz1998,EnockSchwartz1992,BaajSkandalis1993,MasudaNakagami1994,VanDaele1998}.

\subsection{Duality between spaces and algebras}\label{ssec:spacesAlgebras}
One of the most fruitful ideas in mathematics is to study a space $X$ through the algebra of complex functions on $X$. To illustrate this, suppose first that $X$ is a finite set $X = \{1,\ldots,n\}$. Complex functions $f:X\rightarrow\cc$ form the commutative algebra $A=\cc^n$ under pointwise addition and multiplication. Pointwise complex conjugation endows $A$ with a natural $*$-operation $f^*(j) = \overline{f(j)}$. $A$ also comes with the norm $\| f \| = \max_j(|f(j)|)$. The norm automatically satisfies $\|f^* f\| = \|f\|^2$.

Conversely, one can classify all finite-dimensional commutative algebras $A$ over the complex numbers equipped with a $*$-operation and a norm satisfying $\|f^* f\| = \|f\|^2$, a.k.a.~finite-dimensional commutative C*-algebras. It is not hard to prove that in fact we must have $A \cong \cc^n$ up to $*$-isomorphism. In other words, such algebras are in one-to-one correspondence with finite sets. One can thus \emph{identify} such algebras with finite sets. In this work, we explain how Virasoro conformal field theory generalizes this simple idea.

As a next step, one wants to generalize the above correspondence to spaces with infinitely many points. One is then forced to introduce additional structure on $X$. A particularly natural such structure is topology. In this case, one takes the algebra $A$ to consist of all continuous functions vanishing at infinity, denoted $C_0(X)$. $A$ is a commutative C*-algebra. Gelfand and Naimark proved that the space $\leftrightarrow$ algebra correspondence generalizes to this setting, as long as $X$ is sufficiently nice~\cite{GelfandNaimark1943}: There is a one-to-one correspondence between locally compact Hausdorff spaces $X$ (up to homeomorphism), and commutative C*-algebras $A$ (up to $*$-isomorphism). To go from spaces to algebras, take $A = C_0(X)$. To go in the other direction, points of $X$ are reconstructed as characters $A\rightarrow \cc$. Indeed, the characters of $C_0(X)$ are precisely evaluations at points of $X$.

Next, suppose $X$ comes equipped with a measure $\mu$. In the case $X = \{1,\ldots,n\}$, a measure is the assignment of weights to the points, i.e.~a map $w:\{1,\ldots,n\}\rightarrow \rr_{\geq 0}$. One can then integrate elements of the function algebra as
\be
\theta(f) = \int\limits_{X} f d\mu := \sum\limits_{j=1}^{n} f(j) w(j)\,.
\ee
A measure can be axiomatized in the language of the function algebra as a positive linear functional, i.e.~a linear map $\theta : A\rightarrow \cc$ such that $\theta(f^* f) \in \rr_{\geq 0}$ for all $f\in A$. The functional is said to be faithful if $\theta(f^* f) = 0$ implies $f = 0$, which happens if and only if all the weights are strictly positive.

In the setting of general measure spaces $(X,\mu)$, a natural generalization of $A$ is the commutative algebra $L^{\infty}(X,\mu)$, consisting of equivalence classes of essentially bounded measurable functions, with functions deemed equivalent if equal almost everywhere. For sufficiently nice $(X,\mu)$, $L^{\infty}(X,\mu)$ is a von Neumann algebra. Integration with respect to $\mu$ gives a normal semifinite faithful weight $\theta$ on $L^{\infty}(X,\mu)$. The analogue of Gelfand--Naimark duality provides a one-to-one correspondence between localizable measure spaces $(X,\mu)$, and abelian von Neumann algebras equipped with a normal semifinite faithful weight~\cite{BlecherGoldsteinLabuschagne2022}.

Since the concept of a space is seen as equivalent to a suitable commutative topological algebra, it is natural to also drop the commutativity assumption. Thus the study of noncommutative C*-algebras is sometimes called non-commutative geometry, and that of noncommutative von Neumann algebras non-commutative measure theory.

In this article, we will give a new definition of two-dimensional irrational conformal field theory in the language of non-commutative von Neumann algebras. Thus, we can think of an irrational CFT as a noncommutative measure space $\xx$ with a weight $\theta$ satisfying certain additional properties.

It is no secret that rational 2d CFTs have a flavor of non-commutative geometry~\cite{Fuchs:2002cm}. Indeed, the case of rational 2d CFT is a noncommutative generalization of the above situation where $X$ is a finite set, with additional structure provided by a modular tensor category. The relevant algebras are then finite-dimensional. Therefore, the situation can be analyzed purely algebraically. On the other hand, irrational conformal field theory is intrinsically infinite-dimensional. This is analogous to the situation where $X$ has infinitely many points. Correspondingly, one is forced to introduce topology on the algebra, and making progress requires a combination of algebra and analysis.

Let us comment on the choice of von Neumann vs C*-algebraic formulation. The algebras that we will consider come with distinguished weights, and thus the von Neumann algebraic framework is natural. In general, passing from the C*-algebra $C_0(X)$ to the von Neumann algebra $L^{\infty}(X,\mu)$ loses much information about the underlying space, in particular about its topology, as seen for example from $L^{\infty}(S^1) \cong L^{\infty}([0,1])$. However, this deficiency disappears in our setting. This is because the quantum space $\xx$ corresponding to a 2d CFT carries an ergodic action of a locally compact quantum group $\gg$. Consequently, we can organize elements of $L^{\infty}(\xx)$ according to how they transform under $\gg$. In particular, we will focus on those elements which transform irreducibly under $\gg$. In the case of a classical homogeneous space $X = G/\Gamma$, irreducible subspaces of $L^2(G/\Gamma)$ contain a vast supply of \emph{smooth} functions, called automorphic forms. This means that the von Neumann algebra $L^{\infty}(\xx,\theta)$ equipped with the $\gg$ action can be expected to have access to detailed geometric information about $\xx$.

\subsection{Finite groups}
The next step is to reformulate the concept of a group in the algebraic language. Let $G$ be a finite group. $G$ is in particular a finite set, so we can describe it in terms of $C(G)$, the finite-dimensional algebra of complex functions on $G$. To describe the group axioms in terms of $C(G)$, note that group multiplication gives rise to the coproduct
\be
\Delta: C(G) \rightarrow C(G)\otimes C(G)\,,\qquad
(\Delta f)(g_1,g_2) := f(g_1 g_2)\,.
\ee
Associativity of group multiplication is equivalent to coassociativity of the coproduct
\be
(\Delta\otimes \id)\Delta = (\id\otimes \Delta)\Delta\,,
\label{eq:coassociativity}
\ee
where $\id$ stands for the identity map. The group $G$ is abelian if and only if $\Delta$ is cocommutative, i.e.~$\Delta(f) = \Delta^{\text{op}}(f)$, where $\Delta^{\text{op}}$ is the coproduct with the tensor factors reversed. The identity element $e\in G$ corresponds to the existence of the counit map
\be
\varepsilon: C(G) \rightarrow\cc \,,\qquad \varepsilon f := f(e),
\ee
satisfying
\be
(\varepsilon\otimes \id)\Delta = (\id\otimes \varepsilon)\Delta = \id\,.
\ee
Finally, existence of group inverses gives rise to the antipode map
\be
S: C(G) \rightarrow C(G)\,,\qquad
(S f)(g) := f(g^{-1})\,.
\ee
The defining property of group inverses is equivalent to the antipode identity
\be
m(S\otimes\id)\Delta(f) = m(\id \otimes S)\Delta(f) = \varepsilon(f) 1\,,
\ee
where $f\in C(G)$, and $1\in C(G)$ is the multiplicative unit, i.e.~the constant function $g \mapsto 1$. $m$ is the algebra product of $C(G)$, i.e.~pointwise multiplication
\be
m: C(G)\otimes C(G) \rightarrow C(G) \,,\qquad (m f)(g) = f(g,g)\,.
\ee
There is a converse theorem also in this setting: Every finite-dimensional commutative C*-algebra equipped with unital $*$-homomorphisms $\Delta$, $\varepsilon$, and $S$ satisfying the above identities is isomorphic to $C(G)$ for some finite group $G$, unique up to isomorphism.

\subsection{Locally compact groups}
The next step is to generalize the discussion to infinite groups. To get a controlled setting, one introduces topology on the group. We will restrict attention to locally compact groups. Every such group admits left- and right-invariant Haar measures, providing a natural setting for harmonic analysis.

Let $G$ be a locally compact group and let $\mu_L$, $\mu_R$ be a left and a right Haar measure. In the spirit of the preceding discussion, one wants to characterize $G$ using the commutative von Neumann algebra $L^{\infty}(G,\mu_L)$, together with additional structure. Since the null sets of $\mu_L$ and $\mu_R$ agree, we have canonically $L^{\infty}(G):=L^{\infty}(G,\mu_L) = L^{\infty}(G,\mu_R)$.

As in the case of a finite group, the group multiplication gives rise to the coproduct
\be
\Delta: L^{\infty}(G) \rightarrow L^{\infty}(G)\botimes L^{\infty}(G)\,,\qquad (\Delta f)(g_1,g_2) := f(g_1 g_2)\,.
\label{eq:LCGDelta}
\ee
Here $\botimes$ is the spatial tensor product, which is the natural tensor product in the category of von Neumann algebras. It is such that $L^{\infty}(G_1)\botimes L^{\infty}(G_2) \cong L^{\infty}(G_1\times G_2)$. $\Delta$ is then a unital, normal $*$-homomorphism, which is the natural kind of morphism in the category of von Neumann algebras. Furthermore, $\Delta$ is coassociative.

How do we describe the Haar measures? When $G$ is compact, Haar integrals give rise to linear functionals $\ph_{L},\ph_{R}: L^{\infty}(G)\rightarrow \cc$, defined as\footnote{Since $G$ is compact, we could take $\mu_L=\mu_R$.}
\be
\ph_L(f) = \int\limits_G f(g) d\mu_L(g)\,,\qquad
\ph_R(f) = \int\limits_G f(g) d\mu_R(g)\,.
\ee
These functionals are \emph{positive}, i.e.~satisfying $\ph(f^*f) \geq 0$ for all $f\in L^{\infty}(G)$, and \emph{faithful}, meaning $\ph(f^*f) = 0$ implies $f=0$. Furthermore, they are \emph{normal}, which is the right version of continuity in this setting.

More care is needed when $G$ is non-compact, since $\ph_L$ and $\ph_R$ are then not well-defined on all of $L^{\infty}(G)$. Instead, they give rise to weights, defined as follows.
\begin{definition}
Let $M$ be a von Neumann algebra and let $M_+=\{x^*x \,:\, x\in M \}$ be the set of its positive elements. A \emph{weight} $\ph$ on $M$ is a map
\be
\ph: M_+ \rightarrow [0,\infty]
\ee
such that $\ph(0) = 0$ and for all $x,y\in M_+$ and $t>0$, we have
\be
\ph(x+y) = \ph(x) + \ph(y)\,,\qquad \ph(t x) = t\ph(x)\,.
\ee
\end{definition}
Note that a weight $\ph$ is allowed to take infinite values on positive elements. It is particularly useful if the set of elements on which $\ph$ is finite is sufficiently large. Thus, we say that $\ph$ is \emph{semi-finite} if the set
\be
\mathfrak{n}_{\ph} = \{x\in M \,|\, \ph(x^* x) < \infty\}
\ee
is dense in $M$. The Haar integrals on a locally compact group define weights $\ph_{L}$, $\ph_{R}$ which are simultaneously normal, semi-finite, and faithful, or n.s.f.~for short.

The defining property of the Haar measures is that they are respectively left- and right-invariant
\be
\int\limits_{G} f(g_1 g) d\mu_L(g) = \int\limits_{G} f(g) d\mu_L(g)\,,\qquad
\int\limits_{G} f(g g_1) d\mu_R(g) = \int\limits_{G} f(g) d\mu_R(g)
\ee
for all $g_1 \in G$. This has an equivalent formulation as a compatibility between the weights and the coproduct, valid for any $f\in L^{\infty}(G)_+$ with a finite integral:
\be
(\id\otimes \ph_L)\Delta_G(f) = \ph_L(f) 1\,,\qquad (\ph_R\otimes\id)\Delta_G(f) = \ph_R(f) 1\,.
\label{eq:invariancePh}
\ee
Here $1\in L^{\infty}(G)$ is the multiplicative identity.

It turns out that these properties completely characterize locally compact groups up to isomorphism. More precisely, every commutative von Neumann algebra $M$ equipped with a coassociative normal unital $*$-homomorphism $\Delta: M\rightarrow M\botimes M$, and admitting a pair of n.s.f.~weights $\ph_L$, $\ph_R$ satisfying~\eqref{eq:invariancePh} arises from a locally compact group $G$ as $(M,\Delta) \cong (L^{\infty}(G),\Delta_G)$.

In this characterization, the existence of Haar weights takes the place of the counit and antipode axioms. In fact, when $G$ is not discrete, $f\in L^{\infty}(G)$ does not have a well-defined value at $g=e$ since the latter has measure zero, and thus the counit map does not exist.

\subsection{Locally compact quantum groups}
The discussion of the previous section directly motivates the definition of a locally compact quantum group. All we have to do is allow the von Neumann algebra $M$ to be non-commutative. Locally compact quantum groups thus generalize locally compact groups to the world of not-necessarily-commutative geometry. One arrives at the following definition, due to Kustermans and Vaes~\cite{KustermansVaes2003}.
\begin{definition}\label{def:LCQG}
A \emph{locally compact quantum group} $(M,\Delta)$ is a von Neumann algebra $M$ together with a normal unital $*$-homomorphism $\Delta: M\rightarrow M\botimes M$, satisfying coassociativity $(\Delta\otimes\id)\Delta = (\id\otimes \Delta)\Delta$, and such that there exists a pair of n.s.f.~weights $\ph_L$, $\ph_R$ on $M$ subject to the following conditions:
\begin{enumerate}
\item $\ph_L$ is left-invariant: $(\id\otimes \ph_L)\Delta(x) = \ph_L(x) 1_M$ for all $x\in M_+$ such that $\ph_L(x)<\infty$.
\item $\ph_R$ is right-invariant: $(\ph_R \otimes \id )\Delta(x) = \ph_R(x) 1_M$ for all $x\in M_+$ such that $\ph_R(x)<\infty$.
\end{enumerate}
\end{definition}
As discussed in the previous section, this definition is equivalent to the notion of a locally compact group when $M$ is commutative. Given $(M,\Delta)$ satisfying the definition, we will write suggestively $M = L^{\infty}(\gg)$ and call $\gg$ the underlying quantum group, although no such object exists unless $M$ is commutative.

Locally compact quantum groups are relevant for our analysis because they play the role of symmetries of noncommutative measure spaces. Once we represent an irrational CFT as a noncommutative space $\xx$, the role usually played by the Virasoro algebra will be replaced by an action of the locally compact quantum group $\mathrm{SL}^+_b(2,\rr)$ on $\xx$.

Locally compact quantum groups have an intrinsic importance in mathematics. This is in part because they elegantly generalize Pontryagin duality from abelian locally compact groups to the nonabelian case. For any locally compact abelian group $G$, the set of continuous characters $\chi: G\rightarrow \mathrm{U}(1)$ itself forms a locally compact abelian group -- the dual of $G$, denoted $\hat{G}$. Group multiplication in $\hat{G}$ is pointwise multiplication of characters. The Pontryagin duality theorem states that taking the dual of $\hat{G}$ recovers the original group $G$. The relationship between $G$ and $\hat{G}$ is a generalization of the Fourier transform.

To generalize Pontryagin duality to nonabelian $G$, it is convenient to transition to the von Neumann algebra language. Let $G$ be a locally compact group and consider the Hilbert space $L^2(G,\mu_L)$. The von Neumann algebra $L^{\infty}(G)$ acts on $L^2(G,\mu_L)$ by pointwise multiplication. This exhibits $L^{\infty}(G)$ as a von Neumann subalgebra of $\mathrm{B}(L^2(G,\mu_L))$ -- the von Neumann algebra of all bounded operators on $L^2(G,\mu_L)$. The latter contains another natural von Neumann algebra. To define it, let $g\in G$ and let $\lambda_g \in \mathrm{B}(L^2(G,\mu_L))$ be the left translation by $g$, i.e.~$(\lambda_g f)(h) = f(g^{-1} h)$ for $f\in L^2(G,\mu_L)$. The left group von Neumann algebra $\mathrm{VN}_L(G)$ is defined as the smallest von Neumann subalgebra of $\mathrm{B}(L^2(G,\mu_L))$ containing all left translations. Equivalently,
\be
\mathrm{VN}_L(G) := \{\lambda_g : g\in G\}''\subset \mathrm{B}(L^2(G,\mu_L))\,,
\ee
where $''$ stands for the double commutant.

When $G$ is abelian, we can normalize the Haar measure $\hat{\mu}$ on $\hat{G}$ so that the Fourier transform is a unitary isomorphism $L^2(G,\mu) \cong L^2(\hat{G},\hat{\mu})$. Furthermore, the Fourier transform exchanges group translations and multiplication by characters. It follows that it provides isomorphisms $L^{\infty}(G) \cong \mathrm{VN}_L(\hat{G})$ and $L^{\infty}(\hat{G}) \cong \mathrm{VN}_L(G)$. Under the latter isomorphism, the usual coproduct on $L^{\infty}(\hat{G})$ corresponds to the cocommutative coproduct
\ba
\hat\Delta:\mathrm{VN}_L(G)\rightarrow
\mathrm{VN}_L(G)\botimes\mathrm{VN}_L(G),
\qquad
\hat\Delta(\lambda_g)=\lambda_g\otimes\lambda_g \quad \text{for all}\quad g\in G\,.
\ea

When $G$ is non-abelian, neither is $\mathrm{VN}_L(G)$, and thus we cannot have $\mathrm{VN}_L(G) \cong L^{\infty}(\hat{G})$ for any classical group $\hat{G}$. Rather, we can \emph{define} the locally compact \emph{quantum} group $\hat{G}$ so that $L^{\infty}(\hat{G}):=\mathrm{VN}_L(G)$ is its underlying function algebra and $\hat{\Delta}$ its coproduct.

In order to obtain a class of objects closed under duality, we are forced to allow $L^{\infty}(G)$ to be noncommutative as well. This is one way to arrive at Definition~\ref{def:LCQG}. Given $(M,\Delta)$ satisfying this definition, one can always construct a dual locally compact quantum group $(\hat{M},\hat{\Delta})$, generalizing the Pontryagin dual. The dual of $(\hat{M},\hat{\Delta})$ gives back $(M,\Delta)$. $\hat{M}$ is commutative if and only if $\Delta$ is cocommutative, and $M$ is commutative if and only if $\hat{\Delta}$ is cocommutative. There exist many interesting examples in which neither $M$ nor $\hat{M}$ are commutative, with $\sl$ being the principal example in this paper. As is usual in the literature, we will use the suggestive notation $M = L^{\infty}(\gg)$ and $\hat{M} = L^{\infty}(\hat{\gg})$, where $\gg$ is the underlying quantum group and $\hat{\gg}$ the dual quantum group.

In the next subsection, we will review the construction of $\hat{M}$ starting from $(M,\Delta)$. This construction provides a useful point of view on our results. Indeed, $\hat{\sl}$ is generated by Faddeev's modular double $\cU_{q\check{q}}(\mathfrak{sl}(2,\rr))$, which is known to play a fundamental role in 2d CFT.

\subsection{The multiplicative unitary and duality}\label{ssec:W}
Let $(M,\Delta)$ be a locally compact quantum group with Haar weights $\ph_L$, $\ph_R$. The construction of the dual $(\hat{M},\hat{\Delta})$ generalizes the construction of $(\mathrm{VN}_L(G),\hat{\Delta})$ from the previous subsection. The first step is to generalize the Hilbert space $L^2(G,\mu_L)$ and canonically realize $M$ on it. This is accomplished by the GNS construction for $M$ equipped with weight $\ph_L$. The GNS construction consists of the triple $(\cK,\iota,\Lambda)$, where $\cK$ is a Hilbert space, $\iota: M\rightarrow \mathrm{B}(\cK)$ is a faithful normal unital $*$-representation realizing $M$ in $\mathrm{B}(\cK)$, and $\Lambda: \mathfrak{n}_{\ph_L}\rightarrow \cK$ is linear injective with a dense image. They satisfy
\begin{enumerate}
\item $\iota(x)\Lambda(y) = \Lambda(x y)$ for all $x\in M$ and $y\in \mathfrak{n}_{\ph_L}$.
\item $\langle\Lambda(x),\Lambda(y)\rangle = \ph_L(x^* y)$ for all $x,y\in \mathfrak{n}_{\ph_L}$.
\end{enumerate}
Here $\langle\cdot,\cdot\rangle$ is the inner product on $\cK$, linear in the second variable. $\cK$ generalizes $L^2(G,\mu_L)$, $\iota$ generalizes the canonical realization of $L^{\infty}(G)$ on $L^2(G,\mu_L)$ by pointwise multiplication, and $\Lambda$ generalizes the inclusion $L^{\infty}(G)\cap L^2(G,\mu_L) \hookrightarrow L^2(G,\mu_L)$. We will write $\cK = L^2(\gg)$ where $\gg$ is the underlying quantum group. We will also identify $M$ with its image in $\mathrm{B}(\cK)$ under $\iota$.

The notion of left translations by $\gg$ on $L^2(\gg)$ is encoded in the coproduct $\Delta$. This idea is implemented by a unitary operator $\ww$ acting on two copies of the GNS Hilbert space, i.e.~$\ww \in \mathrm{U}(L^2(\gg)\otimes L^2(\gg))$. $\ww$ is called the (left) multiplicative unitary. It is uniquely defined by its action on the image of $\Lambda\otimes\Lambda$, which uses the coproduct as follows
\be
\ww^* (\Lambda(x)\otimes\Lambda(y)) = (\Lambda\otimes\Lambda)(\Delta(y)(x\otimes 1))\,,\qquad x,y\in \mathfrak{n}_{\ph_L}\,.
\label{eq:WDef}
\ee
When $(M,\Delta) = (L^\infty(G),\Delta_G)$ for a locally compact group $G$, we have
\be
(\ww f)(g_1,g_2) = f(g_1,g_1^{-1} g_2)\,,\qquad f\in L^2(G \times G)\,.
\ee
We see that $\ww$ encodes the generalization of the left regular representation of $\gg$ on $L^2(\gg)$, translating the second entry by the first entry. To prove unitarity of $\ww$, one uses left invariance of $\ph_L$ under $\Delta$. Coassociativity of the coproduct implies that $\ww$ satisfies the pentagon equation
\be
\ww_{23}\ww_{12} = \ww_{12}\ww_{13}\ww_{23}\,.
\label{eq:Wpentagon}
\ee
Here both sides are elements of $\mathrm{B}(L^2(\gg)\otimes L^2(\gg)\otimes L^2(\gg))$, and $\ww_{ij}$ is the operator acting like $\ww$ on legs $i$ and $j$.

To construct $\hat{M}$, one takes the smallest von Neumann subalgebra of $\mathrm{B}(L^2(\gg))$ containing ``all the possible ways $\ww$ acts on the second leg''. To make this precise, one first introduces the predual of $M$, denoted $M_*$. $M_*$ is the space of all normal linear functionals $\omega: M\rightarrow \cc$. When $M = L^{\infty}(G)$ for a locally compact group $G$, we have $M_* = L^1(G,\mu_L)$, with $F \in L^1(G,\mu_L)$ acting on $f \in L^{\infty}(G)$ as
\be
f\mapsto \int\limits_{G} F(g) f(g) d\mu_L(g)\,.
\ee
One can show that $\ww \in M \botimes \mathrm{B}(L^2(\gg))$. To construct $\hat{M}$, we apply all possible $\omega \in M_*$ to the first leg of $\ww$ and take the completion
\be
\hat{M} = \{(\omega\otimes \id) \ww \,:\, \omega \in M_* \}'' \subset \mathrm{B}(L^2(\gg))\,.
\label{eq:MhatFromW}
\ee
Indeed, in the classical case, when $\omega$ corresponds to smearing against $F \in L^1(G,\mu_L)$, we have
\be
(\omega\otimes \id) \ww = \int\limits_{G} F(g) \lambda_g \, d\mu_L(g) \in \mathrm{VN}_L(G)\,.
\ee
It turns out that we can recover $(M,\Delta)$ from $\ww$. Firstly, by construction of $\hat{M}$, we have $\ww \in M \botimes \hat{M}$. We can recover $M$ from $\ww$ by the symmetric image of~\eqref{eq:MhatFromW}
\be
M = \{(\id \otimes \omega) \ww \,|\, \omega \in (\hat{M})_* \}'' \subset \mathrm{B}(L^2(\gg))\,.
\label{eq:MFromW}
\ee
The coproduct $\Delta$ is given by
\be
\Delta(x) = \ww^*(1\otimes x)\ww\,,\qquad x\in M\,.
\label{eq:DeltaFromW}
\ee
It remains to construct the coproduct $\hat{\Delta}$ for the dual quantum group $\hat{\gg}$. To that end, let us first introduce
\be
\hat{\ww} = \ww^*_{21} \in \hat{M}\botimes M\,.
\ee
The notation $21$ refers to the leg flip, i.e.~$\hat{\ww} = \Sigma \ww^*\Sigma$, where $\Sigma(v\otimes w):= w\otimes v$. $\hat{\ww}$ is the left multiplicative unitary for $\hat{\gg}$. In particular, it satisfies the same pentagon equation
\be
\hat{\ww}_{23}\hat{\ww}_{12} = \hat{\ww}_{12}\hat{\ww}_{13}\hat{\ww}_{23}\,.
\ee
Correspondingly, the dual coproduct is given by
\be
\hat{\Delta}(x) = \hat{\ww}^*(1\otimes x)\hat{\ww}\,,\qquad x\in\hat{M}\,.
\label{eq:DeltaHatFromW}
\ee
In summary, the single unitary operator $\ww$ simultaneously encodes $(M,\Delta)$ and $(\hat{M},\hat{\Delta})$. A fundamental theorem of~\cite{KustermansVaes2003} states that there exists a pair of n.s.f.~weights $\hat{\ph}_L$, $\hat{\ph}_R$ on $\hat{M}$, respectively left and right invariant under $\hat{\Delta}$. In other words, $(\hat{M},\hat{\Delta})$ is itself a locally compact quantum group. Since $\hat{\raisebox{0pt}[0.85\height]{$\hat{\ww}$}} = \ww$, the double dual recovers the original quantum group.

Conversely, given a Hilbert space $\cK$ and a unitary operator $\ww$ on $\cK\otimes \cK$ satisfying the pentagon equation, one can construct a pair of von Neumann algebras $M,\hat{M}\subset \mathrm{B}(\cK)$ with coassociative coproducts $\Delta$, $\hat{\Delta}$ using \eqref{eq:MhatFromW},~\eqref{eq:MFromW},~\eqref{eq:DeltaFromW},~\eqref{eq:DeltaHatFromW}.\footnote{To indeed construct $M$, $\hat{M}$ from nothing more than $\ww$, one replaces $M_*$ and $(\hat{M})_*$ in~\eqref{eq:MhatFromW} and~\eqref{eq:MFromW} with $\mathrm{B}(\cK)_*$.} However, this does not necessarily mean that $(M,\Delta)$ is a locally compact quantum group. Indeed, the existence of invariant Haar weights is an additional constraint on $\ww$.

\subsection{Right translations}\label{ssec:V}
Having discussed the left regular representation, let us turn to its right-handed analogue. We will assume for simplicity until the end of this subsection that the quantum group $\gg$ is unimodular, meaning $\ph:=\ph_L = \ph_R$. $\sl$ is unimodular.

In the classical world, right translation by $g\in G$ gives rise to the unitary $\rho_g \in \mathrm{B}(L^2(G))$, defined by $(\rho_g f)(h) = f(hg)$ for $f\in L^2(G)$. Right translations generate the right group von Neumann algebra $\mathrm{VN}_R(G)$
\be
\mathrm{VN}_R(G) := \{\rho_g : g\in G\}''\subset \mathrm{B}(L^2(G,\mu))\,.
\ee
$\mathrm{VN}_L(G)$ and $\mathrm{VN}_R(G)$ are commutants of each other. In the quantum world, the generalization of $\mathrm{VN}_R(G)$ is simply $\hat{M}'$. One constructs the right regular representation $\vv$ as the unique unitary in $\mathrm{B}(L^2(\gg)\otimes L^2(\gg))$ satisfying
\be
\vv (\Lambda(x)\otimes\Lambda(y)) = (\Lambda\otimes\Lambda)(\Delta(x)(1\otimes y))\,,\qquad x,y\in \mathfrak{n}_{\ph}\,.
\label{eq:VDef}
\ee
When $(M,\Delta) = (L^\infty(G),\Delta_G)$ for a locally compact group $G$, we have
\be
(\vv f)(g_1,g_2) = f(g_1 g_2,g_2)\,,\qquad f\in L^2(G \times G)\,.
\ee
In other words, $\vv$ encodes a right translation of the first leg by the second leg. We have $\vv\in \hat{M}'\botimes M$, and can recover $\hat{M}'$ as
\be
\hat{M}' = \{(\id\otimes \omega) \vv \,:\, \omega \in M_* \}'' \subset \mathrm{B}(L^2(\gg))\,.
\ee
$\vv$ satisfies the same pentagon equation as $\ww$
\be
\vv_{23}\vv_{12} = \vv_{12}\vv_{13}\vv_{23}\,.
\ee
Finally, $\vv$ encodes the coproduct on $M$ as follows
\be
\Delta(x) = \vv (x\otimes 1)\vv^*\,,\qquad x\in M\,.
\label{eq:DeltaFromV}
\ee

\subsection{Conjugation and group inverses from modular theory}\label{ssec:modularGeneral}
Let us now review how to generalize two natural operations from classical to quantum groups: complex conjugation of functions on the group, and inverses of group elements. At the level of the von Neumann algebra $M = L^{\infty}(\gg)$, the generalization of pointwise complex conjugation is simply the star operation $x\mapsto x^*$. At the level of the Hilbert space $L^2(\gg)$, one would like complex conjugation to be implemented by an antiunitary involution. In order to make it agree with $*$ via the GNS map, one may at first try to define it as the closure $S$ of the densely defined operator
\be
S_0 \Lambda(x)= \Lambda(x^*)\,,\qquad x\in \mathfrak{n}_{\ph_L}\cap \mathfrak{n}_{\ph_L}^*\,.
\ee
$S$ is the Tomita operator for the GNS construction $(\cK,\iota,\Lambda)$. While an antilinear involution, $S$ is in general not an isometry. In fact, $S$ can be unbounded. The resolution is to perform the polar decomposition of $S$, provided by Tomita--Takesaki theory. It takes the form
\be
S = J \nabla^{\frac{1}{2}}\,.
\ee
Here $J:L^2(\gg)\rightarrow L^2(\gg)$ is an antiunitary involution, called the \emph{modular conjugation}, and $\nabla$ is a (generally unbounded) injective positive self-adjoint operator on $\cK$, called the \emph{modular operator}. $J$ and $\nabla$ are unique subject to these conditions. Furthermore, we have
\be
J \nabla J = \nabla^{-1}\,.
\ee
$J$ is the quantum analogue of pointwise complex conjugation on $L^2(\gg)$. For classical (even non-unimodular) groups, we have
\ba
(J f)(g) = \overline{f(g)}\,,\qquad \nabla = 1\,.
\ea
In particular, $\nabla$ should not be confused with the modular character measuring the difference between left and right Haar weights. Rather, $\nabla$ quantifies the failure of $\ph_L$ to be tracial.

We can generalize the operation of taking inverses of group elements by considering modular theory of the dual quantum group $\hat{\gg}$. For example for $G=(\rr,+)$, pointwise complex conjugation in Fourier space acts as $\overline{f(-x)}$ in position space, thus supplying the group inverse $x\mapsto -x$. Let $\hat{\Lambda}:\mathfrak{n}_{\hat{\ph}_L}\rightarrow L^2(\gg)$ be the GNS map for the pair $\hat{M}$, $\hat{\ph}_L$. Let $\hat{S}$ be the Tomita operator for this GNS construction, satisfying
\be
\hat{S}\hat{\Lambda}(x) = \hat{\Lambda}(x^*)\,,\qquad x\in \mathfrak{n}_{\hat{\ph}_L}\cap \mathfrak{n}_{\hat{\ph}_L}^*\,.
\ee
Write its polar decomposition as
\be
\hat{S} = \hat{J} \hat{\nabla}^{\frac{1}{2}}\,,
\ee
where $\hat{J}:L^2(\gg)\rightarrow L^2(\gg)$ is the dual modular conjugation, and $\hat{\nabla}$ the dual modular operator. In the classical case, we have
\be
(\hat{J}f)(g) = \Delta_G(g)^{-\frac{1}{2}}\overline{f(g^{-1})}\,,\qquad
(\hat{\nabla}^{it} f)(g) = (\Delta_G(g))^{i t}f(g)
\label{eq:hatJhatNabla}
\ee
for $f\in L^2(G,\mu_L)$, $t\in\rr$. Here $\Delta_G: G\rightarrow \rr_{>0}$ is the modular character of $G$.\footnote{$\Delta_G$ measures the failure of the left Haar measure to be right-invariant as follows: $\int_G f(g h)d\mu_L(g) = \Delta_G(h)^{-1}\int_G f(g)d\mu_L(g).$} In the quantum setting, the group inverse is implemented on $M = L^{\infty}(\gg)$ through the \emph{unitary antipode} $R$, defined as
\be
R(x) := \hat{J} x^* \hat{J}\,,\qquad x\in M\,.
\ee
A theorem of~\cite{KustermansVaes2003} shows that $R$ maps $M$ onto itself, and is a normal involutive $*$-antiautomorphism. In the classical setting, it follows from~\eqref{eq:hatJhatNabla} that indeed
\be
(R(f))(g) = f(g^{-1})
\ee
for all $f\in L^{\infty}(G)$.

Tomita--Takesaki theory establishes that conjugation by the unitary $\nabla^{it}$ with $t\in \rr$ preserves $M$ (and similarly for $\hat{\nabla}^{it}$ and $\hat{M}$). Thus one defines the modular automorphism groups
\ba
\,&\sigma_t : M\rightarrow M\,,\qquad \sigma_t(x) := \nabla^{it} x \nabla^{- it}\,,\\
&\hat{\sigma}_t : \hat{M}\rightarrow \hat{M}\,,\qquad \hat{\sigma}_t(x) := \hat{\nabla}^{it} x \hat{\nabla}^{- it}\,.
\ea
It turns out that conjugation by $\hat{\nabla}^{it}$ also preserves $M$ (and vice versa conjugation by $\nabla^{it}$ preserves $\hat{M}$). Thus, one defines the so-called scaling group of $(M,\Delta)$ as
\be
\tau_t : M\rightarrow M\,,\qquad \tau_t(x) = \hat{\nabla}^{it} x \hat{\nabla}^{- it}\,.
\ee
For classical $G$, $\tau_t$ is trivial. It is a symmetry of the quantum group structure with no classical analogue. In particular, it preserves the coproduct: $\Delta\circ\tau_t = (\tau_t\otimes\tau_t)\circ\Delta$.

\subsection{Unitary representations as corepresentations}\label{ssec:reps}
Unitary representations play a central role in the study of groups. Let us review how to generalize this concept to locally compact quantum groups. The first step is to reformulate it for classical locally compact groups $G$ in terms of the von Neumann algebra $L^\infty(G)$.

Let $G$ be a locally compact group with left Haar measure $\mu_L$ and let $\cH$ be a separable Hilbert space. The Hilbert space $L^2(G,\mu_L)\otimes\cH$ consists of square-integrable $\cH$-valued functions on $G$. Similarly, $L^{\infty}(G)\botimes \mathrm{B}(\cH)$ is the von Neumann algebra of essentially bounded $\mathrm{B}(\cH)$-valued functions on $G$. Unitaries in $L^{\infty}(G)\botimes \mathrm{B}(\cH)$ are the same as measurable maps $G\rightarrow \mathrm{U}(\cH)$. All maps from $G$ are understood modulo almost-everywhere equality.

Suppose $\pi: G\rightarrow \mathrm{U}(\cH)$ is a strongly continuous unitary representation of $G$ on $\cH$, and let $\uu_\pi \in L^{\infty}(G)\botimes \mathrm{B}(\cH)$ be the corresponding unitary operator provided by the previous paragraph. $\pi$ satisfies the group homomorphism property
\be
\pi(g_1 g_2) = \pi(g_1)\pi(g_2)\,,\qquad g_1,g_2\in G\,.
\ee
By definition of the coproduct~\eqref{eq:LCGDelta}, this is equivalent to the equality in $L^{\infty}(G)\botimes L^{\infty}(G) \botimes \mathrm{B}(\cH)$
\be
(\Delta\otimes \id)\uu_{\pi} = (\uu_{\pi})_{13}(\uu_{\pi})_{23}\,.
\label{eq:corepresentationDef}
\ee
Conversely, any unitary operator $\uu \in L^{\infty}(G)\botimes \mathrm{B}(\cH)$ satisfying~\eqref{eq:corepresentationDef} arises from a unique strongly continuous unitary representation of $G$ on $\cH$. This formulation generalizes seamlessly to the quantum setting in terms of unitary corepresentations. Their definition makes sense for all von Neumann bialgebras:\footnote{A von Neumann bialgebra is a pair $(M,\Delta)$, where $M$ is a von Neumann algebra and $\Delta:M\rightarrow M\botimes M$ is a coassociative, unital, normal $*$-homomorphism. In particular, a locally compact quantum group is a von Neumann bialgebra with invariant weights.}
\begin{definition}\label{def:UnitaryRep}
Let $(M,\Delta)$ be a von Neumann bialgebra and let $\cH$ be a separable Hilbert space. A \emph{unitary corepresentation} of $(M,\Delta)$ on $\cH$ is a unitary operator $\uu \in M\botimes \mathrm{B}(\cH)$ satisfying
\be
(\Delta\otimes \id) \uu = \uu_{13} \uu_{23}\,.
\label{eq:corepresentationId}
\ee
\end{definition}
When $(M,\Delta) = (L^{\infty}(\gg),\Delta_{\gg})$ for a locally compact quantum group $\gg$, we will refer to $\uu$ as a unitary representation of $\gg$. In this case, the identity~\eqref{eq:corepresentationId} is the same as
\be
\uu_{23}\ww_{12}=\ww_{12} \uu_{13} \uu_{23}\,,
\ee
where $\ww$ is the left multiplicative unitary of $\gg$. This equation generalizes the pentagon equation~\eqref{eq:Wpentagon} for $\ww$. Indeed, setting $\cH = L^2(\gg)$ and $\uu=\ww$, we see that $\ww$ defines a unitary representation of $\gg$ on $L^2(\gg)$. This is the left regular representation. Similarly, recall the right multiplicative unitary $\vv$ from Section~\ref{ssec:V}. The unitary $\vv_{21}:=\Sigma\vv\Sigma \in M\botimes \hat{M}'$ is a unitary representation of $\gg$ on $L^2(\gg)$. This is the right regular representation.

Various standard definitions generalize from the classical to the quantum world, starting with the notion of an intertwining map, and an irreducible representation.
\begin{definition}\label{def:intertwinerIrr}
Let $\uu_1$, $\uu_2$ be unitary corepresentations of a von Neumann bialgebra $(M,\Delta)$ on separable Hilbert spaces $\cH_1$, $\cH_2$. A bounded linear operator $\cI: \cH_1\rightarrow \cH_2$ is called an \emph{intertwiner} for $(\uu_1,\uu_2)$ if
\be
(1_M \otimes \cI) \uu_1 = \uu_2(1_M \otimes \cI)\,.
\ee
One says that $\uu_1$, $\uu_2$ are \emph{unitarily equivalent} if there exists a unitary intertwiner $\cI:\cH_1\rightarrow \cH_2$. A unitary corepresentation $\uu$ of $(M,\Delta)$ on a nonzero Hilbert space $\cH$ is said to be \emph{irreducible} if the only intertwiners $\cI:\cH\rightarrow \cH$ for $(\uu,\uu)$ are scalar multiples of the identity.
\end{definition}

Let us consider the notion of conjugate representations and reality. Suppose first $\cH$ is an arbitrary complex Hilbert space. One defines the conjugate Hilbert space $\overline{\cH}$ to have the same underlying set as $\cH$, the same addition, and complex conjugate scalar multiplication and inner product. In other words, the identity of the underlying sets defines a canonical antiunitary map $j:\cH\rightarrow \overline{\cH}$. $\overline{\cH}$ is canonically isomorphic to the Hilbert space $\cH^*$ of bounded linear functionals on $\cH$ via $j(v) \mapsto \langle v,\cdot \rangle_{\cH}$. Given a bounded operator $T\in \mathrm{B}(\cH)$, we define $\cJ(T)\in \mathrm{B}(\overline{\cH})$ as
\be
\cJ(T) = j T^* j^{-1}\,.
\ee
Under the isomorphism $\overline{\cH} \rightarrow \cH^*$, $\cJ(T)$ becomes the transpose of $T$, i.e.~the operator mapping the linear functional $v\mapsto \omega(v)$ to $v\mapsto \omega(T v)$.

\begin{definition}[The conjugate representation]\label{def:conjRep}
Let $\uu$ be a unitary representation of a locally compact quantum group $\gg$ on a Hilbert space $\cH$. The conjugate representation $\uu^c$ of $\gg$ is defined as
\be
\uu^c = (R\otimes \cJ)\uu\,,
\ee
where $R: L^{\infty}(\gg)\rightarrow L^{\infty}(\gg)$ is the unitary antipode of $\gg$.
\end{definition}
$\uu^c$ is a unitary representation of $\gg$ on $\overline{\cH}$.
\begin{definition}[Real structure on a unitary representation]\label{def:realStructure}
We say that a unitary representation $\uu$ of $\gg$ on $\cH$ is real if there exists an antiunitary map $J_{\uu}:\cH\rightarrow \cH$ satisfying $J_{\uu}^2 = 1_{\cH}$ such that
\be
(R\otimes \cJ_{\uu})\uu = \uu\,,
\ee
where $\cJ_{\uu}(T) := J_{\uu} T^* J_{\uu}$. Equivalently,
\be
(\hat{J} \otimes J_{\uu})\uu (\hat{J} \otimes J_{\uu}) = \uu^*\,.
\ee
The map $J_{\uu}$ is called a real structure on $\uu$.
\end{definition}
If $\uu$ is real, then $\uu^c$ is unitarily equivalent to $\uu$, since we can take $j J_{\uu}: \cH\rightarrow \overline{\cH}$ as the unitary intertwiner. The trivial representation $\cH=\cc$ is real, with real structure provided by $J_{\uu}(z)= \overline{z}$.

\subsection{Quantum group actions as coactions}\label{ssec:QMS}
Let us review the central concept used in this work to reformulate conformal field theory: actions of locally compact quantum groups on quantum measure spaces. A quantum measurable space $\xx$ is described by a von Neumann algebra $N$, interpreted as the algebra of essentially bounded functions on $\xx$, denoted $L^{\infty}(\xx):=N$. In particular, a locally compact quantum group is a quantum measurable space equipped with a coproduct and invariant weights. Actions generalize as coactions:
\begin{definition}
Let $N$ be a von Neumann algebra and $(M,\Delta)$ a von Neumann bialgebra. A \emph{left coaction} of $(M,\Delta)$ on $N$ consists of a normal, injective, unital $*$-homomorphism $\alpha: N\rightarrow M\botimes N$ such that
\be
(\Delta\otimes\id_N) \alpha = (\id_M\otimes \alpha)\alpha\,.
\ee
A \emph{right coaction} of $(M,\Delta)$ on $N$ consists of a normal, injective, unital $*$-homomorphism $\beta: N\rightarrow N\botimes M$ such that
\be
(\id_N\otimes \Delta) \beta = (\beta\otimes \id_M)\beta\,.
\ee
When $(M,\Delta) = (L^{\infty}(\gg),\Delta_{\gg})$ for a locally compact quantum group $\gg$, and $N = L^{\infty}(\xx)$ for a quantum measurable space $\xx$, we call such $\alpha$, $\beta$ respectively a left, right action of $\gg$ on $\xx$.
\end{definition}
This generalizes the classical definition of an action: when $G$ acts on a space $X$ on the left $x\mapsto g\cdot x$, resp.~right $x\mapsto x\cdot g$, the above maps $\alpha,\beta$ are realized on $f\in L^{\infty}(X)$ as
\be
(\alpha f)(g, x) = f(g\cdot x)\,,\qquad (\beta f)(x,g) = f(x\cdot g)\,.
\ee
The coproduct of a locally compact quantum group $\gg$ provides both a left and a right action of $\gg$ on itself via $\alpha=\beta=\Delta_{\gg}$. These generalize the action of a classical group $G$ on itself by left and right translations.

Classically, an action is called ergodic if the only measurable functions preserved by it are constant almost everywhere. This definition generalizes seamlessly to the noncommutative case. We use $M$ and $L^{\infty}(\gg)$ interchangeably in the following, and the same goes for $N$ and $L^{\infty}(\xx)$.
\begin{definition}
Let $\alpha: N\rightarrow M\botimes N$ be a left action of a quantum group $\gg$ on a quantum space $\xx$. Define the fixed-point von Neumann subalgebra of $N$ as
\be
N^{\alpha} := \{a\in N\, : \, \alpha(a) = 1_M \otimes a \}\,.
\ee
We say $\alpha$ is ergodic if $N^{\alpha} = \cc\,1_N$. The definition for right actions is analogous, with
\be
N^{\beta} := \{a\in N\, : \, \beta(a) =  a\otimes 1_M \}\,.
\ee
\end{definition}

To generalize the notion of measure space $(X,\mu)$, one equips $N$ with an n.s.f.~weight $\theta : N_{+}\rightarrow [0,\infty]$. We call the pair $(\xx,\theta)$ a quantum measure space. The following definition generalizes the classical notion of invariant measure:
\begin{definition}
Let $\alpha: N\rightarrow M\botimes N$ be a left action of a quantum group $\gg$ on a quantum measure space $(\xx,\theta)$. We say that $\theta$ is \emph{invariant} under $\alpha$ if
\be
(\id\otimes\theta)\alpha(a) = \theta(a) 1_M
\ee
for all $a\in N_+$ such that $\theta(a)<\infty$. Similarly, when $\beta: N\rightarrow N\botimes M$ is a right action, we say that $\theta$ is invariant under $\beta$ if
\be
(\theta\otimes\id)\beta(a) = \theta(a) 1_M
\ee
for all $a\in N_+$ such that $\theta(a)<\infty$.
\end{definition}
The canonical left and right actions of a locally compact quantum group on itself provide an example, respectively with $(\alpha,\theta) = (\Delta_{\gg}, \ph_L)$ and $(\beta,\theta) = (\Delta_{\gg}, \ph_R)$. We have seen in Sections~\ref{ssec:W} and~\ref{ssec:V} that it is useful to implement these actions on the GNS Hilbert space $L^2(\gg)$ through the left and right multiplicative unitaries $\ww\,,\vv\in \mathrm{B}(L^2(\gg)\otimes L^2(\gg))$.

More generally, every action of a locally compact quantum group $\gg$ on a quantum measure space $(\xx,\theta)$ has a canonical implementation as a unitary representation of $\gg$~\cite{Vaes2001UnitaryImplementation}, which we describe now.

Let $(\cK_\theta,\iota_\theta,\Lambda_\theta)$ be the GNS construction for $(N,\theta)$. Here $\cK_{\theta}$ is a Hilbert space generalizing $L^2(X,\mu)$. $\iota_{\theta}: L^{\infty}(\xx)\rightarrow \mathrm{B}(\cK_{\theta})$ realizes $N$ on $\cK_{\theta}$.  $\Lambda_{\theta}: \mathfrak{n}_{\theta}\rightarrow \cK_{\theta}$ is the GNS map. It is injective and its image is dense in $\cK_{\theta}$. Denoting the inner product on $\cK_{\theta}$ by $\langle\cdot,\cdot \rangle_{\theta}$, we have
\begin{align}
\,&a\,\Lambda_{\theta}(b) = \Lambda_{\theta}(a b) &&\qquad a\in N\,, b\in \mathfrak{n}_\theta\,,\\
&\langle\Lambda_{\theta}(a),\Lambda_{\theta}(b)\rangle_{\theta} = \theta(a^* b)&&\qquad a,b\in \mathfrak{n}_{\theta}\,,
\end{align}
where we identified $N$ with its image under $\iota_{\theta}$. We will write $\cK_{\theta}$ and $L^2(\xx,\theta)$ interchangeably.

Recall Definition~\ref{def:UnitaryRep} of a unitary representation of $\gg$. As in Section~\ref{ssec:V}, we will assume for simplicity that the quantum group is unimodular with $\ph:=\ph_L=\ph_R$, and denote by $\Lambda_\ph: \mathfrak{n}_{\ph}\rightarrow L^2(\gg)$ the GNS map for $(M,\ph)$. Left and right actions on $\xx$ preserving $\theta$ then have the following unitary implementations:
\begin{theorem}[Vaes~\cite{Vaes2001UnitaryImplementation}]\label{thm:UImplementation}
Let $\alpha: N\rightarrow M\botimes N$ be a left action of a unimodular quantum group $\gg$ on $\xx$, preserving an n.s.f.~weight $\theta$ on $N$. Then there exists a unique unitary operator $\uu_{\alpha}\in M \botimes \mathrm{B}(\cK_{\theta})$ such that
\be
\uu_{\alpha}^* (\Lambda_\ph(a)\otimes\Lambda_{\theta}(b)) = (\Lambda_\ph\otimes\Lambda_{\theta})(\alpha(b)(a\otimes 1_N))\,,\qquad
a\in \mathfrak{n}_{\ph}\,,\, b\in\mathfrak{n}_{\theta}\,.
\label{eq:UStarDefinition}
\ee
$\alpha$ can be recovered from $\uu_{\alpha}$ as
\be
\alpha(b) = \uu_{\alpha}^*(1_M\otimes b) \uu_{\alpha}\,,\qquad b\in N\,.
\ee
Furthermore, $\uu_{\alpha}$ is a unitary representation of $\gg$ on $\cK_{\theta}$:
\be
(\Delta_\gg\otimes \id)\uu_{\alpha} =  (\uu_{\alpha})_{13} (\uu_{\alpha})_{23}\,.
\ee

Similarly, let $\beta: N\rightarrow N\botimes M$ be a right action of $\gg$ on $\xx$, preserving an n.s.f.~weight $\theta$ on $N$. Then there exists a unique unitary operator $\uu_{\beta}\in \mathrm{B}(\cK_{\theta})\botimes M$ such that
\be
\uu_{\beta} (\Lambda_{\theta}(b)\otimes\Lambda_{\ph}(a)) = (\Lambda_{\theta}\otimes\Lambda_{\ph})(\beta(b)(1_N \otimes a))\,,\qquad
a\in \mathfrak{n}_{\ph}\,,\, b\in\mathfrak{n}_{\theta}\,.
\ee
$\beta$ can be recovered from $\uu_{\beta}$ as
\be
\beta(b) = \uu_{\beta}(b\otimes 1_M) \uu_{\beta}^*\,, \qquad b\in N\,.
\ee
Furthermore, the tensor flip $(\uu_{\beta})_{21} \in M \botimes \mathrm{B}(\cK_{\theta})$ is a unitary representation of $\gg$ on $\cK_{\theta}$:
\be
(\id\otimes \Delta_\gg)\uu_{\beta} =  (\uu_{\beta})_{12} (\uu_{\beta})_{13}\,.
\ee
\end{theorem}
Thus, $\uu_{\alpha}$ generalizes $\ww$ and $\uu_{\beta}$ generalizes $\vv$, see~\eqref{eq:WDef},~\eqref{eq:DeltaFromW},~\eqref{eq:VDef},~\eqref{eq:DeltaFromV}. Theorem~\ref{thm:UImplementation} generalizes the canonical unitary representation of $G$ on the Hilbert space $L^2(X,\mu)$ arising from a measure-preserving action.

\subsection{Real structure and modular flow of quantum measure spaces}\label{ssec:QMSrealmodular}
Every quantum measure space $(\xx,\theta)$ comes with a natural notion of conjugation, generalizing pointwise complex conjugation of functions on classical measure spaces. At the level of the von Neumann algebra $N$, this is the $*$-operation. Let $S_{\theta}$ be the Tomita operator of $(N,\theta)$ on $\cK_{\theta}$
\be
S_{\theta} \Lambda_{\theta}(a) = \Lambda_{\theta}(a^*)\,,\qquad a\in \mathfrak{n}_{\theta}\cap\mathfrak{n}^*_{\theta}\,.
\ee
Let $J_{\theta}$ and $\nabla_{\theta}$ be the modular conjugation and modular operator
\be
S_{\theta} = J_{\theta} \nabla^{\frac{1}{2}}_{\theta}\,.
\ee
$J_{\theta}$ generalizes pointwise complex conjugation at the level of $\cK_{\theta}$.

Suppose $(\xx,\theta)$ comes equipped with a $\theta$-preserving left or right action of a locally compact quantum group. Its canonical unitary implementation on $\cK_{\theta}$ provided by Theorem~\ref{thm:UImplementation} is then automatically compatible with the real structure provided by $J_{\theta}$:
\begin{proposition}[Vaes~\cite{Vaes2001UnitaryImplementation}]\label{prop:UJ}
Under the hypotheses of Theorem~\ref{thm:UImplementation}, $J_{\theta}$ provides a real structure for the unitary representations $\uu_{\alpha}$, $(\uu_{\beta})_{21}$, i.e.
\ba
\,&(\hat{J}\otimes J_{\theta})\uu_{\alpha} (\hat{J}\otimes J_{\theta}) = \uu^*_{\alpha}\,,\\
&(J_{\theta}\otimes \hat{J})\uu_{\beta} (J_{\theta}\otimes \hat{J}) = \uu^*_{\beta}\,.
\ea
\end{proposition}
Similarly, the canonical unitary implementations are compatible with the modular flow of $\theta$, provided we simultaneously apply the scaling group of $\gg$ to the quantum group leg:
\begin{proposition}[Vaes~\cite{Vaes2001UnitaryImplementation}]\label{prop:UNabla}
Under the hypotheses of Theorem~\ref{thm:UImplementation}, the unitary implementation $\uu_{\alpha}$ of the left action $\alpha$ satisfies
\be
(\tau_t\otimes \id) \uu_{\alpha} = (1_M\otimes \nabla^{-i t}_{\theta})\uu_{\alpha}(1_M\otimes \nabla^{i t}_{\theta})\,,\qquad t\in \rr\,.
\ee
Equivalently,
\be
(\hat{\nabla}^{i t}\otimes \nabla^{i t}_{\theta})\uu_{\alpha}(\hat{\nabla}^{-i t}\otimes \nabla^{-i t}_{\theta})=\uu_{\alpha} \,,\qquad t\in \rr\,.
\label{eq:scalingModularFlowL}
\ee
Similarly, the unitary implementation $\uu_{\beta}$ of the right action $\beta$ satisfies
\be
(\id\otimes \tau_t) \uu_{\beta} = (\nabla^{i t}_{\theta}\otimes 1_M)\uu_{\beta} (\nabla^{-i t}_{\theta}\otimes 1_M)\,,\qquad t\in \rr\,.
\ee
Equivalently,
\be
(\nabla^{-i t}_{\theta}\otimes \hat{\nabla}^{i t})\uu_{\beta}(\nabla^{i t}_{\theta}\otimes \hat{\nabla}^{-i t})=\uu_{\beta} \,,\qquad t\in \rr\,.
\label{eq:scalingModularFlowR}
\ee
\end{proposition}
Notice the opposite relative signs in~\eqref{eq:scalingModularFlowR}, with respect to~\eqref{eq:scalingModularFlowL}. This occurs because a right action is equivalent to a left action with opposite coproduct, and switching the coproduct order reverses the scaling group.

In the special case of $\gg$ acting on itself by left and right translations, the above propositions reduce to the identities
\ba
\,&(\hat{J}\otimes J )\ww (\hat{J}\otimes J) = \ww^*\,,\\
& (J \otimes \hat{J})\vv (J \otimes \hat{J}) = \vv^*\,,
\label{eq:JWV}
\ea
and
\ba
\,&(\hat{\nabla}^{i t}\otimes \nabla^{i t})\ww(\hat{\nabla}^{-i t}\otimes \nabla^{-i t})=\ww\,,\\
&(\nabla^{-i t}\otimes \hat{\nabla}^{i t})\vv(\nabla^{i t}\otimes \hat{\nabla}^{-i t})=\vv\,.
\label{eq:nablaWV}
\ea

\section{$\sl$}\label{sec:slb2r}
In this section, we review the locally compact quantum group $\sl$ and its dual. To our knowledge, the relevant structure was first discussed in Section~5 of the classic paper~\cite{PonsotTeschner1999}. The same reference also sketched a duality between $\sl$ and Faddeev's modular double~\cite{Faddeev1999ModularDouble}. The dual object was further discussed in~\cite{PonsotTeschner2001,BytskoTeschner2003}. A detailed description of $\sl$ was worked out in~\cite{Ip2013QuantumPlane}. Finally, reference~\cite{DeCommerSchraderShapiroVoigt2025} constructed $\sl$ and its generalization $\mathrm{SL}^+_b(N,\rr)$, in the regime $Q\geq 2$, rigorously in the sense of Definition~\ref{def:LCQG}.

Our exposition combines elements of~\cite{PonsotTeschner1999,Ip2013QuantumPlane,DeCommerSchraderShapiroVoigt2025}. In addition, we provide evidence that the construction of~\cite{DeCommerSchraderShapiroVoigt2025} can be analytically continued to define a locally compact quantum group also for $0<Q<2$. We also explain in which sense the modular double generates the dual quantum group. Finally, we compute the action of operators arising in the modular theory of $\sl$ and its dual. To keep expressions readable, we will use the notation $\gg_b = \sl$.

\subsection{Von Neumann algebra $L^{\infty}(\sl)$}
References~\cite{Ip2013QuantumPlane,DeCommerSchraderShapiroVoigt2025} construct $\gg_b$ as a quotient of the locally compact quantum group $\mathrm{GL}^+_b(2,\rr)$. The latter is the Drinfeld double of the positive $ax+b$ quantum group, playing the role of positive upper-triangular 2 by 2 matrices. Here we will focus on describing the final result of this construction, and refer to~\cite{Ip2013QuantumPlane,DeCommerSchraderShapiroVoigt2025} for its origin. 

$\gg_b$ is parametrized by
\be
b \in \{e^{i\theta}\,|\, \theta\in (-\tfrac{\pi}{2},\tfrac{\pi}{2})\}\cup \rr_{>0}\,.
\ee
We assume that $b$ lies in this set from now on. We will see that $\gg_b$ is manifestly invariant under $b\leftrightarrow b^{-1}$:
\be
\gg_b = \gg_{1/b}
\ee
as locally compact quantum groups. We can therefore label them uniquely by
\be
Q = b+\frac{1}{b} \in (0,\infty)\,.
\ee
In that light, perhaps a better notation would have been $\mathrm{SL}^+_{Q}(2,\rr)$. We will use $b$ as it simplifies some formulas, and the relationship with the modular double. Note that the inequivalent values of the parameter map one-to-one to the range of central charges $c$ appropriate for unitary irrational 2d CFTs\footnote{Strictly speaking, unitary irrational CFTs also exist at $c=1$, which is just outside the stated range. It would be interesting to develop a quantum group viewpoint also for $c=1$ as a degenerate limit of our analysis. Note that the classification of unitary CFTs at $c=1$ was recently completed in~\cite{Eberhardt:2026qqm}.}
\be
c = 1 + 6Q^2 \in (1,\infty)\,.
\ee

Let us describe the von Neumann algebra $L^{\infty}(\gg_b)$. We will first give its concrete realization on $\cK_0:=L^2(\rr^2)$ and later pass to the realization on the GNS Hilbert space $L^2(\gg_b)\cong L^2(\rr^3)$. The objects and maps in the first realization carry the subscript $0$.

Denote the coordinates on $\rr^2$ by $x_{1},x_{2}$ and let $\xbf_j$, $\pbf_j$ be the unbounded self-adjoint operators on $\cK_0$ initially defined by
\be
\xbf_j f = x_j f\,,\qquad \pbf_j f = \frac{1}{2\pi i} \frac{\partial f}{\partial x_j}\,,\qquad j=1,2\,.
\ee
\begin{definition}
Let $C_0  \subset \mathrm{B}(\cK_0)$ be the von Neumann algebra generated by the unitaries $e^{2\pi i(\xbf_1+\xbf_2+\pbf_1)t}$:
\be
C_0 = \{e^{2\pi i(\xbf_1+\xbf_2+\pbf_1)t} \,:\, t\in \rr \}''\subset \mathrm{B}(\cK_0)\,.
\ee
We define $M_0$ as the commutant of $C_0$:
\be
M_0=C_0'\subset \mathrm{B}(\cK_0)\,.
\ee
$M_0$ provides our initial realization of $L^{\infty}(\gg_b)$.
\end{definition}
\begin{lemma}
$M_0$ is generated by the unitaries $U_{s,t,u} := e^{2\pi i[(-\xbf_1-\pbf_1)s+(\pbf_1-\xbf_2-\pbf_2)t+\xbf_2 u]}$:
\be
M_0 = \{ U_{s,t,u} \, : \, s,t,u\in \rr\}'' \subset \mathrm{B}(\cK_0)\,.
\ee
\end{lemma}
General elements of $M_0$ arise as $\int_{\rr^3} F(s,t,u)U_{s,t,u} ds dt du$, where $F$ is a distribution on $\rr^3$. It will be useful to work with a $*$-subalgebra of $M_0$ in which $F$ ranges over particularly nice functions:
\begin{definition}
Let $\alpha,\beta\geq 0$. The Gelfand--Shilov space $S^{\alpha}_{\beta}(\rr^n)$ consists of all smooth functions $f: \rr^n\rightarrow \cc$ for which there exist constants $A,B,C > 0$ such that
\be
\sup_{x\in\rr^n} |x^p \partial^q f(x)| \leq C A^{|p|} B^{|q|} (p!)^{\beta} (q!)^{\alpha}\quad \text{for all}\quad p,q\in \mathbb{N}^n_0\,,
\ee
where $p! := p_1! \ldots p_n!$ and similarly for $q!$. We define
\be
\sA = S^{1/2}_{1/2}(\rr^3)\,.
\ee
\end{definition}
Every $F \in \sA$ extends to an entire function on $\cc^3$ with entire Fourier transform $\hat{F}$. Moreover, there exist constants $a,b,c>0$, depending on $F$, such that
\be
|F(x+i y)| \leq c \,e^{-a |x|^2 + b |y|^2}\,,\quad 
|\hat{F}(x+i y)| \leq c\, e^{-a |x|^2 + b |y|^2}\quad\text{for all}\quad x,y\in\rr^3\,.
\ee
In particular, both $F$ and $\hat{F}$ exhibit at least Gaussian decay along $\rr^3 \subset \cc^3$. The Fourier transform maps $\sA$ bijectively to itself.

Define the following skew-symmetric bilinear form on $\rr^3$
\be
\Omega: \rr^3\times \rr^3\rightarrow \rr\,,\qquad \Omega(x,x') := (s+u)t' -t(s'+u')\,,
\label{eq:OmegaSkew}
\ee
where $x=(s,t,u)$ and $x'=(s',t',u')$. Introduce the following twisted convolution product on $\sA$
\be
(F_1\cdot F_2)(x)= \int\limits_{\rr^3}e^{i\pi \,\Omega(x,x')} F_1(x-x')F_2(x') dx'\,,
\label{eq:Aproduct}
\ee
where $F_1,F_2\in\sA$. Introduce the $*$-operation on $\sA$
\be
(F^*)(x) =  \overline{F(-x)}\,,\qquad F\in \sA\,.
\label{eq:Astar}
\ee
\begin{lemma}\label{lem:pi0}
\eqref{eq:Aproduct} and~\eqref{eq:Astar} make $\sA$ into a $*$-algebra. Furthermore, the map
\be
\pi_0: \sA \rightarrow M_0\,,\qquad \pi_0(F) :=  \int\limits_{\rr^3} F(s,t,u) U_{s,t,u}\, ds dt du\,,
\ee
is an injective \mbox{$*$-algebra} homomorphism whose image is dense in $M_0$.
\end{lemma}
Thus $\pi_0$ identifies $\sA$ with a dense subalgebra of $M_0$, which we will use for explicit calculations. Whenever convenient, we will extend the domain of $\pi_0$ to tempered distributions $\mathcal{S}'(\rr^3)$. Every element of $M_0$ equals $\pi_0(F)$ for some $F\in \mathcal{S}'(\rr^3)$. For example
\be
\pi_0(\delta_{(s_0,t_0,u_0)}) = U_{s_0,t_0,u_0}\,.
\ee
On the other hand, $\pi_0(F)$ is not always a bounded operator for arbitrary $F\in\mathcal{S}'(\rr^3)$.

In order to start justifying the name $\sl$, let us introduce the following unbounded operators
\ba
A &:= U_{i\frac{b}{2},i\frac{b}{2},i\frac{b}{2}}=e^{\pi b(\xbf_1+\pbf_2)}\,,\\
B &:= U_{i\frac{b}{2},i\frac{b}{2},-i\frac{b}{2}}=e^{\pi b(\xbf_1+2\xbf_2+\pbf_2)}\,,\\
C &:= U_{-i\frac{b}{2},i\frac{b}{2},i\frac{b}{2}}=e^{-\pi b(\xbf_1+2\pbf_1-\pbf_2)}\,,\\
D &:= U_{-i\frac{b}{2},-i\frac{b}{2},-i\frac{b}{2}} + U_{-i\frac{b}{2},i\frac{b}{2},-i\frac{b}{2}} = A^{-1}+ e^{-i\pi b^2}A^{-1}BC\,.
\ea
They satisfy the relations
\ba
BA &= q AB\,,\quad CA = q AC\,,\quad CB = BC\,,\quad D B = q B D\,,\quad D C = q C D\,,\\
AD &= 1+q^{-1}BC\,,\quad DA = 1 + q BC\,,
\label{eq:ABCDrelations}
\ea
where
\be
q:= e^{ i \pi b^2}\,.
\ee
Heuristically, $A,B,C,D$ are the matrix elements of a quantum version of the $\mathrm{SL}(2,\rr)$ group, consisting of matrices
\be
\begin{pmatrix}
A & B\\ C & D
\end{pmatrix}\,.
\ee
$A,B,C,D$ being unbounded operators, i.e.~not belonging to the von Neumann algebra $L^{\infty}(\gg_b)$, is consistent with $\gg_b$ being non-compact. When $Q\geq 2$, $A,B,C,D$ are positive self-adjoint. This justifies the name ``positive quantum $\mathrm{SL}(2,\rr)$ group'', and is responsible for the superscript in $\sl$. $A,B,C$ generate $L^{\infty}(\gg_b)$ in the sense that we can recover the unitaries $U_{s,t,u}$ as
\be
U_{s,t,u} = e^{\frac{i \pi }{2} (s+u) (s-2 t+u)} A^{-\tfrac{i (s+u)}{b}}B^{-\tfrac{i (t-u)}{b}}C^{-\tfrac{i (t-s)}{b}}\,.
\ee

Intriguingly, $\gg_b$ gives rise to a second copy of these generators, obtained by $b\mapsto 1/b$. Indeed, introduce the unbounded operators
\ba
\check{A} &:= U_{i\frac{1}{2b},i\frac{1}{2b},i\frac{1}{2b}}=e^{\pi b^{-1}(\xbf_1+\pbf_2)}\,,\\
\check{B} &:= U_{i\frac{1}{2b},i\frac{1}{2b},-i\frac{1}{2b}}=e^{\pi b^{-1}(\xbf_1+2\xbf_2+\pbf_2)}\,,\\
\check{C} &:= U_{-i\frac{1}{2b},i\frac{1}{2b},i\frac{1}{2b}}=e^{-\pi b^{-1}(\xbf_1+2\pbf_1-\pbf_2)}\,,\\
\check{D} &:= U_{-i\frac{1}{2b},-i\frac{1}{2b},-i\frac{1}{2b}} + U_{-i\frac{1}{2b},i\frac{1}{2b},-i\frac{1}{2b}} = \check{A}^{-1}+ e^{-i\pi b^{-2}}\check{A}^{-1}\check{B}\check{C}\,.
\ea
They satisfy the relations
\ba
\check{B}\check{A} &= \check{q} \check{A}\check{B}\,,\quad \check{C}\check{A} = \check{q} \check{A}\check{C}\,,\quad \check{C}\check{B} = \check{B}\check{C}\,,\quad \check{D} \check{B} = \check{q} \check{B} \check{D}\,,\quad \check{D} \check{C} = \check{q} \check{C} \check{D}\,,\\
\check{A}\check{D} &= 1+\check{q}^{-1}\check{B}\check{C}\,,\quad \check{D}\check{A} = 1 + \check{q} \check{B}\check{C}\,,
\ea
where
\be
\check{q}:= e^{ i \pi b^{-2}}\,.
\ee
When $Q\geq 2$, also the new operators are positive self-adjoint. By the spectral functional calculus for positive self-adjoint operators, we get
\be
\check{A} = A^{1/b^2}\,,\qquad
\check{B} = B^{1/b^2}\,,\qquad
\check{C} = C^{1/b^2}\,,\qquad
\check{D} = D^{1/b^2}\,.
\ee
When $0<Q<2$, i.e.~$b=e^{i\theta}$ with $\theta\in (0,\frac{\pi}{2})$, the generators are no longer self-adjoint. Rather, they are adjoints of each other
\be
A^* = \check{A}\,,\qquad
B^* = \check{B}\,,\qquad
C^* = \check{C}\,,\qquad
D^* = \check{D}\,.
\ee
The unbounded operators $A,B,C,D$ and $\check{A},\check{B},\check{C},\check{D}$ are useful for intuition. Making rigorous statements about $\sl$ requires working with bounded operators in $L^{\infty}(\gg_b)$. Note that the above construction of $L^{\infty}(\gg_b)$ is completely $b$-independent. $b$ only enters if we insist on introducing the coordinates $A,B,C,D$. Nevertheless, as a locally compact quantum group, $\gg_b$ does depend on $b$ because the coproduct depends on it.

\subsection{Coproduct}
Recall from Section~\ref{sec:LCQG} that the coproduct $\Delta$ encodes the group multiplication law of the underlying quantum group. It is a unital normal $*$-homomorphism of von Neumann algebras $\Delta:L^{\infty}(\gg_b)\rightarrow L^{\infty}(\gg_b)\botimes L^{\infty}(\gg_b)$. We will denote the realization of this coproduct on $M_0$ as $\Delta_0: M_0\rightarrow M_0\botimes M_0$.

We will first motivate the form of $\Delta_0$ by its action on the unbounded operators $A,B,C,D$, and later formulate its action on $M_0$ rigorously. Let us initially assume $Q\geq 2$, so that $A,B,C,D$ are positive self-adjoint.

The coproducts of the matrix elements take the usual matrix-multiplication form\footnote{It is at this point where the dependence of the quantum group on $b$ first enters: one demands that the $b$-dependent coordinates $A,B,C$ have the simple coproducts above.}
\ba
\Delta_0(A) &= A\otimes A+B\otimes C\,,\quad \Delta_0(B) = A\otimes B+B\otimes D\,,\\
\Delta_0(C) &= C\otimes A+D\otimes C\,,\quad \Delta_0(D) = C\otimes B+D\otimes D\,.
\ea
Associativity of matrix multiplication means that the coproduct is formally coassociative. The formulas are nontrivially consistent with $\Delta_0$ being an algebra homomorphism, i.e.~they preserve the relations~\eqref{eq:ABCDrelations}.

The coproducts of arbitrary elements of $M_0$ in principle follow by the homomorphism property of $\Delta_0$ and functional calculus. Consider first the family of unitaries $U_{t,t,t}\in M_0$ with $t\in \rr$. We have $U_{t,t,t} = A^{-\frac{2i t}{b}}$ and thus
\be
\Delta_0(U_{t,t,t}) = \Delta_0(A^{-\tfrac{2i t}{b}}) = \Delta_0(A)^{-\tfrac{2i t}{b}} = (A\otimes A+B\otimes C)^{-\tfrac{2i t}{b}} \in M_0\botimes M_0\,.
\label{eq:DeltaUttt}
\ee
We would like to express the unitary on the right-hand side in terms of $U_{s_1,t_1,u_1}\otimes U_{s_2,t_2,u_2}$. To that end, we need an analogue of the binomial theorem for positive self-adjoint operators commuting up to a phase. To state the relevant theorem, we will need to introduce Faddeev's noncompact quantum dilogarithm~\cite{Faddeev1995,KashaevNakanishi2011,FaddeevKashaev1994}:
\begin{definition}[The noncompact quantum dilogarithm]
Let $b \in \{e^{i\theta}\,|\, \theta\in (-\tfrac{\pi}{2},\tfrac{\pi}{2})\}\cup \rr_{>0}$. The noncompact quantum dilogarithm $\phi_b$ is defined by
\be
\phi_b(z) := \exp\!\Big[\int\limits_{\Omega}
\frac{e^{-2i x z}}{4 \sinh \left(b x\right) \sinh \left(x/b\right)}\frac{dx}{x}\Big]\quad\text{for}\quad |\mathrm{Im}(z)| < \frac{Q}{2}\,.
\label{eq:phiIntegral}
\ee
Here $\Omega$ is a contour from $x=-\infty$ to $x=\infty$ avoiding the pole at $x=0$ from just above.
\end{definition}
\begin{lemma}
$\phi_b$ extends to a meromorphic function on $\cc$ with the following analytic structure:
\be
\phi_b(z)\text{ has }\begin{cases}
\text{ poles at }z= i \left(\frac{Q}{2}+ m b+n b^{-1}\right)\text{ with }m,n\in\zz_{\geq 0}\,,\\
\text{ zeros at }z= -i \left(\frac{Q}{2}+ m b+n b^{-1}\right)\text{ with }m,n\in\zz_{\geq 0}\,.
\end{cases}
\ee
$\phi_b$ has the manifest symmetry
\be
\phi_b(z) = \phi_{1/b}(z)\,.
\ee
Finally, for the stated range of $b$, we have
\be
|\phi_b(x)| = 1\,,\qquad x\in \rr\,.
\ee
\end{lemma}
$\phi_b$ satisfies various additional remarkable properties, such as its functional equations and the pentagon identity~\eqref{eq:phiPentagon} stated below; see e.g.~\cite{FaddeevKashaev1994,Faddeev1995,BytskoTeschner2003,KashaevNakanishi2011}. The same function appears in the literature under various names and normalizations, notably in Woronowicz's theory of quantum exponential functions~\cite{Woronowicz2000}. We will find the following version particularly convenient
\be
s_b(z) := e^{-\frac{i\pi}{24}(Q^2-2)}e^{-\frac{i\pi}{2}z^2} \phi_b(z)\,.
\label{eq:sbDefinition}
\ee
\begin{lemma}
We have
\be
s_b(z)s_b(-z) = 1 \,,\qquad s_b(z)\, \overline{s_b(\zb)} = 1\,,\qquad z\in \cc\,.
\label{eq:sIdentities}
\ee
Furthermore,
\be
s_b(z) = S_b(\tfrac{Q}{2} + i z)\,,\qquad z\in\cc\,,
\ee
where $S_b$ is the double sine function, itself related to the Barnes double Gamma function $\Gamma_b$:
\be
S_b(z) = \frac{\Gamma_b(z)}{\Gamma_b(Q- z)}\,.
\ee
\end{lemma}
\begin{definition}[The quantum binomial coefficient]\label{def:QBC}
Let $B_b$ be the following meromorphic kernel
\be
B_b(x,y) = \frac{s_b(x+i \tfrac{Q}{2})s_b(y+i \tfrac{Q}{2})}{s_b(x+y+i \tfrac{Q}{2})} =
\frac{S_b(i x)S_b(i y)}{S_b(ix+iy)}\,,\qquad x,y\in \cc\,.
\label{eq:Bmeromorphic}
\ee
\end{definition}
$B_b(x,y)$ is holomorphic in the domain $\mathrm{Im}(x)<0$, $\mathrm{Im}(y)<0$, $\mathrm{Im}(x+y)>-Q$. We will need its boundary value for $x,y\in \rr$. This is a distribution defined as follows
\be
B_b(x,y) = \lim_{\epsilon_{1},\epsilon_{2}\searrow 0} B_b(x-i\epsilon_1,y-i\epsilon_2)\,.
\ee
We are now ready to state the quantum binomial theorem~\cite[Appendix~B]{BytskoTeschner2003}.
\begin{theorem}[Quantum binomial theorem]\label{thm:qBinomial}
Let $b>0$ and let $X$, $Y$ be strictly positive self-adjoint operators satisfying
\be
X^{i r} Y^{i s} = e^{2\pi i  b^2 r s} Y^{i s} X^{i r}
\ee
for all $r,s\in\rr$. Then $X+Y$ is positive self-adjoint and we have
\be
(X+Y)^{-i\tfrac{2t}{b}} = \int\limits_{\rr} e^{-i\pi (t+z)(t-z)}B_b(t+z,t-z)X^{-i\tfrac{t+z}{b}}Y^{-i\tfrac{t-z}{b}}dz
\label{eq:qBinomial}
\ee
for all $t\in \rr$.\footnote{Strictly speaking, $X+Y$ here is the positive form sum of $X$ and $Y$.}
\end{theorem}
The formula has a similar shape to the ordinary binomial theorem, with sum replaced by integral, and with $s_b$ playing the role of the factorial. The distributional definition of $B_b$ is equivalent to its meromorphic form~\eqref{eq:Bmeromorphic} with the contour prescription where the pole at $z=-t$ is avoided from below and the one at $z=t$ from above. The formula works also for $t=0$ thanks to the identity $B_b(x,-x)=\delta(x)$, arising from pinching of the contour.

Let us finally apply Theorem~\ref{thm:qBinomial} to the right-hand side of~\eqref{eq:DeltaUttt} with $X = A\otimes A$ and $Y = B\otimes C$. We get
\be
\Delta_0(U_{t,t,t}) =  \int\limits_{\rr}
B_b(t+z,t-z)
\,U_{t,t,z}\otimes U_{z,t,t} \, dz\,,\qquad t\in \rr\,.
\ee
By a straightforward generalization of this procedure, we arrive at the coproduct of a general $U_{s,t,u}\in M_0$ 
\ba
\Delta_0(U_{s,t,u})=
&\int\limits_{\rr^3}
B_b(t-z_1,z_1-s)
B_b(t-z_3,z_3-u) B_b(t+z_2,z_1-z_2+z_3-t)\\
&\quad\quad\times U_{s,z_1,z_2}\otimes U_{z_2,z_3,u}\;dz_1 dz_2 dz_3\,.
\label{eq:DeltaUstu}
\ea
Something nice has happened: since $B_b(x,y) = B_{1/b}(x,y)$, the final coproduct formula is manifestly invariant under $b\leftrightarrow 1/b$, although the intermediate steps were not. Furthermore, the formula makes perfect sense also for $Q\in(0,2)$ although we assumed $Q\geq 2$ in the intermediate steps.

\eqref{eq:DeltaUstu} lets us write down the action of the coproduct on $\pi_0(\sA)$. Since $\pi_0(\sA)$ is dense in $M_0$, this defines the coproduct on $M_0$. We expect that the following claim can be proved using the methods of~\cite{DeCommerSchraderShapiroVoigt2025}.
\begin{claim}
Let $Q = b+b^{-1}> 0$. There exists a coassociative, normal, unital $*$-homomorphism $\Delta_0: M_0\rightarrow M_0\botimes M_0$ such that
\be
\Delta_0(\pi_0(F)) = (\pi_0\otimes \pi_0)(D_b F)\,,\qquad F\in \sA\,,
\ee
where $D_b: \sA \rightarrow \mathcal{S}'(\rr^6)$ is given by
\ba
\,&(D_b F)(s_1,t_1,u_1;s_2,t_2,u_2) \\
&=\delta(s_2-u_1)
\int\limits_{\rr}
B_b(z-t_1,t_1-s_1)
B_b(z-t_2,t_2-u_2)
B_b(z+u_1,t_1+t_2-u_1-z)\\
&\qquad\qquad\qquad\qquad\times  F(s_1,z,u_2)\, dz\,.
\label{eq:DeltaF}
\ea
\end{claim}

\subsection{Haar weight}\label{ssec:HaarGb}
In order to establish that $\gg_b$ is a locally compact quantum group, one needs to demonstrate that $(M_0,\Delta_0)$ admits invariant Haar weights $\ph_L$, $\ph_R$. Their existence was proved for $Q\geq 2$ in~\cite{DeCommerSchraderShapiroVoigt2025}. In fact, like its classical counterpart, $\gg_b$ is unimodular, meaning we can take $\ph_L=\ph_R$. We expect the methods of~\cite{DeCommerSchraderShapiroVoigt2025} apply also for $0<Q<2$, leading to the following claim:
\begin{claim}\label{clm:GbLCQG}
Let $Q = b+b^{-1}> 0$. Then $(M_0,\Delta_0)$ is a unimodular locally compact quantum group. Its Haar weight $\ph_0$ (conveniently normalized) satisfies
\be
\ph_0(\pi_0(F)) = F(0,-iQ,0)\,,\qquad F\in \sA\,.
\ee
\end{claim}
Let us check that this formula is consistent with $\ph_0$ being an invariant weight on $M_0$.
Positivity can be proved directly from~\eqref{eq:Aproduct} and~\eqref{eq:Astar}
\ba
\ph_0(\pi_0(F^*\cdot F)) &= 
 \int\limits_{\rr^3}e^{-\pi Q (s+u)}  \overline{F(\bar{s},\bar{t}-i Q,\bar{u})} F(s,t,u) dsdtdu\\
&= 
 \int\limits_{\rr^3}e^{-\pi Q (s+u)} |F(s,t-i\tfrac{Q}{2},u)|^2 dsdtdu\geq 0\,.
 \label{eq:ph0FstarF}
\ea
The second equality follows by shifting the $t$ integration contour by $-iQ/2$. Note that $\ph_0(\pi_0(F^*\cdot F)) = 0$ for $F\in\sA$ only if $F\equiv 0$, which is consistent with $\ph_0$ being faithful. Note also that $\ph_0(\pi_0(F))$ is finite for all $F\in\sA$. Thus, if $\ph_0$ exists, it is semifinite.

Consistency with left and right invariance can be checked using~\eqref{eq:DeltaF}. The main ingredient is the following distributional identity for a boundary value of the quantum binomial coefficient:
\be
\lim_{\epsilon_1,\epsilon_2\searrow 0} B_b(x-i\epsilon_1, -i Q+i\epsilon_1+i\epsilon_2) = \delta(x)\quad\text{for}\quad x\in \rr\,.
\ee
We then get from~\eqref{eq:DeltaF}, for any $F\in \sA$\footnote{To properly define the left-hand side of these identities, we need to deform the $z$ contour in~\eqref{eq:DeltaF} so that no poles are crossed as $t_2$ (respectively $t_1$) is taken from the real axis to $-i Q$.}
\ba
\lim_{\epsilon\searrow 0}(D_b F)(s,t,u;0,-i Q+i \epsilon,0) &= F(0,-i Q,0) \delta(s)\delta(t)\delta(u)\,,\\
\lim_{\epsilon\searrow 0}(D_b F)(0,-i Q+i \epsilon,0;s,t,u) &= F(0,-i Q,0) \delta(s)\delta(t)\delta(u)\,.
\ea
These identities are precisely left invariance: $(\id\otimes \ph_0)\Delta_0(\pi_0(F)) = \ph_0(\pi_0(F)) 1_{M_0}$, and right invariance: $(\ph_0 \otimes \id)\Delta_0(\pi_0(F)) = \ph_0(\pi_0(F)) 1_{M_0}$. Indeed, the multiplicative unit $1_{M_0}$ corresponds to the distribution $F(s,t,u) = \delta(s)\delta(t)\delta(u)$. 

We take the above checks as strong evidence that $\gg_b$ is a locally compact quantum group for all $Q>0$. However, the question remains open at the rigorous level.

\subsection{GNS representation}
We are ready to construct the GNS representation of $(M_0,\ph_0)$.
\begin{definition}
Let $\cK = L^2(\rr^3)$. Define a $*$-representation $\pi:\sA\rightarrow \mathrm{B}(\cK)$ by
\be
(\pi(F) f)(x) := \int\limits_{\rr^3}e^{i\pi \,\Omega(x,x')} F(x-x') f(x') dx'\,,\qquad F\in \sA\,,\, f\in \cK\,.
\label{eq:PiA}
\ee
Here $\Omega$ is the skew-symmetric bilinear form defined in~\eqref{eq:OmegaSkew}. Furthermore, define the linear bijection $T:\sA\rightarrow\sA$
\be
(T F)(s,t,u) := e^{-\frac{\pi Q}{2}(s+u)} F(s,t-i\tfrac{Q}{2},u)\,,\qquad F\in\sA\,.
\ee
\end{definition}
$\sA = T(\sA)$ is a dense subspace of $\cK$ under the natural embedding $S^{1/2}_{1/2}(\rr^3)\hookrightarrow L^2(\rr^3)$.
\begin{lemma}\label{lem:piT}
We have
\be
\pi(F_1) TF_2 = T(F_1\cdot F_2)\,,\qquad F_1,F_2\in\sA\,,
\label{eq:piF1TF2}
\ee
and
\be
\ph_0(\pi_0(F^*_1\cdot F_2)) = \langle T F_1, T F_2 \rangle\,,\qquad F_1,F_2\in \sA\,.
\label{eq:phi0pi0}
\ee
Here $\langle\cdot,\cdot\rangle$ denotes the inner product on $\cK$. 
\end{lemma}
\begin{proof}
\eqref{eq:piF1TF2} follows directly from the definition of the product~\eqref{eq:Aproduct}, and the definitions of $\pi$ and $T$.~\eqref{eq:phi0pi0} is the same as~\eqref{eq:ph0FstarF}.
\end{proof}

\begin{claim}
Let $Q=b+b^{-1}>0$. There exists a unique faithful normal unital $*$-representation $\iota: M_0\rightarrow \mathrm{B}(\cK)$ satisfying
\be
\iota(\pi_0(F)) = \pi(F)\,,\qquad F\in \sA\,.
\ee
Furthermore, there exists a unique linear map $\Lambda_0: \mathfrak{n}_{\ph_0}\rightarrow \cK$ such that
\be
\Lambda_0 (\pi_0(F)) = T F\,,\qquad  F\in\sA
\ee
and such that $(\cK,\iota,\Lambda_0)$ is a GNS construction for $(M_0,\ph_0)$.
\end{claim}
Indeed, for $a,b\in \pi_0(\sA)$, the GNS conditions
\be
\iota(a)\Lambda_0(b) = \Lambda_0(a b)\,,\qquad \langle\Lambda_0(a),\Lambda_0(b)\rangle = \ph_0(a^*b)
\ee
are equivalent to the two parts of Lemma~\ref{lem:piT}. Furthermore, $\Lambda_0(\pi_0(\sA)) = T\sA = \sA$ is dense in $\cK$. For a rigorous proof, this algebraic argument should be supplemented by appropriate analytic reasoning.

We can now finally define the GNS realization of $\gg_b$:
\begin{definition}
Define the von Neumann algebra $M \subset \mathrm{B}(\cK)$ as
\be
M:= \iota(M_0) = \pi(\sA)'' \subset \mathrm{B}(\cK)\,.
\ee
In particular, $\pi:\sA\rightarrow M$ is an injective $*$-algebra homomorphism whose image is dense in $M$. Transport the coproduct and Haar weight from $M_0$ to $M$ by setting
\ba
\Delta &:=(\iota\otimes\iota)\, \Delta_0\, \iota^{-1}\,:\quad M\rightarrow M\botimes M\,,\\
\ph &:= \ph_0\, \iota^{-1}\,: \quad M_+ \rightarrow [0,\infty]\,.
\ea
\end{definition}
$(M,\Delta,\ph)$ provides the GNS realization of $\gg_b$. The corresponding GNS map is
\be
\Lambda:= \Lambda_0 \, \iota^{-1}: \mathfrak{n}_{\ph}\rightarrow \cK\,.
\ee
In particular
\be
\ph(\pi(F)) = F(0,-iQ,0)\,,\qquad F\in\sA
\ee
and
\be
\Lambda(\pi(F))(s,t,u) = e^{-\frac{\pi Q}{2}(s+u)} F(s,t-i\tfrac{Q}{2},u) \,,\qquad F\in\sA\,.
\label{eq:GNSGb}
\ee
In the spirit of the discussion following Lemma~\ref{lem:pi0}, we can parametrize elements of $M$ as $\pi(F)$, with the domain of $\pi$ extended to all tempered distributions $\mathcal{S}'(\rr^3)$ for which $\pi(F)$ is a bounded operator.

\subsection{Multiplicative unitaries}\label{ssec:WVGb}
Let us determine the left multiplicative unitary $\ww$ on $\cK\otimes\cK$. Recall from Section~\ref{ssec:W} that it is characterized by
\begin{equation}
\ww^*(\Lambda(x)\otimes\Lambda(y))
= (\Lambda\otimes\Lambda)(\Delta(y)(x\otimes 1))\,,\qquad x,y\in \mathfrak{n}_{\ph}.
\end{equation}
To state the result, let $\sbf,\tbf,\ubf$ denote multiplication by the corresponding coordinates on $\cK=L^2(\rr^3)$, and introduce the momentum operators
\begin{equation}
\pbf_s=\frac{1}{2\pi i}\frac{\partial}{\partial s}\,,\qquad
\pbf_t=\frac{1}{2\pi i}\frac{\partial}{\partial t}\,,\qquad
\pbf_u=\frac{1}{2\pi i}\frac{\partial}{\partial u}\,.
\end{equation}
All real linear combinations of position and momentum appearing below are understood as self-adjoint operators. It follows from~\eqref{eq:PiA} that the algebra $M$ is generated by
\ba
\mathbf{a}&=\frac{\tbf}{2}-\pbf_s\,,\\
\mathbf{b}&=\frac{\sbf+\ubf-\tbf}{2}+\pbf_s+\pbf_t\,,\\
\mathbf{c}&=\frac{\sbf+\ubf}{2}-\tbf+\pbf_s+\pbf_t+\pbf_u\,.
\label{eq:abc}
\ea
\begin{claim}
Let $Q = b+b^{-1}>0$. Then
\ba
\ww={}&
\phi_b\!\left(\mathbf{a}_1+\pbf_{s,2}+\frac{\tbf_2}{2}\right)^{*}
\phi_b\!\left(\mathbf{a}_1+\pbf_{s,2}+\sbf_2-\frac{\tbf_2}{2}\right)^{*}\\
&\times\phi_b\!\left(\mathbf{b}_1-\pbf_{s,2}-\frac{\tbf_2}{2}\right)
\phi_b\!\left(\mathbf{b}_1-\pbf_{s,2}-\pbf_{t,2}+\frac{\sbf_2+\ubf_2-\tbf_2}{2}\right)\\
&\times\phi_b\!\left(\mathbf{b}_1-\pbf_{s,2}-\pbf_{t,2}+\frac{\sbf_2-\ubf_2+\tbf_2}{2}\right)
\phi_b\!\left(\mathbf{b}_1-\pbf_{s,2}+\sbf_2+\frac{\tbf_2}{2}\right)
e^{2\pi i \mathbf{c}_1\sbf_2}\,.
\label{eq:Wfactorized}
\ea
Subscripts $1,2$ indicate the tensor factor of $\cK\otimes \cK$ on which an operator acts.
\end{claim}
\eqref{eq:Wfactorized} was obtained by substituting the formulas for $\Delta$ and $\Lambda$ into the defining relation for $\ww$, and evaluating the resulting quantum-dilogarithm integrals. The factors are ordered as displayed. Since $|\phi_b(x)|=1$ for $x\in\rr$ throughout our range of $b$, this expression makes the unitarity of $\ww$ manifest. It also manifests the symmetry $b\leftrightarrow b^{-1}$ and the fact that $\ww\in M\botimes \mathrm{B}(\cK)$. The pentagon identity~\eqref{eq:Wpentagon} for $\ww$ follows from the pentagon identity for $\phi_b$:
\be
\phi_b(\mathbf{p})\phi_b(\mathbf{x})=\phi_b(\mathbf{x})\phi_b(\mathbf{x}+\mathbf{p})\phi_b(\mathbf{p})\,,
\qquad [\mathbf{p},\mathbf{x}]=\frac{1}{2\pi i}\,.
\label{eq:phiPentagon}
\ee

Recall from Section~\ref{ssec:V} the right regular multiplicative unitary $\vv$. In the case of $\gg_b$, $\vv$ has a simple relation to $\ww$:
\begin{claim}
Let $P:\cK\to\cK$ exchange the first and last coordinates,
\be
(P f)(s,t,u)=f(u,t,s)\,,
\label{eq:reflectionSU}
\ee
and let $\Sigma$ exchange the two tensor factors of $\cK\otimes\cK$. Then
\be
\vv=\Sigma(P\otimes P)\ww^*(P\otimes P)\Sigma\,.
\label{eq:VWreflection}
\ee
\end{claim}

\subsection{Dual von Neumann algebra}
We can use $\ww$ to describe the dual von Neumann algebra $\hat M=L^{\infty}(\widehat{\sl})$. As explained in Section~\ref{ssec:W}, $\hat{M}$ generalizes the group algebra. Recall that $\hat{M}$ is generated by the second leg of $\ww$ as
\be
\hat M=\{(\omega\otimes\id)\ww:\omega\in M_*\}''\subset \mathrm{B}(\cK)\,.
\label{eq:MHatFromW2}
\ee
To extract the second-leg algebra from~\eqref{eq:Wfactorized}, first introduce two unitaries on $\cK$:
\be
G_e=\phi_b(\tbf-\sbf)\,,\qquad
G_f=\phi_b(\tbf-\ubf)
\phi_b\!\left(\frac{\sbf+\ubf}{2}-\pbf_t\right)
\phi_b(\sbf+\tbf)\,.
\label{eq:GEGF}
\ee
Define the self-adjoint operators $\kbf$, $\ebf$, $\fbf$ on $\cK$ by
\be
\kbf=\frac{\sbf}{2}\,,\qquad G^*_e\,\mathbf{e}\,G_e=\frac{\sbf-\tbf}{2}+\pbf_s\,,\qquad
G^*_f\,\mathbf{f}\,G_f =-\frac{\sbf+\tbf}{2}-\pbf_s\,.
\label{eq:HEHF}
\ee
\begin{claim}
The multiplicative unitary~\eqref{eq:Wfactorized} can be rewritten as
\be
\ww=
\phi_b\Big(\mathbf{a}_1+\kbf_2+\mathbf{e}_2\Big)^{*}
\phi_b\Big(\mathbf{b}_1+\kbf_2+\mathbf{f}_2\Big)
e^{4\pi i \mathbf{c}_1\kbf_2}\,.
\label{eq:WdualGenerators}
\ee
\end{claim}
The sums in the arguments are understood with their self-adjoint closures, which follow from~\eqref{eq:HEHF} and the commutation of $G_e,G_f$ with $\kbf$. The result follows by combining the first two and the last four quantum dilogarithms in~\eqref{eq:Wfactorized} by repeated use of the pentagon identity~\eqref{eq:phiPentagon}. 

\begin{theorem}
For all $Q>0$, the dual von Neumann algebra $\hat{M}$ is generated by $\kbf$, $\ebf$, $\fbf$
\be
\hat M=\{e^{i z \kbf},e^{i z \ebf },e^{i z \fbf}\,:\, z\in\rr\}'' \subset \mathrm{B}(\cK)\,.
\label{eq:MhatLogarithmic}
\ee
\end{theorem}
\begin{proof}[Proof sketch]
The inclusion $\hat{M}\subset \{e^{i z \kbf},e^{i z \ebf },e^{iz\fbf}\,:\, z\in\rr\}''$ follows from~\eqref{eq:MHatFromW2} and~\eqref{eq:WdualGenerators}. To see the converse inclusion, we translate the first-leg coordinates so that $\mathbf{a}$ and $\mathbf{b}$ tend simultaneously to $-\infty$ while $\mathbf{c}$ stays fixed. Since $\phi_b(x)\to1$ as $x\to-\infty$, we recover the second-leg algebra generated by $\kbf$. Removing this factor and repeating the same steps recovers the algebra generated by $\fbf$, and finally also $\ebf$.
\end{proof}

\subsection{Modular double from $\hat{\sl}$}\label{ssec:modularDouble}
Suppose now $b>0$, and introduce the following positive unbounded self-adjoint operators on $\cK$
\be
K=\exp(2\pi b\,\kbf)\,,\qquad E=\exp(2\pi b \,\ebf)\,,\qquad F=\exp(2 \pi b \,\fbf)\,.
\label{eq:KEFspectral}
\ee
$\hat M$ is generated by $K,E,F$
\be
\hat M=
\{K^{-\tfrac{iz}{b}},E^{-\tfrac{iz}{b}},F^{-\tfrac{iz}{b}}:z \in\rr\}''\,.
\label{eq:MhatKEF}
\ee
We have
\ba
K^{-\tfrac{iz}{b}}&=\exp(- i \pi z\,\sbf)\,,\\
E^{-\tfrac{iz}{b}}&=G_e \exp[-i\pi z(\sbf-\tbf+2\pbf_s)]G_e^*\,,\\
F^{-\tfrac{iz}{b}}&=G_f \exp[i\pi z(\sbf+\tbf+2\pbf_s)]G_f^*\,.
\ea
The functional equations of $\phi_b$ give explicit expressions for $K,E,F$ in terms of finite difference operators
\ba
K=&\exp(\pi b\sbf)\,,\\
 E=&\exp[\pi b(\sbf-\tbf+2\pbf_s)]
+\exp[\pi b(-\sbf+\tbf+2\pbf_s)]\,,\\
F=&\exp[\pi b(\sbf+\tbf-2\pbf_s)]
+\exp[\pi b(-\sbf-\tbf-2\pbf_s)]\\
&+\exp[\pi b(\tbf-\ubf-2\pbf_s-2\pbf_t)]
+\exp[\pi b(-\tbf+\ubf-2\pbf_s-2\pbf_t)]\,.
\label{eq:LeftKEF}
\ea
Here the sums are positive form sums, as in the quantum binomial theorem.
In fact, the unbounded operators $K,E,F$ are none other than the generators of Faddeev's modular double~\cite{Faddeev1999ModularDouble}; see also~\cite{PonsotTeschner2001,BytskoTeschner2003,NidaievTeschner2013}. Indeed, the explicit expressions~\eqref{eq:LeftKEF} and canonical commutation relations imply
\be
K E=q E K\,,\qquad K F=q^{-1}FK\,,\qquad
EF-FE=(q^{-1}-q)(K^2-K^{-2})\,,
\label{eq:LeftQuantumGroupRelations}
\ee
where $q=e^{i\pi b^2}$. These identities hold on the dense subspace $\sA \subset \cK$. They are the relations of the quantized enveloping algebra $\cU_q(\mathfrak{sl}(2,\rr))$ in a normalization adapted to positive generators.

Just as was the case for the coordinates $A,B,C,D$, another set of generators $\check{K},\check{E},\check{F}$ is obtained by replacing $b$ with $b^{-1}$. Since $\phi_b=\phi_{1/b}$, the unitaries $G_e,G_f$ and the self-adjoint operators $\kbf,\ebf,\fbf$ are unchanged by this replacement. Consequently,
\be
\check{K}=e^{2\pi b^{-1}\kbf} \,,\qquad \check{E}=e^{2\pi b^{-1} \ebf}\,,\qquad \check{F}=e^{2 \pi b^{-1} \fbf}\,.
\label{eq:CheckedKEF}
\ee
Furthermore, we have
\be
\check{K}= K^{1/b^2}\,,\qquad \check{E}=E^{1/b^2}\,,\qquad \check{F}=F^{1/b^2}\,.
\ee
The new generators satisfy the commutation relations
\be
\check{K} \check{E}=\check{q} \check{E} \check{K}\,,\qquad \check{K} \check{F}=\check{q}^{-1}\check{F}\check{K}\,,\qquad
\check{E}\check{F}-\check{F}\check{E}=(\check{q}^{-1}-\check{q})(\check{K}^2-\check{K}^{-2})\,,
\ee
with $\check{q}=e^{i\pi b^{-2}}$.
The simultaneous appearance of these two quantized enveloping algebras is Faddeev's modular double $\cU_{q\check q}(\mathfrak{sl}(2,\rr))$. This is the sense in which the modular double can be thought of as a generalization of the Lie algebra of $\sl$.

Finally, let us consider $0<Q<2$, where $b=e^{i\theta}$ and $\bar b=b^{-1}$. Operators $G_e,G_f$ remain unitary and operators $\kbf,\ebf,\fbf$ remain self-adjoint. Equations~\eqref{eq:KEFspectral} and~\eqref{eq:CheckedKEF} now define normal operators, and give
\be
K^*=\check K\,,\qquad
E^*=\check E\,,\qquad
F^*=\check F\,.
\label{eq:ModularDoubleStar}
\ee
Thus adjunction exchanges the two halves of the modular double. This is compatible with $\bar q=\check q^{-1}$. The same change of real structure occurs for the coordinate generators $A,B,C,D$ discussed above.

The passage from the modular double $\cU_{q\check q}(\mathfrak{sl}(2,\rr))$ to the locally compact quantum group $\sl$ clarifies various subtle analytic points. For example, the meaning of the modular duality $b\leftrightarrow b^{-1}$ is transparent: it is simply the manifest equality $\sl=\mathrm{SL}^+_{1/b}(2,\rr)$. The advantages become even more evident when we consider unitary representations. It is essential for the applications to conformal field theory that the category of representations includes the trivial representation. Indeed, the latter corresponds to the vacuum module of the Virasoro algebra. The positive representations of $\cU_{q\check q}(\mathfrak{sl}(2,\rr))$ discussed in \cite{PonsotTeschner1999} do not include the trivial representation, so this class must be enlarged for our purposes. The locally compact quantum group formulation provides a natural framework for doing so: its category of unitary corepresentations automatically contains the trivial corepresentation $1_M$ on $\cc$.

\subsection{Dual coproduct}
Recall from Section~\ref{ssec:W} that the coproduct of the dual quantum group $\hat{\Delta} : \hat{M} \rightarrow \hat{M}\botimes\hat{M}$ is obtained as
\be
\hat{\Delta}(x) = \hat{\ww}^*(1\otimes x)\hat{\ww}\,,\qquad x\in \hat{M},
\ee 
where $\hat{\ww} = \Sigma \ww^* \Sigma$ is the dual multiplicative unitary. The dual coproduct of the normal generators $K$, $E$, $F$ of $\hat{M}$, defined in~\eqref{eq:KEFspectral}, takes a particularly simple form.
\begin{claim}\label{clm:DeltaHatKEF}
Let $Q>0$. We have
\ba
\hat{\Delta}(K) &= K\otimes K\,,\\
\hat{\Delta}(E) &= E\otimes K + K^{-1}\otimes E\,,\\
\hat{\Delta}(F) &= F\otimes K + K^{-1}\otimes F\,.
\ea
\end{claim}
This exactly agrees with the coproduct of the modular double generators~\cite{BytskoTeschner2003}. The dual coproduct of $\check{K}$, $\check{E}$, $\check{F}$ takes the same form with $\check{}$ above every letter. One can also find formulas for the dual coproduct of the generating unitaries $e^{i z \kbf},e^{i z \ebf },e^{iz\fbf}$ but they will not be needed in the following.

\subsection{Modular operators of $\sl$ and $\hat{\sl}$}
Let us record the operators appearing in the modular theory of $\gg_b$ and $\hat{\gg}_b$. Recall from Section~\ref{ssec:modularGeneral} the definition of the Tomita operator $S$, and its associated modular conjugation $J$ and modular operator $\nabla$. Their action can be computed directly from the GNS map~\eqref{eq:GNSGb}, with the following result.
\begin{claim}[Modular operators of $\gg_b$]\label{eq:GbModular}
On their respective domains in $\cK \cong L^2(\rr^3_{s,t,u})$, the operators $S$, $J$, $\nabla$ for $\gg_b$ take the form
\ba
(S f)(s,t,u) &= e^{-\pi Q(s+u)} \overline{f(-s,-t,-u)}\,,\\
(J f)(s,t,u) &=  \overline{f(-s,-t,-u)}\,,\\
(\nabla f)(s,t,u) &=  e^{2\pi Q(s+u)}f(s,t,u)\,.
\ea
\end{claim}
$S$ and $\nabla$ are unbounded, while $J$ is an antiunitary involution.

The modular operators $\hat{S}$, $\hat{J}$, $\hat{\nabla}$ of the dual quantum group $\hat{\gg}_b$ make use of the quantum binomial coefficient $B_b(x,y)$ from Definition~\ref{def:QBC}.
\begin{claim}[Modular operators of $\hat{\gg}_b$]
On their respective domains in $\cK \cong L^2(\rr^3_{s,t,u})$, the operators $\hat{S}$, $\hat{J}$, $\hat{\nabla}$ take the form
\ba
(\hat{S} f)(s,t,u) &=  e^{\pi  Q (u-s)} \int\limits_{\rr}
B_b(t',-s-t'-u)\overline{f(u,-t-t',s)}\,dt'\,,\\
(\hat{J} f)(s,t,u) &= \int\limits_{\rr}
B_b(t',-s-t'-u)\overline{f(u,-t-t',s)}\,dt'\,,\\
(\hat{\nabla} f)(s,t,u) &=  e^{2\pi Q(s-u)}f(s,t,u)\,.
\ea
\end{claim}
Antiunitarity of $\hat{J}$ can be seen by Fourier transforming with respect to $t$, whereupon the integral over $t'$ turns into a multiplication by a function of unit modulus. From $\hat{J}$, we can construct the unitary antipode $R: M\rightarrow M$ as $R(x) = \hat{J}x^*\hat{J}$. We can express its action on $\pi(\sA)$ as follows.
\begin{claim}[The unitary antipode of $\gg_b$]\label{clm:UnitaryAntipode}
Let $F\in \sA$. Then $R(\pi(F)) = \pi(r(F))$, where $r: \sA \rightarrow \cS(\rr^3)$ is given by
\be
(r F)(s,t,u) =  \int\limits_{\rr}
B_b(t',-s-t'-u)F(-u,t+t',-s)\,dt'\,.
\ee
\end{claim}
Finally, let us record the action of the modular automorphism group and the scaling group of $\gg_b$. The former is defined as
$\sigma_\alpha: M\rightarrow M$, $\sigma_\alpha(x) = \nabla^{i\alpha} x \nabla^{-i\alpha}$. By a small abuse of notation, we will use the same symbol for its action on $\sA$ induced by $\pi$, i.e.~$\sigma_\alpha(\pi(F)) = \pi(\sigma_\alpha(F))$. Similarly, write the action of the scaling group $\tau_\alpha: M\rightarrow M$, $\tau_\alpha(x) = \hat{\nabla}^{i\alpha} x \hat{\nabla}^{-i\alpha}$ on $F\in\sA$ as $\tau_\alpha(\pi(F)) = \pi(\tau_\alpha(F))$.
\begin{claim}[Modular and scaling groups of $\gg_b$]\label{clm:sigmaTau}
Let $F\in \sA$. Then
\be
(\sigma_\alpha(F))(s,t,u)  = e^{2 \pi  i Q \alpha (s+u)} F(s,t,u)\,,
\ee
and
\be
(\tau_\alpha(F))(s,t,u) = e^{2 \pi  i Q \alpha (s-u)} F(s,t,u)\,.
\ee
\end{claim}
$\sA$ is preserved by these actions even if we take $\alpha\in \cc$.

\section{Unitary representations of $\sl$}\label{sec:reps}
In this section, we discuss unitary representations of $\sl$. Recall from Section~\ref{ssec:reps} that a unitary representation of a locally compact quantum group $\gg$ is defined as unitary corepresentation of the von Neumann bialgebra $(L^{\infty}(\gg),\Delta_{\gg})$. These objects form a well-behaved category closed under tensor products and containing the trivial representation.

We will construct a family of unitary irreducible representations of $\sl$. This family generalizes the positive principal series of the modular double, constructed in~\cite{PonsotTeschner1999} and further discussed in~\cite{PonsotTeschner2001,BytskoTeschner2003,FrenkelIp2014}. Our construction supplements it with the complementary series. Together with the trivial representation, the principal and complementary series are in one-to-one correspondence with unitary irreducible lowest-weight modules of the Virasoro algebra at central charge $c=1+6Q^2$. Based on this evidence, we conjecture that they provide a complete list of unitary irreducible representations of $\sl$.

\subsection{Unitary irreducible representations}\label{ssec:unitaryIrreps}
A unitary representation of $\gg_b$ on a separable Hilbert space $\cH$ is a unitary operator $\uu \in M\botimes \mathrm{B}(\cH)$ such that $(\Delta\otimes \id)\uu = \uu_{13}\uu_{23}$ holds in the von Neumann algebra $M\botimes M\botimes \mathrm{B}(\cH)$. Recall also the notions of intertwiners and irreducibility from Definition~\ref{def:intertwinerIrr}.

The simplest unitary irreducible representation of $\gg_b$ is the trivial one, with $\cH = \cc$ and norm given by $\|z\|^2 = |z|^2$. Its underlying unitary is $\uu = 1_M$.

The principal and complementary series are infinite-dimensional representations realized as completions of spaces of functions on the real line. The representations will be labelled by $\nu$, with $\nu\in\rr$ corresponding to the principal series and $\nu \in i(-Q/2,Q/2)$ to the complementary series. We use the following convention for the Fourier transform
\be
\hat{f}(p) = \int\limits_{\rr} e^{-2\pi i p x} f(x) dx\,,\qquad f(x) = \int\limits_{\rr} e^{2\pi i p x} \hat{f}(p) dp\,.
\ee
\begin{definition}[Inner products]
Let $Q=b+b^{-1}>0$, and let $\nu \in \cc$ with $|\mathrm{Im}(\nu)|<Q/2$. Define the following family of inner products $\langle\cdot,\cdot\rangle_{\nu}$ on functions on the real line
\be
\langle f_1,f_2 \rangle_{\nu} := \int\limits_{\rr} |s_b(p+\nu)|^2 \,\overline{\hat{f}_1(p)} \hat{f}_2(p) dp \label{eq:IPcomplementary}
\ee
with $s_b$ defined in~\eqref{eq:sbDefinition}. Let $\cH_{\nu}$ be the Hilbert space completion of
\be
\{f\in L^2(\rr):\, \|f\|_{\nu}<\infty\}
\ee
with respect to $\|\cdot \|_{\nu}$.
\end{definition}
For $\nu\in \rr$, we have $|s_b(p+\nu)|^2 = 1$, and thus $\cH_{\nu} = L^2(\rr)$ in that case. For general $\nu$ such that $|\mathrm{Im}(\nu)|<Q/2$, let us define
\be
V_{\nu}: \cH_{\nu}\rightarrow L^2(\rr)\,,\qquad \hat{V_{\nu} f}(p) := s_b(p+\nu) \hat{f}(p)\,.
\label{eq:Vnudef}
\ee
$V_{\nu}$ is a unitary isomorphism by construction.

Recall the realization of $M$ on $L^2(\gg_b)\cong L^2(\rr^3)$. We will construct the principal and complementary series representations as unitary operators on $L^2(\gg_b)\otimes \cH_{\nu}$. Recall from~\eqref{eq:abc} the unbounded self-adjoint operators $\mathbf{a},\mathbf{b},\mathbf{c}$ on $L^2(\gg_b)$. They are affiliated with $M$ and generated it. Let $\xbf$ be the standard coordinate multiplication operator on $L^2(\rr)$ and let $\pbf$ be its Fourier dual.
\begin{lemma}\label{lem:Theta}
Define
\be
\Theta_\nu(p):= s_b(p+\nu)s_b(p-\nu)\,.
\ee
$\Theta_\nu(\pbf)$ is a unitary operator on $L^2(\rr)$ for all $\nu \in \rr\cup i(-\frac{Q}{2},\frac{Q}{2})$.
\end{lemma}
\begin{proof}
Suppose $p\in\rr$. It follows from~\eqref{eq:sIdentities} that for all $|\mathrm{Im}(\nu)|<Q/2$
\be
\overline{\Theta_\nu(p)} = \frac{1}{s_b(p+\overline{\nu})s_b(p-\overline{\nu})}\,.
\ee
Thus $\overline{\Theta_\nu(p)} = \Theta_\nu(p)^{-1}$ as long as $\nu \in \rr\cup i(-\frac{Q}{2},\frac{Q}{2})$.
\end{proof}
\begin{definition}
Let $Q=b+b^{-1}>0$ and let $\nu \in \rr\cup i(-Q/2,Q/2)$. Define $\pp_{\nu}\in M\botimes \mathrm{B}(\cH_{\nu})$ by
\be
(1_M\otimes V_{\nu})\pp_{\nu}(1_M\otimes V^*_{\nu})
=
\phi_b\Big(\mathbf{a}-\xbf+\frac{\pbf}{2}\Big)^{*}
\Theta_{\nu}(\pbf)
\phi_b\Big(\mathbf{b}+\xbf+\frac{\pbf}{2}\Big)
\Theta_{\nu}(\pbf)^*
e^{2\pi i\,\mathbf{c}\, \mathbf{p}}\,.
\label{eq:Pnu}
\ee
\end{definition}
Since $\mathbf{a},\mathbf{b},\mathbf{c},\mathbf{x},\mathbf{p}$ are self-adjoint operators on $L^2(\rr^4_{s,t,u,x})$, the right-hand side of~\eqref{eq:Pnu} is a unitary element of $M\botimes \mathrm{B}(L^2(\rr))$ by Lemma~\ref{lem:Theta}. Unitarity of $\pp_{\nu}$ and its belonging to $M\botimes \mathrm{B}(\cH_{\nu})$ then follow by construction of $V_{\nu}$ in~\eqref{eq:Vnudef}.

\begin{claim}\label{clm:PnuRep}
Let $Q=b+b^{-1}>0$ and $\nu \in \rr\cup i (-Q/2,Q/2)$. Then $\pp_{\nu}$ defines a unitary irreducible representation of $\gg_b$ on $\cH_\nu$. In particular, $\pp_{\nu}$ satisfies
\be
(\Delta\otimes\id)\pp_{\nu} = (\pp_{\nu})_{13}(\pp_{\nu})_{23}\,.
\ee
\end{claim}

We will refer to the unitary representation $(\pp_{\nu},\cH_{\nu})$ with $\nu \in \rr$ as the principal series and with $\nu \in i(-Q/2,Q/2)$ as the complementary series.

For calculations, it is convenient to represent $\pp_{\nu}$ from~\eqref{eq:Pnu} as a distributional kernel.
\begin{claim}\label{clm:PnuKernel}
Let $Q=b+b^{-1}>0$ and  $\nu \in \rr\cup i (-Q/2,Q/2)$. In the coefficient realization $\pi$ of $M$, the unitary $\pp_{\nu}$ has the distributional kernel
\be
\mathscr{P}_\nu(s,t,u;x_1,x_2)
= e^{2 \pi i (u x_2-s x_1)}e^{\frac{\pi}{2}Q(u-s)}
\frac{s_b(s-\nu )s_b(u+\nu)s_b(t-u+i\tfrac{Q}{2}-i0^+)}{s_b(t-\nu) s_b(t+\nu) s_b(s-t-i\tfrac{Q}{2}+i0^+)}\,.
\ee
More explicitly, write $z=(s,t,u)$ and let $U_z = \pi(\delta_z)$ be the corresponding Weyl unitary. Then
\be
[\pp_{\nu}(\xi \otimes f)](x_1) = \int\limits_{\rr^3}\int\limits_{\rr} \mathscr{P}_\nu(z;x_1,x_2) (U_z \xi) f(x_2) dx_2 dz\,,\qquad \xi\in \cH\,,\; f\in \cH_{\nu}\,.
\ee
\end{claim}

Let us define the family of unitary isomorphisms $\cI_{\nu}:\cH_{\nu}\rightarrow \cH_{-\nu}$ as the compositions $\cI_{\nu}:= V^*_{-\nu}V_{\nu}$. We have
\be
\cI_{\nu}: \cH_{\nu}\rightarrow \cH_{-\nu}\,,\qquad \hat{\cI_{\nu}(f)}(p) = \frac{s_b (p+\nu)}{s_b (p-\nu )} \hat{f}(p)\,.
\label{eq:ISignFlip}
\ee
Note that we have $\cI^*_{\nu} = \cI_{-\nu}$.
\begin{claim}
Representations $\pp_{\nu}$ and $\pp_{-\nu}$ are unitarily equivalent. $\cI_{\nu}$ provides the corresponding intertwining operator:
\be
(1_M \otimes \cI_{\nu}) \pp_{\nu} = \pp_{-\nu}(1_{M}\otimes \cI_{\nu})\,.
\ee
\end{claim}

The inequivalent unitary irreducible representations constructed so far are in one-to-one correspondence with unitary irreducible lowest-weight modules of the Virasoro algebra at central charge $c=1+6Q^2$. The trivial representation maps to the Virasoro vacuum module, and the representation $\pp_{\nu}$ maps to the Virasoro module with lowest weight
\be
h = \frac{Q^2}{4} + \nu^2 \in (0,\infty)\,.
\ee
As we will discuss later, this correspondence extends to an equivalence of appropriate monoidal structures. The correspondence motivates the following conjecture:
\begin{conjecture}\label{conj:unitaryDual}
The complete list of inequivalent unitary irreducible representations of $\sl$ consists of:
\begin{enumerate}
\item The trivial representation $(1_M,\cc)$.
\item The complementary series $(\pp_{\nu},\cH_{\nu})$, with $\nu \in i(0,\frac{Q}{2})$.
\item The principal series $(\pp_{\nu},\cH_{\nu})$, with $\nu \in \rr_{\geq 0}$.
\end{enumerate}
\end{conjecture}

\subsection{The trivial representation as a limit of the complementary series}\label{ssec:CfromP}
Consider the unitary lowest-weight module of the Virasoro algebra at central charge $c>1$ and conformal weight $h>0$. As we take $h\rightarrow 0^{+}$, the module develops a null submodule, generated by $L_{-1}v$, where $v$ the lowest-weight vector. Quotienting by this submodule recovers the unitary $h=0$ module.

This mechanism has an analogue in the language of the unitary irreducible representations $\pp_{\nu}$ of $\gg_b$. The limit $h\rightarrow 0^+$ corresponds to $\nu \rightarrow i Q/2$ from below. To study the inner product of the complementary series~\eqref{eq:IPcomplementary} in this limit, let us first rewrite it in position space.
\begin{claim}\label{lem:IPComp}
For $\nu \in i(0,Q/2)$, the inner product~\eqref{eq:IPcomplementary} takes the following form:
\be
\langle f_1,f_2 \rangle_{\nu} = \int\limits_{\rr^2} \frac{s_b(x_1-x_2-\nu+ i\tfrac{Q}{2})}{s_b(i\tfrac{Q}{2}-2\nu)s_b(x_1-x_2+\nu - i\tfrac{Q}{2})}\overline{f_1(x_1)} f_2(x_2)dx_1 dx_2\,.
\ee
\end{claim}
Setting $\nu = i(\frac{Q}{2}-\epsilon)$, the two $x$-dependent dilogarithms terms cancel in the $\epsilon\rightarrow 0^+$ limit, with the remaining dilogarithm producing a pole. In fact, for any fixed $f\in \cS(\rr)$
\be
\|f\|^2_{i\frac{Q}{2}-i\epsilon}  = \frac{1}{4\pi\epsilon}\Big|\int\limits_{\rr} f(x)dx\Big|^2+ O(1)\quad\text{as}\quad \epsilon\rightarrow 0^+\,.
\label{eq:NormId}
\ee
We see that at the leading order at small $\epsilon$, the norm is only sensitive to the total integral of $f$. The null space consists of functions with $\int_\rr f (x) dx = 0$. Quotienting by this subspace yields the one-dimensional Hilbert space $\cc$. This quotient can be identified with the trivial representation $\uu = 1_M$ since
\be
\lim_{\nu\rightarrow i\frac{Q}{2}}\int\limits_{\rr} \mathscr{P}_{\nu}(s,t,u;x_1,x_2) dx_1 = \delta(s)\delta(t)\delta(u)\,.
\ee
The limit is taken through the complementary series and understood distributionally in $(s,t,u)$.

It will be convenient to incorporate the trivial representation uniformly into the notation $\pp_{\nu}$, $\cH_{\nu}$ as the special case $\nu = i \frac{Q}{2}$. Thus, let us define
\be
\cH_{i\frac{Q}{2}} := \cc\,,\qquad \pp_{i\frac{Q}{2}} := 1_M\,.
\ee

\subsection{Real structure and scaling group action}\label{ssec:JNablaP}
Recall Definitions~\ref{def:conjRep},~\ref{def:realStructure} of conjugate representation and real structure. The following result can be checked by appealing to the action of the unitary antipode of $\gg_b$, recorded in Claim~\ref{clm:UnitaryAntipode}.
\begin{claim}\label{clm:Jnu}
Let $Q>0$. The unitary irreducible representations $\pp_{\nu}$ of $\gg_b$ are real for all $\nu\in\rr\,\cup i (-Q/2,Q/2)$. Indeed, writing $C: L^2(\rr)\rightarrow L^2(\rr)$ for pointwise conjugation $(C f)(x) = \overline{f(x)}$ in position space, $\pp_{\nu}$ admits the real structure:
\be
J_{\nu} : \cH_{\nu}\rightarrow \cH_{\nu}\,,\qquad J_{\nu}:= V^*_{\nu} C V_{\nu}\,.
\ee
Explicitly, we have
\ba
\hat{(J_\nu f)}(p) &=
\frac{s_b(p-\nu)}{s_b(p+\nu)} \overline{\hat{f}(-p)} \quad &&\text{for}\quad \nu\in\rr\,,\\
\hat{(J_\nu f)}(p) &=\overline{\hat{f}(-p)} \quad&&\text{for}\quad \nu\in i(-Q/2,Q/2)\,.
\ea
In particular, $J_{\nu}$ is an antiunitary involution and
\be
(\hat{J}\otimes J_{\nu}) \pp_{\nu}(\hat{J}\otimes J_{\nu}) = \pp^*_{\nu}\,.
\label{eq:JPnu}
\ee
\end{claim}
This situation is closely analogous to the real structure of the principal and complementary series of $\mathrm{PSL}(2,\rr)$: complex conjugation in position space turns $\pp_{\nu}$ into $\pp_{-\overline{\nu}}$, and thus provides real structure for $\nu\in i(-Q/2,Q/2)$. For $\nu\in \rr$, we need to postcompose it with $\cI_{-\nu}$.

Let us incorporate the trivial representation $(\pp_{i\frac{Q}{2}},\cH_{i\frac{Q}{2}})$ into the notation by denoting its natural real structure as:
\be
J_{i\frac{Q}{2}}: \cc\rightarrow \cc\,,\qquad 
J_{i\frac{Q}{2}}(z) := \zb\,.
\ee
We will hereafter identify $\cH_{\nu}$ with $\overline{\cH_\nu}$ via the map $j J_{\nu}$, which lets us write $\pp^c_{\nu} = \pp_\nu$.

Let us recall the scaling group $\tau_\alpha: M\rightarrow M$, discussed in Section~\ref{ssec:modularGeneral}. For $\gg_b$, the action of $\tau_{\alpha}$ on $M$ is described in Claim~\ref{clm:sigmaTau}. $\tau_\alpha$ preserves the coproduct $\Delta\circ\tau_\alpha = (\tau_\alpha\otimes\tau_\alpha)\circ\Delta$. Thus if $\uu$ is a unitary representation, then also $\uu_{\alpha}:=(\tau_{\alpha}\otimes \id)\uu$ with $\alpha\in\rr$ is one. The following claim asserts that the scaling group acts on nontrivial $\pp_{\nu}$ as conjugation by a translation. It follows easily by combining Claim~\ref{clm:sigmaTau} and Claim~\ref{clm:PnuKernel}.
\begin{claim}\label{clm:tauNu}
Let $Q>0$ and $\nu\in \rr \cup i (-Q/2,Q/2)$. Define a positive self-adjoint operator on $\cH_{\nu}$ by the formula
\be
\nabla_{\nu} = e^{2\pi Q \pbf}\,.
\label{eq:nablaNu}
\ee
We have
\be
(\tau_\alpha \otimes \id)\pp_{\nu} = (1 \otimes \nabla_{\nu}^{-i \alpha}) \pp_{\nu} (1 \otimes \nabla_{\nu}^{i \alpha})\,, \qquad \alpha\in \rr\,.
\ee
Equivalently,
\be
(\hat{\nabla}^{i \alpha}\otimes  \nabla_{\nu}^{i \alpha})\pp_{\nu}(\hat{\nabla}^{-i \alpha}\otimes \nabla_{\nu}^{-i \alpha}) = \pp_{\nu}\,,\qquad \alpha\in \rr\,,
\label{eq:tauPnu}
\ee
where $\hat{\nabla}$ is the modular operator of $\hat{M}$.
\end{claim}
The scaling group action on the trivial representation is trivial: $\tau_{\alpha}(1_M) = 1_M$. Let us therefore define $\nabla_{i\frac{Q}{2}}$ to be the identity map
\be
\nabla_{i\frac{Q}{2}} : \cc \rightarrow \cc \,,\qquad \nabla_{i\frac{Q}{2}}(z) := z\,.
\ee
Note that for all $\nu \in \rr \cup i(-Q/2,Q/2]$, the operators $J_{\nu}$ and $\nabla_{\nu}$ satisfy the same compatibility as modular conjugation and modular operator
\be
J_{\nu}\nabla_{\nu}J_{\nu} = \nabla^{-1}_{\nu}\qquad\Leftrightarrow \qquad J_{\nu}\nabla^{i\alpha}_{\nu} = \nabla^{i\alpha}_{\nu}J_{\nu}\,,\quad \alpha\in\rr\,.
\ee

\subsection{Positive representations of the modular double}\label{ssec:posRepMD}
Let us explain the relationship between the unitary representations $\pp_{\nu}$ and the positive principal series representations of the modular double, considered in~\cite{PonsotTeschner1999,PonsotTeschner2001,BytskoTeschner2003,FrenkelIp2014}.
\begin{claim}\label{clm:kefnu}
Let $Q>0$, $\nu\in\rr\,\cup i (-Q/2,Q/2)$. Let $\xbf,\pbf$ be the canonical pair of self-adjoint operators on $L^2(\rr)$. Define the following self-adjoint operators on $\cH_{\nu}$:
\ba
\kbf_\nu &= \tfrac{1}{2}V^*_{\nu}\pbf V_{\nu}\,,\\
\ebf_\nu &= -V^*_{\nu}\xbf V_{\nu}\,,\\
\fbf_{\nu}&= V^*_{\nu}\Theta_{\nu}(\pbf) \xbf \Theta_{\nu}(\pbf)^* V_{\nu}\,.
\ea
$\pp_{\nu}$ then admits the equivalent presentation:
\be
\pp_{\nu} =
\phi_b\Big(\mathbf{a}+\mathbf{k}_\nu+\mathbf{e}_\nu\Big)^{*}
\phi_b\Big(\mathbf{b}+\mathbf{k}_\nu+\mathbf{f}_\nu\Big)
e^{4\pi i\, \mathbf{c}\, \kbf_\nu}\,.
\label{eq:Pkef}
\ee
\end{claim}
The point of this rewriting is that $\pp_{\nu}$ now takes the same form as the left multiplicative unitary $\ww$ in~\eqref{eq:WdualGenerators}, provided we identify $\kbf,\ebf,\fbf$ in~\eqref{eq:WdualGenerators} with $\kbf_\nu,\ebf_\nu,\fbf_\nu$ in~\eqref{eq:Pkef}.

This means that $\pp_{\nu}$ gives rise to a unitary representation of the modular double. Take $b>0$ and define the positive self-adjoint operators $K_\nu$, $E_\nu$, $F_\nu$ on $\cH_{\nu}$ as in~\eqref{eq:KEFspectral}:
\be
K_\nu=\exp(2\pi b\,\kbf_\nu)\,,\qquad E_\nu=\exp(2\pi b \,\ebf_\nu)\,,\qquad F_\nu=\exp(2 \pi b \,\fbf_\nu)\,.
\ee
They satisfy the commutation relations~\eqref{eq:LeftQuantumGroupRelations} of $\cU_q(\mathfrak{sl}(2,\rr))$. Furthermore, they are equal to the following finite difference operators
\ba
K_\nu &= \exp(\pi b \pbf)\,,\\
E_\nu &= \exp[\pi b(\pbf-2\xbf+\nu)] + \exp[\pi b(-\pbf-2\xbf-\nu)]\,,\\
F_\nu &= \exp[\pi b(-\pbf+2\xbf+\nu)] + \exp[\pi b(\pbf+2\xbf-\nu)]\,.
\ea
For $\nu\in\rr$, this agrees with the positive principal series representations of the modular double. The special case $\nu = i/(4 b)$ of the complementary series was constructed in~\cite{IpMan2024}. For the remaining values of $\nu$, our construction seems to be novel.

The modular double possesses a quadratic Casimir element, built bilinearly from $E$, $F$ and $K$:
\be
C = FE - qK^2 - q^{-1}K^{-2} = EF - q^{-1}K^2 - qK^{-2}\,.
\label{eq:Casimir}
\ee
The two expressions agree by~\eqref{eq:LeftQuantumGroupRelations}, and a short computation using the same relations shows that $C$ commutes with $K$, $E$ and $F$. Up to the normalization of $E$ and $F$ and an additive constant, $C$ is the Casimir element of~\cite[Section~3.1]{PonsotTeschner2001}, which generates the center of $\cU_q(\mathfrak{sl}(2,\rr))$ for generic $q$. Its counterpart $\check{C}$ in the other half of the modular double is obtained by replacing $b$ with $b^{-1}$, i.e.~$q$ with $\check{q} = e^{i\pi/b^2}$. Since $\check{K}$ anticommutes with $E$ and $F$, while $\check{E}$ and $\check{F}$ commute with $E$ and $F$ and anticommute with $K$, the Casimir $C$ commutes with $\check{K}$, $\check{E}$, $\check{F}$ as well, and similarly for $\check{C}$. For $b>0$, $C$ and $\check{C}$ are formally self-adjoint, while for $|b|=1$,~\eqref{eq:ModularDoubleStar} implies $C^* = \check{C}$. Using the finite difference operators above, one finds that on $\cH_\nu$
\be
C = 2\cosh(2\pi b\nu)\,,\qquad \check{C} = 2\cosh(2\pi b^{-1}\nu)\,,
\label{eq:CasimirEigenvalue}
\ee
for $\nu\in\rr\cup i(-Q/2,Q/2)$, where for $|b|=1$ the identities hold on analytic vectors. On the trivial representation, where $K=1$ and $E=F=0$, we get $C = -q-q^{-1}$ and $\check{C} = -\check{q}-\check{q}^{-1}$, which is~\eqref{eq:CasimirEigenvalue} at $\nu = iQ/2$, in agreement with~\cref{ssec:CfromP}. The pair $(C,\check{C})$ plays the role of a quantum Laplace operator. Classically, the quadratic Casimir of $\mathfrak{sl}(2,\rr)$ acts on the principal and complementary series of $\mathrm{PSL}(2,\rr)$ by $\frac14+\nu^2$, which becomes the Laplace eigenvalue on a hyperbolic surface. At fixed $b$, the eigenvalues~\eqref{eq:CasimirEigenvalue} are functions of $h = \frac{Q^2}{4}+\nu^2$, but need not determine $h$ uniquely. We will always work directly with $h$.

\subsection{Plancherel theorem for $\sl$}\label{ssec:Plancherel}
Let $G$ be a classical unimodular locally compact group. $L^2(G)$ is then a unitary representation of $G\times G$, with the left and right factors acting by left and right translations. When $G$ is sufficiently nice, the Plancherel theorem presents $L^2(G)$ as a direct integral of unitary irreducible representations of $G\times G$
\be
L^2(G) \cong \int_{\hat{G}}^{\oplus} \cH_{\pi}\otimes \overline{\cH_{\pi}}\, d\mu_{\mathrm{P}}(\pi)\,.
\ee
Here $\hat{G}$ is the unitary dual of $G$, and $\mu_{\mathrm{P}}$ is the Plancherel measure. This result has a precise analogue for $\gg_b$, to which we turn now. The following two definition-lemmas are standard:
\begin{deflem}
For $i=1,2$, let $\gg_i$ be a locally compact quantum group, with $M_{i} = L^{\infty}(\gg_{i})$, and coproduct $\Delta_{i}$. Set $M = M_1 \botimes M_2$ and $\Delta =\sigma_{23}\circ (\Delta_1\otimes \Delta_2)$. Then $(M,\Delta)$ is a locally compact quantum group, representing the direct product $\gg_1 \times \gg_2$.
\end{deflem}

\begin{deflem}\label{deflem:ExtTenProd}
For $i=1,2$, let $\uu_i$ be a unitary representation of a locally compact quantum group $\gg_i$ on a Hilbert space $\cH_i$. Define
\be
\uu_1 \boxtimes \uu_2 := (\uu_1)_{13}(\uu_2)_{24} \in M_1\botimes M_2 \botimes \mathrm{B}(\cH_1)\botimes \mathrm{B}(\cH_2)\,.
\ee
Then $\uu_1\boxtimes \uu_2$ is a unitary representation of $\gg_1\times \gg_2$ on $\cH_1\otimes \cH_2$. We call $\uu_1\boxtimes \uu_2$ the external tensor product of $\uu_1$ and $\uu_2$.
\end{deflem}

Recall the left and right multiplicative unitaries $\ww$, $\vv$ of $\gg_b$, described in Section~\ref{ssec:WVGb}. Both $\ww \in M\botimes \hat{M}$ and $\vv_{21}=\Sigma\vv\Sigma \in M\botimes \hat{M}'$ are unitary representations of $\gg_b$ on the same GNS Hilbert space $L^2(\gg_b)$, with commuting slices. We can thus define
\be
\ww_{LR} := \ww_{13}\vv_{32} = \vv_{32}\ww_{13} \in M\botimes M\botimes \mathrm{B}(L^2(\gg_b))\,.
\ee
\begin{lemma}
$\ww_{LR}$ is a unitary representation of $\gg_b\times\gg_b$ on $L^2(\gg_b)$.
\end{lemma}

The Plancherel theorem for $\gg_b$ decomposes $L^2(\gg_b)$ as a direct integral of $\cH_{\nu}\otimes\cH_{\nu}$ with $\nu \geq 0$, in such a way that the unitary representation $\ww_{LR}$ of $\gg_b\times\gg_b$ is intertwined with $\pp_\nu\boxtimes\pp_\nu$.
\begin{claim}[Plancherel theorem for $\sl$]\label{clm:plancherel}
Let $Q=b+b^{-1}>0$. Define the Plancherel Hilbert space as
\be
\cH_{\mathrm{P}} := \int_{\rr_+}^{\oplus} \cH_{\nu}\otimes \cH_{\nu} \, d\nu  \cong L^2(\rr_{+,\nu}\times \rr_{x_L}\times\rr_{x_R})\,.
\ee
There is then a unitary isomorphism of representations of $\gg_b\times \gg_b$
\be
L^2(\gg_b) \cong  \cH_{\mathrm{P}}\,,\qquad
\ww_{LR} \simeq  \int_{\rr_+}^{\oplus} \pp_{\nu}\boxtimes \pp_{\nu} \, d\nu\,.
\label{eq:PlancherelSLb}
\ee
The unitary intertwiner realizing this isomorphism takes the form
\ba
\,&\psi :  \cH_{\mathrm{P}} \rightarrow L^2(\gg_b)\\
&(\psi(f))(s,t,u) = \int\limits_{0}^{\infty}
\frac{\sqrt{\rho_b(\nu)}s_b(t-s)s_b(t-u) s_b(s+\nu)s_b(u+\nu)}{s_b(t+\nu -i\tfrac{Q}{2}+i 0^+)s_b(t-\nu-i\tfrac{Q}{2}+i0^+)}\hat{f}(\nu,s,-u)\,d\nu\,,
\label{eq:psiPlancherel}
\ea
where
\be
\rho_b(\nu):= \big|s_b(2\nu - i\tfrac{Q}{2}) \big|^2 = 4\sinh(2\pi b \nu) \sinh(2\pi b^{-1} \nu)
\ee
and
\be
\hat{f}(\nu,p_L,p_R) := \int\limits_{\rr^2} e^{-2\pi i(p_L x_L+p_R x_R)}f(\nu,x_L,x_R)dx_L dx_R\,.
\ee
In particular,
\be
\psi^*\psi = 1_{\cH_{\mathrm{P}}}\,,\qquad \psi \psi^* = 1_{L^2(\gg_b)}\,,
\label{eq:psiUnitarity}
\ee
and
\be
\ww_{LR}(1_M\otimes 1_M\otimes \psi) = (1_M\otimes 1_M\otimes \psi)  \int_{\rr_+}^{\oplus} \pp_{\nu}\boxtimes \pp_{\nu} \, d\nu\,.
\label{eq:psiEquivariance}
\ee
\end{claim}
We see that the Plancherel decomposition is supported precisely on the unitary principal series. The trivial representation and complementary series do not appear. The measure $\rho_b(\nu)d\nu$ is sometimes called the Plancherel measure of $\sl$. However, in contrast with classical unimodular groups, there is no canonical Plancherel measure in the quantum setting -- only a canonical measure class. Indeed, the $\nu$-dependent normalization of the measure must be specified together with that of Duflo-Moore operators. This is familiar from classical non-unimodular group.\footnote{For classical unimodular groups, one can take the Duflo-Moore operators to be the identity, which makes the Plancherel measure canonical.} In this paper, we normalize the measure to be $d\nu$ rather than $\rho_b(\nu)d\nu$. This is natural because the spectral density of the Liouville CFT is constant in $\nu$, as can be seen from the torus partition function.

The Plancherel theorem for $\sl$ was first stated in~\cite{PonsotTeschner1999} in the language of the positive principal series representations of the modular double. A more detailed argument was given in~\cite{Ip2013QuantumPlane}. Claim~\ref{clm:plancherel} represents a reformulation in terms of unitary corepresentations. The unitarity of the integral transform is based on Kashaev’s orthogonality and completeness relations~\cite{Kashaev2001}.

Unitarity~\eqref{eq:psiUnitarity}, and equivariance~\eqref{eq:psiEquivariance} fix the distributional kernel of $\psi$ up to a $\nu$-dependent phase. The remaining phase in~\eqref{eq:psiPlancherel} is fixed (up to a sign) by requiring that $\psi$ intertwines the natural real structures. Indeed, the modular conjugation $J: L^2(\gg_b)\rightarrow L^2(\gg_b)$ provides a real structure for $\ww_{LR}$.
\begin{claim}\label{clm:PlancherelReal}
Let $J:L^2(\gg_b)\rightarrow L^2(\gg_b)$ be the modular conjugation of $(M,\ph)$, given in~\eqref{eq:GbModular}, and let $J_{\nu}:\cH_{\nu}\rightarrow\cH_{\nu}$ be the real structure of the representation $(\pp_{\nu},\cH_{\nu})$, given in Claim~\ref{clm:Jnu}. Define the real structure on the Plancherel Hilbert space as
\be
J_{\mathrm{P}}: \cH_{\mathrm{P}}\rightarrow \cH_{\mathrm{P}}\,,\qquad J_{\mathrm{P}} = \int_{\rr_+}^{\oplus} J_{\nu}\otimes J_{\nu}\, d\nu\,.
\ee
The Plancherel isomorphism $\psi$, given by~\eqref{eq:psiPlancherel}, satisfies
\be
J \psi = \psi  J_{\mathrm{P}}\,.
\ee
\end{claim}
In particular, $\psi$ intertwines the reality identity~\eqref{eq:JPnu} satisfied by $\pp_{\nu}$ with the reality identities~\eqref{eq:JWV} satisfied by $\ww$, $\vv$.

The analogous comments apply to the modular flow. Indeed, recall the modular operator $\nabla$ of $(M,\ph)$, given in~\eqref{eq:GbModular}, as well as the family~\eqref{eq:nablaNu} of positive self-adjoint operators $\nabla_{\nu}$ on $\cH_{\nu}$ implementing the action of the scaling group on $\pp_{\nu}$.

\begin{claim}\label{clm:PlancherelModular}
Define the modular flow on the Plancherel Hilbert space as
\be
\nabla^{i\alpha}_{\mathrm{P}}: \cH_{\mathrm{P}}\rightarrow \cH_{\mathrm{P}}\,,\qquad
\nabla^{i\alpha}_{\mathrm{P}} = \int_{\rr_+}^{\oplus} \nabla^{i\alpha}_{\nu}\otimes \nabla^{-i\alpha}_{\nu}\, d\nu\,,\qquad\alpha\in\rr\,.
\ee
The Plancherel isomorphism $\psi$ then satisfies
\be
\nabla^{i\alpha}\psi = \psi \nabla^{i\alpha}_{\mathrm{P}}\,,\qquad \alpha\in \rr\,.
\ee
\end{claim}
Notice the opposite signs on the left and right tensor factors, arising because a switch from a left to a right action reverses the scaling group. In particular, $\psi$ intertwines the scaling group identity~\eqref{eq:tauPnu} satisfied by $\pp_{\nu}$ with the scaling group identities~\eqref{eq:nablaWV} satisfied by $\ww$, $\vv$.

\section{Invariant forms}\label{sec:invariantForms}
The main goal of this section is to describe invariant trilinear forms on the irreducible representations $\pp_{\nu}$. Our construction is closely related to the Clebsch--Gordan kernels for the positive principal series of Ponsot and Teschner~\cite{PonsotTeschner2001}, and their interpretation as invariant trilinear forms by Teschner and Vartanov~\cite{TeschnerVartanov2014}. Here we generalize these structures to continuous trilinear functionals on spaces of analytic test vectors for unitary irreducible representations of $\mathbb G_b$.

\subsection{Tensor products of unitary corepresentations}
Unitary representations of locally compact quantum groups admit a natural notion of tensor product, generalizing the classical definition:
\begin{deflem}
Let $\uu_1$, $\uu_2$ be unitary representations of a locally compact quantum group $\gg = (M,\Delta)$ on Hilbert spaces $\cH_{1}$, $\cH_{2}$. Let us define the unitary
\be
\uu_1\times \uu_2 := (\uu_2)_{13}(\uu_1)_{12} \in M \botimes \mathrm{B}(\cH_{1}) \botimes \mathrm{B}(\cH_{2})\,.
\label{eq:UboxV}
\ee
Then $\uu_1\times\uu_2$ is a unitary representation of $\gg$ on $\cH_{1}\otimes\cH_{2}$. $\uu_1\times \uu_2$ is called the tensor product of $\uu_1$ and $\uu_2$.
\end{deflem}
We use the notation $\uu_1\times\uu_2$ since the operator tensor product already has the meaning $\uu_1\otimes \uu_2 = (\uu_1)_{12}(\uu_2)_{34}$. Contrast the tensor product $\uu_1\times \uu_2$, which is a representation of $\gg$, with the external tensor product $\uu_1\boxtimes \uu_2$ of Definition-Lemma~\ref{deflem:ExtTenProd}, which is a representation of $\gg\times\gg$. Another valid definition for $\uu_1\times \uu_2$ would have been $(\uu_1)_{12} (\uu_2)_{13}$. It is equivalent to~\eqref{eq:UboxV} under simultaneously reversing the order of the corepresentations and flipping their Hilbert-space factors. For classical groups, the distinction disappears because $M$ is commutative. The reason we prefer~\eqref{eq:UboxV} is that it is compatible with the dual coproduct in the sense
\be
(\id\otimes \hat{\Delta})\ww = \ww\times\ww\,.
\label{eq:DeltaHatTimes}
\ee
\begin{lemma}
The tensor product is associative, i.e.
\be
(\uu_1\times\uu_2)\times \uu_3 = \uu_1\times(\uu_2\times \uu_3)
\ee
as representations of $\gg$ on $\cH_{1}\otimes\cH_{2}\otimes\cH_{3}$.
\end{lemma}

Let us consider tensor products $\pp_{\nu_1}\times\pp_{\nu_2}$ of unitary irreducible representations of $\gg_b$, described in Section~\ref{sec:reps}. We will only need the case where both belong to the principal series $\nu_1,\nu_2\in \rr$. The following is a reformulation of the Clebsch--Gordan decomposition discussed in~\cite{PonsotTeschner2001,NidaievTeschner2013,Ip2015TensorProduct} in terms of unitary corepresentations of $\gg_b$. We normalize the spectral measure $\mu_b$ as follows
\be
d\mu_b(\nu) = \big|s_b(2\nu - i\tfrac{Q}{2}) \big|^2 d\nu\,.
\ee
\begin{claim}[Clebsch--Gordan decomposition for principal series]\label{clm:CGPrincipal}
Let $Q>0$ and let $\nu_1,\nu_2\in \rr$. Define the Clebsch--Gordan Hilbert space $\cH_{\mathrm{CG}}$ as
\be
\cH_{\mathrm{CG}} := \int_{\rr_+}^{\oplus} \cH_{\nu_3}\,d\mu_b(\nu_3) \cong L^2(\rr_{+,\nu_3},\mu_b) \otimes  L^2(\rr_{x_3})\,.
\ee
There is then a unitary isomorphism of representations of $\gg_b$
\be
\cH_{\nu_1}\otimes \cH_{\nu_2} \cong \cH_{\mathrm{CG}}\,,\qquad
\pp_{\nu_1}\times \pp_{\nu_2} \simeq \int_{\rr_+}^{\oplus} \pp_{\nu_3}d\mu_b(\nu_3)\,.
\label{eq:Hnu1Hnu2Dec}
\ee
The unitary intertwiner realizing this isomorphism takes the form
\ba
\,&C_{\nu_1,\nu_2} : \cH_{\nu_1}\otimes \cH_{\nu_2} \rightarrow \cH_{\mathrm{CG}}\,,\\
&(C_{\nu_1,\nu_2}(F))(\nu_3,x_3) = \int\limits_{\rr^2}\mathscr{C}_{\nu_1,\nu_2}(\nu_3,x_3;x_1,x_2) F(x_1,x_2) dx_1 dx_2\,,
\ea
with the distributional kernel $\mathscr{C}_{\nu_1,\nu_2}$ given by
\ba
\,&\mathscr{C}_{\nu_1,\nu_2}(\nu_3,x_3;x_1,x_2) = \frac{s_b(\nu_1-\nu_2-\nu_3)}{s_b(\nu_1-\nu_2+\nu_3)}\\
&\quad\times\frac{
s_b(x_{12}-\frac{\nu_1+\nu_2+2\nu_3}{2}+i\tfrac Q2-i0)
s_b(x_{31}-\frac{\nu_1-2\nu_2-\nu_3}{2})
s_b(x_{32}-\frac{\nu_2-\nu_3}{2}+i\tfrac Q2-i0)
}{
s_b(x_{12}+\frac{\nu_1+\nu_2}{2})
s_b(x_{31}+\frac{\nu_1-\nu_3}{2}-i\tfrac Q2+i0)
s_b(x_{32}-\frac{2\nu_1-\nu_2+\nu_3}{2})
}\,,
\label{eq:Cnu1nu2}
\ea
where $x_{ij}:=x_i-x_j$. In particular, 
\be
C^*_{\nu_1,\nu_2} C_{\nu_1,\nu_2} = 1_{\cH_{\nu_1}\otimes \cH_{\nu_2}}\,,\qquad C_{\nu_1,\nu_2} C^*_{\nu_1,\nu_2} = 1_{\cH_{\mathrm{CG}}}\,,
\ee
and
\be
\int_{\rr_+}^{\oplus} \pp_{\nu_3} d\mu_b(\nu_3) \circ (1_M\otimes C_{\nu_1,\nu_2}) = (1_M\otimes C_{\nu_1,\nu_2}) \circ(\pp_{\nu_1}\times\pp_{\nu_2})\,.
\label{eq:CGIntertwine}
\ee
\end{claim}
The prefactor $s_b(\nu_1-\nu_2-\nu_3)/s_b(\nu_1-\nu_2+\nu_3)$ in the kernel $\mathscr{C}_{\nu_1,\nu_2}$ is a pure phase, and removing it would not affect the validity of the claim. It will be useful to keep it because it partially restores symmetry between $\nu_1,\nu_2,\nu_3$. To make contact with the modular double formulation, we use the relation~\eqref{eq:DeltaHatTimes} between the tensor product and the dual coproduct. Recall the dual coproduct of the modular double generators from Claim~\ref{clm:DeltaHatKEF} and their realization on $\cH_{\nu}$ from Section~\ref{ssec:posRepMD}. The relation~\eqref{eq:CGIntertwine} then implies
\ba
C^*_{\nu_1,\nu_2}\Big(\int_{\rr_+}^{\oplus} K_{\nu_3} d\mu_b(\nu_3) \Big) C_{\nu_1,\nu_2}&= K_{\nu_1}\otimes K_{\nu_2} \,,\\
C^*_{\nu_1,\nu_2}\Big(\int_{\rr_+}^{\oplus} E_{\nu_3} d\mu_b(\nu_3) \Big) C_{\nu_1,\nu_2}&=E_{\nu_1}\otimes K_{\nu_2}+K^{-1}_{\nu_1}\otimes E_{\nu_2}\,,\\
C^*_{\nu_1,\nu_2}\Big(\int_{\rr_+}^{\oplus} F_{\nu_3} d\mu_b(\nu_3) \Big) C_{\nu_1,\nu_2}&=F_{\nu_1}\otimes K_{\nu_2}+K^{-1}_{\nu_1}\otimes F_{\nu_2}
\ea
on a suitable analytic domain in $\cH_{\nu_1}\otimes \cH_{\nu_2}$.

\subsection{Space of analytic test vectors}\label{ssec:testVectors}
It should be stressed that the Clebsch--Gordan maps $C_{\nu_1,\nu_2}$ do not give rise to bounded intertwiners $\cH_{\nu_1}\otimes\cH_{\nu_2}\rightarrow \cH_{\nu_3}$ between Hilbert spaces. Indeed, no such intertwiner can exist since otherwise $\cH_{\nu_3}$ would embed in $\cH_{\nu_1}\otimes\cH_{\nu_2}$ as a discrete summand, contradicting~\eqref{eq:Hnu1Hnu2Dec}. In fact, the image of $F\in \cH_{\nu_1}\otimes \cH_{\nu_2}$ under $C_{\nu_1,\nu_2}$ is only defined up to equality almost everywhere on $\rr_{+,\nu_3} \times \rr_{x_3}$ and thus cannot be evaluated at a fixed $\nu_3$.

Similarly, there exist no nonzero bounded invariant maps of the form $\cH_{\nu}\otimes \cH_{\nu}\rightarrow \cc$ and $\cH_{\nu_1}\otimes\cH_{\nu_2}\otimes\cH_{\nu_3}\rightarrow \cc$. All of this is already familiar from the setting of automorphic forms on $\mathrm{PSL}(2,\rr)$, reviewed in Section~\ref{sec:bootstrapClassical}. The classical setting suggests that one should instead work with continuous intertwiners between suitable spaces of test vectors. The spaces of intertwiners can then be nontrivial.

The idea is to restrict $\cH_{\nu}$ to a dense subspace $\cE_{\nu}$ of ``analytic vectors'', equipped with a Fr\'{e}chet topology. The completed projective tensor product $\cE_{\nu_1}\hat{\otimes}_{\pi} \cE_{\nu_2}$ embeds into $\cH_{\nu_1}\otimes\cH_{\nu_2}$. We will see that for vectors in this subspace, the image under the Clebsch--Gordan map $C_{\nu_1,\nu_2}$, interpreted through its distributional kernel, can be evaluated at a fixed spectral parameter $\nu_3$. This evaluation gives rise to a continuous equivariant linear map $\cE_{\nu_1}\hat{\otimes}_{\pi} \cE_{\nu_2}\rightarrow \cE_{\nu_3}$. Similarly, we will construct nontrivial continuous invariant linear maps $\cE_{\nu}\hat{\otimes}_{\pi} \cE_{\nu}\rightarrow \cc$ and $\cE_{\nu_1}\hat{\otimes}_{\pi} \cE_{\nu_2}\hat{\otimes}_{\pi}\cE_{\nu_3}\rightarrow \cc$.

In the classical setting, one can take $\cE_{\nu}$ to be the space of smooth vectors in $\cH_{\nu}$. Since we are not aware of a universal definition of smooth vectors in the quantum setting, we will proceed differently. We will give an ad hoc definition of $\cE_{\nu}$ as an explicit subspace of $\cH_{\nu}$, and check that it satisfies various required properties. We do not claim that our definition of $\cE_{\nu}$ is the unique one possible.
\begin{deflem}[Space of analytic test vectors]\label{def:ESpace}
We define the space of analytic vectors to consist of entire functions that decay rapidly in all horizontal strips. More precisely, for $f:\cc\rightarrow \cc$ entire, define the family of seminorms
\be
\| f \|_{a,m} = \sup_{\substack{x,y\in \rr\\ |y|\leq a}} |x^m f(x+i y)|\,,\qquad a,m\in\zz_{\geq 0}\,,
\ee
and let
\be
\cE = \{f: \cc\rightarrow \cc \,: \, f \text{ is entire},\, \| f \|_{a,m} <\infty\text{ for all }a,m\in \zz_{\geq 0} \}\,.
\ee
Equipped with the topology generated by these seminorms, $\cE$ is a Fr\'{e}chet space. Restriction to the real axis defines a continuous inclusion $\cE\hookrightarrow \cH_{\nu}$ with dense image for all $\nu$ in the complementary and principal series. We will write $\cE_{\nu}$ for the image of this inclusion.
\end{deflem}
It will be convenient to extend the notation to the trivial representation by defining
\be
\cE_{i \frac{Q}{2}} := \cH_{i \frac{Q}{2}} = \cc\,.
\ee


For the rest of this section, $Q>0$ is fixed and all representation parameters $\nu$ are understood to belong to $\rr\cup i(-Q/2,Q/2]$, unless stated otherwise.

Recall the real structure $J_{\nu}:\cH_{\nu}\rightarrow \cH_{\nu}$ from Claim~\ref{clm:Jnu} and the scaling group operator $\nabla_{\nu}$ on $\cH_{\nu}$ from Claim~\ref{clm:tauNu}. For $\nu\neq i\frac{Q}{2}$, recall also the unitary equivalence $\cI_{\nu}:\cH_{\nu}\rightarrow \cH_{-\nu}$ from~\eqref{eq:ISignFlip}.
\begin{lemma}\label{eq:JnuNablanucE}
$J_{\nu}$ restricts to a continuous antilinear involution of $\cE_{\nu}$. Furthermore, for every $z\in \cc$, we have $\cE_{\nu}\subset\mathrm{Dom}(\nabla^{i z}_{\nu})$ and $\nabla^{i z}_{\nu}$ restricts to a continuous linear automorphism of $\cE_{\nu}$. For $\nu\neq i\frac{Q}{2}$, $\cI_{\nu}$ restricts to a continuous linear isomorphism $\cE_{\nu}\rightarrow\cE_{-\nu}$.
\end{lemma}
\begin{proof}
For the first part, we go to Fourier space, where $\cE$ has the following characterization:
\be
\hat{\cE} = \{\hat{f} \in C^{\infty}(\rr) \,: \, \sup_{p\in\rr}| e^{a |p|} \partial^m_p\hat{f}(p) |<\infty\text{ for all }a,m\in \zz_{\geq 0} \}\,.
\ee
The family of suprema reproduces the original Fr\'{e}chet topology on $\cE$. Complex conjugation preserves $\cE$ continuously. For $\nu \in \rr\cup i(-Q/2,Q/2)$, the double-sine asymptotics imply that the functions
\be
\frac{s_b(p-\nu)}{s_b(p+\nu)}
\qquad\text{and}\qquad
\frac{s_b(p+\nu)}{s_b(p-\nu)}
\ee
are smooth on $\rr$, with all derivatives growing at most exponentially. Multiplication by either function therefore
preserves $\hat{\cE}$ continuously. The explicit formulas for $J_\nu$ and $\cI_\nu$ then show that both preserve $\cE$ continuously. Since $\cI_{-\nu}=\cI_\nu^{-1}$, the latter restricts to a continuous linear isomorphism $\cE_\nu\rightarrow\cE_{-\nu}$. For the second part, we have $(\nabla^{i z}_{\nu} f)(x) = f(x+Q z)$ for any $z\in\cc$. This is indeed a continuous linear automorphism of $\cE_\nu$. For the trivial representation, i.e.~$\nu = i\frac{Q}{2}$, the proof is immediate. 
\end{proof}

In order to formulate the notion of an invariant multilinear form on $\cE_{\nu}$, we will realize the action of $\gg_b$ on $\cH_{\nu}$ through normal slices $(\omega \otimes \id)\pp_{\nu}\in \mathrm{B}(\cH_{\nu})$ and $(\omega \otimes \id)\pp^*_{\nu}\in \mathrm{B}(\cH_{\nu})$, where $\omega \in M_*$. The following claim asserts that any vector in $\cE_{\nu}$ can be obtained from any nonzero vector in $\cH_{\nu}$ by acting with either type of normal slice:
\begin{claim}\label{clm:EsliceReachability}
Let $f\in \cH_{\nu}$ with $f\neq 0$. Then
\be
\cE_{\nu} \subset \{(\omega\otimes \id)\pp_{\nu} f \,:\, \omega \in M_*\}\quad\text{and}\quad
\cE_{\nu} \subset \{(\omega\otimes \id)\pp^*_{\nu} f \,:\, \omega \in M_*\}\,.
\ee
\end{claim}
Note that irreducibility of $\pp_\nu$ only implies that the space generated from $f\neq 0$ by all normal slices is dense in $\cH_{\nu}$. The above claim is strictly stronger and follows from the explicit form of $\pp_\nu$ in Claim~\ref{clm:PnuKernel}.

We want to restrict to $\omega\in M_*$ such that the corresponding normal slices preserve $\cE_{\nu}$. Since $M$ acts on $L^2(\gg_b)$ in standard form, every $\omega \in M_*$ can be represented as
\be
\omega_{\eta,\xi}(a) = \langle \eta, a \xi \rangle
\ee
for some pair of vectors $\eta,\xi \in L^2(\gg_b)$, where $\langle\cdot,\cdot\rangle$ is the inner product on $L^2(\gg_b)$. Let us recall the dense Gelfand--Shilov subspace $\sA := S^{1/2}_{1/2}(\rr^3) \subset L^2(\rr^3)= L^2(\gg_b)$, introduced in Section~\ref{sec:slb2r}.
\begin{claim}\label{clm:SliceContinuity}
Let $\eta,\xi \in \sA \subset L^2(\gg_b)$. Then the normal slices produce continuous maps
\be
(\omega_{\eta,\xi} \otimes \id)\pp_{\nu}\,,\;(\omega_{\eta,\xi} \otimes \id)\pp^*_{\nu} : \;\cH_{\nu}\rightarrow \cE_{\nu}\,.
\ee
In particular, they restrict to continuous endomorphisms of $\cE_{\nu}$.
\end{claim}

Let us pick an orthonormal basis $(e_p)_{p=1}^{\infty}$ for $L^2(\gg_b)$ with $e_p \in \sA$ for all $p$. Let us denote the corresponding normal slice actions on $\cH_{\nu}$ as
\be
A^{(\nu)}_{pq} := (\omega_{e_p,e_q} \otimes \id)\pp_{\nu}\,,\qquad
B^{(\nu)}_{pq} := (\omega_{e_p,e_q} \otimes \id)\pp^*_{\nu}\,.
\label{eq:ABdef}
\ee
Each $A^{(\nu)}_{pq}$ and $B^{(\nu)}_{pq}$ is a continuous endomorphism of $\cE_{\nu}$.

\subsection{Invariance of forms}
We are interested in multilinear forms on $\cE_{\nu}$. By definition, these are multilinear maps
\be
T:\cE_{\nu_1}\times \ldots \times \cE_{\nu_n} \rightarrow \cc
\ee
jointly continuous in the Fr\'{e}chet topologies. Such maps are equivalent to continuous linear maps
\be
T: \cE_{\nu_1}\hat{\otimes}_{\pi} \ldots \hat{\otimes}_{\pi}  \cE_{\nu_n} \rightarrow \cc\,.
\label{eq:formT}
\ee
The following definition-lemma describes the completed projective tensor products explicitly.
\begin{deflem}\label{deflem:HatOtimesE}
Suppose $F:\cc^n \rightarrow \cc$ is entire, and define the family of seminorms
\be
\| F \|_{a,m} = \sup_{\substack{x,y\in \rr^n \\ \|y\|_{\infty}\leq a}} \|x\|^m |F(x+i y)|\,,\qquad a,m\in\zz_{\geq 0}\,.
\ee
Let
\be
\cE^{(n)} = \{F: \cc^n \rightarrow \cc \,: \, F \text{ is entire},\, \| F \|_{a,m} <\infty\text{ for all }a,m\in \zz_{\geq 0} \}\,.
\ee
Equipped with the topology generated by these seminorms, $\cE^{(n)}$ is a Fr\'{e}chet space. We have the canonical isomorphism
\be
\underbrace{\cE  \hat{\otimes}_{\pi}\ldots  \hat{\otimes}_{\pi}\cE}_{n\text{-times}} \cong \cE^{(n)}\,,\qquad
f_1\otimes\ldots\otimes f_n\mapsto f_1(z_1)\ldots f_n(z_n)\,.
\ee
Let $(\nu_1,\ldots,\nu_n)$ be representation labels and let $n'$ be the number of nontrivial representations in it. Restriction to $\rr^{n'}$ defines a continuous inclusion $\cE^{(n')}\hookrightarrow \cH_{\nu_1}\otimes\ldots\otimes \cH_{\nu_n}$. We will identify $\cE_{\nu_1}  \hat{\otimes}_{\pi}\ldots  \hat{\otimes}_{\pi}\cE_{\nu_n}$ with its image.
\end{deflem}

We would like to formulate a notion of invariance of forms~\eqref{eq:formT} under the action of $\gg_b$. The action is inherited from the tensor product representation $\pp_{\nu_1}\times\ldots\times \pp_{\nu_n}$ via the embedding $\cE_{\nu_1}  \hat{\otimes}_{\pi}\ldots  \hat{\otimes}_{\pi}\cE_{\nu_n} \subset \cH_{\nu_1}\otimes\ldots \otimes \cH_{\nu_n}$. If it were the case that $T$ extends to a bounded intertwiner $T : \cH_{\nu_1}\otimes\ldots \otimes \cH_{\nu_n}\rightarrow \cc$, invariance could be expressed as either of the equivalent conditions
\be
(1_M\otimes T)  (\pp_{\nu_1}\times\ldots\times \pp_{\nu_n}) = (1_M\otimes T)\,,\qquad
(1_M\otimes T)  (\pp_{\nu_1}\times\ldots\times \pp_{\nu_n})^* = (1_M\otimes T)
\,.
\label{eq:invarianceT0}
\ee
However, these equalities do not make sense for $T$ of the form~\eqref{eq:formT}. Even if we apply both sides to a factorized vector $\xi \otimes f_1 \otimes \ldots \otimes f_n$ with $\xi \in L^2(\gg_b)$ and $f_i \in \cE_{\nu_i}$, the left-hand side is not well-defined since the tensor product representation takes us outside the domain of $1_M\otimes T$. To make progress, let us first take a normal slice of~\eqref{eq:invarianceT0} with $\omega_{\eta,\xi}$:
\be
T (\omega_{\eta,\xi}\otimes \id)(\pp_{\nu_1}\times\ldots\times \pp_{\nu_n})= \langle\eta,\xi\rangle \, T\,,\qquad
T (\omega_{\eta,\xi}\otimes \id)(\pp_{\nu_1}\times\ldots\times \pp_{\nu_n})^*= \langle\eta,\xi\rangle \, T\,.
\label{eq:invarianceT1}
\ee
By itself, this does not cure the domain problem. Let us proceed formally and act with \eqref{eq:invarianceT1} on a factorized vector $f_1 \otimes \ldots \otimes f_n \in \cE_{\nu_1} \hat{\otimes}_{\pi}\ldots  \hat{\otimes}_{\pi} \cE_{\nu_n}$. Use completeness of the basis $(e_r)_{r=1}^{\infty}$ to write the left-hand sides as:
\ba
\,&T (\omega_{e_p,e_q}\otimes \id)(\pp_{\nu_1}\times\ldots\times \pp_{\nu_n})(f_1 \otimes \ldots \otimes f_n)\\
&\qquad\qquad= \sum\limits_{r_1,\ldots,r_{n-1}=1}^{\infty} T(A^{(\nu_1)}_{r_{1} q} f_1 \otimes A^{(\nu_2)}_{r_{2} r_{1}}f_2\otimes\ldots\otimes A^{(\nu_n)}_{p r_{n-1}}f_n)\,,\\
\,&T (\omega_{e_p,e_q}\otimes \id)(\pp_{\nu_1}\times\ldots\times \pp_{\nu_n})^*(f_1 \otimes \ldots \otimes f_n)\\
&\qquad\qquad= \sum\limits_{r_1,\ldots,r_{n-1}=1}^{\infty} T(B^{(\nu_1)}_{pr_1} f_1\otimes B^{(\nu_2)}_{r_1 r_2}f_2\otimes\ldots\otimes B^{(\nu_n)}_{r_{n-1} q}f_n)\,,
\ea
where we used
\ba
\pp_{\nu_1}\times\ldots\times \pp_{\nu_n} &= (\pp_{\nu_n})_{1 n+1}\ldots  (\pp_{\nu_2})_{13}(\pp_{\nu_1})_{12}\,,\\
(\pp_{\nu_1}\times\ldots\times \pp_{\nu_n})^* &= (\pp^*_{\nu_1})_{12}(\pp^*_{\nu_2})_{13}\ldots (\pp^*_{\nu_n})_{1 n+1}\,.
\ea
Since all $A^{(\nu)}_{pq}$ and $B^{(\nu)}_{pq}$ are endomorphisms of $\cE_{\nu}$, each term in the infinite sum is now well-defined. The above formal manipulations directly motivate our definition of invariance:
\begin{definition}[Invariant form]\label{def:InvariantForm}
Let $T:\cE_{\nu_1}  \hat{\otimes}_{\pi}\ldots  \hat{\otimes}_{\pi}\cE_{\nu_n}\rightarrow \cc$ be linear and continuous. We say that $T$ is $\gg_b$-invariant with respect to the orthonormal basis $(e_p)_{p=1}^{\infty}$ of $L^2(\gg_b)$ if
\ba
\lim_{R\rightarrow \infty }\sum\limits_{r_1,\ldots,r_{n-1}=1}^{R} T(A^{(\nu_1)}_{r_{1} q} f_1 \otimes A^{(\nu_2)}_{r_{2} r_{1}}f_2\otimes\ldots\otimes A^{(\nu_n)}_{p r_{n-1}}f_n) = \delta_{p q}\,T(f_1 \otimes \ldots \otimes f_n)
\label{eq:defInvA}
\ea
and
\ba
\lim_{R\rightarrow \infty }\sum\limits_{r_1,\ldots,r_{n-1}=1}^{R} T(B^{(\nu_1)}_{pr_1} f_1\otimes B^{(\nu_2)}_{r_1 r_2}f_2\otimes\ldots\otimes B^{(\nu_n)}_{r_{n-1} q}f_n) = \delta_{p q}\,T(f_1 \otimes \ldots \otimes f_n)
\label{eq:defInvB}
\ea
for all $p,q \in \zz_{>0}$ and all $f_1 \in \cE_{\nu_1},\ldots, f_n \in \cE_{\nu_n}$. Here $A^{(\nu)}_{pq}$, $B^{(\nu)}_{pq}$ are endomorphisms of $\cE_\nu$ defined in~\eqref{eq:ABdef}.
\end{definition}
We expect that whether a form is invariant or not is independent of the choice of the basis but do not have a proof. From now on, we will call any $T$ which obeys Definition~\ref{def:InvariantForm} an invariant form.

\begin{claim}[Invariant 1-forms]\label{clm:inv1form}
Let $T:\cE_{\nu}\rightarrow \cc$ be $\gg_b$-invariant. Then $T=0$ unless $\nu$ is the trivial representation.
\end{claim}
\begin{proof}
Invariance gives $T=T\circ B^{(\nu)}_{pp}$ for any $p$. Since by Claim~\ref{clm:SliceContinuity}, $B^{(\nu)}_{pp}:\cH_\nu\rightarrow\cE_\nu$ is continuous, $T$ extends to a bounded functional on $\cH_\nu$. The invariance identities extend by density, making this extension a bounded intertwiner from $\pp_\nu$ to the trivial representation. Schur's lemma then proves the claim.
\end{proof}

\subsection{Bilinear forms}
We will now discuss some important examples of invariant forms. Unless stated otherwise, we keep assuming $Q>0$ and $\nu\in \rr\cup i(-Q/2,Q/2]$. Let us first define an analogue of the Tomita operator for the representation $\pp_{\nu}$.
\begin{definition}
Set
\be
S_{\nu}:=J_{\nu} \nabla^{\frac{1}{2}}_{\nu} =  \nabla^{-\frac{1}{2}}_{\nu} J_{\nu}\,.
\ee
\end{definition}
It follows immediately from Lemma~\ref{eq:JnuNablanucE} that $S_{\nu}: \cE_{\nu}\rightarrow \cE_{\nu}$ is a continuous antilinear involution. For $\nu\neq i\frac{Q}{2}$, we have $S_{-\nu} = \cI_{\nu}S_{\nu}\cI^*_{\nu}$. We can use $S_{\nu}$ to define distinguished bilinear forms.
\begin{definition}\label{def:bilinearForm}
Let us define a family $\cD_{\nu_1,\nu_2}$ of bilinear forms on $\cE_{\nu_1}\times \cE_{\nu_2}$ by the following expressions, with $f_1\in \cE_{\nu_1}$, $f_2\in \cE_{\nu_2}$
\ba
 \cD_{\nu_1,\nu_1}(f_1, f_2) &:= \langle S_{\nu_1} f_1, f_2 \rangle_{\nu_1}\,,\\
 \cD_{\nu_1,-\nu_1}(f_1, f_2) &:= \cD_{\nu_1,\nu_1}(f_1, \cI_{-\nu_1}f_2)\,,\quad &&\nu_1\neq i\tfrac{Q}{2}\,,\\
 \cD_{\nu_1,\nu_2}(f_1,f_2) &:= 0\,,\qquad &&\nu_1\neq \pm \nu_2\,.
 \label{eq:DnuDef}
\ea
\end{definition}
$\cD_{\nu,\nu}$ plays the role of an invariant bilinear pairing on $\cE_{\nu}$.
\begin{proposition}
$\cD_{\nu_1,\nu_2}$ extend to unique continuous linear maps $\cD_{\nu_1,\nu_2}: \cE_{\nu_1}\hat{\otimes}_{\pi}\cE_{\nu_2}\rightarrow \cc$. For the principal and complementary series, we have
\ba
\cD_{\nu,\nu}(f_1,f_2) &= \int\limits_{\rr}  e^{\pi Q p}\frac{s_b(p+\nu)}{s_b(p-\nu)} \hat{f_1}(p) \hat{f_2}(-p) dp\,,\\
\cD_{\nu,-\nu}(f_1,f_2) &= \int\limits_{\rr}  e^{\pi Q p}\hat{f_1}(p) \hat{f_2}(-p) dp\,.
\ea
For the trivial representation, we have
\be
\cD_{i\frac{Q}{2},i\frac{Q}{2}}(z_1, z_2) = z_1 z_2\,,\qquad z_1,z_2\in\cc\,.
\label{eq:DTrivial}
\ee
The bilinear forms $\cD_{\nu,\nu}$, $\cD_{\nu,-\nu}$ are nondegenerate.
\end{proposition}
\begin{proof}
Continuity follows from continuity of $S_\nu$ and the inclusion $\mathcal E_\nu\hookrightarrow\cH_\nu$. The unique extension then follows from the defining property of the completed projective tensor product. The integral formula follows from the explicit expressions for the inner product and $S_\nu$. Nondegeneracy follows from $\cD_{\nu,\nu}(f,S_{\nu} f) = \|S_{\nu} f\|^2_{\nu}$ and $\cD_{\nu,\nu}(S_{\nu} f,f) = \| f\|^2_{\nu}$. The corresponding statements for $\cD_{\nu,-\nu}$ follow by transport through the continuous isomorphism $\cI_{-\nu}$.
\end{proof}

The following two assertions express the key invariance and uniqueness properties.
\begin{claim}[Invariance of bilinear forms]
$\cD_{\nu_1,\nu_2}$ are invariant in the sense of Definition~\ref{def:InvariantForm}.
\end{claim}
\begin{claim}[Uniqueness of bilinear forms]\label{clm:bilinearUniqueness}
Every invariant form on $\cE_{\nu_1}\hat{\otimes}_{\pi}\cE_{\nu_2}$ is proportional to $\cD_{\nu_1,\nu_2}$. In particular, if $\nu_1 \neq \pm \nu_2$, every invariant form on $\cE_{\nu_1}\hat{\otimes}_{\pi}\cE_{\nu_2}$ vanishes.
\end{claim}

$\cD_{\nu_1,\nu_2}$ are symmetric up to a twist by the modular operator and satisfy a reality property.
\begin{lemma}[KMS symmetry and reality of bilinear forms]
We have
\be
\cD_{\nu_1,\nu_2}(f_1, f_2)  =\cD_{\nu_2,\nu_1}(\nabla^{-1}_{\nu_2}f_2, f_1)= \cD_{\nu_2,\nu_1}(f_2,\nabla_{\nu_1}f_1)\,,
\ee
and
\be
\overline{\cD_{\nu_1,\nu_2}(f_1 , f_2)} = \cD_{\nu_2,\nu_1}(S_{\nu_2}f_2 ,S_{\nu_1} f_1)\,.
\ee
\end{lemma}
\begin{proof}
If $\nu_1\neq\pm\nu_2$, all terms vanish by definition. Otherwise, since $\cI_\nu$ intertwines both $S_\nu$ and $\nabla_\nu$, transport through $\cI_\nu$ reduces the assertion to $\nu_1=\nu_2=\nu$. In this case
\ba
\cD_{\nu,\nu}(f_1,f_2)
&=\langle S_\nu^*f_2,f_1\rangle_\nu\\
&=\langle S_\nu\nabla_\nu^{-1}f_2,f_1\rangle_\nu
=\cD_{\nu,\nu}(\nabla_\nu^{-1}f_2,f_1)\\
&=\langle S_\nu f_2,\nabla_\nu f_1\rangle_\nu
=\cD_{\nu,\nu}(f_2,\nabla_\nu f_1)\,,
\ea
where we used $S_\nu^*=S_\nu\nabla_\nu^{-1}=\nabla_\nu S_\nu$. The proof of reality is immediate from the definition.
\end{proof}
Finally, we will check that the limit of $\cD_{\nu,\nu}$ as $\nu\rightarrow i\frac{Q}{2}$ agrees with the trivial representation formula~\eqref{eq:DTrivial}, in the spirit of Section~\ref{ssec:CfromP}. Indeed, $\cD_{\nu,\nu}$ admits the position-space representation
\be
\cD_{\nu,\nu}(f_1,f_2) = \iint \frac{s_b(x_{12}-\nu + i Q)}{s_b(i \tfrac{Q}{2}-2\nu) s_b(x_{12}+\nu)} f_1(x_1) f_2(x_2) dx_1 dx_2\,.
\ee
For $\nu\in i(0,Q/2)$, the contours can be taken as $x_1 \in \rr - i \frac{Q}{2}$, $x_2 \in \rr$. It follows
\be
\cD_{i\frac{Q}{2}-i \epsilon,i\frac{Q}{2}-i \epsilon}(f_1,f_2) = \frac{1}{4\pi \epsilon}\int f_{1}(x_1) dx_1\int f_{2}(x_2) dx_2 +O(1)\quad\text{as}\quad\epsilon\rightarrow 0^+\,.
\ee
At the leading order, this precisely reproduces $\cD_{i\frac{Q}{2},i\frac{Q}{2}}(z_1,z_2)=z_1 z_2$ including the overall normalization, set by~\eqref{eq:NormId}.

\subsection{Trilinear forms}\label{ssec:trilinear}
We will define invariant trilinear forms using the Clebsch--Gordan kernel $\mathscr{C}_{\nu_1,\nu_2}$, recorded in~\eqref{eq:Cnu1nu2}. The normalization used in the Clebsch--Gordan decomposition is adapted to unitarity. For continuous maps between the test spaces, a different normalization will be convenient. Let
\be
\cN_b(\nu_1,\nu_2,\nu_3) := \sqrt{\frac{\Gamma_b(Q)^3}{\Gamma_b(2Q)}}\,
\frac{\sqrt{\prod\limits_{k=1}^{3}\Gamma_b(Q+2i\nu_k)\Gamma_b(Q-2i\nu_k)}}{\Gamma_b(\tfrac{Q}{2}-i\nu_{\Sigma})\prod\limits_{k=1}^{3}\Gamma_b(\tfrac{Q}{2}-i\nu_{\Sigma}+2i\nu_k)}\,,
\label{eq:3PtN}
\ee
where $\nu_{\Sigma}:= \nu_1+\nu_2+\nu_3$. Note that the argument of each square root is positive whether $\nu_k$ is on the principal or complementary series. $\cN_b$ is fully symmetric in its arguments, and has a number of other useful properties, discussed below.

Let $\nu_1,\nu_2,\nu_3 \in \rr\cup i(-Q/2,Q/2)$ and take $F\in \cE_{\nu_2}\hat{\otimes}_{\pi}\cE_{\nu_3}$. Define $\cT^{\nu_1}_{\nu_2,\nu_3}(F)$ by the formula
\be
(\cT^{\nu_1}_{\nu_2,\nu_3}(F))(x_1):= \cN_b(\nu_1,\nu_2,\nu_3) \iint \mathscr{C}_{\nu_2,\nu_3}(\nu_1,x_1;x_2,x_3) F(x_2,x_3) dx_2 dx_3\,.
\label{eq:2To1Def}
\ee
For $\nu_1,\nu_2,\nu_3\in \rr$ and $x_1\in\rr$, the integral is taken over
\be
x_2 \in \rr -i\epsilon\,,\qquad x_3\in \rr + i \epsilon\,,\qquad 0<\epsilon < \tfrac{Q}{4}\,.
\ee
For complementary-series parameters, it is defined by analytic continuation of the full normalized integral from the principal series. The integration contours are deformed so that the poles remain on their original sides. At removable singularities, the value is defined by taking the limit of the full expression.

\begin{claim}
For all $\nu_1,\nu_2,\nu_3 \in \rr\cup i(-Q/2,Q/2)$, we have
\be
\cT^{\nu_1}_{\nu_2,\nu_3}:\cE_{\nu_2}\hat{\otimes}_{\pi}\cE_{\nu_3} \rightarrow \cE_{\nu_1}\,.
\ee
Moreover, $\cT^{\nu_1}_{\nu_2,\nu_3}$ is continuous with respect to the corresponding Fr\'{e}chet topologies.
\end{claim}
\begin{proof}[Proof sketch]
Write $x_2=x_1+v$ and $x_3=x_1+w$, so that the kernel depends only on $v,w$. The horizontal-strip asymptotics of $s_b$ give exponential decay along the integration contours, implying that the output is entire and satisfies seminorm estimates
\be
\|\cT^{\nu_1}_{\nu_2,\nu_3}F\|_{a,m}
\leq C_{a,m}\|F\|_{a',m'}
\ee
for suitable $a',m'$. For complementary-series parameters, deform the contours and include the resulting residue terms. These satisfy analogous estimates by Cauchy's inequalities, while the full normalization cancels the possible contour-pinching singularities. The resulting seminorm bounds prove the claimed continuity.
\end{proof}

We extend $\cT^{\nu_1}_{\nu_2,\nu_3}$ in the natural way to the cases where at least one representation is trivial:
\ba
\cT^{i\frac{Q}{2}}_{\nu_1,\nu_2} &:= \cD_{\nu_1,\nu_2}\,,\\
\cT^{\nu}_{i\frac{Q}{2},\nu} &=\cT^{\nu}_{\nu,i\frac{Q}{2}} := 1_{\nu}\,,\\
\cT^{-\nu}_{i\frac{Q}{2},\nu} &=\cT^{-\nu}_{\nu,i\frac{Q}{2}} := \cI_{\nu}\,,\qquad &&\nu\neq i \tfrac{Q}{2}\,,\\
\cT^{\nu_1}_{i\frac{Q}{2},\nu_2} &=\cT^{\nu_1}_{\nu_2,i\frac{Q}{2}} := 0\,,\qquad &&\nu_1\neq \pm \nu_2\,.
\ea
Here $1_{\nu}: \cE_{\nu}\rightarrow \cE_{\nu}$ is the identity map. The normalization factor $\cN_{b}(\nu_1,\nu_2,\nu_3)$ is such that
\be
\cT^{\nu_1}_{\nu,\nu} \rightarrow \cD_{\nu,\nu}\,,\qquad \cT^{\nu}_{\nu_1,\nu}\rightarrow 1_{\nu}\,,\qquad \cT^{\nu}_{\nu,\nu_1}\rightarrow 1_{\nu}\qquad\text{as}\qquad\nu_1\rightarrow i\tfrac{Q}{2}
\ee
 in the sense of Section~\ref{ssec:CfromP}. These limiting relations fix the overall multiplicative constant in~\eqref{eq:3PtN}.

We can now define trilinear forms by composing $\cT^{\nu_1}_{\nu_2,\nu_3}$ with $\cD_{\nu_1,\nu_1}$.
\begin{definition}\label{def:trilinearForm}
Let $\nu_1,\nu_2,\nu_3 \in \rr\cup i(-Q/2,Q/2]$. Define the trilinear form $\cT_{\nu_1,\nu_2,\nu_3}$ on $\cE_{\nu_1}\times \cE_{\nu_2}\times \cE_{\nu_3}$ by
\be
\cT_{\nu_1,\nu_2,\nu_3}(f_1,f_2,f_3) := \cD_{\nu_1,\nu_1}(f_1,\cT^{\nu_1}_{\nu_2,\nu_3}(f_2\otimes f_3))\,,\qquad 
f_1\in \cE_{\nu_1}\,,\,f_2\in \cE_{\nu_2}\,,\,f_3\in \cE_{\nu_3}\,.
\ee
\end{definition}
It follows from continuity of $\cD_{\nu_1,\nu_1}$ and $\cT^{\nu_1}_{\nu_2,\nu_3}$ that $\cT_{\nu_1,\nu_2,\nu_3}$ extends uniquely to a continuous linear map
\be
\cT_{\nu_1,\nu_2,\nu_3}: \cE_{\nu_1}\hat{\otimes}_{\pi}\cE_{\nu_2}\hat{\otimes}_{\pi}\cE_{\nu_3}\rightarrow \cc\,.
\ee
The explicit form of the Clebsch--Gordan kernel leads to an integral expression for $\cT_{\nu_1,\nu_2,\nu_3}$:
\begin{proposition}
Let $\nu_1,\nu_2,\nu_3 \in \rr\cup i(-Q/2,Q/2)$ and $F\in \cE_{\nu_1}\hat{\otimes}_{\pi}\cE_{\nu_2}\hat{\otimes}_{\pi}\cE_{\nu_3}$. We have
\be
\cT_{\nu_1,\nu_2,\nu_3}(F) = \iiint \mathscr{T}_{\nu_1,\nu_2,\nu_3}(x_1,x_2,x_3) F(x_1,x_2,x_3) dx_1dx_2 dx_3\,,
\ee
where
\ba
\,&\mathscr{T}_{\nu_1,\nu_2,\nu_3}(x_1,x_2,x_3) = \cN_b(\nu_1,\nu_2,\nu_3)\\
&\qquad\times \frac{s_b(x_{12}-\frac{\nu_1+\nu_2}{2}+\nu_3+i\tfrac Q2)s_b(x_{13}-\frac{\nu_1+\nu_3}{2}+iQ)s_b(x_{23}-\frac{\nu_2+\nu_3}{2}+\nu_1+i\tfrac Q2)}{s_b(x_{12}+\frac{\nu_1+\nu_2}{2})s_b(x_{13}+\frac{\nu_1+\nu_3}{2}-\nu_2+i\tfrac Q2)s_b(x_{23}+\frac{\nu_2+\nu_3}{2})} \,.
\ea
For $\nu_1,\nu_2,\nu_3\in \rr$, the integration contours can be taken as
\be
x_1\in\rr-i\eta\,,\qquad x_2\in\rr\,,\qquad
x_3\in\rr+i\eta\,,\qquad \tfrac{Q}{4}<\eta<\tfrac{Q}{2}\,.
\ee
The integral is absolutely convergent and independent of $\eta$ in this range. For complementary-series parameters, it is defined by analytic continuation of the full normalized integral, deforming the contours so that the pole families remain on their original sides. At removable singularities, the value is obtained by taking the limit of the full expression.
\end{proposition}

In the special cases where at least one representation is trivial, Definition~\ref{def:trilinearForm} reduces to
\be
\cT_{i\frac{Q}{2},\nu_1,\nu_2}(z,f_1,f_2)
=\cT_{\nu_1,i\frac{Q}{2},\nu_2}(f_1,z,f_2) 
=\cT_{\nu_1,\nu_2,i\frac{Q}{2}}(f_1,f_2,z)
= z\, \cD_{\nu_1,\nu_2}(f_1,f_2)\,.
\ee
For $\nu_1=\pm\nu_2$, these identities agree with the normalized degeneration of the generic trilinear form to the trivial representation, including the overall normalization, in the sense of Section~\ref{ssec:CfromP}.

The following two assertions express the key invariance and uniqueness properties of $\cT_{\nu_1,\nu_2,\nu_3}$.
\begin{claim}[Invariance of trilinear forms]
$\cT_{\nu_1,\nu_2,\nu_3}$ is invariant in the sense of Definition~\ref{def:InvariantForm}.
\end{claim}
\begin{conjecture}[Uniqueness of trilinear forms]\label{conj:trilinearUniqueness}
Every invariant form on $\cE_{\nu_1}\hat{\otimes}_{\pi}\cE_{\nu_2}\hat{\otimes}_{\pi}\cE_{\nu_3}$ is proportional to $\cT_{\nu_1,\nu_2,\nu_3}$.
\end{conjecture}

It follows from the uniqueness conjecture that $\cT_{\nu_1,\nu_2,\nu_3}\circ(\cI_{-\nu_1} \otimes 1_{\nu_2} \otimes 1_{\nu_3})$ must be proportional to $\cT_{-\nu_1,\nu_2,\nu_3}$, and similarly for the other two entries. The normalization~\eqref{eq:3PtN} ensures that the proportionality constant is one:
\begin{lemma}
We have
\ba
\cT_{\nu_1,\nu_2,\nu_3}\circ(\cI_{-\nu_1} \otimes 1_{\nu_2} \otimes 1_{\nu_3}) &=  \cT_{-\nu_1,\nu_2,\nu_3}\,,\\
\cT_{\nu_1,\nu_2,\nu_3}\circ(1_{\nu_1}\otimes \cI_{-\nu_2} \otimes 1_{\nu_3}) &=  \cT_{\nu_1,-\nu_2,\nu_3}\,,\\
\cT_{\nu_1,\nu_2,\nu_3}\circ(1_{\nu_1}\otimes 1_{\nu_2} \otimes \cI_{-\nu_3}) &= \cT_{\nu_1,\nu_2,-\nu_3}\,.
\ea
\end{lemma}
The trilinear forms satisfy KMS cyclic symmetry and reality properties analogous to those satisfied by the bilinear forms. Again, the normalization~\eqref{eq:3PtN} ensures that the proportionality constants are one:
\begin{lemma}[KMS symmetry and reality of trilinear forms]
For any $\nu_1,\nu_2,\nu_3 \in \rr\cup i(-Q/2,Q/2]$, we have
\be
\cT_{\nu_1,\nu_2,\nu_3}(f_1, f_2,f_3)  =\cT_{\nu_3,\nu_1,\nu_2}(\nabla^{-1}_{\nu_3}f_3, f_1,f_2)
\label{eq:TKMS}
\ee
and
\be
\overline{\cT_{\nu_1,\nu_2,\nu_3}(f_1,f_2,f_3)} = \cT_{\nu_3,\nu_2,\nu_1}(S_{\nu_3}f_3,S_{\nu_2}f_2 ,S_{\nu_1} f_1)\,.
\label{eq:TReality}
\ee
\end{lemma}
The previous two lemmas follow from integral identities satisfied by the double sine function.

It is instructive to ask: To what extent do the above properties of $\cT_{\nu_1,\nu_2,\nu_3}$ determine the overall normalization $\cN_b$? Write $h_k = \frac{Q^2}{4}+\nu^2_k$. The remaining freedom is multiplication by a nowhere-vanishing function $\kappa_b: \rr^{3}_{>0}\rightarrow \cc$ satisfying
\ba
\kappa_b(h_1,h_2,h_3) &=  \kappa_b(h_3,h_1,h_2)\,,\\
\overline{\kappa_b(h_1,h_2,h_3)} &=  \kappa_b(h_3,h_2,h_1)\,,\\
\lim_{h_1\searrow 0}\kappa_b(h_1,h,h) &= 1\,.
\label{eq:freedomN}
\ea
This is precisely the freedom one has in defining conformal three-point functions in boundary CFT.

\subsection{Quadrilinear forms and 6j symbols}\label{ssec:quadrilinear}
One can generalize the construction of $\cT_{\nu_1,\nu_2,\nu_3}$ to invariant $n$-linear forms by composing the maps $\cT^{\nu_i}_{\nu_j,\nu_k}$ and contracting the result with $\cD_{\nu,\nu}$. These coupling schemes are encoded by planar trivalent trees with $n$ cyclically ordered leaves, labelled by $\nu_1,\ldots,\nu_n$, and internal edges labelled by intermediate representations. For $n\geq 3$, there are $n-3$ internal representation labels. For $n=4$, there exist two inequivalent couplings:
\begin{definition}\label{def:quadrilinarForms}
Let $\nu_1,\nu_2,\nu_3,\nu_4\in \rr\cup i(-Q/2,Q/2]$. For each $\nu_5\in  \rr\cup i(-Q/2,Q/2]$, define the pair of quadrilinear forms on $\cE_{\nu_1}\times \cE_{\nu_2}\times \cE_{\nu_3}\times \cE_{\nu_4}$ as follows:
\begin{align}
 \cB^s_{\nu_1,\nu_2,\nu_3,\nu_4;\nu_5}(f_1,f_2,f_3,f_4)
 &= \cT_{\nu_1,\nu_2,\nu_5}(f_1,f_2,\cT^{\nu_5}_{\nu_3,\nu_4}(f_3,f_4))\,,\\[5pt]
 \cB^t_{\nu_1,\nu_2,\nu_3,\nu_4;\nu_5}(f_1,f_2,f_3,f_4)
 &=  \cT_{\nu_1,\nu_5,\nu_4}(f_1,\cT^{\nu_5}_{\nu_2,\nu_3}(f_2,f_3),f_4)\,,
\end{align}
where $f_k \in \cE_{\nu_k}$.
\end{definition}
Thanks to continuity of trilinear forms, $\cB^s_{\nu_1,\nu_2,\nu_3,\nu_4;\nu_5}$ and $\cB^t_{\nu_1,\nu_2,\nu_3,\nu_4;\nu_5}$ both extend to unique continuous linear maps $\cE_{\nu_1}\hat{\otimes}_{\pi}\cE_{\nu_2}\hat{\otimes}_{\pi}\cE_{\nu_3}\hat{\otimes}_{\pi}\cE_{\nu_4}\rightarrow \cc$. They are expected to be invariant in the sense of Definition~\ref{def:InvariantForm}.

\begin{claim}[Trivial intermediate representation]\label{clm:BFactorization}
For nontrivial principal- or complementary-series labels
$\nu,\lambda$, the quadrilinear forms satisfy
\ba
\lim_{\mu\rightarrow iQ/2}
\cB^s_{\nu,\nu,\lambda,\lambda;\mu}(f_1,f_2,f_3,f_4)
&=\cD_{\nu,\nu}(f_1,f_2)\,
  \cD_{\lambda,\lambda}(f_3,f_4)\,,\\
\lim_{\mu\rightarrow iQ/2}
\cB^t_{\nu,\lambda,\lambda,\nu;\mu}(f_1,f_2,f_3,f_4)
&=\cD_{\nu,\nu}(f_1,f_4)\,
  \cD_{\lambda,\lambda}(f_2,f_3)\,,
\ea
with the limits taken through the complementary series. Thus, for pairwise equal external labels, these forms are continuous at the trivial intermediate representation.
\end{claim}

The KMS symmetry~\eqref{eq:TKMS} relates the two forms:
\be
 \cB^t_{\nu_1,\nu_2,\nu_3,\nu_4;\nu_5}(f_1,f_2,f_3,f_4) =   \cB^s_{\nu_4,\nu_1,\nu_2,\nu_3;\nu_5}(\nabla^{-1}_{\nu_4}f_4,f_1,f_2,f_3)\,.
\ee
A more interesting relation between the two follows from associativity of the tensor product of unitary representations
\be
(\pp_{\nu_2}\times \pp_{\nu_3})\times \pp_{\nu_4} = \pp_{\nu_2}\times (\pp_{\nu_3}\times \pp_{\nu_4})\,.
\ee
Suppose $\nu_2,\nu_3,\nu_4$ are all in the principal series. Composing the Clebsch--Gordan decomposition described in Claim~\ref{clm:CGPrincipal} in two different ways provides the following two unitary isomorphisms of representations
\ba
(\cH_{\nu_2}\otimes \cH_{\nu_3}) \otimes \cH_{\nu_4} &\cong  \int^{\oplus}_{\rr_{+}} \cH_{\nu_t}\otimes \cH_{\nu_4}  \,d\mu_b(\nu_t) 
\cong  \int^{\oplus}_{\rr_{+}}\int^{\oplus}_{\rr_{+}} \cH_{\nu_1}  \,d\mu_b(\nu_1) \,d\mu_b(\nu_t)\\
&\cong  \int^{\oplus}_{\rr_{+}}[L^2(\rr_{+},\mu_b(\nu_t))\otimes \cH_{\nu_1}]  \,d\mu_b(\nu_1)\,,
\label{eq:CGT}
\ea
and
\ba
\cH_{\nu_2}\otimes( \cH_{\nu_3} \otimes \cH_{\nu_4}) &\cong  \int^{\oplus}_{\rr_{+}} \cH_{\nu_2}\otimes \cH_{\nu_s}  \,d\mu_b(\nu_s) 
\cong  \int^{\oplus}_{\rr_{+}}\int^{\oplus}_{\rr_{+}} \cH_{\nu_1}  \,d\mu_b(\nu_1) \,d\mu_b(\nu_s)\\
&\cong  \int^{\oplus}_{\rr_{+}}[L^2(\rr_{+},\mu_b(\nu_s))\otimes \cH_{\nu_1}]  \,d\mu_b(\nu_1)\,.
\label{eq:CGS}
\ea
They are implemented by the unitary intertwiners
\ba
\,&C^t_{\nu_2,\nu_3,\nu_4}: \cH_{\nu_2}\otimes \cH_{\nu_3} \otimes \cH_{\nu_4} \rightarrow 
\int^{\oplus}_{\rr_{+}}[L^2(\rr_{+},\mu_b(\nu_t))\otimes \cH_{\nu_1}]  \,d\mu_b(\nu_1)\,,\\
&C^t_{\nu_2,\nu_3,\nu_4} = \mathfrak{F}_t\Big(\int_{\rr_+}^{\oplus} C_{\nu_t,\nu_4} d\mu_b(\nu_t)\Big) (C_{\nu_2,\nu_3}\otimes 1_{\nu_4})\,,
\ea
and
\ba
\,& C^s_{\nu_2,\nu_3,\nu_4}: \cH_{\nu_2}\otimes \cH_{\nu_3} \otimes \cH_{\nu_4} \rightarrow
 \int^{\oplus}_{\rr_{+}}[L^2(\rr_{+},\mu_b(\nu_s))\otimes \cH_{\nu_1}]  \,d\mu_b(\nu_1)\,,\\
& C^s_{\nu_2,\nu_3,\nu_4} = \mathfrak{F}_s\Big(\int_{\rr_+}^{\oplus} C_{\nu_2,\nu_s} d\mu_b(\nu_s)\Big) (1_{\nu_2}\otimes C_{\nu_3,\nu_4})\,.
\ea
Here $\mathfrak{F}_t,\mathfrak{F}_s$ are the canonical Fubini unitaries implementing the final rearrangements in~\eqref{eq:CGT} and~\eqref{eq:CGS}. Now, $C^s_{\nu_2,\nu_3,\nu_4}(C^t_{\nu_2,\nu_3,\nu_4})^*$ is a unitary intertwiner between the two decompositions. By uniqueness of the principal-series spectral decomposition and
Schur's lemma, it follows that
\be
C^s_{\nu_2,\nu_3,\nu_4}
=
\left[
\int_{\rr_+}^{\oplus}
\bigl(F_{\nu_1,\nu_2,\nu_3,\nu_4}\otimes 1_{\nu_1}\bigr)
\,d\mu_b(\nu_1)
\right]
C^t_{\nu_2,\nu_3,\nu_4}\,,
\label{eq:recoupling}
\ee
where $\nu_1\mapsto F_{\nu_1,\nu_2,\nu_3,\nu_4}$ is a measurable field of unitary operators
\be
F_{\nu_1,\nu_2,\nu_3,\nu_4} : L^2(\rr_{+},\mu_b(\nu_t))\rightarrow L^2(\rr_{+},\mu_b(\nu_s))\,.
\ee
$F_{\nu_1,\nu_2,\nu_3,\nu_4}$ was determined in~\cite{PonsotTeschner1999,PonsotTeschner2001}. See also~\cite{TeschnerVartanov2014} for more details. The result is:
\begin{claim}\label{clm:6jsymbol}
Let $\nu_1,\nu_2,\nu_3,\nu_4\in \rr$, and let $f\in L^2(\rr_{+},\mu_b(\nu_t))$. Then
\be
(F_{\nu_1,\nu_2,\nu_3,\nu_4} f )(\nu_s) = \int_0^{\infty}\mathscr{F}_{\nu_1,\nu_2,\nu_3,\nu_4}(\nu_s,\nu_t) f(\nu_t)\, d\mu_b(\nu_t)\,,\qquad \nu_s \in \rr_{+}\,,
\ee
where
\ba
\,&\mathscr{F}_{\nu_1,\nu_2,\nu_3,\nu_4}(\nu_s,\nu_t)\\
&\qquad= 
\frac{
s_b(\nu_s-\nu_1-\nu_2)\,
s_b(\nu_t+\nu_1+\nu_4)\,
s_b(\nu_t+\nu_1-\nu_4)\,
s_b(\nu_t+\nu_2-\nu_3)
}{
s_b(\nu_s+\nu_1+\nu_2)\,
s_b(\nu_s+\nu_1-\nu_2)\,
s_b(\nu_s-\nu_3+\nu_4)\,
s_b(\nu_t-\nu_1-\nu_4)
}\\
&\qquad\qquad\times\int\limits_{\rr}  \prod\limits_{\eta\in\{-1,1\}}\frac{s_b(x+\nu_1+\eta\, \nu_2+i\tfrac{Q}{4})s_b(x-\nu_3+\eta \,\nu_4+i\tfrac{Q}{4})}{s_b(x+\eta\,\nu_s-i\tfrac{Q}{4})s_b(x+\nu_1-\nu_3+\eta\, \nu_t-i\tfrac{Q}{4})}dx\,.
\ea
\end{claim}
The integral kernel $\mathscr{F}_{\nu_1,\nu_2,\nu_3,\nu_4}(\nu_s,\nu_t)$ is the 6j symbol for the unitary principal series representations of $\sl$.

Recall the construction~\eqref{eq:2To1Def} of the map $\cT^{\nu_1}_{\nu_2,\nu_3}$ from the Clebsch--Gordan map $C_{\nu_2,\nu_3}$. The quadrilinear invariant forms $\cB^s_{\nu_1,\nu_2,\nu_3,\nu_4;\nu_s}$,  $\cB^t_{\nu_1,\nu_2,\nu_3,\nu_4;\nu_t}$ arise by composing $\cT^{\nu_i}_{\nu_j,\nu_k}$ in the same order as the unitary intertwiners $C^s_{\nu_2,\nu_3,\nu_4}$, $C^t_{\nu_2,\nu_3,\nu_4}$ arise by composing $C_{\nu_j,\nu_k}$. The $\cB^s$ and $\cB^t$ are therefore expected to satisfy the analogous recoupling relation, differing only by the normalization $\cN_b$:
\begin{claim}
Let $\nu_1,\nu_2,\nu_3,\nu_4,\nu_s\in \rr$ and let $f\in\cE_{\nu_1}\hat{\otimes}_{\pi}\cE_{\nu_2}\hat{\otimes}_{\pi}\cE_{\nu_3}\hat{\otimes}_{\pi}\cE_{\nu_4}$. Then
\be
\cB^s_{\nu_1,\nu_2,\nu_3,\nu_4;\nu_s}(f) = \int_0^{\infty}\mathscr{F}^{\mathrm{CFT}}_{\nu_1,\nu_2,\nu_3,\nu_4}(\nu_s,\nu_t) \,\cB^t_{\nu_1,\nu_2,\nu_3,\nu_4;\nu_t}(f)\, d\mu_b(\nu_t)\,.
\label{eq:recouplingB}
\ee
Here
\be
\mathscr{F}^{\mathrm{CFT}}_{\nu_1,\nu_2,\nu_3,\nu_4}(\nu_s,\nu_t) = 
\frac{\cN_b(\nu_1,\nu_2,\nu_s)\cN_b(\nu_3,\nu_4,\nu_s)}{\cN_b(\nu_4,\nu_1,\nu_t)\cN_b(\nu_2,\nu_3,\nu_t)}
\mathscr{F}_{\nu_1,\nu_2,\nu_3,\nu_4}(\nu_s,\nu_t)\,.
\ee
\end{claim}
What is missing to prove this claim? Suppose $f_k\in \cE_{\nu_k}$ with $k=1,\ldots,4$. Apply the recoupling identity~\eqref{eq:recoupling} to $f_2\otimes f_3 \otimes f_4$ and contract both sides with $\cD_{\nu_1,\nu_1}(f_1,\cdot)$, which is a bounded functional on $\cH_{\nu_1}$. This gives the equality~\eqref{eq:recouplingB} in $L^2(\rr^2_{+},\mu_b(\nu_s)\times \mu_b(\nu_1))$, hence equality almost everywhere in $(\nu_s,\nu_1)$. To obtain the stated identity at each fixed $(\nu_s,\nu_1)$ requires additional analysis.

The recoupling identity is stated in~\eqref{eq:recouplingB} for principal-series parameters. It is expected to hold also after analytic continuation to the complementary series, provided we deform the integration contour so that no poles cross it, or, equivalently, keep track of residues arising from poles crossing the real axis.

\section{Quantum measure spaces and boundary conformal field theory}\label{sec:boundary}
In this section, we define a natural class of quantum measure spaces equipped with a measure-preserving action of the locally compact quantum group $\sl$. We then argue that they are equivalent to unitary Virasoro boundary conditions with $c>1$.

\subsection{Quantum $\gg_b$-spaces}\label{ssec:Gspaces}
Let us recall the notion of an action of a locally compact quantum group on a quantum measure space, reviewed in Section~\ref{ssec:QMS}. A quantum measurable space $\xx$ is simply a von Neumann algebra $N:=L^{\infty}(\xx)$. When $N$ comes equipped with a normal semifinite faithful weight $\theta: N_+\rightarrow [0,\infty]$, we call the pair $(\xx,\theta)$ a quantum measure space. Let $\gg$ be a locally compact quantum group with underlying von Neumann algebra $M:=L^{\infty}(\gg)$ and coproduct $\Delta:M\rightarrow M\botimes M$. A (left) action of $\gg$ on $\xx$ is defined as a normal, injective, unital $*$-homomorphism $\alpha: N\rightarrow M\botimes N$ such that
\be
(\Delta\otimes \id_N)\alpha = (\id_M\otimes\alpha)\alpha\,.
\label{eq:actionDef}
\ee
We say the action $\alpha$ preserves the weight $\theta$ if
\be
(\id_M\otimes \theta)\alpha(a) = \theta(a) 1_M\,,\qquad \text{for all }a\in N_+\text{ such that }\theta(a)<\infty\,.
\ee
It will be convenient to give a name to the above situation.
\begin{definition}[Quantum $\gg$-spaces]\label{def:quantumGspace}
Let $\gg$ be a locally compact quantum group with $M = L^{\infty}(\gg)$. A \emph{quantum $\gg$-space} is a triple $(N,\theta,\alpha)$, where
\begin{itemize}
\item[-] $N$ is a von Neumann algebra,
\item[-] $\theta:N_+\rightarrow[0,\infty]$ is a n.s.f.~weight on $N$,
\item[-] $\alpha: N\rightarrow M\botimes N$ is a $\theta$-preserving left action.
\end{itemize}
\end{definition}
Let us also repeat the definition of ergodicity.
\begin{definition}[Ergodicity]
We say that a quantum $\gg$-space $(N,\theta,\alpha)$ is \emph{ergodic} if the fixed-point von Neumann algebra $N^{\alpha}$ is trivial, i.e.~if
\be
N^{\alpha}:= \{a\in N\,:\, \alpha(a) = 1_M\otimes a\} = \cc 1_N\,.
\ee
\end{definition}
Let us fix $Q>0$ and $b$ such that $Q=b+b^{-1}$. In the rest of this section, we will specialize to $\gg = \gg_b := \sl$. We will use the notation $M := L^{\infty}(\gg_b)$, $\Delta := \Delta_{\gg_b}$.

Let $(N,\theta,\alpha)$ be a quantum $\gg_b$-space, not necessarily ergodic. Consider the GNS construction for $(N,\theta)$ and denote the GNS Hilbert space $\cK_{\theta}:=L^2(\xx,\theta)$. From now on, $N$ is realized as a von Neumann subalgebra of $\mathrm{B}(\cK_{\theta})$. Let $\Lambda_{\theta}: \mathfrak{n}_{\theta}\rightarrow \cK_{\theta}$ be the GNS map. As reviewed in Section~\ref{ssec:QMS}, the action $\alpha$ then admits a canonical unitary implementation $\uu$: a unitary representation of $\gg_b$ on $\cK_{\theta}$. In particular, $\uu \in M\botimes \mathrm{B}(\cK_{\theta})$ is a unitary operator satisfying the corepresentation identity
\be
(\Delta\otimes\id)\uu=\uu_{13}\uu_{23}\,.
\ee
Starting from $\alpha$, the operator $\uu^*$ is fully specified by the action of its slices as follows
\be
[(\omega\otimes \id)\uu^* ]\Lambda_{\theta}(a) = \Lambda_{\theta}((\omega\otimes \id)\alpha(a))\,,\qquad 
\omega\in M_*\,,\, a\in \mathfrak{n}_{\theta}\,,
\label{eq:slicedUStar}
\ee
which is just a restatement of~\eqref{eq:UStarDefinition}. Conversely, $\uu$ implements the action as follows
\be
\alpha(a) = \uu^{*}(1_M\otimes a)\uu\,,\qquad a\in N\,.
\ee

We are interested in the \emph{spectrum} of the quantum $\gg_b$-space: the decomposition of $(\uu,\cK_{\theta})$ into unitary irreducible representations of $\gg_b$. Unitary irreducible representations of $\gg_b$ were discussed in Section~\ref{sec:reps}. In particular, Conjecture~\ref{conj:unitaryDual} asserts that the unitary dual of $\gg_b$ is exhausted by the trivial representation, the complementary series, and the principal series, i.e.~$(\pp_{\nu},\cH_{\nu})$ with $\nu\in \rr_{\geq 0}\cup i(0,Q/2]$. Unitary representations of $\gg_b$ are equivalent to nondegenerate representations of the universal dual C*-algebra $C^u_0(\hat{\gg}_b)$. If this is a type-I C*-algebra, every unitary representation decomposes as an essentially unique direct integral of unitary irreducible representations.

While we expect $C^u_0(\hat{\gg}_b)$ is type I, we do not prove it here. Instead, we will simply restrict attention to quantum $\gg_b$-spaces such that $(\uu,\cK_{\theta})$ does decompose into the irreducibles featuring in Conjecture~\ref{conj:unitaryDual}. In other words, 
\be
\cK_{\theta} \cong \int^{\oplus}_{\cS} ( \cH_{\nu}\otimes \cM_{\nu}) d\mu_{\uu}(\nu)\,,\qquad
\uu\simeq \int^{\oplus}_{\cS}( \pp_{\nu}\otimes 1_{\cM_\nu}) d\mu_{\uu}(\nu)\,.
\label{eq:spectralDecomposition0}
\ee
Here $\cS = \rr_{\geq 0}\cup i(0,Q/2]$, and $\mu_{\uu}$ is a spectral measure of $\uu$. $(\cM_{\nu})_{\nu\in\cS}$ is a measurable field of Hilbert spaces, accounting for multiplicity at spectral parameter $\nu$. The decomposition can be phrased in terms of the quantum Laplace operators of~\eqref{eq:Casimir}. Acting on $\cK_\theta$ through the representation $\uu$, the Casimirs $C$ and $\check{C}$ become unbounded operators commuting with $\uu$, which act on the summand $\cH_\nu\otimes\cM_\nu$ by the scalars~\eqref{eq:CasimirEigenvalue}. The spectrum of irreducible representations in $\cK_\theta$ is therefore encoded in the joint spectrum of $C$ and $\check{C}$, the quantum analogue of the Laplace spectrum of a hyperbolic surface, up to the coincidences discussed below~\eqref{eq:CasimirEigenvalue}.

Let us illustrate the preceding discussion by constructing several examples of quantum $\gg_b$-spaces, all admitting the decomposition~\eqref{eq:spectralDecomposition0}.
\begin{example}[The trivial quantum $\gg_b$-space]\label{ex:trivialGSpace}
The simplest example is the trivial $\gg_b$-space. It has $N = \cc$, with weight $\theta(z)=z$ and action $\alpha(z) = z 1_M$. Its GNS Hilbert space is one-dimensional $\cK_{\theta}=\cc$, with GNS map $\Lambda_\theta(z) = z$. The unitary implementation is given by $\uu = 1_M$. It follows that the spectral decomposition~\eqref{eq:spectralDecomposition0} of $(N,\theta,\alpha)$ is supported at the trivial representation with unit multiplicity, i.e.
\be
\cK_{\theta} \cong \cH_{i\frac{Q}{2}}\,,\qquad \uu \simeq \pp_{i\frac{Q}{2}}\,.
\ee
Under the correspondence between quantum $\gg_b$-spaces and Virasoro boundary conditions, this example is equivalent to the elementary ZZ boundary condition of Liouville CFT~\cite{ZZ2001}.
\end{example}
\begin{example}[Matrix algebras]\label{ex:ZZstack}
Let us generalize the previous example to $N = M_r(\cc)$, the algebra of $r\times r$ matrices. The weight is given by $\theta(a) =\mathrm{tr}(a)/r$, and the action is still trivial $\alpha(a) = 1_M\otimes a$. The GNS Hilbert space $\cK_{\theta}$ coincides with $M_r(\cc)$ with the (normalized) Hilbert--Schmidt norm, i.e.
\be
\cK_{\theta} = \mathrm{HS}(\cc^r) \cong \cc^{r^2}\,.
\ee
$N$ acts on $\cK_{\theta}$ by left multiplication, and the GNS map is the canonical identification $M_r(\cc)= \mathrm{HS}(\cc^r)$. The unitary implementation of $\alpha$ is again trivial $\uu = 1_M\otimes 1_{\cK_{\theta}}$. The spectral decomposition~\eqref{eq:spectralDecomposition0} is thus supported on the trivial representation, with $M_r(\cc)$ serving as the multiplicity space
\be
\cK_{\theta} \cong \cH_{i\frac{Q}{2}}\otimes M_r(\cc) \,,\qquad \uu \simeq \pp_{i\frac{Q}{2}} \otimes 1_{\cK_{\theta}}\,.
\ee
This example is equivalent to a stack of $r$ coincident ZZ branes in the Liouville CFT. The matrix entries represent open strings stretching between a pair of branes. It is ergodic if and only if $r=1$, in which case it reduces to the previous example.
\end{example}
The following gives a simple class of examples with infinite-dimensional $N$.
\begin{example}\label{ex:innerGSpace}
Let $\sigma \in \rr_{\geq 0}\cup i(0,Q/2]$, and set $N_{\sigma} = \mathrm{B}(\cH_{\sigma})$, where $\cH_{\sigma}$ is the underlying Hilbert space of the irreducible representation $\pp_\sigma$. Define a n.s.f.~weight $\theta_{\sigma}$ by
\be
\theta_{\sigma}(a) := \mathrm{tr}(\nabla^{\frac{1}{2}}_{\sigma}a\nabla^{\frac{1}{2}}_{\sigma})\,,\qquad a \in (N_{\sigma})_+\,,
\ee
where $\nabla_{\sigma}$ is the modular operator on $\cH_{\sigma}$~\eqref{eq:nablaNu}. Define the action $\alpha_\sigma$ of $\gg_b$ on $N_\sigma$ by
\be
\alpha_{\sigma}(a) := \pp^*_{\sigma}(1_M \otimes a) \pp_{\sigma}\,,\qquad a\in N_{\sigma}\,.
\ee
$\alpha_{\sigma}$ satisfies~\eqref{eq:actionDef} since $\pp_{\sigma}$ is a corepresentation.
Invariance of $\theta_{\sigma}$ under $\alpha_{\sigma}$ follows from the scaling group action on $\pp_{\sigma}$, i.e.~\eqref{eq:tauPnu}, together with the antipode identity
\be
(R\tau_{-i/2}\otimes \id )\pp_{\sigma} = \pp^*_{\sigma}\,,
\ee
where $R$ and $\tau$ are the unitary antipode and the scaling group of $\gg_b$. The GNS Hilbert space then canonically coincides with the space of Hilbert--Schmidt operators on $\cH_{\sigma}$
\be
\cK_{\theta_{\sigma}} \cong \mathrm{HS}(\cH_{\sigma}) \cong \cH_{\sigma}\otimes \overline{\cH_{\sigma}} \cong \cH_{\sigma}\otimes\cH_{\sigma}\,.
\ee
The GNS map is $\Lambda_{\theta_\sigma}(a)=a\nabla_\sigma^{1/2}$, and $N_\sigma$ acts by left multiplication. The unitary implementation $\uu_{\sigma}$ of $\alpha_{\sigma}$ coincides with the tensor product representation
\be
\uu_{\sigma} = \pp_{\sigma}\times \pp_{\sigma}\,.
\ee
It follows that the spectral decomposition~\eqref{eq:spectralDecomposition0} of the $\gg_b$-space $(N_\sigma,\theta_\sigma,\alpha_\sigma)$ is nothing other than the Clebsch--Gordan decomposition of the tensor square. When $\sigma$ is in the principal series, Claim~\ref{clm:CGPrincipal} implies that the spectral decomposition~\eqref{eq:spectralDecomposition0} is supported on the principal series with unit multiplicity
\be
\cK_{\theta_\sigma} \cong \int^{\oplus}_{\rr_+} \cH_{\nu} \,d\mu_b(\nu)\,,\qquad
\uu_\sigma\simeq \int^{\oplus}_{\rr_+}\pp_{\nu} \,d\mu_b(\nu)\,.
\ee
For $\sigma$ in the complementary series, the Clebsch--Gordan decomposition can contain atomic complementary series terms.

The quantum $\gg_b$-spaces $(N_\sigma,\theta_\sigma,\alpha_\sigma)$ with $\sigma\in \rr$ are equivalent to the FZZT boundary conditions in the Liouville CFT~\cite{FZZ2000,Teschner2000Boundary}. Their continuation to the complementary series was considered in~\cite{Teschner2000Boundary,Teschner2004BoundaryPerturbations}. For $\sigma = i\frac{Q}{2}$, this example reduces to the trivial quantum $\gg_b$-space, and thus the ZZ boundary condition.
\end{example}
The previous three examples admit a common generalization obtained by replacing $(\pp_\sigma,\cH_{\sigma})$ with any finite direct sum of trivial, complementary-series and principal-series representations. Call the resulting representation $(\pp,\cH)$, take $N = \mathrm{B}(\cH)$, and generalize $\theta_\sigma$ and $\alpha_\sigma$ in the obvious way. The resulting quantum $\gg_b$-space is equivalent to the composite boundary condition in Liouville CFT obtained by a finite stack of ZZ and FZZT branes. It is ergodic if and only if $(\pp,\cH)$ is irreducible, i.e.~if the boundary condition is elementary.

\subsection{Spectral compactness}\label{ssec:spectralCompactness}
As discussed in Section~\ref{sec:bootstrapClassical}, compact hyperbolic surfaces provide a particularly pleasant setting for spectral theory of automorphic forms on $\mathrm{PSL}(2,\rr)$. Their spectrum is discrete, contains the zero eigenvalue, and each multiplicity space is finite-dimensional. The same spectral properties are also desirable in conformal field theory, where they correspond to the notion of compact CFT.

We will therefore focus on quantum $\gg_b$-spaces with precisely these spectral properties. They can be thought of as quantum analogues of homogeneous spaces $\mathrm{PSL}(2,\rr)/\Gamma$, where $\Gamma < \mathrm{PSL}(2,\rr)$ is a cocompact lattice. We will see that they correspond to compact Virasoro boundary conditions.

\begin{definition}[Spectral compactness]\label{def:spectralCompactness}
Let $(N,\theta,\alpha)$ be a quantum $\gg_b$-space admitting a spectral decomposition of the form~\eqref{eq:spectralDecomposition0}. We say $(N,\theta,\alpha)$ is \emph{spectrally compact} if $\theta$ is a state and the spectral measure $\mu_\uu$ is purely atomic, its set of atoms has no accumulation points in $\cS$, and the multiplicity space at each atom is finite-dimensional.
\end{definition}

From now on, we will assume that $(N,\theta,\alpha)$ is a spectrally compact quantum $\gg_b$-space.\footnote{However, much of the ensuing discussion can be adapted to the case of general decomposition~\eqref{eq:spectralDecomposition0}.} Let us write the spectral decomposition as
\be
\cK_{\theta} \cong \bigoplus\limits_{i \in I} \cH_{\nu_i}\,,\qquad
\uu \simeq  \bigoplus\limits_{i \in I} \pp_{\nu_i}\,.
\label{eq:spectralDecomposition1}
\ee
Here $I$ is a countable indexing set and $\nu_i \in \cS $ for all $i\in I$. We call the indexed family
$(\nu_i)_{i\in I}$ the \emph{spectrum} of $(N,\theta,\alpha)$, with repetitions recording multiplicities.
 
\begin{lemma}
A spectrally compact quantum $\gg_b$-space is ergodic if and only if the trivial representation appears in $\cK_{\theta}$ with unit multiplicity.
\end{lemma}
\begin{proof}
Let $\cK^\uu_\theta \subseteq \cK_{\theta}$ be the subspace of $\gg_b$-invariant vectors. $\mathrm{dim}(\cK^\uu_\theta)$ is precisely the multiplicity of the trivial representation in $\cK_\theta$. Since $\theta$ is a faithful normal invariant state, we have
\be
\cK^\uu_\theta = \overline{\Lambda_{\theta}(N^{\alpha})}\,.
\ee
Since $1_N \in N^{\alpha}$, we have $\mathrm{dim}(\cK^\uu_\theta)\geq 1$, with equality if and only if $\alpha$ is ergodic. In that case, $\cK^\uu_\theta$ is spanned by $\Lambda_{\theta}(1_N)$.
\end{proof}

We will see that ergodic spectrally compact quantum $\gg_b$-spaces correspond to elementary compact Virasoro boundary conditions, i.e.~precisely those that feature the Virasoro vacuum module with unit multiplicity. Conversely, dropping the ergodicity assumption naturally incorporates non-elementary boundary conditions in our framework. We have already encountered these in Example~\ref{ex:ZZstack}.

From now on, assume that $(N,\theta,\alpha)$ is ergodic and spectrally compact. We label the unique trivial summand by $0\in I$, so that
$\nu_0=i\frac{Q}{2}$, while $\nu_i\in\rr_{\geq 0}\cup i(0,Q/2)$ for all $i\neq 0$. The spectral decomposition~\eqref{eq:spectralDecomposition1} is equivalent to the existence of equivariant embeddings, together with their adjoints
\be
\psi_i : \cH_{\nu_i}\rightarrow \cK_{\theta}\,,\qquad 
\psi^*_i: \cK_{\theta} \rightarrow \cH_{\nu_i}\,,\qquad i \in I\,.
\ee
The embeddings $\psi_i$ are isometric, mutually orthogonal, and complete:
\be
\psi_{i}^*\psi_i = 1_{\nu_i},\quad i\in I; \qquad \psi_{i}^*\psi_j = 0,\quad \; i\neq j\,;\qquad \sum\limits_{i\in I} \psi_i \psi^*_i = 1_{\cK_{\theta}}\,,
\ee
where $1_{\nu_i}$ denotes the identity on $\cH_{\nu_{i}}$, and the sum over $i \in I$ converges in the strong operator topology. Finally, $\psi_i$ intertwines $\pp_{\nu_{i}}$ with $\uu$:
\be
\uu(1_M\otimes \psi_i) = (1_M\otimes \psi_i) \pp_{\nu_{i}}\,, \qquad  i\in I\,.
\label{eq:psiIntertwinesU}
\ee
There is some freedom in choosing $\psi_i$, equivalent to the choice of an orthonormal basis for each multiplicity space $\cM_{\nu_i}$. We will soon discuss how this freedom is reduced by imposing compatibility with the natural real structures. For the trivial summand, we fix the remaining phase by requiring
\be
\psi_0(1) = \Omega_{\theta} := \Lambda_{\theta}(1_N)\,.
\label{eq:psi0Def}
\ee
We have
\be
\Lambda_{\theta}(a) =  a \Omega_{\theta}\qquad\text{for all}\qquad a\in N\,.
\ee

\subsection{Real structure and modular flow}\label{ssec:realmodularPsi}
Every quantum $\gg_b$-space comes equipped with a natural real structure, generalizing pointwise complex conjugation of functions. As reviewed in Section~\ref{ssec:QMSrealmodular}, it is implemented by the modular conjugation $J_{\theta}: \cK_{\theta}\rightarrow \cK_{\theta}$. $J_{\theta}$ is a canonical antilinear involution arising from the polar decomposition of the Tomita operator $S_{\theta}\Lambda_\theta(a) = \Lambda_\theta(a^*)$ as follows:
\be
S_{\theta} = J_{\theta}\nabla^{\frac{1}{2}}_{\theta}\,,
\ee
where $\nabla_{\theta}$ is the modular operator. Propositions~\ref{prop:UJ} and~\ref{prop:UNabla} then state that both $J_{\theta}$ and $\nabla_{\theta}$ are compatible with the unitary implementation $\uu$ of $\alpha$:
\begin{align}
\,&(\hat{J}\otimes J_{\theta}) \uu (\hat{J}\otimes J_{\theta}) = \uu^*\,,\\
&(\hat{\nabla}^{i t}\otimes \nabla^{i t}_{\theta}) \uu (\hat{\nabla}^{-i t}\otimes \nabla^{-i t}_{\theta})= \uu\,,\qquad t\in\rr\,.
\label{eq:tauU}
\end{align}
The first equation means that $(\uu,\cK_{\theta})$ is a real representation, with $J_{\theta}$ providing the real structure in the sense of Definition~\ref{def:realStructure}.

The trivial, complementary-series, and principal-series representations $(\pp_\nu,\cH_{\nu})$ of $\gg_b$ satisfy very similar compatibility conditions. Indeed, Claims~\ref{clm:Jnu} and~\ref{clm:tauNu} state that
\begin{align}
\,&(\hat{J}\otimes J_{\nu}) \pp_{\nu} (\hat{J}\otimes J_{\nu}) = \pp^*_{\nu}\,,\\
&(\hat{\nabla}^{i t}\otimes \nabla^{i t}_{\nu}) \pp_{\nu} (\hat{\nabla}^{-i t}\otimes \nabla^{-i t}_{\nu})= \pp_{\nu}\,,\qquad t\in\rr\,.
\label{eq:tauP}
\end{align}
In particular, $(\pp_\nu,\cH_{\nu})$ is a real representation, with $J_{\nu}$ providing the real structure. $J_{\nu}$ and $\nabla_{\nu}$ are described explicitly in Section~\ref{ssec:JNablaP}.

The following proposition states that, much like their classical counterparts~\eqref{eq:psiClassical}, the embeddings $\psi_i$ can be chosen to be compatible with the real structure.
\begin{proposition}
We can choose the embeddings $\psi_i: \cH_{\nu_i}\rightarrow \cK_{\theta}$ so that
\be
J_{\theta}\psi_i = \psi_i J_{\nu_i}\quad\text{for all}\quad i \in I\,.
\label{eq:psiReal}
\ee
\end{proposition}
\begin{proof}
Fix $\nu$ in the spectrum and consider the finite-dimensional multiplicity space $\cM_\nu
:= \operatorname{Hom}_{\gg_b}(\cH_\nu,\cK_\theta)$. By Schur's lemma, it carries the inner product determined by
$\psi^*\chi=\langle\psi,\chi\rangle\,1_{\cH_\nu}$.
The reality identities for $\uu$ and $\pp_\nu$ imply that $c_\nu(\psi):=J_\theta\psi J_\nu$ preserves $\cM_\nu$. Moreover, $c_\nu$ is an antiunitary involution. We can therefore choose an orthonormal basis of $\cM_\nu$ fixed by $c_\nu$. Its elements are mutually orthogonal isometric intertwining embeddings satisfying $J_\theta\psi_i=\psi_i J_\nu$. The definition~\eqref{eq:psi0Def} of $\psi_0$ satisfies $J_{\theta}\psi_0 = \psi_0 J_{i\frac{Q}{2}}$ because $J_{\theta}\Omega_{\theta}=\Omega_{\theta}$ and $J_{i\frac{Q}{2}}(z) = \zb$.
\end{proof}
From now on, we assume that the embeddings $\psi_i$ satisfy~\eqref{eq:psiReal}. For each nontrivial spectral label $\nu$, the remaining freedom consists of orthogonal changes of basis in the multiplicity space $\cM_\nu$, parametrized by $\mathrm{O}(\dim\cM_\nu,\rr)$.

The modular flow $\nabla^{it}_{\theta}:\cK_{\theta}\rightarrow\cK_{\theta}$ induces a one-parameter group on each multiplicity space. Compatibility with the real structure forces this group to act by real orthogonal matrices. The following proposition makes this precise:
\begin{proposition}\label{prop:modularFlowPsi}
Let $I_{\nu} = \{i\in I: \nu_i = \nu\}$ and suppose that the embeddings $\psi_i$ satisfy~\eqref{eq:psiReal}. For each $\nu$ in the spectrum, there exists a continuous homomorphism
\be
A_{\nu}:(\rr,+)\rightarrow \mathrm{SO}(|I_{\nu}|,\rr)
\ee
such that
\be
\nabla^{i t}_\theta \,  \psi_j = \sum\limits_{k\in I_{\nu}}(A_{\nu}(t))_{kj} \psi_k \, \nabla^{i t}_\nu\,,\qquad
j\in I_\nu\,,\, t\in\rr\,.
\ee
In particular, if the spectrum is multiplicity-free, then
\be
\nabla^{i t}_\theta \,  \psi_j = \psi_j \, \nabla^{it}_\nu\,,\qquad j\in I_\nu\,,t\in\rr\,.
\ee
\end{proposition}
\begin{proof}
Modular covariance~\eqref{eq:tauU},~\eqref{eq:tauP} implies that $a_\nu(t)\psi:=\nabla_\theta^{it}\psi \nabla^{-it}_\nu$ defines a continuous homomorphism from $(\rr,+)$ to the unitary group of $\cM_\nu$. The identities
\be
J_\theta\nabla_\theta^{it}J_\theta=\nabla_\theta^{it},
\qquad
J_\nu \nabla^{it}_\nu J_\nu= \nabla^{it}_\nu
\ee
show that $a_\nu(t)$ commutes with $c_\nu$. Its matrix $A_\nu(t)$ in the basis satisfying~\eqref{eq:psiReal} is therefore real orthogonal. Since $a_{\nu}(0)=1$, continuity implies $\det A_{\nu}(t)= 1$. For multiplicity one, $\mathrm{SO}(1,\rr) = \{1\}$ gives the final assertion. For the trivial representation, the statement also follows from preservation of the vacuum vector by the modular group, $\nabla^{i t}_\theta \Omega_{\theta} = \Omega_{\theta}$.
\end{proof}

The previous proposition has an important structural consequence:
\begin{theorem}\label{thm:typeIII}
Suppose $(N,\theta,\alpha)$ is an ergodic spectrally compact quantum $\gg_b$-space. If $N$ is nontrivial, it is a type III$_1$ factor.
\end{theorem}
Let $N_\theta$ denote the centralizer of $\theta$, i.e.~the fixed-point algebra of its modular automorphism group $\sigma_t^\theta$. The proof of Theorem~\ref{thm:typeIII} will follow from:
\begin{lemma}[Lemma~2.1 of~\cite{MarrakchiVaes:2024ErgodicStates}]
A nontrivial von Neumann algebra admitting a faithful normal state with trivial centralizer is a type~III$_1$ factor.
\end{lemma}
\begin{proof}[Proof of Theorem~\ref{thm:typeIII}]
It remains to show that $N_\theta=\cc1_N$. On the $\nu$-isotypic component of~$\cK_{\theta}$, Proposition~\ref{prop:modularFlowPsi} gives $\nabla_\theta^{it}\simeq \nabla_\nu^{it}\otimes A_\nu(t)$. Since the multiplicity space is finite-dimensional, $A_\nu(t)$ can be diagonalized into characters $e^{it\lambda_r}$. For every nontrivial $\nu$, the operator $\log\nabla_\nu=2\pi Q\pbf$ has no eigenvalues. Thus none of the groups $e^{it\lambda_r}\nabla_\nu^{it}$ has a nonzero invariant vector. Ergodicity implies that
the trivial representation occurs with unit multiplicity. It follows that
\be
\{\xi\in\cK_\theta:\nabla_\theta^{it}\xi=\xi
\text{ for all }t\in\rr\}=\cc\Omega_\theta\,.
\ee
For $a\in N_\theta$, we have $\nabla_\theta^{it}\Lambda_\theta(a)=\Lambda_\theta(\sigma_t^\theta(a))=\Lambda_\theta(a)$. Consequently, $\Lambda_\theta(a)\in\cc\Omega_\theta$, and faithfulness of $\theta$ implies $a\in\cc1_N$. Thus $N_\theta=\cc1_N$.
\end{proof}

Let us give a name to quantum $\gg_b$-spaces whose modular flow acts trivially on the multiplicity spaces. This condition plays an important role in the correspondence with conformal boundary conditions. As we will see, it ensures the expected cyclicity of boundary correlation functions, without additional mixing of equal-weight fields as they are transported around the disk. An analogous condition appears under the name ``symmetry'' in the FRS construction of RCFTs~\cite{Fuchs:2002cm}. We adopt the same terminology:
\begin{definition}[Symmetry]\label{def:symmetry}
An ergodic spectrally compact quantum $\gg_b$-space $(N,\theta,\alpha)$ is called \emph{symmetric} if
\be
\nabla^{it}_\theta\psi_j
=\psi_j\nabla^{it}_{\nu_j}
\qquad\text{for all }j\in I,\ t\in\rr\,,
\ee
or, equivalently, if $A_\nu(t)=1$ for every spectral label $\nu$.
\end{definition}
If the spectrum is multiplicity-free, symmetry is automatic.

 \subsection{Quantum automorphic forms}\label{ssec:automorphicForms}
 In analogy with the classical case, one expects that compatibility of the spectral decomposition~\eqref{eq:spectralDecomposition1} with associativity of algebra multiplication in $N$ imposes strong constraints on the possible spectra of quantum $\gg_b$-spaces. To implement this idea, one wants to work with elements of $N$ which transform irreducibly under the action $\alpha$. In other words, one wants to lift elements of $\psi_i(\cH_{\nu_i})\subset \cK_{\theta}$ to $N$ via the inverse GNS map $\Lambda^{-1}_{\theta}$.
 
For $i \in I$ and general $f\in \cH_{\nu_i}$, we do not expect $\psi_i(f)$ to belong to the image of $\Lambda_{\theta}$. This is familiar from the classical discussion, where $\psi_i(f)$ is not necessarily a bounded function on $\mathrm{PSL}(2,\rr)/\Gamma$. Thus, let us define the subspaces of $\cH_{\nu_i}$ whose elements can be lifted to the algebra:
\be
\cV_{i}:= \{f\in \cH_{\nu_i} \,:\, \psi_i(f) \in \Lambda_{\theta}(N) \}\,,\qquad i\in I\,.
\ee
Restricted to $\cV_i$, the embeddings $\psi_i$ admit the lifts
\be
\Psi_i : \cV_{i} \rightarrow N\,,\qquad \Psi_i(f) := \Lambda^{-1}_{\theta}\psi_i(f)\,,\qquad i\in I\,.
\ee
The definition determines the action of $\Psi_i(f)$ on the vacuum:
\be
\Psi_i(f)\Omega_\theta = \psi_i(f)\,,\qquad i\in I\,, \,f\in\cV_{i}\,.
\ee
The following lemma describes the sense in which $\Psi_i$ is equivariant.
\begin{lemma}\label{lem:PsiEquivariance}
Let $\omega \in M_*$ and $i\in I$. The slice action $(\omega\otimes\id)\pp^*_{\nu_i}$ preserves $\cV_{i}$. Furthermore, we have:
\be
\Psi_i([(\omega\otimes\id)\pp^*_{\nu_i}]f)=(\omega\otimes\id)\alpha(\Psi_i(f))\,,\qquad \omega\in M_*\,,\,i\in I\,,\, f\in \cV_{i}\,.
\ee
\end{lemma}
\begin{proof}
Let $f \in \cV_{i}$ and let $a\in N$ be such that $\Lambda_{\theta}(a) = \psi_i(f)$. Let $\omega\in M_*$. The intertwining property of $\psi_i$~\eqref{eq:psiIntertwinesU} and the defining property of $\uu^*$~\eqref{eq:slicedUStar} imply:
\be
\psi_i([(\omega\otimes\id)\pp^*_{\nu_i}]f) = [(\omega\otimes\id)\uu^*]\Lambda_{\theta}(a) =
\Lambda_{\theta}((\omega\otimes\id)\alpha(a))\,.
\label{eq:actionUpsiI}
\ee
It follows that $[(\omega\otimes\id)\pp^*_{\nu_i}]f\in \cV_i$. The second part follows by applying $\Lambda^{-1}_{\theta}$ to~\eqref{eq:actionUpsiI}.
\end{proof}

The discussion so far in principle allows the unfortunate possibility that $\cV_i$ contains no nonzero vectors. Happily, this does not happen for classical compact quotients $\mathrm{PSL}(2,\rr)/\Gamma$. Indeed, in that case, restricting $f$ to the dense spaces of $K$-finite or smooth vectors ensures that their images are smooth, hence bounded, functions on the compact quotient. Taking $f$ to be $K$-finite then directly leads to the classical definition of an automorphic form on $\mathrm{PSL}(2,\rr)/\Gamma$.

In the quantum setting, we replace the spaces of $K$-finite or smooth vectors with the dense spaces of analytic test vectors $\cE_{\nu}\subset\cH_{\nu}$ introduced in Definition-Lemma~\ref{def:ESpace}. For a fixed quantum $\gg_b$-space, we use the shorthand $\sE_i:=\cE_{\nu_i}\subset\cH_{\nu_i}$, $i\in I$. These spaces have the following useful property:
\begin{claim}\label{clm:EautomaticInclusion}
Let $i \in I$ and suppose $\cV_i \neq \{0\}$. Then $\sE_{i}\subseteq \cV_{i}$.
\end{claim}
\begin{proof}
Let $f$ be a nonzero element of $\cV_i$. By Lemma~\ref{lem:PsiEquivariance}, we have
\be
\{[(\omega\otimes \id)\pp^*_{\nu_i}]f \,: \, \omega \in M_*\} \subseteq \cV_i\,.
\ee
The conclusion then follows from Claim~\ref{clm:EsliceReachability}.
\end{proof}

 The discussion so far motivates the following definition.
 \begin{definition}[Quantum automorphic forms]\label{def:quantumAutomorphicForms}
 Let $(N,\theta,\alpha)$ be an ergodic spectrally compact quantum $\gg_b$-space. Define the vector subspaces of $N$
 \be
 \cA_i =\Psi_{i}(\cV_i \cap \sE_{i}) = \{a\in N\,:\, \Lambda_{\theta}(a) \in \psi_i(\sE_{i })\} \,,\qquad i\in I\,.
 \ee
 We call elements of $\cA_i$ \emph{quantum automorphic forms}.
 \end{definition}

Claim~\ref{clm:EautomaticInclusion} implies that $\cA_i \neq \{0\}$ if and only if $\cV_i \neq \{0\}$. In analogy with the classical setting, we expect that $\cV_i \neq \{0\}$ for all $i\in I$. Since we do not have a proof, we instead give a definition and a conjecture:
\begin{definition}[Regularity]\label{def:regularity}
We say an ergodic spectrally compact quantum $\gg_b$-space is \emph{regular} if $\sE_i \subseteq \cV_i $ for all $i \in I$.
\end{definition}
Regularity is independent of the choice of embeddings $\psi_i$ -- it is a genuine property of $(N,\theta,\alpha)$.
\begin{conjecture}\label{conj:regularity}
Every ergodic spectrally compact quantum $\gg_b$-space is regular.
\end{conjecture}

From now on, we will assume that $(N,\theta,\alpha)$ is regular (as well as ergodic and spectrally compact). Thus $\cA_i = \Psi_i(\sE_{i})$ for all $i\in I$. We will view $\Psi_i$ as a linear map $\sE_{i}\rightarrow N$ by restriction.
\begin{lemma}\label{lem:PsiCts}
Suppose $i\in I$. Then $\Psi_i:\sE_{i} \rightarrow N$ is continuous with respect to the Fr\'{e}chet topology on $\sE_{i}$ and the operator norm on $N$.
\end{lemma}
\begin{proof}
We apply the closed graph theorem to the linear map $\Psi_i: \sE_{i}\rightarrow N$. Suppose $f_n\rightarrow f$ in $\sE_{i}$ and $\Psi_i(f_n)\rightarrow a$ in the norm on $N$. Since the embedding $\sE_{i}\hookrightarrow \cH_{\nu_{i}}$ is continuous and $\psi_i$ is an isometry, we have $\|\psi_i(f_n)-\psi_i(f)\|_{\theta}\rightarrow 0$. At the same time,
\be
\|\psi_i(f_n)-\Lambda_{\theta}(a)\|_{\theta}=\|\Lambda_{\theta}(\Psi_i(f_n)-a)\|_{\theta} \leq \|\Psi_i(f_n)-a\| \rightarrow 0\,.
\ee
It follows that $\psi_i(f) = \Lambda_{\theta}(a)$, and hence $\Psi_i(f) = a$ by injectivity of $\Lambda_{\theta}$. The closed graph theorem implies $\Psi_i$ is continuous.
\end{proof}

\subsection{Correlations}\label{ssec:correlations}
The correspondence with conformal field theory suggests that we study $(N,\theta,\alpha)$ through the lens of correlations. These are defined by applying the state $\theta$ to a product of quantum automorphic forms:
\begin{definition}[$n$-point correlations]
Let $(N,\theta,\alpha)$ be a regular, ergodic, spectrally compact quantum $\gg_b$-space. Let $n\geq 1$, and $i_1,\ldots,i_n \in I$. The \emph{$n$-point correlation} is defined as the following multilinear form on $\sE_{i_1}\times\ldots \times \sE_{i_n}$
\ba
\cC^{(n)}_{i_1,\ldots,i_n}
&: \sE_{i_1}\times\cdots\times\sE_{i_n}\rightarrow\cc\,,\\
\cC^{(n)}_{i_1,\ldots,i_n}(f_1,\ldots,f_n)
&:= \theta\bigl(\Psi_{i_1}(f_1)\cdots\Psi_{i_n}(f_n)\bigr)\\
&= \bigl\langle \Omega_\theta,\,
\Psi_{i_1}(f_1)\cdots\Psi_{i_n}(f_n)\Omega_\theta
\bigr\rangle_\theta\,,
\ea
where $\Psi_{i_1}(f_1)\ldots \Psi_{i_n}(f_n)$ is the algebra product of $\Psi_{i_1}(f_1),\ldots,\Psi_{i_n}(f_n)\in N$, and $\langle\cdot,\cdot\rangle_{\theta}$ is the inner product on $\cK_{\theta}$.
\end{definition}
Since $\theta$ is a state, all correlations are finite. In fact, more is true:
\begin{proposition}[Continuity of correlations]
$\cC^{(n)}_{i_1,\ldots,i_n}: \sE_{i_1}\times\ldots \times \sE_{i_n}\rightarrow \cc$ is jointly continuous with respect to the Fr\'{e}chet topologies.
\end{proposition}
\begin{proof}
Since $\theta$ is a state, we have $|\theta(a)|\leq \|a\|$, where $\|\cdot\|$ is the operator norm. Thus
\ba
|\cC^{(n)}_{i_1,\ldots, i_n}(f_1,\ldots,f_n)| &= |\theta(\Psi_{i_1}(f_1)\ldots\Psi_{i_n}(f_n))|\\
&\leq \|\Psi_{i_1}(f_1)\ldots\Psi_{i_n}(f_n)\| \leq \prod\limits_{k=1}^{n} \|\Psi_{i_k}(f_k)\|\,.
\ea
The proposition then follows from Lemma~\ref{lem:PsiCts}.
\end{proof}
It follows that each correlation extends uniquely to a continuous linear form
\be
\cC^{(n)}_{i_1,\ldots, i_n} : \sE_{i_1}\hat{\otimes}_{\pi}\ldots\hat{\otimes}_{\pi}\sE_{i_n} \rightarrow \cc\,.
\ee

Recall the notion of $\gg_b$-invariance of a continuous multilinear form from Definition~\ref{def:InvariantForm}. The following important proposition is a consequence of invariance of $\theta$ under $\alpha$ and equivariance of $\Psi_{i}$.
\begin{proposition}[Invariance of correlations]
All $n$-point correlations are $\gg_b$-invariant.
\end{proposition}
\begin{proof}
Invariance of $\theta$ under $\alpha$ means
\be
(\omega\otimes\theta)\alpha(a) = \omega(1_M)\theta(a)\,,\qquad \omega\in M_*,\,a\in N\,.
\ee
Let $i_1,\ldots,i_n\in I$ and $f_k \in \sE_{i_k}$ with $k=1,\ldots,n$. Write $b_k:=\Psi_{i_k}(f_k)$. By multiplicativity of $\alpha$ and invariance of $\theta$, we have
\be
(\omega\otimes\theta)[\alpha(b_1)\ldots \alpha(b_n)] = \omega(1_M)\theta(b_1\ldots b_n) = \omega(1_M)\, \cC^{(n)}_{i_1,\ldots,i_n}
(f_1\otimes\ldots\otimes f_n)\,.
\label{eq:corInv0}
\ee
Let $(e_p)_{p=1}^{\infty}$ be an orthonormal basis of $L^2(\gg_b)$ with $e_p\in \sA$, and let $B^{(k)}_{pq} := (\omega_{e_p,e_q}\otimes \id)\pp^*_{\nu_{i_k}}$, where $\omega_{\eta,\xi}(b):=\langle\eta, b\xi\rangle$. Let $\pi_R$ be the orthogonal projection onto $\operatorname{span}\{e_1,\ldots,e_R\}$ and set $\Pi_R:=\pi_R\otimes 1_N$. By Lemma~\ref{lem:PsiEquivariance}, we have
\ba
&\sum_{r_1,\ldots,r_{n-1}=1}^{R}
\cC^{(n)}_{i_1,\ldots,i_n}
\bigl(B^{(1)}_{pr_1}f_1,\ldots,
B^{(n)}_{r_{n-1}q}f_n\bigr)\\
&\qquad=
\left\langle e_p\otimes\Omega_\theta,\,
\alpha(b_1)\Pi_R\alpha(b_2)\ldots \Pi_R\alpha(b_n)
(e_q\otimes\Omega_\theta)\right\rangle.
\ea
Since $\pi_R\rightarrow 1$ strongly as $R\rightarrow \infty$, the right-hand side converges to
\be
(\omega_{e_p,e_q}\otimes\theta)\alpha(b_1\ldots b_n)
=  \delta_{pq}\, \cC^{(n)}_{i_1,\ldots,i_n}
(f_1\otimes\ldots\otimes f_n)\,,
\ee
where we used~\eqref{eq:corInv0}. This proves the $B$-slice part of invariance, i.e.~\eqref{eq:defInvB}.

To prove the $A$-slice part of invariance, i.e.~\eqref{eq:defInvA}, first note that
\be
\overline{\cC^{(n)}_{i_1,\ldots,i_n}(f_1\otimes \ldots\otimes f_n)}
=
\cC^{(n)}_{i_n,\ldots,i_1}
(J_{\nu_{i_n}}f_n\otimes \ldots\otimes J_{\nu_{i_1}}f_1).
\label{eq:corJReality}
\ee
Indeed, setting $a_k:=\Psi_{i_k}(J_{\nu_{i_k}}f_k)$, compatibility with the real structures gives $a_k\Omega_\theta=(J_\theta b_kJ_\theta)\Omega_\theta$. Since $J_\theta b_kJ_\theta\in N'$ commutes with every $a_\ell$, we obtain $a_n\cdots a_1\Omega_\theta=J_\theta b_1\cdots b_n\Omega_\theta$. Taking the vacuum expectation proves~\eqref{eq:corJReality}.

Set $A^{(k)}_{pq}:= (\omega_{e_p,e_q}\otimes \id)\pp_{\nu_{i_k}}$ and use the orthonormal basis $\tilde e_p:=\hat J e_p$. The corresponding slices satisfy
\be
\widetilde B^{(k)}_{pq}
:=(\omega_{\tilde e_p,\tilde e_q}\otimes\id)\pp_{\nu_{i_k}}^*
=
J_{\nu_{i_k}}A^{(k)}_{pq}J_{\nu_{i_k}},
\ee
and therefore preserve $\sE_{i_k}$. By~\eqref{eq:corJReality}, each finite $A$-slice sum is the complex conjugate of the $\widetilde B$-slice sum for the reversed correlation with inputs $J_{\nu_{i_n}}f_n,\ldots,J_{\nu_{i_1}}f_1$. The preceding $B$-slice argument applies to this basis and, using~\eqref{eq:corJReality} again, gives the required limit.
\end{proof}

From now on, we will impose the symmetry condition of Definition~\ref{def:symmetry}. The KMS condition for the weight $\theta$ then translates into a twisted cyclic symmetry of the correlations:

\begin{proposition}[KMS condition for correlations]
Let $(N,\theta,\alpha)$ be a regular, symmetric, ergodic,
spectrally compact quantum $\gg_b$-space. For $n\geq 2$,
$i_1,\ldots,i_n\in I$, and $f_k\in\sE_{i_k}$, we have
\be
\cC^{(n)}_{i_1,\ldots,i_n}(f_1,\ldots,f_n)
=
\cC^{(n)}_{i_n,i_1,\ldots,i_{n-1}}
(\nabla_{\nu_{i_n}}^{-1}f_n,f_1,\ldots,f_{n-1})\,.
\ee
\end{proposition}
\begin{proof}
Recall the modular automorphism group $\sigma_t^\theta$ of
$(N,\theta)$. Symmetry and injectivity of $\Lambda_\theta$ imply
\be
\sigma_t^\theta(\Psi_j(f))
=\Psi_j(\nabla_{\nu_j}^{it}f)\,,
\qquad t\in\rr,\quad f\in\sE_j\,.
\ee
Since $z\mapsto\nabla_{\nu_j}^{iz}f$ is entire with values in
$\sE_j$, and $\Psi_j:\sE_j\rightarrow N$ is continuous,
$\Psi_j(f)$ is an entire analytic element for the modular flow.
In particular,
\be
\sigma_i^\theta(\Psi_j(f))
=\Psi_j(\nabla_{\nu_j}^{-1}f)\,.
\ee
Write $a_k:=\Psi_{i_k}(f_k)$. The KMS condition for $\theta$ gives
\be
\theta(a_1\cdots a_n)
=\theta(\sigma_i^\theta(a_n)a_1\cdots a_{n-1})\,.
\ee
Substituting the preceding identity proves the proposition.
\end{proof}

Let us study $n$-point correlations with $n=1,2,3$. For $n=1$, we have
\be
\cC^{(1)}_{i}(f) = \theta(\Psi_i(f)) = \langle \Omega_{\theta},\psi_i(f) \rangle_{\theta} =  \langle \psi_0(1),\psi_i(f) \rangle_{\theta}\,.
\ee
Thus $\cC^{(1)}_{i}=0$ unless $i=0$. This can also be seen from Claim~\ref{clm:inv1form}, asserting that invariant 1-forms on nontrivial representations vanish.
\begin{lemma}[Two-point correlations]\label{lem:2ptcor}
Let $(N,\theta,\alpha)$ be a regular, symmetric, ergodic,
spectrally compact quantum $\gg_b$-space. Then
\be
\cC^{(2)}_{ij}(f_1,f_2)
=\delta_{ij}\,\cD_{\nu_i,\nu_i}(f_1,f_2)\,,
\qquad i,j\in I,\quad f_1\in\sE_i,\ f_2\in\sE_j\,.
\ee
Here $\cD_{\nu,\nu}$ is the invariant 2-form from Definition~\ref{def:bilinearForm}.
\end{lemma}
\begin{proof}
Recall the Tomita operator $S_{\theta}$ on $\cK_{\theta}$ and the antilinear involution $S_{\nu} := J_{\nu}\nabla^{\frac{1}{2}}_{\nu}$ on $\cE_{\nu}$. Compatibility of the embeddings with the real structures and modular flows gives
\be
S_\theta\psi_i(f)=\psi_i(S_{\nu_i}f)\,,
\qquad f\in\sE_i\,.
\ee
Using the GNS construction and orthogonality of the embeddings,
we obtain
\ba
\cC^{(2)}_{ij}(f_1,f_2)
&=\langle \Psi_i(f_1)^*\Omega_\theta,
          \Psi_j(f_2)\Omega_\theta\rangle_\theta\\
&=\langle S_\theta\psi_i(f_1),\psi_j(f_2)\rangle_\theta\\
&=\langle \psi_i(S_{\nu_i}f_1),\psi_j(f_2)\rangle_\theta
=\delta_{ij}\,\cD_{\nu_i,\nu_i}(f_1,f_2)\,.
\ea
\end{proof}
Without symmetry, two-point correlations still vanish when $\nu_i\neq\nu_j$, but for $i,j\in I_\nu$ the formula becomes
\be
\cC^{(2)}_{ij}(f_1,f_2)
=(A_\nu(i/2))_{ij}\,\cD_{\nu,\nu}(f_1,f_2)\,,
\ee
where $A_\nu(z)$ denotes the entire extension of the finite-dimensional one-parameter group $A_\nu(t)$ featuring in Proposition~\ref{prop:modularFlowPsi}. These results are consistent with the assertion that all invariant bilinear forms are proportional to $\cD_{\nu_1,\nu_2}$, as asserted in Claim~\ref{clm:bilinearUniqueness}.

For the rest of this section, assume Conjecture~\ref{conj:trilinearUniqueness}.
\begin{claim}[Three-point correlations]\label{clm:threePoint}
Let $(N,\theta,\alpha)$ be a regular, symmetric, ergodic,
spectrally compact quantum $\gg_b$-space. There exist constants
$c_{ijk}\in\cc$ such that
\be
\cC^{(3)}_{i,j,k}(f_1,f_2,f_3)
=c_{ijk}\,\cT_{\nu_i,\nu_j,\nu_k}(f_1,f_2,f_3)\,,
\qquad f_1\in\sE_i,\ f_2\in\sE_j,\ f_3\in\sE_k\,.
\ee
Here $\cT_{\nu_1,\nu_2,\nu_3}$ is the invariant 3-form from
Definition~\ref{def:trilinearForm}. The constants satisfy
\be
c_{ijk}=c_{jki}=c_{kij}\,,\qquad
\overline{c_{ijk}}=c_{kji}\,,\qquad
c_{0ij}=\delta_{ij}\,.
\ee
Whenever $\cT_{\nu_i,\nu_j,\nu_k}$ vanishes identically,\footnote{Recall that this happens if and only if at least one representation is the identity, and the other two are not equivalent.}
we set $c_{ijk}=0$.
\end{claim}
\begin{proof}
The existence of $c_{ijk}$ follows from invariance of correlations
and Conjecture~\ref{conj:trilinearUniqueness}.
The KMS condition for correlations gives
\be
\cC^{(3)}_{i,j,k}(f_1,f_2,f_3)
=\cC^{(3)}_{k,i,j}(\nabla_{\nu_k}^{-1}f_3,f_1,f_2)\,.
\ee
Comparing with the KMS symmetry~\eqref{eq:TKMS} of
$\cT_{\nu_i,\nu_j,\nu_k}$ gives
$c_{ijk}=c_{kij}$, hence cyclic symmetry.

As in the proof of the two-point formula, compatibility with
the real structures and modular flows implies
$\Psi_i(f)^*=\Psi_i(S_{\nu_i}f)$.
Taking adjoints inside the state therefore gives
\be
\overline{\cC^{(3)}_{i,j,k}(f_1,f_2,f_3)}
=\cC^{(3)}_{k,j,i}
(S_{\nu_k}f_3,S_{\nu_j}f_2,S_{\nu_i}f_1)\,.
\ee
Comparison with the reality property~\eqref{eq:TReality} of
$\cT_{\nu_i,\nu_j,\nu_k}$ yields
$\overline{c_{ijk}}=c_{kji}$.

Finally, $\Psi_0(1)=1_N$ implies
$\cC^{(3)}_{0,i,j}(1,f_1,f_2)=\cC^{(2)}_{i,j}(f_1,f_2)$.
The two-point formula and the normalization
$\cT_{i\frac{Q}{2},\nu_i,\nu_j}(1,f_1,f_2)
=\cD_{\nu_i,\nu_j}(f_1,f_2)$ then give
$c_{0ij}=\delta_{ij}$.
\end{proof} 
 
 \subsection{Product expansion and four-point correlations}\label{ssec:productExpansion}
We continue assuming that $(N,\theta,\alpha)$ is regular,
symmetric, ergodic, and spectrally compact. The spectral
decomposition~\eqref{eq:spectralDecomposition1} of $\cK_\theta$ allows us to expand the vector
$\Psi_j(f_1)\psi_k(f_2)$ into irreducible components. The
coefficients are precisely the structure constants appearing
in the three-point correlations.

\begin{proposition}[Product expansion]\label{prop:productExpansion}
Let $j,k\in I$, $f_1\in\sE_j$, and $f_2\in\sE_k$. Then
\be
\Psi_j(f_1)\psi_k(f_2)
=
\sum_{i\in I}c_{ijk}\,
\psi_i\bigl(
\cT^{\nu_i}_{\nu_j,\nu_k}(f_1,f_2)
\bigr)\,,
\label{eq:productExpansion}
\ee
where the sum converges in the norm on $\cK_\theta$. Here $\cT^{\nu_i}_{\nu_j,\nu_k}:\cE_{\nu_j}\times \cE_{\nu_k}\rightarrow \cE_{\nu_i}$ are the continuous equivariant bilinear maps described in Section~\ref{ssec:trilinear}.
\end{proposition}
\begin{proof}
Let $i\in I$ and $h\in\sE_i$. Using the GNS construction
and the three-point formula, we obtain
\ba
\langle\psi_i(h),\Psi_j(f_1)\psi_k(f_2)\rangle_\theta
&=\cC^{(3)}_{i,j,k}(S_{\nu_i}h,f_1,f_2)\\
&=c_{ijk}\,\cT_{\nu_i,\nu_j,\nu_k}
(S_{\nu_i}h,f_1,f_2)\\
&=c_{ijk}\,
\langle h,\cT^{\nu_i}_{\nu_j,\nu_k}(f_1,f_2)\rangle_{\nu_i}\,.
\ea
The last equality follows from the definition of $\cT_{\nu_i,\nu_j,\nu_k}$ through
$\cD_{\nu_i,\nu_i}$ and the identity $S_{\nu_i}^2=1$.
Since $\sE_i$ is dense in $\cH_{\nu_i}$, it follows that
\be
\psi_i^*\bigl(\Psi_j(f_1)\psi_k(f_2)\bigr)
=c_{ijk}\,\cT^{\nu_i}_{\nu_j,\nu_k}(f_1,f_2)\,.
\ee
The expansion and its norm convergence now follow from
orthogonality and completeness of the embeddings $\psi_i$.
\end{proof}

We can use the product expansion to express four-point
correlations in terms of the quadrilinear forms
$\cB^s$ and $\cB^t$ introduced earlier, in Definition~\ref{def:quadrilinarForms}.
\begin{proposition}[Four-point correlations]\label{prop:fourPoint}
Let $i,j,k,\ell\in I$ and let
$f_1\in\sE_i$, $f_2\in\sE_j$, $f_3\in\sE_k$, and
$f_4\in\sE_\ell$. The four-point correlation admits the
$s$-channel expansion
\be
\cC^{(4)}_{i,j,k,\ell}(f_1,f_2,f_3,f_4)
=
\sum_{m\in I}c_{ijm}\,c_{k\ell m}\,
\cB^s_{\nu_i,\nu_j,\nu_k,\nu_\ell;\nu_m}
(f_1,f_2,f_3,f_4)
\label{eq:fourPointS}
\ee
and the $t$-channel expansion
\be
\cC^{(4)}_{i,j,k,\ell}(f_1,f_2,f_3,f_4)
=
\sum_{m\in I}c_{\ell i m}\,c_{jkm}\,
\cB^t_{\nu_i,\nu_j,\nu_k,\nu_\ell;\nu_m}
(f_1,f_2,f_3,f_4)\,.
\label{eq:fourPointT}
\ee
Both sums converge absolutely.
\end{proposition}
\begin{proof}
Write the correlation as an inner product
\be
\cC^{(4)}_{i,j,k,\ell}(f_1,f_2,f_3,f_4)
=\langle X,Y\rangle_\theta\,,
\ee
where
\be
X:=\Psi_j(S_{\nu_j}f_2)\psi_i(S_{\nu_i}f_1)\,,
\qquad
Y:=\Psi_k(f_3)\psi_\ell(f_4)\,.
\ee
For $g\in\sE_m$, the three-point formula gives
\be
\langle X,\psi_m(g)\rangle_\theta
=\cC^{(3)}_{i,j,m}(f_1,f_2,g)
=c_{ijm}\,\cT_{\nu_i,\nu_j,\nu_m}(f_1,f_2,g)\,.
\ee
Applying Proposition~\ref{prop:productExpansion} to $Y$
therefore yields~\eqref{eq:fourPointS}, by the definition
of $\cB^s$ and cyclic symmetry $c_{mk\ell}=c_{k\ell m}$.

To prove absolute convergence, set $P_m:=\psi_m\psi_m^*$.
The $m$th summand is $\langle P_mX,P_mY\rangle_\theta$,
so orthogonality and Cauchy--Schwarz give
\be
\sum_{m\in I}
|\langle P_mX,P_mY\rangle_\theta|
\leq
\left(\sum_{m\in I}\|P_mX\|_\theta^2\right)^{1/2}
\left(\sum_{m\in I}\|P_mY\|_\theta^2\right)^{1/2}
=\|X\|_\theta\|Y\|_\theta\,.
\ee

For the $t$-channel expansion, first apply the KMS condition
\be
\cC^{(4)}_{i,j,k,\ell}(f_1,f_2,f_3,f_4)
=
\cC^{(4)}_{\ell,i,j,k}
(\nabla_{\nu_\ell}^{-1}f_4,f_1,f_2,f_3)\,.
\ee
The $s$-channel expansion of the right-hand side gives
\eqref{eq:fourPointT}, using the KMS identity
\be
\cB^s_{\nu_\ell,\nu_i,\nu_j,\nu_k;\nu_m}
(\nabla_{\nu_\ell}^{-1}f_4,f_1,f_2,f_3)
=
\cB^t_{\nu_i,\nu_j,\nu_k,\nu_\ell;\nu_m}
(f_1,f_2,f_3,f_4)\,.
\ee
Its absolute convergence follows from the $s$-channel result,
since $\nabla_{\nu_\ell}^{-1}f_4\in\sE_\ell$.
\end{proof}
 
 \subsection{Associativity and reconstruction}\label{ssec:boundaryAssociativity}
The quantum $\gg_b$-spaces considered above provide our proposed reformulation of the notion of elementary conformal boundary conditions. It will be convenient to give them a shorter name.
\begin{definition}[Quantum boundary spaces]\label{def:quantumBoundarySpace}
A \emph{quantum boundary space} for $\gg_b$ is a regular, symmetric, ergodic, spectrally compact quantum $\gg_b$-space.
\end{definition}
Thus a quantum boundary space comes with a faithful normal invariant state, a unique trivial summand, and the quantum automorphic forms constructed above.

Comparing the two expansions of the four-point correlation gives the following constraints on its spectral data.
\begin{theorem}[Associativity equations]\label{thm:boundaryAssociativity}
The spectral data $(\nu_i)_{i\in I}$, $(c_{ijk})_{i,j,k\in I}$ of a quantum boundary space satisfy
\be
\sum_{m\in I}c_{ijm}c_{k\ell m}
\cB^s_{\nu_i,\nu_j,\nu_k,\nu_\ell;\nu_m}(F)
=
\sum_{m\in I}c_{\ell i m}c_{jkm}
\cB^t_{\nu_i,\nu_j,\nu_k,\nu_\ell;\nu_m}(F)
\label{eq:BAssoc2}
\ee
for all $i,j,k,\ell\in I$ and
$F\in\sE_i\hat\otimes_\pi\sE_j\hat\otimes_\pi\sE_k\hat\otimes_\pi\sE_\ell$.
Both sums converge absolutely.
\end{theorem}
\begin{proof}
For elementary tensors, both sides equal the same four-point correlation. The Cauchy--Schwarz estimates used in the channel expansions bound each sum of absolute values by a product of continuous seminorms of the four inputs. The identity and absolute convergence therefore extend to the completed projective tensor product.
\end{proof}
We retain the convention $0\in I$, so the vacuum contribution is already included in these sums. Recall that $\cB^s_{\nu_1,\nu_1,\nu_3,\nu_3;\nu_5}$ and $\cB^t_{\nu_1,\nu_2,\nu_2,\nu_1;\nu_5}$ are continuous at $\nu_5=i\frac{Q}{2}$, where they reduce to products of invariant bilinear forms, see Claim~\ref{clm:BFactorization}. Therefore, the vacuum contribution does not require special treatment.

We conjecture that quantum boundary spaces satisfy the following converse theorem, analogous to the converse theorem for hyperbolic surface spectra~\cite{Adve:2025sld}:
\begin{conjecture}[Reconstruction of quantum boundary spaces]\label{conj:boundaryReconstruction}
Fix $Q>0$. Let $I$ be a countable indexing set containing $0$, and let $(\nu_i)_{i\in I}$ be a discrete spectrum with finite multiplicities, where
\be
\nu_0=i\tfrac Q2\,,\qquad
\nu_j\in\rr_{\geq0}\cup i(0,Q/2)\quad\text{for }j\neq0\,.
\ee
Let $(c_{ijk})_{i,j,k\in I}$ satisfy
\be
c_{ijk}=c_{jki}=c_{kij}\,,\qquad
\overline{c_{ijk}}=c_{kji}\,,\qquad
c_{0ij}=\delta_{ij}\,.
\ee
Suppose these data satisfy the associativity equations~\eqref{eq:BAssoc2}, in the sense that the sums on both sides are absolutely convergent for all $i,j,k,\ell\in I $ and $F\in\sE_i\hat\otimes_\pi\sE_j\hat\otimes_\pi\sE_k\hat\otimes_\pi\sE_\ell$, and define the same continuous linear functional on this test space. Then there exists a quantum boundary space realizing these data, unique up to an equivariant, state-preserving isomorphism.
\end{conjecture}

The idea behind reconstruction is simple. Start with
\be
\cK=\bigoplus_{i\in I}\cH_{\nu_i}\,,\qquad
\uu=\bigoplus_{i\in I}\pp_{\nu_i}\,,\qquad
\Omega=\psi_0(1)\,,
\ee
where $\psi_i$ are the canonical inclusions. The product expansion prescribes the action of the prospective quantum automorphic forms:
\be
\Psi_j(f)\psi_k(g)
=\sum_{i\in I}c_{ijk}\,
\psi_i\bigl(\cT^{\nu_i}_{\nu_j,\nu_k}(f,g)\bigr)\,.
\ee
If these prescriptions define bounded operators, one can take the von Neumann algebra they generate and the vector state given by $\Omega$. The representation $\uu$ supplies the candidate action. The associativity equations express the compatibility of these constructions. The main challenge is to show that the associativity equations imply boundedness of the operators $\Psi_i(f)$, and that the resulting algebra, state, and action have the required analytic properties.

\subsection{Virasoro boundary bootstrap}\label{ssec:VirBoundaryBootstrap}

We now compare these constraints with the bootstrap equations for an elementary compact unitary Virasoro-preserving boundary condition $a$ at central charge $c>1$. Its Hilbert space on an interval, with boundary condition $a$ at both endpoints, decomposes as
\be
\cH_{aa}\cong\bigoplus_{i\in I}\cP(h_i,c)\,,\qquad
h_0=0\,,\quad h_i>0\quad\text{for }i\neq0\,.
\ee
Here $\cP(h,c)$ denotes the unitary irreducible lowest-weight Virasoro module of weight $h$. The spectrum is discrete with finite multiplicities, and the vacuum occurs once.

Choose Hermitian boundary primaries $\cO_i$ with $\cO_0=\mathbf1$, normalized by
\be
\langle\cO_i(x_1)\cO_j(x_2)\rangle
=\frac{\delta_{ij}}{|x_{12}|^{2h_i}}\,.
\ee
For $x_1>x_2>x_3$, write
\be
\langle\cO_i(x_1)\cO_j(x_2)\cO_k(x_3)\rangle
=\frac{c_{ijk}}
{x_{12}^{h_i+h_j-h_k}x_{13}^{h_i+h_k-h_j}x_{23}^{h_j+h_k-h_i}}\,.
\label{eq:BCFT3pt}
\ee
Cyclic symmetry, reflection positivity, and the identity operator imply
\be
c_{ijk}=c_{jki}=c_{kij}\,,\qquad
\overline{c_{ijk}}=c_{kji}\,,\qquad c_{0ij}=\delta_{ij}\,.
\ee
These are precisely the relations obeyed by the structure constants of a quantum boundary space.

To compare four-point functions, use the conformal frame
$(x_1,x_2,x_3,x_4)=(\infty,1,\chi,0)$, with $0<\chi<1$, where
$\cO_i(\infty):=\lim_{x\to+\infty}x^{2h_i}\cO_i(x)$.
In a general ordered configuration, we have $\chi=(x_{12}x_{34})/(x_{13}x_{24})$.
The two OPEs give the \emph{Virasoro crossing equations}
\begin{align}
\langle\cO_i(\infty)\cO_j(1)\cO_k(\chi)\cO_\ell(0)\rangle
&=\sum_{m\in I}c_{ijm}c_{k\ell m}
\cG_{h_i,h_j,h_k,h_\ell;h_m}(\chi)\notag\\
&=\sum_{m\in I}c_{\ell i m}c_{jkm}
\cG_{h_i,h_\ell,h_k,h_j;h_m}(1-\chi)\,.
\label{eq:VirBoot}
\end{align}
Here $\cG$ is the Virasoro conformal block, with the central charge suppressed from the notation. Our normalization is fixed by
\be
\cG_{h_1,h_2,h_3,h_4;h}(\chi)
=\chi^{h-h_3-h_4}\bigl[1+O(\chi)\bigr]\qquad(\chi\to0^+)\,.
\label{eq:VirBlockRed}
\ee
When $h_1=h_2$ and $h_3=h_4$, the limit $h\rightarrow 0$ of $\cG_{h_1,h_1,h_3,h_3;h}$ continuously reproduces the Virasoro block of the irreducible vacuum module, in perfect correspondence with what happens for the quantum group forms $\cB^s,\cB^t$.

The resemblance between~\eqref{eq:BAssoc2} and~\eqref{eq:VirBoot} is striking: the coefficients $c_{ijk}$ have identical symmetry properties, while the quantum group forms $\cB^s,\cB^t$ are replaced by Virasoro conformal blocks. This is much more than a formal analogy. As we will argue,~\eqref{eq:BAssoc2} and~\eqref{eq:VirBoot} impose precisely the same constraints on the spectral data.

For this purpose, it is useful to express crossing through linear functionals. Introduce the uniformizing variables
\be
\tau(\chi)=i\frac{K(1-\chi)}{K(\chi)}\,,\qquad
w=\frac{i-\tau}{i+\tau}\,,\qquad
\chi(w)=\lambda\!\left(i\frac{1-w}{1+w}\right)\,,
\label{eq:wVar}
\ee
where $K$ is the complete elliptic integral and $\lambda$ is the modular lambda function. The upper half-plane in $\tau$ parametrizes the universal cover of the punctured cross-ratio plane, and $w$ maps it to the open unit disk $\mathbb D$. The crossing-symmetric point is $w=0$, and $\chi(-w)=1-\chi(w)$.

After lifting to $\mathbb D$, the Virasoro blocks and the two channel sums are holomorphic, and the sums converge uniformly on compact subsets. Consequently, crossing is equivalent to equality of all derivatives at $w=0$. More generally, one may apply
\be
\omega_{a,N}(g):=\sum_{n=0}^N a_n g^{(n)}(0)
\label{eq:functional}
\ee
term by term to both sides. A functional that can be exchanged with the OPE sums is called \emph{swappable}~\cite{Qiao:2017lkv}. Finite derivative functionals are swappable. Infinite derivative expansions $\omega_{a,\infty}$ are also useful when they converge and satisfy this swapping property.

\subsection{Equivalence with conformal boundary conditions}\label{ssec:boundaryEquivalence}

The relation between representation labels and conformal weights is
\be
c=1+6Q^2\,,\qquad h_i=\frac{Q^2}{4}+\nu_i^2\,.
\ee
The trivial, complementary-series, and principal-series representations correspond respectively to $h=0$, $0<h<Q^2/4$, and $h\geq Q^2/4$.
\begin{conjecture}[Boundary correspondence]\label{conj:equivalence1}
For $c=1+6Q^2$, there is a one-to-one correspondence between:
\begin{enumerate}
\item quantum boundary spaces for $\gg_b$, up to equivariant, state-preserving isomorphism,
\item solutions $(h_i)_{i\in I}$, $(c_{ijk})_{i,j,k\in I}$ of the full unitary Virasoro boundary bootstrap, with discrete spectrum, finite multiplicities, and a unique vacuum,
\item elementary compact unitary Virasoro-preserving conformal boundary conditions at central charge $c$.
\end{enumerate}
The correspondence identifies $h_i=Q^2/4+\nu_i^2$ and preserves the structure constants $c_{ijk}$. In the second item, data are identified under relabelling and real orthogonal changes of basis among primaries of equal weight, with the vacuum fixed. In the third item, boundary conditions are understood up to equivalence, including their bulk CFTs, and satisfy the sewing constraints at every genus.
\end{conjecture}

Here the \emph{full boundary bootstrap} means~\eqref{eq:VirBoot} for every ordered quadruple of external labels, together with the normalization, cyclicity, conjugation, and convergence properties described above. The conjecture therefore includes a realization statement: every such solution should determine an actual conformal boundary condition and its compatible bulk CFT.

Two ingredients would establish the correspondence between the first two items. The first is the reconstruction conjecture stated above as Conjecture~\ref{conj:boundaryReconstruction}. The second is equivalence of the quantum group associativity equations~\eqref{eq:BAssoc2} and the Virasoro crossing equations~\eqref{eq:VirBoot}. We now explain the evidence for this second assertion.

The main heuristic motivation is that the two families of blocks have the same fusion kernel. In the normalization used here, the Ponsot--Teschner transformation~\cite{PonsotTeschner1999,PonsotTeschner2001,TeschnerVartanov2014} reads
\begin{align}
&\cG_{h_1,h_2,h_3,h_4;h_s}(\chi)\notag\\
&\quad=\int_0^\infty
\mathscr{F}^{\mathrm{CFT}}_{\nu_1,\nu_2,\nu_3,\nu_4}(\nu_s,\nu_t)
\cG_{h_1,h_4,h_3,h_2;h_t}(1-\chi)\,d\mu_b(\nu_t)
\label{eq:crossingVir}
\end{align}
for principal-series parameters. This transformation takes precisely the same form as the recoupling relation~\eqref{eq:recouplingB} for $\cB^s$ and $\cB^t$, with the same kernel $\mathscr{F}^{\mathrm{CFT}}_{\nu_1,\nu_2,\nu_3,\nu_4}$ and the same measure $d\mu_b(\nu)$. Thus the quantum group and Virasoro blocks describe the same change between the two coupling schemes.

Demanding that the fusion kernels for Virasoro blocks and quantum group forms are identical eliminates the freedom in the normalization of the invariant trilinear forms $\cT$, discussed around~\eqref{eq:freedomN}, and requires $\kappa_b(h_1,h_2,h_3)=1$. The freedom to choose $\kappa_b$ is equivalent to the freedom to use a different normalization for $c_{ijk}$ in~\eqref{eq:BCFT3pt}.

The common fusion kernel suggests a simple formal explanation of
the equivalence. Formally substitute~\eqref{eq:crossingVir} into the $s$-channel
of~\eqref{eq:VirBoot}, exchange the spectral sum with the fusion
integral, and compare the coefficients of the $t$-channel blocks.
This gives the formal identity between distributions in the $\nu$ variable
\be
\sum_{m\in I} c_{ijm}c_{k\ell m}\,
\mathscr{F}^{\mathrm{CFT}}_{\nu_i,\nu_j,\nu_k,\nu_\ell}(\nu_m,\nu)
=
\sum_{m\in I} c_{\ell i m}c_{jkm}\,
\rho_b(\nu_m)^{-1}\delta(\nu-\nu_m)
\qquad\text{(formally)}\,.
\label{eq:formalCrossing}
\ee
Starting from quantum group associativity~\eqref{eq:BAssoc2}
and using~\eqref{eq:recouplingB} leads to exactly the same identity.
Thus quantum group associativity and Virasoro crossing appear to impose the same constraints
on the spectrum and structure constants in both descriptions.

This argument captures the intuition behind the conjectured
equivalence, but the infinite sums make it insufficient as a proof.
Exchanging the spectral sum with the fusion integral requires
justification, as does comparing the coefficients of the resulting
continuous expansion. Furthermore, the vacuum and complementary
series contributions require analytic continuation and contour
residues, whereas their appearance in~\eqref{eq:formalCrossing} is schematic.

On the other hand, both original sets of equations~\eqref{eq:BAssoc2} and~\eqref{eq:VirBoot} are perfectly well-defined. We tested their equivalence directly, without resorting to the fusion kernel argument. In particular, if the equivalence holds, then any known solution of Virasoro crossing must also satisfy quantum group associativity. For this purpose, we considered Neumann boundary conditions for compact free bosons at $c=19$ and $c=31$, probing both the $|b|=1$ and $b>0$ regimes. Decomposing exact four-point functions of vertex operators into Virasoro blocks in both channels gives the exchanged spectra and OPE products, which can then be inserted into the quantum group associativity equations. We used several configurations of the external scaling dimensions, covering equal and unequal dimensions in the principal and complementary series, including a mixed example. Taking $F$ to be a product of Gaussian wavepackets in the four variables, numerical evaluation of $\cB^s(F)$ and $\cB^t(F)$ yields an excellent agreement between the channel sums, with discrepancies decreasing as the cutoff increases. The identity representation makes a nontrivial contribution: omitting it spoils the agreement, including in examples where it occurs in only one channel. The precise setup and results are described in Appendix~\ref{app:freeBosonNumerics}.

Another way of checking the equivalence is to relate the ways in which the two families of blocks are tested. Fix nontrivial unitary labels
$\nu_1,\nu_2,\nu_3,\nu_4\in\rr_{\geq0}\cup i(0,Q/2)$.
For these fixed external labels, abbreviate
\ba
\cB^s_{\nu_s}(F)&:=\cB^s_{\nu_1,\nu_2,\nu_3,\nu_4;\nu_s}(F)\,,\qquad
&&\cB^t_{\nu_t}(F):=\cB^t_{\nu_1,\nu_2,\nu_3,\nu_4;\nu_t}(F)\,,\\
\cG^s_{h_s}(w)&:=\cG_{h_1,h_2,h_3,h_4;h_s}(\chi(w))\,,
&&\cG^t_{h_t}(w):=\cG_{h_1,h_4,h_3,h_2;h_t}(1-\chi(w))\,.
\ea
We identify the completed tensor product of the four test spaces with $\cE^{(4)}$. The following two conjectures would establish equivalence of the bootstrap equations by implementing the correspondence for each choice of external labels, simultaneously in both channels.

\begin{conjecture}[Virasoro crossing from quantum group associativity]\label{conj:equivalence2}
For all $w\in\mathbb{D}$, there exists a test function $F_w\in\cE^{(4)}$ such that
\be
\cB^s_{\nu_s}(F_w)=\cG^s_{h_s}(w)\,,\qquad
\cB^t_{\nu_t}(F_w)=\cG^t_{h_t}(w)
\label{eq:GfromB}
\ee
for every nontrivial unitary intermediate label $\nu_s,\nu_t$. The identities also hold for the trivial intermediate representation whenever the corresponding channel admits a vacuum contribution.
\end{conjecture}
Evaluating quantum group associativity on this single test function $F_w$ would give the Virasoro crossing equation at $\chi(w)$, with all intermediate representations included.

\begin{conjecture}[Quantum group associativity from Virasoro crossing]\label{conj:equivalence3}
For all $F\in\cE^{(4)}$, there exists a swappable functional $\omega_F$, expressible as an infinite derivative expansion of the form~\eqref{eq:functional}, such that
\be
\cB^s_{\nu_s}(F)=\omega_F(\cG^s_{h_s})\,,\qquad
\cB^t_{\nu_t}(F)=\omega_F(\cG^t_{h_t})
\label{eq:BfromG}
\ee
for every nontrivial unitary intermediate label $\nu_s,\nu_t$. The identities also hold for the trivial intermediate representation whenever the corresponding channel admits a vacuum contribution.
\end{conjecture}
Applying the same functional $\omega_F$ to both sides of Virasoro crossing would then reproduce quantum group associativity evaluated on $F$. The requirement that one test function, or one functional, work in both channels is essential.

We tested Conjecture~\ref{conj:equivalence3} numerically as follows. For Gaussian test tensors at $c=19$ and $c=31$, with external representations in the complementary and principal series respectively, we approximated $\omega_F$ by a finite derivative functional~\eqref{eq:functional}. Its coefficients were fitted using only $s$-channel values of $\cB^s(F)$. The same functional was then applied to the $t$-channel Virasoro blocks. At moderate derivative order, the predictions agree well with the independently evaluated $t$-channel quantum group forms $\cB^t(F)$. The setup and results are described in Appendix~\ref{app:functionalNumerics}. This simultaneous agreement provides substantial independent evidence for the proposed correspondence between quantum group test functions and functionals on Virasoro blocks. It would be interesting to construct this correspondence explicitly.

The passage from boundary CFT data to an actual conformal boundary condition is the remaining part of Conjecture~\ref{conj:equivalence1}. Its intuition comes from sewing: boundary OPEs determine amplitudes in open channels, while alternative decompositions of the same surfaces reveal bulk data. For example, the annulus with boundary insertions should determine the bulk spectrum and bulk-to-boundary couplings, and a sphere with three disks removed should determine bulk three-point coefficients. The conjecture asserts that full boundary crossing makes this reconstruction consistent. It is this expected bulk completion that motivates the study of quantum $\gg_b$-bispaces in the next section.
 
\section{Quantum bispaces and bulk conformal field theory}\label{sec:bulkObjects}

\subsection{Quantum $\gg_b$-bispaces}\label{ssec:bulkObjectsDef}
We now introduce the quantum measure spaces that will play the role of bulk conformal field theories. The main novelty compared with the previous section is that they carry commuting left and right actions and satisfy braided commutativity. We use the notions of action and invariant weight from Section~\ref{ssec:QMS}, and follow the boundary discussion whenever the arguments are unchanged.
\begin{definition}[Quantum $\gg$-bispaces]\label{def:BulkO}
Let $\gg$ be a locally compact quantum group, with $M=L^{\infty}(\gg)$. A \emph{quantum $\gg$-bispace} is a quadruple $(N,\theta,\alpha_L,\alpha_R)$, where
\begin{itemize}
\item[-] $N$ is a von Neumann algebra,
\item[-] $\theta:N_+\rightarrow[0,\infty]$ is a n.s.f.\ weight,
\item[-] $\alpha_L:N\rightarrow M\botimes N$ is a $\theta$-preserving left action,
\item[-] $\alpha_R:N\rightarrow N\botimes M$ is a $\theta$-preserving right action,
\end{itemize}
and the two actions commute:
\be
(\alpha_L\otimes\id_M)\alpha_R
=(\id_M\otimes\alpha_R)\alpha_L\,.
\ee
The bispace is called \emph{ergodic} if the common fixed-point algebra is trivial:
\be
N^{\alpha_L}\cap N^{\alpha_R}=\cc1_N\,.
\ee
\end{definition}
Thus ergodicity refers to the pair of actions and neither action is required to be ergodic separately. We will see that this corresponds to allowing Virasoro primary conserved currents in the bulk CFT spectrum, while requiring a unique vacuum.
\begin{example}[The trivial bispace]
The trivial example of a quantum $\gg$-bispace is obtained by setting $N = \cc$, $\theta(z)=z$, $\alpha_L(z) = \alpha_R(z) = z 1_M$. It is ergodic. When $\gg=\gg_b$, this example is equivalent to the genus-zero bulk CFT of central charge $c=1+6Q^2$ containing only the identity operator and its Virasoro descendants.
\end{example}
\begin{example}[The canonical bispace]\label{ex:canonicalBispace}
If $\gg$ is unimodular, a canonical example of an ergodic quantum $\gg$-bispace is provided by the quantum group acting on itself by left and right translations. It corresponds to setting
\be
N=M\,,\qquad \theta=\ph\,,\qquad \alpha_L=\alpha_R=\Delta\,,
\ee
where $\Delta$ is the coproduct and $\ph$ the Haar weight. Coassociativity of $\Delta$ implies commutativity of the actions, and preservation of $\ph$ follows from left and right Haar invariance. We will explain that for $\gg=\gg_b$, this example is precisely equivalent to the bulk Liouville CFT at central charge $c=1+6Q^2$, elaborating on Section 5 of~\cite{PonsotTeschner1999}.
\end{example}

Fix $Q=b+b^{-1}>0$. Throughout this section, we specialize to $\gg=\gg_b=\sl$ and write $M=L^{\infty}(\gg_b)$.

We use the same GNS notation as in the boundary case: $\cK_\theta$, $\Lambda_\theta$. When $\theta$ is a state, we write $\Omega_\theta=\Lambda_\theta(1_N)$. We realize $N$ on $\cK_\theta$. Let $\uu_L \in M\botimes \mathrm{B}(\cK_{\theta})$ and $\uu_R\in \mathrm{B}(\cK_{\theta})\botimes M$ be the canonical unitary implementations of $\alpha_L$ and $\alpha_R$ provided by Theorem~\ref{thm:UImplementation}. They satisfy
\be
\alpha_L(a)=\uu_L^*(1_M\otimes a)\uu_L\,,\qquad
\alpha_R(a)=\uu_R(a\otimes1_M)\uu_R^*\,,\qquad a\in N\,.
\ee
$\uu_L$ and $(\uu_R)_{21}$ are unitary representations of $\gg_b$. Commutativity of the actions implies
\be
(\uu_L)_{12}(\uu_R)_{23}=(\uu_R)_{23}(\uu_L)_{12}\,.
\ee
Consequently,
\be
\uu_{LR}:=(\uu_L)_{13}(\uu_R)_{32}
=(\uu_R)_{32}(\uu_L)_{13}
\in M\botimes M\botimes\mathrm{B}(\cK_\theta)
\ee
is a unitary representation of $\gg_b\times\gg_b$. Its spectral decomposition replaces the decomposition of $\uu$ in the boundary case. Recall the external tensor product from Definition-Lemma~\ref{deflem:ExtTenProd}:
\be
\pp_{\nu^L}\boxtimes\pp_{\nu^R}
:=(\pp_{\nu^L})_{13}(\pp_{\nu^R})_{24}\,,\qquad
\nu^L,\nu^R\in\cS:=\rr_{\geq0}\cup i(0,Q/2]\,.
\label{eq:irrepsBulk}
\ee

\begin{definition}[Spectral compactness]\label{def:bulkSpectralCompactness}
A quantum $\gg_b$-bispace is \emph{spectrally compact} if $\theta$ is a state and
\be
\cK_\theta\cong\bigoplus_{i\in I}
\cH_{\nu_i^L}\otimes\cH_{\nu_i^R}\,,\qquad
\uu_{LR}\simeq\bigoplus_{i\in I}
\pp_{\nu_i^L}\boxtimes\pp_{\nu_i^R}\,,
\label{eq:bulkDecomposition}
\ee
with a discrete spectrum and finite multiplicities. Here $I$ is countable, $\nu_i^L,\nu_i^R\in\cS$, and
\be
\#\{i\in I:|\nu_i^L|^2+|\nu_i^R|^2<A\}<\infty
\qquad\text{for every }A>0\,.
\ee
We call the indexed family $((\nu_i^L,\nu_i^R))_{i\in I}$ the \emph{spectrum} of the bispace.
\end{definition}
As in the boundary case, the spectrum can be phrased in terms of quantum Laplace operators. The Casimirs~\eqref{eq:Casimir} of the left and right actions give two commuting pairs $(C_L,\check{C}_L)$ and $(C_R,\check{C}_R)$ of operators on $\cK_\theta$. On the summand $\cH_{\nu_i^L}\otimes\cH_{\nu_i^R}$, they act by~\eqref{eq:CasimirEigenvalue} with $\nu = \nu_i^L$ and $\nu = \nu_i^R$, respectively. The spectrum $((\nu_i^L,\nu_i^R))_{i\in I}$ is thus encoded in their joint spectrum, up to the coincidences discussed below~\eqref{eq:CasimirEigenvalue}.

From now on, assume that the bispace is ergodic and spectrally compact. As in the boundary case, the joint trivial representation occurs with unit multiplicity. We label it by $0\in I$, so that
\be
(\nu_0^L,\nu_0^R)=(i\tfrac Q2,i\tfrac Q2)\,.
\ee
The other summands may be trivial on one side, but not on both. The presence of such summands distinguishes joint ergodicity from separate ergodicity, as the following proposition shows.
\begin{proposition}[Chiral currents and separate ergodicity]
\label{prop:separateErgodicity}
Let $(N,\theta,\alpha_L,\alpha_R)$ be an ergodic spectrally compact
quantum $\gg_b$-bispace. Then $N^{\alpha_R}\neq\cc 1_N$ if and only if
the spectrum contains a summand with $\nu_j^R=i\tfrac Q2$ and
$\nu_j^L\neq i\tfrac Q2$. The same holds with $L$ and $R$ exchanged.
In particular, the bispace is separately ergodic if and only if its
spectrum contains no half-trivial summands.
\end{proposition}
\begin{proof}
Since $\theta$ is a faithful normal invariant state, the subspace of
right-invariant vectors in $\cK_\theta$ is
$\overline{\Lambda_\theta(N^{\alpha_R})}$. By
\eqref{eq:bulkDecomposition}, this is precisely the direct sum of
the summands with $\nu_j^R=i\tfrac Q2$: a nontrivial irreducible
representation has no nonzero invariant vectors. Joint ergodicity
implies that exactly one of these summands is also trivial on the
left, namely $\cc\Omega_\theta$. Thus $N^{\alpha_R}\neq\cc 1_N$
if and only if there is another summand with trivial right label.
The argument for $\alpha_L$ is identical.
\end{proof}
In the CFT dictionary, a nonvacuum summand with trivial right label
corresponds to a holomorphic Virasoro primary, and one with trivial
left label to an antiholomorphic Virasoro primary. Thus separate
ergodicity means that neither chiral algebra extends beyond the
Virasoro vacuum algebra.

Choose isometric equivariant embeddings
\be
\psi_i:\cH_{\nu_i^L}\otimes\cH_{\nu_i^R}\rightarrow\cK_\theta\,,\qquad i\in I\,,
\ee
with orthogonal ranges spanning $\cK_\theta$. Thus $\psi_i^*\psi_i=1$, $\psi_i^*\psi_j=0$ for $i\neq j$, and $\sum_i\psi_i\psi_i^*=1$ strongly. Equivariance means
\be
\uu_{LR}(1_M\otimes1_M\otimes\psi_i)
=(1_M\otimes1_M\otimes\psi_i)
(\pp_{\nu_i^L}\boxtimes\pp_{\nu_i^R})\,.
\label{eq:bulkPsiEquivariance}
\ee
We fix the vacuum embedding by $\psi_0(1\otimes1)=\Omega_\theta$.

\subsection{Real structure and modular flow}
Proposition~\ref{prop:UJ} applies to both actions, and the multiplicity-space argument of Section~\ref{ssec:realmodularPsi} applies to each joint spectral label. In particular, the embeddings can be chosen to preserve the real structures:
\begin{proposition}\label{claim:realityBulk}
We can choose $\psi_i$ so that
\be
J_\theta\psi_i=\psi_i(J_{\nu_i^L}\otimes J_{\nu_i^R})\,,\qquad i\in I\,.
\ee
\end{proposition}
We make this choice from now on. The remaining freedom consists of real orthogonal changes of basis within each joint multiplicity space, with the vacuum fixed.

Combining the left and right modular identities~\eqref{eq:scalingModularFlowL},~\eqref{eq:scalingModularFlowR}, we obtain
\be
(\hat\nabla^{it}\otimes\hat\nabla^{-it}\otimes\nabla_\theta^{it})\uu_{LR}
=\uu_{LR}(\hat\nabla^{it}\otimes\hat\nabla^{-it}\otimes\nabla_\theta^{it})\,.
\ee
Notice the opposite signs on the two quantum group legs, reflecting the difference between left and right actions. The multiplicity-space argument of Proposition~\ref{prop:modularFlowPsi} gives
\be
\nabla_\theta^{it}\psi_j
=\sum_{k\in I_{(\nu^L,\nu^R)}}
(A_{\nu^L,\nu^R}(t))_{kj}\,
\psi_k(\nabla_{\nu^L}^{it}\otimes\nabla_{\nu^R}^{-it})\,,
\label{eq:modularFlowBulk}
\ee
where $j\in I_{(\nu^L,\nu^R)}:=\{j:(\nu_j^L,\nu_j^R)=(\nu^L,\nu^R)\}$ and
$A_{\nu^L,\nu^R}:\rr\rightarrow\mathrm{SO}(|I_{(\nu^L,\nu^R)}|,\rr)$ is a continuous homomorphism.

The structural consequence of Theorem~\ref{thm:typeIII} extends to bispaces.
\begin{theorem}\label{thm:typeIIIBulk}
If an ergodic spectrally compact quantum $\gg_b$-bispace has $N\neq\cc$, then $N$ is a type~III$_1$ factor.
\end{theorem}
\begin{proof}
On each joint isotypic component, the modular flow is the tensor product of
$\nabla_{\nu^L}^{it}\otimes\nabla_{\nu^R}^{-it}$ with the finite-dimensional group $A_{\nu^L,\nu^R}(t)$.
For a nontrivial joint label, its generator is unitarily equivalent to multiplication by $2\pi Q(p_L-p_R)$ in Fourier space, omitting a momentum when the corresponding representation is trivial, together with the finitely many shifts coming from $A_{\nu^L,\nu^R}$.
These operators have no eigenvalues. Thus the only modular-invariant vectors are multiples of $\Omega_\theta$, and the centralizer of $\theta$ is $\cc1_N$.
The result follows from the same argument and the Marrakchi--Vaes lemma used in Theorem~\ref{thm:typeIII}.
\end{proof}

\begin{definition}[Symmetry]\label{def:symmetricBulkObj}
An ergodic spectrally compact quantum $\gg_b$-bispace is \emph{symmetric} if
\be
\nabla_\theta^{it}\psi_j
=\psi_j(\nabla_{\nu_j^L}^{it}\otimes\nabla_{\nu_j^R}^{-it})
\qquad\text{for all }j\in I,\ t\in\rr\,.
\ee
Equivalently, every $A_{\nu^L,\nu^R}(t)$ is the identity.
\end{definition}
Every ergodic spectrally compact bispace with multiplicity-free joint spectrum is automatically symmetric.

\subsection{Quantum automorphic forms}
Assume from now on that the bispace is also symmetric. As in the boundary case, define the subspaces whose vectors admit bounded algebra lifts:
\ba
\,&\cV_i:=\{F\in\cH_{\nu_i^L}\otimes\cH_{\nu_i^R}:
\psi_i(F)\in\Lambda_\theta(N)\}\,,\\
&\Psi_i:\cV_i\rightarrow N\,,\quad
\Psi_i(F):=\Lambda_\theta^{-1}\psi_i(F)\,.
\ea
In particular, $\Psi_i(F)\Omega_\theta=\psi_i(F)$.

The slice argument of Lemma~\ref{lem:PsiEquivariance} applies separately to the two actions and gives
\begin{lemma}[Equivariance of the algebra lifts]\label{clm:LRcoactions}
For $F\in\cV_i$ and $\omega_L,\omega_R\in M_*$, the sliced vectors on the right-hand sides below belong to $\cV_i$, and
\ba
(\omega_L\otimes\id)\alpha_L(\Psi_i(F))
&=\Psi_i\bigl([((\omega_L\otimes\id)\pp_{\nu_i^L}^*)\otimes1_{\nu_i^R}]F\bigr)\,,\\
(\id\otimes\omega_R)\alpha_R(\Psi_i(F))
&=\Psi_i\bigl([1_{\nu_i^L}\otimes((\omega_R\otimes\id)\pp_{\nu_i^R})]F\bigr)\,.
\label{eq:BulkEquivariance}
\ea
\end{lemma}
The adjoint on the left representation, and its absence on the right, will account for the reversed ordering of right-moving invariant forms.

Using the analytic test spaces $\cE_\nu$ introduced in Definition-Lemma~\ref{def:ESpace}, set
\be
\sE_i:=\cE_{\nu_i^L}\hat\otimes_\pi\cE_{\nu_i^R}
\subset\cH_{\nu_i^L}\otimes\cH_{\nu_i^R}\,.
\ee

The definition of regularity and the related conjecture translate from boundary to bulk without change:
\begin{definition}[Regularity]\label{def:bulkRegularity}
An ergodic spectrally compact quantum $\gg_b$-bispace is \emph{regular} if $\sE_i\subseteq \cV_i $ for every $i\in I$.
\end{definition}
\begin{conjecture}\label{conj:bulkRegularity}
Every ergodic spectrally compact quantum $\gg_b$-bispace is regular.
\end{conjecture}

From now on, assume regularity as well. The elements of
\be
\cA_i:=\Psi_i(\sE_i)\subset N
\ee
will again be called \emph{quantum automorphic forms}. The closed graph argument of Lemma~\ref{lem:PsiCts} shows that $\Psi_i:\sE_i\rightarrow N$ is continuous.

Let us introduce the shorthand notation
\be
\nabla_i^{LR}:=\nabla_{\nu_i^L}\otimes\nabla_{\nu_i^R}^{-1}\,,\qquad
S_i^{LR}:=(J_{\nu_i^L}\otimes J_{\nu_i^R})(\nabla_i^{LR})^{1/2}
=S_{\nu_i^L}\otimes S_{\nu_i^R}^*\,.
\ee
By Lemma~\ref{eq:JnuNablanucE}, $S_i^{LR}$ is a continuous antilinear involution of $\sE_i$, and all complex powers of $\nabla_i^{LR}$ act continuously on this space.
\begin{lemma}[The star operation and modular flow]\label{clm:starBulk}
For $F\in\sE_j$ and $z\in\cc$, we have
\be
\Psi_j(F)^*=\Psi_j(S_j^{LR}F)\,,\qquad
\sigma_z^\theta(\Psi_j(F))
=\Psi_j((\nabla_j^{LR})^{iz}F)\,.
\ee
In particular, quantum automorphic forms are entire analytic for the modular group.
\end{lemma}
\begin{proof}
Reality and symmetry give $S_\theta\psi_j=\psi_jS_j^{LR}$ and
$\nabla_\theta^{it}\psi_j=\psi_j(\nabla_j^{LR})^{it}$.
Apply these identities to $\Psi_j(F)\Omega_\theta=\psi_j(F)$ and use injectivity of $\Lambda_\theta$. The modular identity extends to complex $z$ by continuity of $\Psi_j$ and analyticity of the test vectors.
\end{proof}

\subsection{Correlations}
\begin{definition}[$n$-point correlations]
Let $n\geq1$ and $i_1,\ldots,i_n\in I$. Define
\ba
\cC^{(n)}_{i_1,\ldots,i_n}:&\ \sE_{i_1}\times\cdots\times\sE_{i_n}\rightarrow\cc\,,\\
\cC^{(n)}_{i_1,\ldots,i_n}(F_1,\ldots,F_n)
&:=\theta(\Psi_{i_1}(F_1)\cdots\Psi_{i_n}(F_n))\,.
\ea
\end{definition}
The operator-norm estimate used in the boundary case proves joint continuity and hence a unique extension to the completed projective tensor product of the $\sE_{i_k}$.

The equivariance identities~\eqref{eq:BulkEquivariance} involve
$\pp^*$ for the left action and $\pp$ for the right action.
This difference accounts for the opposite tensor-product orders
in the two chiral sectors.
\begin{proposition}[Invariance of bulk correlations]
\label{prop:bulkCorrelationInvariance}
The $n$-point correlations are $\gg_b\times\gg_b$-invariant with respect to the
ordered tensor products
\be
\pp_{\nu^L_{i_1}}\times\ldots\times\pp_{\nu^L_{i_n}}\,,
\qquad
\pp_{\nu^R_{i_n}}\times\ldots\times\pp_{\nu^R_{i_1}}
\ee
in the left and right sectors, respectively.

More precisely, on factorized inputs $F_k=f_k^L\otimes f_k^R$,
the correlation defines an invariant form in
$(f_1^L,\ldots,f_n^L)$ with the right-moving arguments fixed,
and an invariant form in $(f_n^R,\ldots,f_1^R)$ with the
left-moving arguments fixed. Invariance is understood in the
sense of Definition~\ref{def:InvariantForm}.
\end{proposition}
\begin{proof}
Set $a_k:=\Psi_{i_k}(F_k)$. Multiplicativity of the actions
and invariance of $\theta$ give
\ba
(\id\otimes\theta)
[\alpha_L(a_1)\ldots\alpha_L(a_n)]
&=\theta(a_1\ldots a_n)1_M\,,\\
(\theta\otimes\id)
[\alpha_R(a_1)\ldots\alpha_R(a_n)]
&=\theta(a_1\ldots a_n)1_M\,.
\ea
Using~\eqref{eq:BulkEquivariance}, the first identity gives the
adjoint slice condition for the ordered left tensor product.
The second gives the nonadjoint slice condition for the reversed
right tensor product. The rest of the proof is the same as in the boundary case.
\end{proof}

The KMS symmetry and reality properties follow from the modular
structure of $(N,\theta)$, just as in the boundary case.
\begin{proposition}[KMS symmetry and reality of bulk correlations]
\label{prop:bulkCorrelationKMSReality}
For all $F_k\in\sE_{i_k}$, we have
\ba
\cC^{(n)}_{i_1,\ldots,i_n}(F_1,\ldots,F_n)
&=\cC^{(n)}_{i_n,i_1,\ldots,i_{n-1}}
\bigl((\nabla_{i_n}^{LR})^{-1}F_n,F_1,\ldots,F_{n-1}\bigr)\,,\\[4pt]
\overline{\cC^{(n)}_{i_1,\ldots,i_n}(F_1,\ldots,F_n)}
&=\cC^{(n)}_{i_n,\ldots,i_1}
\bigl(S_{i_n}^{LR}F_n,\ldots,S_{i_1}^{LR}F_1\bigr)\,.
\ea
\end{proposition}

To keep the left and right arguments together, we use $T_L\boxtimes T_R$ for the external product of multilinear forms, defined on factorized inputs by
\be
(T_L\boxtimes T_R)(f_1^L\otimes f_1^R,\ldots,f_n^L\otimes f_n^R)
:=T_L(f_1^L,\ldots,f_n^L)T_R(f_1^R,\ldots,f_n^R)\,.
\ee
For continuous forms, this extends to the completed test spaces. We use the same notation for vector-valued multilinear maps, with multiplication of the outputs replaced by their tensor product.

Starting from Definition~\ref{def:bilinearForm}, Definition~\ref{def:trilinearForm}, and the maps~\eqref{eq:2To1Def}, define the reverse-order invariant forms and maps by
\ba
\wt{\cD}_{\nu_1,\nu_2}(f_1,f_2)
&:=\cD_{\nu_2,\nu_1}(f_2,f_1)\,,\\
\wt{\cT}_{\nu_1,\nu_2,\nu_3}(f_1,f_2,f_3)
&:=\cT_{\nu_3,\nu_2,\nu_1}(f_3,f_2,f_1)\,,\\
\wt{\cT}^{\nu_1}_{\nu_2,\nu_3}(f_2,f_3)
&:=\cT^{\nu_1}_{\nu_3,\nu_2}(f_3,f_2)\,.
\ea
These conventions also apply when a label is trivial.

\begin{lemma}[One- and two-point correlations]
The one-point correlations $\cC^{(1)}_i $vanish for $i\neq0$, while $\cC^{(1)}_0(z)=z$. The two-point correlations are
\be
\cC^{(2)}_{ij}
=\delta_{ij}\,
\cD_{\nu_i^L,\nu_i^L}\boxtimes\wt{\cD}_{\nu_i^R,\nu_i^R}\,.
\label{eq:twoPointBulk}
\ee
\end{lemma}
\begin{proof}
The vacuum is orthogonal to every nontrivial joint summand. For two-point correlations, use
\be
\cC^{(2)}_{ij}(F_1,F_2)
=\langle S_\theta\psi_i(F_1),\psi_j(F_2)\rangle_\theta
=\delta_{ij}\langle S_i^{LR}F_1,F_2\rangle\,.
\ee
Since $\langle S_\nu^*f,g\rangle_\nu=\cD_{\nu,\nu}(g,f)$, this is precisely~\eqref{eq:twoPointBulk}.
\end{proof}

As in the boundary case, the following factorization uses the uniqueness conjecture for invariant trilinear forms in Section~\ref{ssec:trilinear}. We assume that conjecture in the discussion of structure constants and their expansions below. We write $C_{ijk}$ for bulk structure constants to distinguish them from the boundary coefficients $c_{ijk}$.
\begin{claim}[Three-point correlations]\label{prop:cBulkSym}
There are constants $C_{ijk}\in\cc$ such that
\be
\cC^{(3)}_{ijk}
=C_{ijk}\,
\cT_{\nu_i^L,\nu_j^L,\nu_k^L}
\boxtimes\wt{\cT}_{\nu_i^R,\nu_j^R,\nu_k^R}\,.
\label{eq:threePointBulk}
\ee
They satisfy
\be
C_{ijk}=C_{jki}=C_{kij}\,,\qquad
\overline{C_{ijk}}=C_{kji}\,,\qquad
C_{0ij}=\delta_{ij}\,.
\ee
\end{claim}
\begin{proof}
Invariance and trilinear uniqueness applied successively to the two chiral sectors give~\eqref{eq:threePointBulk}. Proposition~\ref{prop:bulkCorrelationKMSReality} and the trilinear identities~\eqref{eq:TKMS},~\eqref{eq:TReality} give cyclicity and conjugation of the coefficients. The vacuum normalization follows from~\eqref{eq:twoPointBulk}.
\end{proof}
If either chiral trilinear form vanishes identically, we set $C_{ijk}=0$. In particular, if one left label is trivial, the other two left labels must coincide for $C_{ijk}$ to be nonzero, and the same holds on the right.

\subsection{Product expansion and four-point correlations}
The spectral decomposition of a product is obtained by exactly the same argument as in Proposition~\ref{prop:productExpansion}.
\begin{proposition}[Product expansion]\label{thm:prodExpBulk}
Let $j,k\in I$, $F_1\in\sE_j$, and $F_2\in\sE_k$. Then
\be
\Psi_j(F_1)\psi_k(F_2)
=\sum_{i\in I}C_{ijk}\,
\psi_i\Bigl[
\bigl(\cT^{\nu_i^L}_{\nu_j^L,\nu_k^L}
\boxtimes\wt{\cT}^{\nu_i^R}_{\nu_j^R,\nu_k^R}\bigr)(F_1,F_2)
\Bigr]\,,
\label{eq:productExpansionBulk}
\ee
where the sum converges in the norm on $\cK_\theta$.
\end{proposition}
\begin{proof}
For $H\in\sE_i$, the three-point formula gives
\ba
\langle\psi_i(H),\Psi_j(F_1)\psi_k(F_2)\rangle_\theta
&=\cC^{(3)}_{ijk}(S_i^{LR}H,F_1,F_2)\\
&=C_{ijk}\left\langle H,
\bigl(\cT^{\nu_i^L}_{\nu_j^L,\nu_k^L}
\boxtimes\wt{\cT}^{\nu_i^R}_{\nu_j^R,\nu_k^R}\bigr)(F_1,F_2)
\right\rangle\,.
\ea
Density of $\sE_i$ identifies the $i$th spectral component. Norm convergence follows from orthogonality and completeness of the embeddings.
\end{proof}

Recall the quadrilinear forms $\cB^s$ and $\cB^t$ from Definition~\ref{def:quadrilinarForms}. Define their reverse-order versions by
\be
\wt{\cB}^{a}_{\nu_1,\nu_2,\nu_3,\nu_4;\nu_5}(f_1,f_2,f_3,f_4)
:=\cB^{a}_{\nu_4,\nu_3,\nu_2,\nu_1;\nu_5}(f_4,f_3,f_2,f_1)\,,
\qquad a=s,t\,.
\ee
\begin{proposition}[Four-point correlations]\label{prop:bulkFourPoint}
For $i,j,k,\ell\in I$, the four-point correlation admits the expansions
\begin{align}
\cC^{(4)}_{ijk\ell}
&=\sum_{m\in I}C_{ijm}C_{k\ell m}\,
\cB^s_{\nu_i^L,\nu_j^L,\nu_k^L,\nu_\ell^L;\nu_m^L}
\boxtimes\wt{\cB}^s_{\nu_i^R,\nu_j^R,\nu_k^R,\nu_\ell^R;\nu_m^R}\,,
\label{eq:bulkSchannel}\\
\cC^{(4)}_{ijk\ell}
&=\sum_{m\in I}C_{\ell im}C_{jkm}\,
\cB^t_{\nu_i^L,\nu_j^L,\nu_k^L,\nu_\ell^L;\nu_m^L}
\boxtimes\wt{\cB}^t_{\nu_i^R,\nu_j^R,\nu_k^R,\nu_\ell^R;\nu_m^R}\,.
\label{eq:bulkTchannel}
\end{align}
Both sums converge absolutely on
$\sE_i\hat\otimes_\pi\sE_j\hat\otimes_\pi\sE_k\hat\otimes_\pi\sE_\ell$.
\end{proposition}
\begin{proof}
For factorized bulk inputs, write $b_r=\Psi_{i_r}(F_r)$, with $(i_1,i_2,i_3,i_4)=(i,j,k,\ell)$, and set
$X=(b_1b_2)^*\Omega_\theta$, $Y=b_3b_4\Omega_\theta$.
The correlation equals $\langle X,Y\rangle_\theta$. Apply~\eqref{eq:productExpansionBulk} to both vectors and use the reality identities for the trilinear maps. This gives the $s$-channel expansion, with absolute convergence bounded by
\be
\sum_m|\langle\psi_m\psi_m^*X,\psi_m\psi_m^*Y\rangle_\theta|
\leq\|X\|_\theta\|Y\|_\theta\,.
\ee
The $t$-channel follows by moving $F_4$ to the front using the KMS condition. On the left this inserts $\nabla_{\nu_\ell^L}^{-1}$; on the right it inserts $\nabla_{\nu_\ell^R}$. The corresponding KMS identities for $\cB$ and $\wt{\cB}$ give~\eqref{eq:bulkTchannel}. The same continuity estimates as in the boundary case extend both expansions and absolute convergence to the completed test space.
\end{proof}

Equality of the two expansions gives the \emph{bulk associativity equations}:
\begin{align}
&\sum_{m\in I}C_{ijm}C_{k\ell m}
\bigl(\cB^s_{\nu_i^L,\nu_j^L,\nu_k^L,\nu_\ell^L;\nu_m^L}
\boxtimes\wt{\cB}^s_{\nu_i^R,\nu_j^R,\nu_k^R,\nu_\ell^R;\nu_m^R}\bigr)(F)
\notag\\
&\qquad=\sum_{m\in I}C_{\ell im}C_{jkm}
\bigl(\cB^t_{\nu_i^L,\nu_j^L,\nu_k^L,\nu_\ell^L;\nu_m^L}
\boxtimes\wt{\cB}^t_{\nu_i^R,\nu_j^R,\nu_k^R,\nu_\ell^R;\nu_m^R}\bigr)(F)\,.
\label{eq:assocBulk}
\end{align}
They hold for all external labels and all $F$ in the corresponding completed test space. The vacuum contribution is included as the $m=0$ term.

\subsection{The R-matrix and quasitriangular structure}\label{ssec:Rmatrix}
To correspond to a local CFT, a quantum bispace must satisfy an additional constraint, called \emph{braided commutativity}. Its intuitive meaning is that exchanging two factors in the algebra product is compensated by braiding their left- and right-moving representations in opposite directions. This is formulated using an R-matrix.

\begin{definition}[R-matrix]\label{def:RMatrix}
Let $\gg=(M,\Delta)$ be a locally compact quantum group, and $\hat{\gg}=(\hat M,\hat\Delta)$ its dual. An \emph{R-matrix} is a unitary operator $\cR\in\hat M\botimes\hat M$ satisfying
\ba
(\hat\Delta\otimes\id)\cR&=\cR_{13}\cR_{23}\,,\\
(\id\otimes\hat\Delta)\cR&=\cR_{13}\cR_{12}\,,\\
\cR\hat\Delta(x)\cR^*&=\hat\Delta^{\mathrm{op}}(x)\,,\qquad x\in\hat M\,.
\label{eq:Rmatrix}
\ea
Here $\hat\Delta^{\mathrm{op}}=\sigma\hat\Delta$, where $\sigma$ flips the tensor legs. When $\cR$ satisfying the above properties exists, one calls $(M,\Delta,\cR)$ a \emph{quasitriangular locally compact quantum group}.
\end{definition}
The last relation says that the dual coproduct is ``cocommutative up to conjugation by $\cR$''. The first two express compatibility of $\cR$ with tensor products of representations. See~\cite{MeyerRoyWoronowicz:2016TwistedII,DeCommerKrajczok:2024Braided} for discussions of R-matrices in the operator-algebraic setting.

It can be shown~\cite{DeCommerKrajczok:2024Braided} that whenever an R-matrix exists, it induces a natural family of braiding isomorphism $\uu_i\times \uu_j\rightarrow \uu_j\times \uu_i$ on tensor products of representations. Indeed, let $\uu_i$, $\uu_j$ be unitary representations of $\gg$ on Hilbert spaces $\cH_i$, $\cH_j$ and recall the tensor-product convention~\eqref{eq:UboxV}, $\uu_i\times\uu_j=(\uu_j)_{13}(\uu_i)_{12}$, and its relation to the dual coproduct~\eqref{eq:DeltaHatTimes}. The R-matrix then induces a canonical unitary operator $\cR_{i,j}\in\mathrm{U}(\cH_i\otimes\cH_j)$ such that the following identity holds in $M\botimes \mathrm{B}(\cH_i)\botimes \mathrm{B}(\cH_j)$
\be
(\cR_{i,j})_{23}(\uu_j)_{13}(\uu_i)_{12}
=(\uu_i)_{12}(\uu_j)_{13}(\cR_{i,j})_{23}\,.
\label{eq:braidingIJ}
\ee
Equivalently $\Sigma\cR_{i,j}:\cH_i\otimes\cH_j\rightarrow\cH_j\otimes\cH_i$ intertwines the two tensor product representations:
\be
(\uu_j\times\uu_i)(1_M\otimes \Sigma\cR_{i,j})= (1_M\otimes \Sigma\cR_{i,j})(\uu_i\times\uu_j)\,.
\ee
When $\cH_i = \cH_j = L^2(\gg)$ and $\uu_i = \uu_j = \ww$ is the left regular representation, the corresponding operator $\cR_{\ww,\ww}$ coincides with $\cR$ from Definition~\ref{def:RMatrix}. The identity~\eqref{eq:braidingIJ} then reads
\be
\cR_{23}\ww_{13}\ww_{12}
=\ww_{12}\ww_{13}\cR_{23}\,,
\ee
which is the same as the last line of~\eqref{eq:Rmatrix}. This family $\cR_{i,j}$ is natural: for bounded intertwiners $\cI_{k,i}:\cH_i\rightarrow\cH_k$ and $\cI_{\ell,j}:\cH_j\rightarrow\cH_\ell$,
\be
(\cI_{k,i}\otimes\cI_{\ell,j})\cR_{i,j}
=\cR_{k,\ell}(\cI_{k,i}\otimes\cI_{\ell,j})\,.
\label{eq:intertwineR}
\ee
It is also compatible with tensor products:
\be
\cR_{i\times j,k}=(\cR_{i,k})_{13}(\cR_{j,k})_{23}\,,\qquad
\cR_{i,j\times k}=(\cR_{i,k})_{13}(\cR_{i,j})_{12}\,.
\ee

For $\gg_b$, the R-matrix has a particularly simple expression in terms of the self-adjoint operator $\kbf,\ebf,\fbf$ defined in~\eqref{eq:HEHF}, which generate $L^{\infty}(\hat{\gg}_b)$ as in~\eqref{eq:MhatLogarithmic}:
\begin{claim}[The R-matrix of $\gg_b$]\label{clm:RmatrixGb}
The quantum group $\gg_b$ is quasitriangular, with R-matrix
\be
\cR=e^{4\pi i\kbf_1\kbf_2}\,
\phi_b(\fbf_1+\ebf_2)^*\,
e^{4\pi i\kbf_1\kbf_2}\,.
\label{eq:RmatrixGb}
\ee
This expression is valid for all $Q>0$.
\end{claim}
This is the R-operator of the modular double constructed by Bytsko and Teschner~\cite{BytskoTeschner2003}, written in our conventions; see also~\cite{Faddeev1999ModularDouble}.
The formula is manifestly unitary: generators on different legs commute, and $|\phi_b(x)|=1$ for real $x$.

\subsection{Braided commutativity}\label{ssec:braidedCommutativity}
Let us return to the setting of a quantum $\gg_b$-bispace with left and right actions implemented through the unitary representations $\uu_L$ and $(\uu_R)_{21}$. The quasitriangular structure of $\gg_b$ then supplies a natural unitary operator $\cK_{\theta}\otimes\cK_{\theta}$:
\begin{definition}[The braiding operator of a quantum bispace]
Let $(N,\theta,\alpha_L,\alpha_R)$ be a quantum $\gg_b$-bispace. Let $\cR_{L,L}$ and $\cR_{R,R}$ be the R-matrices for the pairs $(\uu_L,\uu_L)$ and $((\uu_R)_{21},(\uu_R)_{21})$, respectively. Define
\be
\beta_N:=\Sigma\,\cR_{L,L}\cR_{R,R}^*
\in\mathrm{U}(\cK_\theta\otimes\cK_\theta)\,.
\ee
We call $\beta_N$ the \emph{braiding operator}.
\end{definition}
The adjoint on the right reflects its reversed tensor-product order. Indeed,~\eqref{eq:braidingIJ} implies
\be
(\cR_{i,j}^*)_{23}(\uu_i)_{12}(\uu_j)_{13}
=(\uu_j)_{13}(\uu_i)_{12}(\cR_{i,j}^*)_{23}\,.
\ee
Thus the two chiral sectors exchange in opposite directions.

We would like to impose the additional constraint that the quantum $\gg_b$-bispace be braided commutative. Formally, this is implemented as
\be
\mu_N\beta_N=\mu_N\,,
\label{eq:BC}
\ee
where $\mu_N$ stands for multiplication, initially defined as
\be
\mu_N:\Lambda_\theta(N)\odot\Lambda_\theta(N)\rightarrow\Lambda_\theta(N)\,,\qquad
\mu_N(\Lambda_\theta(a)\otimes\Lambda_\theta(b)):=\Lambda_\theta(ab)\,.
\ee
on the algebraic tensor product $\Lambda_\theta(N)\odot\Lambda_\theta(N)$. The difficulty is that $\beta_N$ need not preserve this space. It is possible to give~\eqref{eq:BC} a precise analytic meaning as follows. First, suppose that the quantum $\gg_b$-bispace is ergodic, spectrally compact, symmetric, and regular. We will impose~\eqref{eq:BC} on the dense subspace
\be
\mathcal{L}_{\mathrm{br}}:=\operatorname{span}\{
(\psi_k\otimes\psi_\ell)(F_1\otimes F_2):
k,\ell\in I,\ F_1\in\sE_k,\ F_2\in\sE_\ell\}
\subset\cK_\theta\otimes\cK_\theta\,.
\label{eq:braidingDomain}
\ee
Naturality~\eqref{eq:intertwineR} of the R-matrix gives
\be
\beta_N(\psi_k\otimes\psi_\ell)(F_1\otimes F_2)
=(\psi_\ell\otimes\psi_k)
(\cR_{\nu_k^L,\nu_\ell^L})_{31}
(\cR_{\nu_k^R,\nu_\ell^R}^*)_{42}
(F_2\otimes F_1)\,.
\label{eq:braidingJK}
\ee
The leg order on the right is $L_\ell,R_\ell,L_k,R_k$.

The R-matrix on irreducible representations has the following distributional kernel.
\begin{claim}\label{clm:Rkernel}
Let $\nu_1,\nu_2\in\rr_{\geq0}\cup i(0,Q/2)$ and $F\in\cE_{\nu_1}\hat\otimes_\pi\cE_{\nu_2}$. For real $x_1,x_2$,
\be
(\cR_{\nu_1,\nu_2}F)(x_1,x_2)
=\iint_{\mathscr C_1\times\mathscr C_2}
\mathscr R_{\nu_1,\nu_2}(x_1,x_2;y_1,y_2)F(y_1,y_2)\,dy_1\,dy_2\,,
\ee
where
\ba
&\mathscr R_{\nu_1,\nu_2}(x_1,x_2;y_1,y_2)\\
&\quad=e^{-2\pi i(x_1-y_1)(x_2-y_2)}
\frac{
\phi_b\!\left(x_1-y_2+\frac{\nu_1-\nu_2}{2}+i\frac Q2\right)
\phi_b\!\left(y_1-x_2-\frac{\nu_1-\nu_2}{2}+i\frac Q2\right)
}{
\phi_b\!\left(x_1-x_2-\frac{\nu_1+\nu_2}{2}\right)
\phi_b\!\left(y_1-y_2+\frac{\nu_1+\nu_2}{2}\right)
}\,.
\ea
A consistent contour prescription is
\be
\mathscr C_1=\rr-i\epsilon\,,\qquad
\mathscr C_2=\rr+i\epsilon\,,\qquad
\epsilon=\frac{Q+2|\operatorname{Im}\nu_1-\operatorname{Im}\nu_2|}{8}\,.
\ee
If either representation is trivial, the R-matrix is the identity under the natural identification with the other representation space.
\end{claim}

The trilinear maps diagonalize the braiding in the following sense.\footnote{The compositions below are understood through their distributional kernels and contour prescriptions; preservation of the analytic test space by $\cR$ itself is not required.}
\begin{claim}[Braiding of the invariant maps]\label{prop:Rnunu}
Let $\nu_1,\nu_2,\nu_3\in\cS$ and set $h_r=Q^2/4+\nu_r^2$. The following compositions are well-defined on $\cE_{\nu_1}\hat\otimes_\pi\cE_{\nu_2}$ and satisfy
\ba
\wt{\cT}^{\nu_3}_{\nu_1,\nu_2}\cR_{\nu_1,\nu_2}
&=e^{-i\pi(h_1+h_2-h_3)}\cT^{\nu_3}_{\nu_1,\nu_2}\,,\\
\cT^{\nu_3}_{\nu_1,\nu_2}\cR_{\nu_1,\nu_2}^*
&=e^{i\pi(h_1+h_2-h_3)}\wt{\cT}^{\nu_3}_{\nu_1,\nu_2}\,.
\label{eq:TRT}
\ea
For the bilinear forms, this gives
\be
\wt{\cD}_{\nu,\nu}\cR_{\nu,\nu}
=e^{-2\pi ih}\cD_{\nu,\nu}\,,\qquad
\cD_{\nu,\nu}\cR_{\nu,\nu}^*
=e^{2\pi ih}\wt{\cD}_{\nu,\nu}\,,
\quad h=Q^2/4+\nu^2\,.
\label{eq:DRD}
\ee
\end{claim}
The bilinear identities are the trivial-output cases of~\eqref{eq:TRT}. The formulas also agree with the prescribed maps when an input representation is trivial.

Define the conformal weights and spin associated with the $j$th joint spectral label by
\be
h_j^L:=\frac{Q^2}{4}+(\nu_j^L)^2\,,\qquad
h_j^R:=\frac{Q^2}{4}+(\nu_j^R)^2\,,\qquad
s_j:=h_j^L-h_j^R=(\nu_j^L)^2-(\nu_j^R)^2\,.
\ee
For fixed $k,\ell,F_1,F_2$, abbreviate the spectral vectors in the product expansion by
\be
v_m:=\psi_m\Bigl[
\bigl(\cT^{\nu_m^L}_{\nu_k^L,\nu_\ell^L}
\boxtimes\wt{\cT}^{\nu_m^R}_{\nu_k^R,\nu_\ell^R}\bigr)(F_1,F_2)
\Bigr]\,.
\ee
Combining~\eqref{eq:braidingJK},~\eqref{eq:TRT}, and the product expansion gives
\begin{align}
\mu_N(\psi_k(F_1)\otimes\psi_\ell(F_2))
&=\sum_{m\in I}C_{k\ell m}\,v_m\,,
\label{eq:OPENobraid}\\
\mu_N\beta_N(\psi_k(F_1)\otimes\psi_\ell(F_2))
&=\sum_{m\in I}e^{i\pi(s_m-s_k-s_\ell)}C_{\ell km}\,v_m\,.
\label{eq:OPEBraid}
\end{align}
We use the norm-convergent series in~\eqref{eq:OPEBraid} to define the notation $\mu_N\beta_N$ on $\mathcal{L}_{\mathrm{br}}$. Indeed, the $v_m$ are mutually orthogonal and $C_{\ell km}=\overline{C_{k\ell m}}$, so the two series have summands of the same norm. This gives a linear map on the stated test domain without assuming that the original algebraic multiplication map acts on its image under $\beta_N$.

\begin{definition}[Braided commutativity]\label{def:bulkBraidedCommutativity}
A regular, symmetric, ergodic, spectrally compact quantum $\gg_b$-bispace is \emph{braided commutative} if
\be
\mu_N\beta_N(\psi_k\otimes\psi_\ell)(F_1\otimes F_2)
=\mu_N(\psi_k\otimes\psi_\ell)(F_1\otimes F_2)
\ee
for all $k,\ell\in I$, $F_1\in\sE_k$, and $F_2\in\sE_\ell$.
\end{definition}

\begin{proposition}[Braided commutativity in spectral terms]\label{prop:bulkBraidingCriterion}
Braided commutativity is equivalent to
\be
s_i\in\zz\qquad\text{for all }i\in I
\label{eq:spinIntegrality}
\ee
and
\be
C_{ijk}=(-1)^{s_i+s_j+s_k}C_{jik}\qquad\text{for all }i,j,k\in I\,.
\ee
\end{proposition}
\begin{proof}
Compare~\eqref{eq:OPENobraid} and~\eqref{eq:OPEBraid} in each orthogonal summand. For $k=\ell$ and $m=0$, the nonzero two-point pairing and $C_{kk0}=1$ give $e^{-2\pi is_k}=1$. For general labels, comparison gives
\be
C_{k\ell m}=e^{i\pi(s_m-s_k-s_\ell)}C_{\ell km}
=(-1)^{s_k+s_\ell+s_m}C_{\ell km}\,.
\ee
Conversely, these relations make the two series termwise equal.
\end{proof}

\subsection{Associativity and reconstruction}\label{ssec:bulkAssociativity}
To avoid repeating the full list of conditions, we introduce a shorter name, parallel to Definition~\ref{def:quantumBoundarySpace}:
\begin{definition}[Quantum bulk spaces]\label{def:quantumBulkSpace}
A \emph{quantum bulk space} for $\gg_b$ is a regular, symmetric, ergodic, spectrally compact quantum $\gg_b$-bispace that is braided commutative.
\end{definition}
We have seen that its spectral data necessarily satisfy spin integrality, the bulk associativity equations~\eqref{eq:assocBulk}, and
\be
C_{ijk}=C_{jki}=C_{kij}\,,\quad
C_{ijk}=(-1)^{s_i+s_j+s_k}C_{jik}\,,\quad
\overline{C_{ijk}}=C_{kji}\,,\quad C_{0ij}=\delta_{ij}\,.
\label{eq:cSymmetryGeneral}
\ee
In particular, $C_{ijk}$ is real when $s_i+s_j+s_k$ is even and imaginary when it is odd.

It is natural to conjecture a converse theorem, parallel to Conjecture~\ref{conj:boundaryReconstruction} and the classical reconstruction result of~\cite{Adve:2025sld}.
\begin{conjecture}[Reconstruction of quantum bulk spaces]\label{conj:converseTheorem}
Fix $Q>0$. Let $I$ be countable with $0\in I$, and let
$((\nu_i^L,\nu_i^R))_{i\in I}\subset\cS^2$ be a discrete spectrum with finite multiplicities and a unique trivial pair, labeled by $0$. Let $(C_{ijk})_{i,j,k\in I}$ satisfy~\eqref{eq:spinIntegrality},~\eqref{eq:cSymmetryGeneral}, and the chiral vacuum selection rules stated after~\eqref{eq:threePointBulk}. Suppose both sides of all bulk associativity equations~\eqref{eq:assocBulk} converge absolutely and define continuous forms on the corresponding test spaces. Then there exists a quantum bulk space realizing these data, unique up to an isomorphism preserving the state and both actions.
\end{conjecture}

The idea behind reconstruction is the same as in the boundary case. The main analytic problem is to establish boundedness of the prescribed operators starting from the associativity equations.

\subsection{The bulk Virasoro bootstrap}\label{ssec:VirBulkBootstrap}
We now compare these constraints with those on bosonic unitary 2d CFTs with $c_L=c_R=c>1$. As in the boundary case, we restrict to discrete spectra with finite multiplicities and a unique vacuum. The Hilbert space on the circle decomposes as
\be
\cH(S^1)\cong\bigoplus_{i\in I}
\cP(h_i^L,c)\otimes\cP(h_i^R,c)\,,\qquad
(h_0^L,h_0^R)=(0,0)\,.
\label{eq:bulkHilbertSpace}
\ee
Locality requires integer spin $s_i=h_i^L-h_i^R$, while unitarity gives $h_i^L,h_i^R\geq0$. The scaling dimension is $\Delta_i=h_i^L+h_i^R$, so $\Delta_i\geq|s_i|$.

The state-operator correspondence associates a Virasoro primary $\cO_i$ to each summand, with $\cO_0=1$. Choose a basis compatible with the adjoint and normalize the two-point functions as
\be
\langle\cO_i(z_1,\zb_1)\cO_j(z_2,\zb_2)\rangle
=\frac{\delta_{ij}}{z_{12}^{2h_i^L}\bar z_{12}^{2h_i^R}}\,.
\label{eq:bulk2Pt}
\ee
The one-point functions are $\langle\cO_j\rangle=\delta_{j0}$. Three-point functions take the form
\ba
\,&\langle\cO_i(z_1,\zb_1)\cO_j(z_2,\zb_2)\cO_k(z_3,\zb_3)\rangle\\
&\qquad\qquad=\frac{C_{ijk}}{z_{12}^{h^L_i+h^L_j-h^L_k}z_{13}^{h^L_i+h^L_k-h^L_j}z_{23}^{h^L_j+h^L_k-h^L_i}\zb_{12}^{h^R_i+h^R_j-h^R_k}\zb_{13}^{h^R_i+h^R_k-h^R_j}\zb_{23}^{h^R_j+h^R_k-h^R_i}}\,.
\label{eq:bulk3Pt}
\ea
The powers are defined by continuation from $z_1>z_2>z_3$ on the real axis. Integer spins make the full expression single-valued away from coincident points.

Permutation symmetry and conjugation give precisely~\eqref{eq:cSymmetryGeneral}. The chiral vacuum selection rules agree as well: if $h_i^L=0$, then $C_{ijk}=0$ unless $h_j^L=h_k^L$, and similarly for the other entries and the right-moving sector. The identity normalization is
\be
C_{0ij}=C_{i0j}=C_{ij0}=\delta_{ij}\,.
\label{eq:cIdentity}
\ee
Thus the spectra and three-point constants obey the same elementary constraints in both descriptions.

For four-point functions, use the frame $(\infty,1,\chi,0)$ and the reduced Virasoro blocks normalized in~\eqref{eq:VirBlockRed}. The s-channel and t-channel OPEs give
\begin{align}
&\langle\cO_i(\infty)\cO_j(1)\cO_k(\chi,\widetilde\chi)\cO_\ell(0)\rangle
\notag\\
&\quad=\sum_{m\in I}C_{ijm}C_{k\ell m}
\cG_{h_i^L,h_j^L,h_k^L,h_\ell^L;h_m^L}(\chi)
\cG_{h_i^R,h_j^R,h_k^R,h_\ell^R;h_m^R}(\widetilde\chi)
\label{eq:CBExpBulkS}\\
&\quad=\sum_{m\in I}C_{\ell im}C_{jkm}
\cG_{h_i^L,h_\ell^L,h_k^L,h_j^L;h_m^L}(1-\chi)
\cG_{h_i^R,h_\ell^R,h_k^R,h_j^R;h_m^R}(1-\widetilde\chi)\,.
\label{eq:CBExpBulkT}
\end{align}
On the physical Euclidean locus, $\widetilde\chi=\bar\chi$. Equality of the two channels is the bulk Virasoro crossing equation
\begin{align}
&\sum_{m\in I}C_{ijm}C_{k\ell m}\,
\cG_{h_i^L,h_j^L,h_k^L,h_\ell^L;h_m^L}(\chi)
\cG_{h_i^R,h_j^R,h_k^R,h_\ell^R;h_m^R}(\widetilde\chi)
\notag\\
&\qquad=\sum_{m\in I}C_{\ell im}C_{jkm}\,
\cG_{h_i^L,h_\ell^L,h_k^L,h_j^L;h_m^L}(1-\chi)
\cG_{h_i^R,h_\ell^R,h_k^R,h_j^R;h_m^R}(1-\widetilde\chi)\,.
\label{eq:VirBootBulk}
\end{align}
Each side of~\eqref{eq:VirBootBulk} is a holomorphic function of two complex variables $\chi$, $\wt{\chi}$, each of which can be continued independently to the universal covers of $\cc\setminus\{0,1\}$. The channel sums converge uniformly on compact sets.

\begin{definition}[Genus-zero compact unitary CFT data]\label{def:genusZeroBulkCFT}
Fix $c>1$. By \emph{genus-zero compact unitary CFT data}, we mean spectral data $((h_i^L,h_i^R))_{i\in I}$ and $(C_{ijk})_{i,j,k\in I}$ such that:
\begin{itemize}
\item[-] $I$ is countable and contains $0$; $h_i^L,h_i^R\geq0$, and $(h_i^L,h_i^R)=(0,0)$ precisely when $i=0$;
\item[-] the scaling dimensions form a discrete spectrum with finite multiplicities, and $s_i=h_i^L-h_i^R\in\zz$;
\item[-] the structure constants satisfy~\eqref{eq:cSymmetryGeneral}, the chiral vacuum selection rules, and $C_{0ij}=\delta_{ij}$;
\item[-] the crossing equations~\eqref{eq:VirBootBulk} hold for every ordered quadruple of external labels and $0<\chi,\widetilde\chi<1$, with the convergence properties described above.
\end{itemize}
\end{definition}
This is the bulk analogue of the boundary bootstrap problem. It concerns consistency on the sphere, consistency at higher genus has not yet been imposed.

\subsection{Equivalence with genus-zero bulk CFT}\label{ssec:bulkEquivalence}
The proposed correspondence identifies
\be
c=1+6Q^2\,,\qquad
h_j^L=\frac{Q^2}{4}+(\nu_j^L)^2\,,\qquad
h_j^R=\frac{Q^2}{4}+(\nu_j^R)^2\,.
\ee
Under this identification, the quantum group and CFT conditions on the spectrum, identity, permutation symmetry, and conjugation coincide. Their remaining constraints are~\eqref{eq:assocBulk} and~\eqref{eq:VirBootBulk}, respectively.

\begin{conjecture}[Bulk correspondence]\label{conj:bulkEquivalence}
For $c=1+6Q^2$, there is a one-to-one correspondence between quantum bulk spaces for $\gg_b$, and genus-zero compact unitary CFT data of central charge $c$. The correspondence identifies the weights as above and preserves $C_{ijk}$.
\end{conjecture}

Quantum bulk spaces are considered up to isomorphisms preserving the state and both actions. Spectral data are considered up to relabeling and real orthogonal changes of basis among primaries of equal left and right weights, with the vacuum fixed.

The evidence for this conjecture parallels the boundary case. The fusion kernel exhibits an exact symmetry under the reversal of the external labels:
\be
\mathscr{F}^{\mathrm{CFT}}_{\nu_4,\nu_3,\nu_2,\nu_1}(\nu_s,\nu_t) = \mathscr{F}^{\mathrm{CFT}}_{\nu_1,\nu_2,\nu_3,\nu_4}(\nu_s,\nu_t)\,.
\ee
It follows that the reversed quantum group forms $\wt{\cB}^{s}$, $\wt{\cB}^{t}$ satisfy the same recoupling identity~\eqref{eq:recouplingB} as their non-reversed counterparts. Therefore, the bulk associativity equations~\eqref{eq:assocBulk}, and the bulk Virasoro crossing equations~\eqref{eq:VirBootBulk} are both formally equivalent to the following identity between distributions in the pair of variables $\nu,\wt{\nu}$
\ba
\,&\sum_{m\in I} C_{ijm}C_{k\ell m}\,
\mathscr{F}^{\mathrm{CFT}}_{\nu^L_i,\nu^L_j,\nu^L_k,\nu^L_\ell}(\nu^L_m,\nu)
\mathscr{F}^{\mathrm{CFT}}_{\nu^R_i,\nu^R_j,\nu^R_k,\nu^R_\ell}(\nu^R_m,\wt{\nu})\\
&\qquad=
\sum_{m\in I} C_{\ell i m}C_{jkm}\,
\rho_b(\nu^L_m)^{-1}\rho_b(\nu^R_m)^{-1}\delta(\nu-\nu^L_m)\delta(\wt{\nu}-\nu^R_m)
\qquad\text{(formally)}\,.
\label{eq:formalCrossingBulk}
\ea

As in the boundary case, the formal identity~\eqref{eq:formalCrossingBulk}
does not replace the original, well-defined associativity and crossing
equations. A direct route to their equivalence is provided by the boundary
Conjectures~\ref{conj:equivalence2} and~\ref{conj:equivalence3}, applied
separately to the two chiral sectors. Taking tensor products of the test
functions supplied by Conjecture~\ref{conj:equivalence2} would reproduce
products of left- and right-moving Virasoro blocks in both channels
simultaneously. Evaluating quantum group bulk associativity on these tensors would then give bulk Virasoro crossing.

Conversely, products of the chiral functionals supplied by
Conjecture~\ref{conj:equivalence3} would recover quantum group associativity
on factorized bulk test tensors, provided their interchange with the bulk
OPE sums is justified. By linearity, the equality would then hold for finite sums of factorized
bulk test tensors. Every element of the completed bulk test space can be
approximated by such finite sums. Provided both channel sums depend
continuously on the test tensor, the equality would therefore extend to
the whole space. Thus the bulk argument follows from the boundary construction by taking tensor
products, with the remaining analytic task of establishing the required
swapping properties. Together with the reconstruction
Conjecture~\ref{conj:converseTheorem}, this would establish
the correspondence Conjecture~\ref{conj:bulkEquivalence}.

Since there exists a vast collection of compact unitary 2d CFTs with
$c>1$, our results provide substantial evidence for the existence of a
correspondingly rich variety of quantum bulk spaces. It will be important
to construct the associated von Neumann algebraic data
$(N,\theta,\alpha_L,\alpha_R)$ rigorously.

Modular invariance should select a distinguished class of quantum bulk
spaces corresponding to CFTs consistent at every genus. This is a genuine
additional requirement. For example, the trivial bispace $N=\cc$ satisfies
the quantum bulk-space axioms, but its Virasoro spectrum contains only
the vacuum pair, and its partition function
\be
Z(\tau,\bar\tau)=|\chi_{\mathrm{vac}}(\tau)|^2
\ee
is not modular invariant for $c>1$.

The comparison with RCFT suggests that modular invariance should admit
an intrinsic formulation as a maximality property of the quantum bispace,
without reference to its spectral data. More concretely, we expect the
modular-invariant bispaces to be precisely those obtained from suitable
quantum boundary spaces by an analytic analogue of the full center
construction. Developing this construction would provide a direct route
from the von Neumann algebraic data of a boundary condition to its bulk
CFT, with braided commutativity and modular invariance built in.

\section{Liouville CFT}\label{sec:Liouville}

The main purpose of this paper was to reformulate \emph{compact} CFTs and conformal boundary conditions as quantum analogues of compact hyperbolic surfaces. There is, however, a distinguished \emph{noncompact} example in which the preceding constructions become particularly explicit: Liouville theory. It plays the role of the homogeneous model underlying hyperbolic geometry. More precisely, the Liouville CFT is the quantum analogue of $\mathrm{PSL}(2,\rr)$ acting on itself by left and right translations, which one may regard as the frame-bundle version of the hyperbolic plane. Its quantum counterpart is simply $\gg_b$ acting on itself. From this point of view, the solvability of the Liouville CFT reflects the fact that its spectral theory is the harmonic analysis of the quantum group itself.

This connection has a long history. Ponsot and Teschner identified the Virasoro fusion transformation with the Racah--Wigner transformation for the positive representations of the modular double~\cite{PonsotTeschner1999,PonsotTeschner2001}; see~\cite{Teschner2001Liouville,Teschner2004Vertex} for reviews. Section~5 of~\cite{PonsotTeschner1999} already introduced the positive quantum group and stated its left--right Plancherel decomposition. The harmonic analysis was developed further in~\cite{Kashaev2001,BytskoTeschner2003,Ip2013QuantumPlane,NidaievTeschner2013,TeschnerVartanov2014}, while the relation between boundary Liouville structure constants and fusion coefficients was established in~\cite{PonsotTeschner2002Boundary}; see also~\cite{Teschner2000Boundary,Teschner2004BoundaryPerturbations}. The relation to quantum Teichm\"uller theory was explored in~\cite{Teschner2004Teichmuller,NidaievTeschner2013}. The discussion below organizes these constructions in the language of quantum $\gg_b$-spaces and bispaces, introduced in the previous two sections. In particular, the DOZZ and FZZT formulas for bulk and boundary structure constants arise by applying invariant weights to products of bounded operators in explicit von Neumann algebras.

Throughout this section, let $Q=b+b^{-1}>0$ and $c=1+6Q^2$.

\subsection{The canonical bispace and the Liouville spectrum}\label{ssec:LiouvilleSpectrum}

Recall the canonical quantum $\gg_b$-bispace from Example~\ref{ex:canonicalBispace}:
\be
(N,\theta,\alpha_L,\alpha_R)=(M,\ph,\Delta,\Delta)\,,
\qquad M=L^\infty(\gg_b)\,.
\label{eq:LiouvilleBispace}
\ee
Its GNS Hilbert space is $L^2(\gg_b)$, and the two unitary implementations are the multiplicative unitaries $\ww$ and $\vv$. Claim~\ref{clm:plancherel} gives its joint spectral decomposition:
\be
L^2(\gg_b)\cong\int_{\rr_+}^\oplus
\cH_\nu\otimes\cH_\nu\,d\nu\,,\qquad
\ww_{LR}\simeq\int_{\rr_+}^\oplus
\pp_\nu\boxtimes\pp_\nu\,d\nu\,.
\ee
Replacing quantum-group representations with Virasoro modules gives precisely the Liouville Hilbert space:
\be
\cH_{\mathrm{Liouville}}(S^1)
\cong\int_{\rr_+}^\oplus
\cP\bigl(\tfrac{Q^2}{4}+\nu^2,c\bigr)
\otimes\cP\bigl(\tfrac{Q^2}{4}+\nu^2,c\bigr)\,d\nu\,.
\ee
Thus the spectrum is diagonal and supported on the principal series:
\be
h^L_\nu=h^R_\nu=\frac{Q^2}{4}+\nu^2\,,\qquad \nu\geq0\,.
\label{eq:LiouvilleSpec}
\ee
The vacuum representation does not appear in this direct integral. On the quantum-group side, this has a simple interpretation: the algebra $M$ has an identity, but $\ph(1_M)=\infty$, so $1_M$ does not give a vector in the GNS Hilbert space. It is the analogue of the constant unit function on a space of infinite volume.

The Plancherel transform $\psi$ in~\eqref{eq:psiPlancherel} replaces the discrete family of embeddings $\psi_i$ used for compact quantum bispaces. Its compatibility with the real structures and modular flows was recorded in Claims~\ref{clm:PlancherelReal} and~\ref{clm:PlancherelModular}. In particular, the modular operator on the fiber $\cH_\nu\otimes\cH_\nu$ is $\nabla_\nu\otimes\nabla_\nu^{-1}$, with the same relative sign as in our construction~\eqref{eq:modularFlowBulk} of quantum \mbox{$\gg_b$-bispaces}. The main change from the compact setting is therefore the continuous spectrum and the use of a semifinite weight instead of a state.

Let us construct the corresponding algebra lifts $\Psi$ of states which transform irreducibly, aka quantum automorphic forms. Write
\be
\cE^{(2)}:=\cE\hat\otimes_\pi\cE\,,\qquad
\mathscr E_{\mathrm P}:=C_c^\infty((0,\infty),\cE^{(2)})\,.
\ee
\begin{claim}[Algebra lifts for the canonical bispace]\label{clm:LiouvilleLifts}
Let $\nu>0$ and $f\in\cE^{(2)}$. Define
\be
a_{\nu,f}(s,t,u)
=\sqrt{\rho_b(\nu)}e^{\frac{\pi Q}{2}(s+u)}
\frac{s_b(s+\nu)s_b(u+\nu)s_b(-t-\nu)s_b(-t+\nu)}
{s_b(s-t-i\tfrac Q2+i0)s_b(u-t-i\tfrac Q2+i0)}
\hat f(s,-u)\,.
\label{eq:LiouvilleLiftKernel}
\ee
The distributional expression
\be
\Psi_\nu(f):=\pi(a_{\nu,f})
\ee
defines a bounded element of $M$. For $F\in\mathscr E_{\mathrm P}$, its spectral integral
\be
\Psi(F):=\int_0^\infty\Psi_\nu(F_\nu)\,d\nu
\label{eq:LiouvilleWavepacket}
\ee
belongs to $\mathfrak n_\ph\cap\mathfrak n_\ph^*$ and satisfies
\be
\Lambda(\Psi(F))=\psi(F)\,,\qquad
\ph(\Psi(F)^*\Psi(F))
=\int_0^\infty\|F_\nu\|^2_{\cH_\nu\otimes\cH_\nu}\,d\nu\,.
\label{eq:LiouvilleGNSLift}
\ee
Moreover,
\be
\Psi(F)^*=\Psi(S_{\mathrm P}F)\,,\qquad \text{where}\qquad
(S_{\mathrm P}F)_\nu=(S_\nu\otimes S_\nu^*)F_\nu\,.
\label{eq:LiouvilleLiftStar}
\ee
\end{claim}
The kernel~\eqref{eq:LiouvilleLiftKernel} is obtained by combining the explicit GNS map~\eqref{eq:GNSGb} with the Plancherel transform~\eqref{eq:psiPlancherel}. The star operation agrees with Claim~\ref{clm:starBulk}. The bounds on these lifts, locally uniform in $\nu$, also give continuity of~\eqref{eq:LiouvilleWavepacket} on the indicated test space.

These operators are analogous to matrix coefficients of classical principal-series representations: a fixed spectral label gives a bounded operator, while square integrability requires smearing in the spectral parameter. In particular, the formula $\Lambda\Psi=\psi$ applies to wavepackets, rather than to individual fibers of the direct integral.

\subsection{Bulk correlations and the DOZZ formula}\label{ssec:DOZZ}

The Haar weight $\ph$ now takes the place of the state in the definition of correlations. $\ph$ extends uniquely to a linear functional on
\be
\mathfrak m_\ph
:= \operatorname{span}\{a\in M_+:\ph(a)<\infty\}\,.
\ee
If $F_1,\ldots,F_n\in\mathscr E_{\mathrm P}$ and $n\geq2$, then
\be
\Psi(F_1)\cdots\Psi(F_n)\in\mathfrak m_\ph\,.
\ee
Indeed, the first and last factors belong to $\mathfrak n_\ph^*$ and $\mathfrak n_\ph$, respectively, and the intervening factors are bounded. Thus we can define the correlations as
\be
\cC^{(n)}(F_1,\ldots,F_n)
:=\ph\bigl(\Psi(F_1)\cdots\Psi(F_n)\bigr)\,.
\label{eq:LiouvilleCorrelations}
\ee
These are continuous multilinear functionals. The useful estimate is
\be
|\ph(a_1\cdots a_n)|
\leq\|\Lambda(a_1^*)\|\,\|\Lambda(a_n)\|
\prod_{k=2}^{n-1}\|a_k\|\,.
\ee
We denote their spectral distribution kernels by $\cC^{(n)}_{\nu_1,\ldots,\nu_n}$. Expressions involving these kernels are understood after smearing the spectral labels.

\begin{claim}[Liouville two- and three-point correlations]\label{clm:LiouvilleCorrelations}
The two-point correlation is
\be
\cC^{(2)}_{\nu_1,\nu_2}
=\delta(\nu_1-\nu_2)
\bigl(\cD_{\nu_1,\nu_1}\boxtimes
\wt{\cD}_{\nu_1,\nu_1}\bigr)\,.
\label{eq:LiouvilleTwoPoint}
\ee
The three-point correlation is
\be
\cC^{(3)}_{\nu_1,\nu_2,\nu_3}
=c_{\mathrm{DOZZ}}(\nu_1,\nu_2,\nu_3)
\bigl(\cT_{\nu_1,\nu_2,\nu_3}\boxtimes
\wt{\cT}_{\nu_1,\nu_2,\nu_3}\bigr)\,,
\label{eq:LiouvilleThreePoint}
\ee
where
\be
c_{\mathrm{DOZZ}}(\nu_1,\nu_2,\nu_3)
=\frac{\sqrt{\rho_b(\nu_1)\rho_b(\nu_2)\rho_b(\nu_3)}}
{|\cN_b(\nu_1,\nu_2,\nu_3)|^2}\,.
\label{eq:DOZZfromN}
\ee
\end{claim}
The reversed forms $\wt{\cD}$, $\wt{\cT}$ and the product $\boxtimes$ have the same meaning as in the preceding section on bulk spaces. The two-point formula follows directly from~\eqref{eq:LiouvilleGNSLift} and~\eqref{eq:LiouvilleLiftStar}, just as in the compact case. The three-point calculation uses the Clebsch--Gordan decomposition in the left and right sectors. Passing from its unitary normalization to the forms $\cT$ and $\wt{\cT}$ supplies the factor $|\cN_b|^{-2}$, while the three algebra lifts supply the factors $\sqrt{\rho_b}$. Equivalently, one can substitute~\eqref{eq:LiouvilleLiftKernel} into the twisted product~\eqref{eq:Aproduct} and evaluate the resulting double-sine integrals.

Substituting~\eqref{eq:3PtN} into~\eqref{eq:DOZZfromN} gives
\be
c_{\mathrm{DOZZ}}(\nu_1,\nu_2,\nu_3)
=\frac{\Gamma_b(2Q)}{\Gamma_b(Q)^3}
\frac{\displaystyle\prod_{\epsilon_1,\epsilon_2,\epsilon_3=\pm1}
\Gamma_b\bigl(\tfrac Q2+i\epsilon_1\nu_1+i\epsilon_2\nu_2+i\epsilon_3\nu_3\bigr)}
{\displaystyle\sqrt{\prod_{j=1}^3\prod_{\epsilon=\pm1}
\bigl[\Gamma_b(2i\epsilon\nu_j)\Gamma_b(Q+2i\epsilon\nu_j)\bigr]}}\,.
\label{eq:cDOZZ}
\ee
Here we used
\be
\rho_b(\nu)
=\frac{\Gamma_b(Q+2i\nu)\Gamma_b(Q-2i\nu)}
{\Gamma_b(2i\nu)\Gamma_b(-2i\nu)}\,.
\ee
This is the DOZZ structure constant~\cite{DornOtto1994,ZamolodchikovZamolodchikov1996,Teschner1995ThreePoint}, with the primary fields normalized to have the delta-function two-point function
\be
\langle\cO_{\nu_1}(z_1)\cO_{\nu_2}(z_2)\rangle
=\frac{\delta(\nu_1-\nu_2)}{|z_{12}|^{4h_{\nu_1}}}\,.
\ee
The overall convention is fixed here by the normalization of $\ph$ and of the invariant forms. The corresponding three-point function is
\be
\langle\cO_{\nu_1}(z_1)\cO_{\nu_2}(z_2)\cO_{\nu_3}(z_3)\rangle
=\frac{c_{\mathrm{DOZZ}}(\nu_1,\nu_2,\nu_3)}
{|z_{12}|^{2(h_1+h_2-h_3)}
 |z_{13}|^{2(h_1+h_3-h_2)}
 |z_{23}|^{2(h_2+h_3-h_1)}}\,.
\ee
Thus the canonical bispace reproduces both the Liouville spectrum and its three-point data. It provides the basic noncompact example of the correspondence proposed in Conjecture~\ref{conj:bulkEquivalence}.

The four-point associativity equations~\eqref{eq:assocBulk} involve an integral over the diagonal spectrum in both channels:
\ba
\,&\int_0^{\infty} c_{\mathrm{DOZZ}}(\nu_1,\nu_2,\nu_s)c_{\mathrm{DOZZ}}(\nu_3,\nu_4,\nu_s)\,
(\cB^{s}_{\nu_1,\nu_2,\nu_3,\nu_4;\nu_s}\boxtimes
\wt{\cB}^{s}_{\nu_1,\nu_2,\nu_3,\nu_4;\nu_s})(F) \,d\nu_s\\
&=\int_0^{\infty} c_{\mathrm{DOZZ}}(\nu_4,\nu_1,\nu_t)c_{\mathrm{DOZZ}}(\nu_2,\nu_3,\nu_t)
(\cB^{t}_{\nu_1,\nu_2,\nu_3,\nu_4;\nu_t}\boxtimes
\wt{\cB}^{t}_{\nu_1,\nu_2,\nu_3,\nu_4;\nu_t})(F)\,d\nu_t\,,
\label{eq:LiouvilleAssociativity}
\ea
The relation~\eqref{eq:DOZZfromN} makes the mechanism transparent: after changing to unitary Clebsch--Gordan normalizations, the identity becomes orthogonality of the recoupling transformation. This is the quantum-group explanation of Liouville crossing symmetry described in~\cite{PonsotTeschner1999,PonsotTeschner2001}. Braided commutativity is immediate: the spectrum has zero spin, and $c_{\mathrm{DOZZ}}$ is fully symmetric, as required by Proposition~\ref{prop:bulkBraidingCriterion}.

In fact, braided commutativity is a general property of the canonical
bispace of any unimodular locally compact quantum group equipped with an
R-matrix. For $(M,\ph,\Delta,\Delta)$, it can be expressed directly as the identity
\be
\cR_{23}\ww_{13}\ww_{12}
=\ww_{12}\ww_{13}\cR_{23}\,.
\ee
Indeed, since $(\id\otimes\hat\Delta)\ww=\ww_{13}\ww_{12}$,
this identity is precisely the defining relation
$\cR\hat\Delta(x)\cR^*=\hat\Delta^{\mathrm{op}}(x)$ applied to the
second leg of $\ww$. The two occurrences of the R-matrix implement
the opposite braidings associated with the left and right actions.

\subsection{Boundary conditions and their structure constants}\label{ssec:LiouvilleBoundaries}
The three examples of quantum $\gg_b$-spaces introduced earlier acquire a concrete interpretation in Liouville theory. The trivial space of Example~\ref{ex:trivialGSpace}, with $N=\cc$, describes the elementary $(1,1)$ ZZ boundary condition~\cite{ZZ2001}. Its boundary spectrum consists only of the Virasoro vacuum module. The matrix algebra of Example~\ref{ex:ZZstack} describes a stack of $r$ such branes: the multiplicity space $M_r(\cc)$ records open strings between the individual branes.

The infinite-dimensional Example~\ref{ex:innerGSpace} reproduces the FZZT family of boundary conditions~\cite{FZZ2000,Teschner2000Boundary}. This example assigns to any principal or complementary series representation $\pp_{\sigma}$ of $\gg_b$ the following quantum $\gg_b$-space
\ba
N_\sigma&=\mathrm{B}(\cH_\sigma)\,,\\
\theta_\sigma(a)&=\operatorname{Tr}
\bigl(\nabla_\sigma^{1/2}a\nabla_\sigma^{1/2}\bigr)\,,
\qquad a\in(N_\sigma)_+\,,\\
\alpha_\sigma(a)&=\pp_\sigma^*(1_M\otimes a)\pp_\sigma\,.
\label{eq:FZZTspace}
\ea
The weight is invariant and the action is ergodic. Its GNS representation is the Hilbert space of Hilbert--Schmidt operators on $\cH_\sigma$
\be
\cK_{\theta_\sigma}\cong\mathrm{HS}(\cH_\sigma)
\cong\cH_\sigma\otimes\cH_\sigma\,.
\label{eq:FZZTGNS}
\ee
The last identification uses $J_\sigma$, sending $|v\rangle\langle  w|$ to $v\otimes (J_\sigma w)$. The GNS action of $N_{\sigma}$ is by left multiplication on $\mathrm{HS}(\cH_\sigma)$. The GNS map is
\be
\Lambda_\sigma(a)=a\nabla_\sigma^{1/2}\,.
\ee
The unitary implementation of $\alpha_\sigma$ is $\pp_\sigma\times\pp_\sigma$. The spectral decomposition of $\cK_{\theta_\sigma}$ is equivalent to the Clebsch--Gordan decomposition of $\cH_\sigma\otimes\cH_\sigma$. For principal-series $\sigma$, this is given in Claim~\ref{clm:CGPrincipal}:
\be
\cK_{\theta_\sigma} \cong\int_{\rr_+}^\oplus \cH_{\nu}\, d\mu_b(\nu)\,,\qquad
\pp_{\sigma}\times\pp_{\sigma}\simeq \int_{\rr_+}^\oplus \pp_{\nu}\,d\mu_b(\nu)\,.
\ee
It reproduces the representation content of Liouville theory on the interval with identical FZZT boundary conditions:
\be
\cH^{\mathrm{FZZT}}_{\sigma\sigma}
\cong\int_{\rr_+}^\oplus
\cP\bigl(\tfrac{Q^2}{4}+\nu^2,c\bigr)\,d\nu\,.
\ee
The change from $d\mu_b(\nu)$ to $d\nu$ is absorbed into the normalization of the spectral maps below.\footnote{This is a choice of normalization for the direct integral, distinct from the density of states on the interval needed for consistency on the annulus.} As in the bulk, the identity exists in the operator algebra but is not normalizable: $\theta_\sigma(1)=\infty$.

The next step is to identify the algebra lifts $\Psi_\nu$ of the irreducible spectral components. They are provided by the equivariant maps $\cT^{\sigma}_{\nu, \sigma}$ introduced in~\eqref{eq:2To1Def}. For $f\in\cE_\nu$, define $L^\sigma_\nu(f)\in N_\sigma$ by its action on $\cE_{\sigma} \subset \cH_\sigma$:
\be
L^\sigma_\nu(f)g:=\cT^\sigma_{\nu,\sigma}(f,g)\,,
\qquad g\in\cE_\sigma\,.
\label{eq:FZZTUnnormalizedField}
\ee
These are the quantum automorphic forms for the FZZT space, interpreted as generalized algebra lifts because the spectrum is continuous. Assume for simplicity that the boundary labels $\sigma$ belong to the principal series. We return to complementary-series $\sigma$ at the end.

The algebra lifts are normalized as follows
\be
n_\sigma(\nu):=|\cN_b(\nu,\sigma,\sigma)|\,,\qquad
\Psi^\sigma_\nu(f):=
\frac{\sqrt{\rho_b(\nu)}}{n_\sigma(\nu)}L^\sigma_\nu(f)\,,
\qquad \nu>0\,.
\label{eq:FZZTNormalizedField}
\ee
This normalization is chosen so that their two-point function has unit coefficient:\footnote{All correlation formulas are understood distributionally in the continuous spectral labels.}
\be
\theta_\sigma\bigl(\Psi^\sigma_{\nu_1}(f_1)
\Psi^\sigma_{\nu_2}(f_2)\bigr)
=\delta(\nu_1-\nu_2)\,
\cD_{\nu_1,\nu_1}(f_1,f_2)\,.
\label{eq:FZZTNormalization}
\ee
Indeed, the KMS and reality identities for $\cT$ identify the generalized GNS vectors of $L^\sigma_\nu$ with the adjoint Clebsch--Gordan kernel. Its orthogonality then gives~\eqref{eq:FZZTNormalization}: $n_\sigma$ removes the absolute value of the trilinear normalization, while $\sqrt{\rho_b}$ converts the Plancherel measure into $d\nu$. The same identities give
\be
\Psi^\sigma_\nu(f)^*=\Psi^\sigma_\nu(S_\nu f)\,,\qquad
\nabla_\sigma^{it}\Psi^\sigma_\nu(f)\nabla_\sigma^{-it}
=\Psi^\sigma_\nu(\nabla_\nu^{it}f)\,.
\ee

\begin{claim}[FZZT structure constants]\label{clm:FZZTStructureConstants}
The three-point functions are
\be
\theta_\sigma\bigl(\Psi^\sigma_{\nu_1}(f_1)
\Psi^\sigma_{\nu_2}(f_2)\Psi^\sigma_{\nu_3}(f_3)\bigr)
=c_\sigma(\nu_1,\nu_2,\nu_3)
\cT_{\nu_1,\nu_2,\nu_3}(f_1,f_2,f_3)\,,
\ee
where
\be
c_\sigma(\nu_1,\nu_2,\nu_3)
=\sqrt{\rho_b(\nu_1)\rho_b(\nu_2)\rho_b(\nu_3)}
\frac{n_\sigma(\nu_3)}{n_\sigma(\nu_1)n_\sigma(\nu_2)}
\mathscr F^{\mathrm{CFT}}_{\sigma,\nu_1,\nu_2,\sigma}(\sigma,\nu_3)\,.
\label{eq:FZZTStructureConstants}
\ee
They satisfy cyclicity and reality,
\be
c_\sigma(\nu_1,\nu_2,\nu_3)
=c_\sigma(\nu_2,\nu_3,\nu_1)\,,\qquad
\overline{c_\sigma(\nu_1,\nu_2,\nu_3)}
=c_\sigma(\nu_3,\nu_2,\nu_1)\,,
\ee
as well as the continuous quantum-group associativity equations.
\end{claim}
The reasoning is as follows: composing two lifts gives a nested pair of Clebsch--Gordan maps. Recoupling this pair and using the two-point normalization gives~\eqref{eq:FZZTStructureConstants}. Cyclicity and reality follow from the KMS condition and the adjoint operation. Associativity is the corresponding specialization of the pentagon identity, with the normalization factors cancelling on internal edges. This recovers the familiar description of boundary Liouville structure constants in terms of fusion coefficients~\cite{PonsotTeschner2002Boundary}, now directly from composition in $N_\sigma$.

\subsection{Boundary changing operators and Morita equivalence}\label{ssec:Morita}
The construction generalizes to boundary-changing operators. For general boundary labels $\sigma_a,\sigma_b$, the algebra $N_\sigma$ is replaced by the space of bounded operators $\cH_{\sigma_b}\rightarrow \cH_{\sigma_a}$:
\be
N_{ab}:=\mathrm{B}(\cH_{\sigma_b},\cH_{\sigma_a})\,.
\ee
For $a\neq b$, this is a bimodule rather than an algebra. The corresponding Hilbert space is obtained by completing the operators of finite norm $\theta_{\sigma_b}(X^*X)^{1/2}$. The generalized GNS map $X\mapsto X\nabla_{\sigma_b}^{1/2}$ identifies it with
\be
\mathrm{HS}(\cH_{\sigma_b},\cH_{\sigma_a})
\cong\cH_{\sigma_a}\otimes\overline{\cH_{\sigma_b}}
\cong\cH_{\sigma_a}\otimes\cH_{\sigma_b}\,.
\ee
Its representation is $\pp_{\sigma_a}\times\pp_{\sigma_b}$, so the spectrum of boundary-changing operators is given by the general Clebsch--Gordan decomposition.

The quantum automorphic forms now use the general case of the equivariant maps $\cT^{\nu_1}_{\nu_2,\nu_3}$. For $f\in \cE_{\nu}$, we define $\Psi^{ab}_\nu(f) \in N_{ab}$ by its action on $\cE_{\sigma_b}\subset \cH_{\sigma_b}$:
\be
\Psi^{ab}_\nu(f)g:=
\frac{\sqrt{\rho_b(\nu)}}{n_{ab}(\nu)}
\cT^{\sigma_a}_{\nu,\sigma_b}(f,g)\,,
\qquad g\in\cE_{\sigma_b}\,,\qquad n_{ab}(\nu):=|\cN_b(\nu,\sigma_a,\sigma_b)|\,.
\label{eq:FZZTChangingLifts}
\ee
Their adjoints reverse the boundary labels,
$\Psi^{ab}_\nu(f)^*=\Psi^{ba}_\nu(S_\nu f)$. The normalization ensures that the two-point functions take the form
\be
\theta_{\sigma_a}\bigl(
\Psi^{ab}_{\nu_1}(f_1)\Psi^{ba}_{\nu_2}(f_2)\bigr)
=\delta(\nu_1-\nu_2)\,
\cD_{\nu_1,\nu_1}(f_1,f_2)\,.
\ee
Correlations are defined by composing operators with matching inputs and outputs. A closed chain belongs to a diagonal algebra, where the corresponding weight can be applied. In particular,
\be
\theta_{\sigma_a}\bigl(
\Psi^{ab}_{\nu_1}(f_1)
\Psi^{bc}_{\nu_2}(f_2)
\Psi^{ca}_{\nu_3}(f_3)\bigr)
=c_{abc}(\nu_1,\nu_2,\nu_3)
\cT_{\nu_1,\nu_2,\nu_3}(f_1,f_2,f_3)\,,
\ee
with the general six-label coefficient
\be
c_{abc}(\nu_1,\nu_2,\nu_3)
=\sqrt{\rho_b(\nu_1)\rho_b(\nu_2)\rho_b(\nu_3)}
\frac{n_{ac}(\nu_3)}{n_{ab}(\nu_1)n_{bc}(\nu_2)}
\mathscr F^{\mathrm{CFT}}_{\sigma_a,\nu_1,\nu_2,\sigma_c}
(\sigma_b,\nu_3)\,.
\label{eq:FZZTChangingStructureConstants}
\ee
The same recoupling argument applies, now to
$\cT^{\sigma_a}_{\nu_1,\sigma_b}
(f_1,\cT^{\sigma_b}_{\nu_2,\sigma_c}(f_2,g))$.
In particular,
\be
c_{abc}(\nu_1,\nu_2,\nu_3)
=c_{cab}(\nu_3,\nu_1,\nu_2)\,,\qquad
\overline{c_{abc}(\nu_1,\nu_2,\nu_3)}
=c_{acb}(\nu_3,\nu_2,\nu_1)\,.
\ee
Setting $\sigma_a=\sigma_b=\sigma_c=\sigma$ recovers~\eqref{eq:FZZTStructureConstants}.

Complementary-series boundary labels are treated by continuing the Clebsch--Gordan kernels, giving rise to additional discrete terms when poles cross the spectral contour. This is the quantum-group counterpart of the discrete boundary states appearing in the continuation of FZZT boundary conditions~\cite{Teschner2000Boundary,Teschner2004BoundaryPerturbations}.

Finally, the preceding discussion provides a concrete example of the Morita picture sketched in
Section~\ref{ssec:BulkFromBoundary}. Two von Neumann
algebras are called Morita equivalent if they can be realized as
complementary full corners of a larger von Neumann algebra, called
a linking algebra. Here fullness means that both corner
projections have central support equal to one. The off-diagonal
corners are the bimodules implementing the equivalence. In our case,
the linking algebra is simply
\be
\mathrm{B}(\cH_a\oplus\cH_b)
=
\begin{pmatrix}
\mathrm{B}(\cH_a) & \mathrm{B}(\cH_b,\cH_a)\\
\mathrm{B}(\cH_a,\cH_b) & \mathrm{B}(\cH_b)
\end{pmatrix},
\ee
whose off-diagonal corners are precisely the boundary-changing
operator spaces. Taking $\cH_a,\cH_b$ to be $\cc$, $\cc^r$, or
$\cH_\sigma$ shows that the underlying von Neumann algebras of
our three examples belong to the Morita class of $\cc$. Fullness is automatic because the linking algebra is a factor and both corner projections are nonzero.

This equivalence is compatible with the quantum-group
actions. Indeed, the direct-sum corepresentation on $\cH_a\oplus\cH_b$
induces an action on the linking algebra. The latter restricts to the
original actions on its diagonal corners and equips the
boundary-changing bimodules with their equivariant structure.
The invariant weights likewise extend by summing their values
on the diagonal corners. Thus the construction refines ordinary
Morita equivalence to an equivariant Morita equivalence between
the quantum $\gg_b$-spaces.

In the CFT interpretation, Liouville theory provides the common
bulk for this trivial Morita class. A suitable analytic and $\gg_b$-equivariant version of the full center
construction should therefore assign the canonical $\gg_b$-bispace
to the trivial quantum $\gg_b$-space.

\section{Discussion}\label{sec:discussion}
\subsection{Summary}\label{ssec:discussionSummary}
We have proposed a reformulation of unitary two-dimensional conformal field theory with $c>1$ in terms of quantum measure spaces carrying actions of the locally compact quantum group $\gg_b=\sl$, with $c = 1+6(b+b^{-1})^2$. Conformal boundary conditions correspond to quantum $\gg_b$-spaces, and bulk CFTs to quantum $\gg_b$-bispaces. Let us summarize the logical structure of the proposal, referring to the precise statements in the text. Their status is collected in~\cref{tab:status}.

\paragraph{Harmonic analysis on $\gg_b$.} In~\cref{sec:slb2r}, we reviewed the construction of $\gg_b$ as a von Neumann algebraic quantum group, whose dual is generated by Faddeev's modular double~\eqref{eq:MhatLogarithmic}. For $b>0$ it is a locally compact quantum group by~\cite{DeCommerSchraderShapiroVoigt2025}, and we gave evidence that the same holds for $|b|=1$ (\cref{clm:GbLCQG}). In~\cref{sec:reps}, we complemented the principal series of Ponsot and Teschner~\cite{PonsotTeschner1999,PonsotTeschner2001} by the complementary series of unitary representations $\pp_{\nu}$, $\nu\in i(0,\frac{Q}{2})$ (\cref{clm:PnuRep}). We showed how the trivial representation arises as their limit (\cref{ssec:CfromP}), and conjectured that these representations exhaust the unitary dual (\cref{conj:unitaryDual}). Under~\eqref{eq:introHnu}, they match the unitary Virasoro modules with $h = 0$, $0<h<\frac{c-1}{24}$, and $h\geq\frac{c-1}{24}$, respectively (\cref{tab:RepCor}). In~\cref{sec:invariantForms}, we introduced spaces of analytic test vectors (\cref{def:ESpace}) and constructed continuous invariant forms on them: the bilinear forms $\cD_\nu$, the trilinear forms $\cT_{\nu_1,\nu_2,\nu_3}$ (\cref{ssec:trilinear}), and the quadrilinear forms $\cB^s$, $\cB^t$, whose relation~\eqref{eq:recouplingB} is governed by the normalized Virasoro fusion kernel.

\paragraph{Boundary conditions.} In~\cref{sec:boundary}, we defined quantum $\gg_b$-spaces $(N,\theta,\alpha)$ (\cref{def:quantumGspace}) and proved that every nontrivial ergodic, spectrally compact one is a type III$_1$ factor (\cref{thm:typeIII}). We defined quantum automorphic forms as bounded lifts of analytic test vectors (\cref{def:quantumAutomorphicForms}). Their existence is the regularity condition (\cref{def:regularity}), which we conjecture to be automatic (\cref{conj:regularity}). We then derived the properties of their correlations: the two-point correlations (\cref{lem:2ptcor}), the three-point correlations together with the cyclicity and reality of the structure constants (\cref{clm:threePoint}), the product expansion convergent in the GNS Hilbert-space norm (\cref{prop:productExpansion}), and the two absolutely convergent expansions of four-point correlations (\cref{prop:fourPoint}). For quantum boundary spaces (\cref{def:quantumBoundarySpace}), these culminate in the quantum associativity equations~\eqref{eq:BAssoc2} (\cref{thm:boundaryAssociativity}). We conjectured that every solution of these equations comes from a unique quantum boundary space (\cref{conj:boundaryReconstruction}). We also conjectured that the equations are equivalent to the Virasoro boundary crossing equations~\eqref{eq:VirBoot}, and that quantum boundary spaces are equivalent to elementary, compact, unitary, Virasoro-preserving conformal boundary conditions (Conjectures~\ref{conj:equivalence1}, \ref{conj:equivalence2}, and~\ref{conj:equivalence3}). The evidence consists of the formal argument~\eqref{eq:formalCrossing} based on the identification of $6j$ symbols with Virasoro fusion kernels, together with two numerical tests (\cref{app:freeBosonNumerics,app:functionalNumerics}).

\paragraph{Bulk CFT.} In~\cref{sec:bulkObjects}, we extended these constructions to quantum $\gg_b$-bispaces (\cref{def:BulkO}), which, when nontrivial, are again type III$_1$ factors under the analogous assumptions (\cref{thm:typeIIIBulk}). Using the R-matrix of $\gg_b$ (\cref{clm:RmatrixGb}), we defined braided commutativity (\cref{def:bulkBraidedCommutativity}) and proved that it is equivalent to spin integrality and the (anti)symmetry of the bulk structure constants (\cref{prop:bulkBraidingCriterion}). The spectral data of quantum bulk spaces (\cref{def:quantumBulkSpace}) satisfy the bulk associativity equations~\eqref{eq:assocBulk}. We conjectured that quantum bulk spaces can be reconstructed from their spectral data (\cref{conj:converseTheorem}) and that they are equivalent to compact unitary bosonic CFTs at genus zero (\cref{conj:bulkEquivalence}).

\paragraph{Liouville theory.} In~\cref{sec:Liouville}, we showed that the canonical bispace of $\gg_b$ (\cref{ex:canonicalBispace}) reproduces the Liouville spectrum~\eqref{eq:LiouvilleSpec} and the DOZZ structure constants,~\eqref{eq:cDOZZ}. The ZZ brane, stacks of ZZ branes, and the FZZT branes correspond to the quantum $\gg_b$-spaces of Examples~\ref{ex:trivialGSpace}, \ref{ex:ZZstack}, and~\ref{ex:innerGSpace}, and the FZZT boundary structure constants follow from~\cref{clm:FZZTStructureConstants}. Boundary-changing operators are described by bimodules, which exhibit all these boundary conditions as equivariantly Morita equivalent, with Liouville theory as their common bulk (\cref{ssec:Morita}).

\paragraph{Boundaries first.} Finally, the analogy with the construction of Fuchs, Runkel and Schweigert (\cref{ssec:QGmodularInvariance,ssec:BulkFromBoundary}) led us to conjecture that every solution of the boundary four-point Virasoro bootstrap with a discrete spectrum and a unique vacuum is an elementary boundary condition of a unique bulk CFT, consistent at all genera (\cref{conj:bulkFromBoundary}). In this picture, quantum boundary spaces are the fundamental objects of 2d CFT, and bulk theories are derived from them.

\begin{table}[t]
\centering
\small
\renewcommand{\arraystretch}{1.2}
\begin{tabular}{>{\raggedright\arraybackslash}p{7.2cm} >{\raggedright\arraybackslash}p{4.4cm} >{\raggedright\arraybackslash}p{2.8cm}}
\toprule
statement & reference & status\\
\midrule
$\gg_b$ is a locally compact quantum group & \cite{DeCommerSchraderShapiroVoigt2025}; \cref{clm:GbLCQG} & $b>0$: proved; $|b|=1$: claim\\
complementary series are unitary irreducible representations & \cref{clm:PnuRep} & claim\\
trivial, complementary, principal series exhaust the unitary dual & \cref{conj:unitaryDual} & conjecture\\
uniqueness of invariant trilinear forms & \cref{conj:trilinearUniqueness} & conjecture\\
type III$_1$ property for $N\neq\cc$ & \cref{thm:typeIII,thm:typeIIIBulk} & theorem\\
regularity is automatic & Conjectures~\ref{conj:regularity} and~\ref{conj:bulkRegularity} & conjecture\\
quantum associativity equations & \cref{thm:boundaryAssociativity,prop:bulkFourPoint} & theorem\\
braided commutativity $\Leftrightarrow$ locality & \cref{prop:bulkBraidingCriterion} & proposition\\
reconstruction from spectral data & Conjectures~\ref{conj:boundaryReconstruction} and~\ref{conj:converseTheorem} & conjecture\\
equivalence with Virasoro BCFT & Conjectures~\ref{conj:equivalence1}, \ref{conj:equivalence2}, and~\ref{conj:equivalence3} & conjecture; numerical evidence\\
equivalence with genus-zero bulk CFT & \cref{conj:bulkEquivalence} & conjecture\\
boundary data determine the bulk CFT & \cref{conj:bulkFromBoundary} & conjecture\\
DOZZ and FZZT structure constants & Claims~\ref{clm:LiouvilleCorrelations} and~\ref{clm:FZZTStructureConstants} & claim\\
\bottomrule
\end{tabular}
\caption{Status of the main statements of this paper. Theorems and propositions are proved assuming the Claims and explicitly stated conjectural inputs they rely on.}
\label{tab:status}
\end{table}

\subsection{Outlook}\label{ssec:outlook}
\paragraph{Quantum lattices and chaotic spectra.} A central problem in quantum field theory is to understand the landscape of generic compact unitary 2d CFTs: theories with no Virasoro primary conserved currents, whose spectra of scaling dimensions are expected to be highly irregular. The framework presented here suggests a new approach to this problem, modeled on the construction of hyperbolic surfaces. Classically, a compact hyperbolic surface (or orbifold) is obtained as a quotient $\hh/\Gamma$ by a cocompact lattice $\Gamma<\mathrm{PSL}(2,\rr)$, and a generic lattice gives rise to an irregular Laplace spectrum with random-matrix statistics~\cite{BohigasGiannoniSchmit1984,Bogomolny1997Arithmetical}. The direct quantum analogue would be a quotient of $\gg_b$ by a discrete quantum subgroup of finite covolume, a \emph{quantum lattice} in $\gg_b$. The quotient would be an ergodic quantum $\gg_b$-space~\cite{Vaes2005Induction,Kasprzak2012QuantumHomogeneousSpaces,KasprzakSoltan2014Embeddable}. Compactness of the quotient (in the C*-algebraic sense) should guarantee spectral compactness of the quantum $\gg_b$-space. By analogy with the classical case, it is plausible that a generic quantum lattice produces a quantum boundary space with a chaotic spectrum, i.e.~a boundary condition of a generic Virasoro CFT. Moreover, cocompact lattices in $\mathrm{PSL}(2,\rr)$ come in continuous families parametrized by Teichm\"uller space. Whether quantum lattices are rigid or deformable may thus bear on the question of whether generic CFTs are isolated. Lattices in locally compact quantum groups have been studied in~\cite{BrannanChirvasituViselter2020}, but to our knowledge lattices in $\gg_b$ have not. We regard the construction of quantum cocompact lattices in $\gg_b$, or more generally compact quantum homogeneous spaces for $\gg_b$, as one of the most important problems raised by this work.

\paragraph{An analytic full center.} Quantum lattices would produce quantum boundary spaces. To obtain bulk CFTs, one needs an analytic analogue of the full center construction. This would associate to a quantum boundary space $(N,\theta,\alpha)$ a modular-invariant quantum $\gg_b$-bispace $Z(N,\theta,\alpha)$, for which the boundary space is an elementary boundary condition. In rational CFT, the full center of a special symmetric Frobenius algebra produces a modular-invariant bulk theory~\cite{Fuchs:2002cm,FjelstadEtAl:2006Sewing,Fjelstad:2006aw,Kong:2009inh}. Modular invariance then appears as a maximality condition, the saturation of the bound~\eqref{eq:dimBbound} (\cref{ssec:QGmodularInvariance}). We expect the analytic full center to satisfy an analogous maximality property. Liouville theory provides a first test: its canonical $\gg_b$-bispace should be the full center of the trivial quantum $\gg_b$-space $\cc$, whose Morita class contains all the Liouville boundary conditions of~\cref{ssec:LiouvilleBoundaries} (\cref{ssec:Morita}). Braided tensor products of operator algebras over quantum groups~\cite{MeyerRoyWoronowicz:2016TwistedII,DeCommerKrajczok:2024Braided} are likely to play a role. An analytic full center would also provide a route to~\cref{conj:bulkFromBoundary}. Together with quantum lattices, it would complete a geometric construction of 2d CFTs: quantum lattices would produce quantum boundary spaces, and the full center would turn them into modular-invariant quantum $\gg_b$-bispaces describing unitary CFTs. Compactness of the bulk would require an additional hypothesis.

\paragraph{Known CFTs as quantum measure spaces.} There is also a less ambitious, but concrete, direction. Many compact unitary CFTs with $c>1$ and their boundary conditions are known: theories of several compact free bosons, WZW models of compact groups, tensor products of minimal models, their orbifolds. Assuming the reconstruction conjectures, each of them defines a quantum bulk space, and each of its elementary boundary conditions a quantum boundary space. Our framework therefore provides an algorithm that produces, from CFT data, type III$_1$ factors carrying ergodic actions of a noncompact locally compact quantum group with an invariant state. Such actions are of independent interest in operator algebras; see, e.g.,~\cite{MarrakchiVaes:2024ErgodicStates} for related questions about ergodic states on type III$_1$ factors. It would be very interesting to carry out this construction explicitly in examples, and to describe the resulting algebras, states, and actions directly. One would like to know, for instance, whether the resulting quantum $\gg_b$-spaces are of quotient type or embeddable in the sense of~\cite{KasprzakSoltan2014Embeddable}.

\paragraph{Boundedness and conformal nets.} The framework of this paper shares its operator-algebraic language with the approach to CFT based on nets of von Neumann algebras associated with spacetime regions~\cite{BrunettiGuidoLongo:1993Modular,LongoRehren:1995Nets,KawahigashiLongoMuger:2001Modularity,Kawahigashi:2015Conformal}, but the two differ in an important respect. In a conformal net, a smeared primary field is typically unbounded; taking bounded functions of it, for instance by exponentiation, gives bounded operators but mixes different conformal families. In a quantum boundary or bulk space, by contrast, quantum automorphic forms are bounded operators transforming irreducibly under the quantum group: boundedness and irreducibility coexist. The price is that boundedness must be established. The regularity conjectures (Conjectures~\ref{conj:regularity} and~\ref{conj:bulkRegularity}) and the existence part of the reconstruction conjectures (Conjectures~\ref{conj:boundaryReconstruction} and~\ref{conj:converseTheorem}) amount to showing that the operators reconstructed from CFT data are bounded. We expect this to require quantitative control over the growth of structure constants, in the spirit of sup-norm estimates for classical automorphic forms~\cite{BernsteinReznikov2004}. It would also be interesting to relate the two frameworks directly. In a conformal net, the modular flow of the vacuum on the algebra of an interval acts geometrically~\cite{BrunettiGuidoLongo:1993Modular}. In a quantum measure space, the modular flow of $\theta$ is implemented by the scaling group of $\gg_b$ (\cref{prop:UNabla,prop:modularFlowPsi}). Understanding whether these facts are related could clarify how the algebra $N$ is encoded in the local algebras of the net, and vice versa.

\paragraph{Foundations.} Several mathematical questions remain open on the quantum group side. These include proving that $\gg_b$ is a locally compact quantum group for $|b|=1$ (\cref{clm:GbLCQG}), establishing the Plancherel theorem in this regime (\cref{clm:plancherel}), and classifying the unitary dual (\cref{conj:unitaryDual}). The decomposition of $L^2(\xx,\theta)$ into the representations of~\cref{conj:unitaryDual} was assumed throughout, and a proof would require establishing type I-ness of the relevant universal C$^*$-algebra. Another open question is uniqueness of invariant trilinear forms (\cref{conj:trilinearUniqueness}), on which the definition of structure constants relies. On the CFT side, the central open problem is to prove the equivalence between quantum associativity and Virasoro crossing (Conjectures~\ref{conj:equivalence2} and~\ref{conj:equivalence3}). These conjectures are completeness statements about families of linear functionals and~\cref{app:functionalNumerics} provides a first numerical exploration.

\paragraph{Chiral algebras and other generalizations.} Chiral algebras were treated implicitly: Virasoro primary conserved currents appear as summands of $L^2(\xx,\theta)$ that are trivial for one of the two actions. In particular, generic Virasoro CFTs correspond precisely to quantum bulk spaces on which $\alpha_L$ and $\alpha_R$ are separately ergodic (\cref{prop:separateErgodicity}). This gives the search for them a clean structural formulation. It would also be interesting to find analogous reformulations for theories that are irrational with respect to a larger chiral algebra. For $W_N$ algebras, natural candidates are the quantum groups $\mathrm{SL}^+_b(N,\rr)$~\cite{DeCommerSchraderShapiroVoigt2025}, whose representation theory has been studied in~\cite{FrenkelIp2014}. Fermionic and superconformal theories may admit similar descriptions.

\paragraph{Beyond two dimensions and beyond conformal symmetry.} Finally, it is tempting to view our results as an instance of a broader dictionary between quantum field theories and noncommutative spaces, summarized in~\cref{tab:speculative}. In this dictionary, a conformal field theory becomes a quantum measure space with a symmetry, its operators become functions on the space, its primary operators become automorphic forms, and its OPE coefficients become integrals of triple products. The conformal bootstrap becomes associativity of multiplication. Two directions seem particularly natural. First, in higher-dimensional CFTs the conformal bootstrap is likewise an expression of the associativity of the operator product expansion. Its reformulation in terms of harmonic analysis and 6j-symbols was proposed in~\cite{Gadde2017}. However, it is not clear what should replace $\gg_b$, which in two dimensions encodes the representation theory of the Virasoro algebra together with the braiding and fusion of conformal blocks. Second, away from conformality, one may ask how renormalization group flows act on the corresponding noncommutative spaces. In two-dimensional sigma models, the leading-order RG flow is the Ricci flow of the target-space metric~\cite{Friedan1980}. This suggests that RG flows between CFTs might be described as a form of noncommutative Ricci flow on quantum measure spaces.

\begin{table}[t]
\centering
\renewcommand{\arraystretch}{1.2}
\begin{tabular}{c c}
\toprule
quantum field theory & noncommutative space $\xx$\\
conformal field theory & $\xx$ with a noncompact quantum symmetry\\
operators & $L^{\infty}(\xx)$\\
conformal primaries & automorphic forms on $\xx$\\
scaling dimensions & spectrum of irreducible representations\\
OPE coefficients & integrals of triple products\\
conformal bootstrap & associativity\\
RG flow & noncommutative Ricci flow?\\
\bottomrule
\end{tabular}
\caption{A speculative dictionary between quantum field theories and noncommutative spaces. The entries above the last row are made precise in this paper for two-dimensional CFTs with $c>1$.}
\label{tab:speculative}
\end{table}

\acknowledgments
The author thanks Anshul Adve, Jan Albert, Yichul Choi, Cl\'{e}ment Delcamp, Petr Kravchuk, Sridip Pal, Eric Perlmutter, Leonardo Rastelli, Sylvain Ribault, Yiannis Tsiares, Jakub Vo\v{s}mera, Zhuo Yang, and Xiang Zhao for discussions. He is grateful to Abhijit Gadde for writing~\cite{Gadde2017}, which served as early inspiration for this work.

This work was supported by the European Research Council (ERC) under the European Union's Horizon Europe research and innovation programme (grant agreement No.~101231994, HARMONICON).

\newpage

\appendix
\section{A free-boson test of quantum-group associativity}
\label{app:freeBosonNumerics}

This appendix describes a direct numerical test of the agreement between
boundary Virasoro crossing and quantum-group associativity. We start with
exact free-boson four-point functions, extract their Virasoro OPE data, and
use the same data in the channel expansions~\eqref{eq:BAssoc2} of the quantum-group
quadrilinear forms. The latter are evaluated from their integral kernels.

The numerical calculations were carried out using Mathematica and commercially-available AI tools.

\subsection{Exact CFT input and choice of parameters}

Consider $d$ compact free bosons with a Neumann boundary condition, so that
$c=d$. Normalize boundary vertex operators $V_q$ at momentum $q\in \rr^d$ by
\begin{equation}
 h(V_q)=\frac{q\cdot q}{2},\qquad
 \left\langle\prod_j V_{q_j}(x_j)\right\rangle
 =\delta_{\sum_jq_j,0}\prod_{j<k}|x_j-x_k|^{q_j\cdot q_k},
\end{equation}
and set $\cO_q=(V_q+V_{-q})/\sqrt2$. The momenta used below lie on one or two
orthogonal axes; the compactification radii can be chosen to make these belong to the lattice of allowed momenta. The other bosons contribute to the central charge even
though the external operators do not depend on them.

We use the ordered conformal frame $(\infty,1,\chi,0)$, with $0<\chi<1$.
For four identical fields of weight $h$, the exact correlator is
\begin{equation}
 G_h(\chi)=\frac12\left[
 \chi^{-2h}(1-\chi)^{-2h}
 +\chi^{2h}(1-\chi)^{-2h}
 +\chi^{-2h}(1-\chi)^{2h}\right].
 \label{eq:numericsEqualBoson}
\end{equation}
For $A=\cO_q$, $B=\cO_r$ with $q\cdot r=0$ and weights $h_A,h_B$, we instead have
\begin{equation}
 G_{AABB}(\chi)
 :=\langle A(\infty)A(1)B(\chi)B(0)\rangle
 =\chi^{-2h_B}.
 \label{eq:numericsUnequalBoson}
\end{equation}
The identity has coefficient one in both channels of
\eqref{eq:numericsEqualBoson}, and in the $s$ channel of
\eqref{eq:numericsUnequalBoson}. It does not appear in the $t$ channel of \eqref{eq:numericsUnequalBoson}.

We use
\begin{equation}
\begin{array}{c|c|c|c}
c&Q&b&Q^2/4\\ \hline
19&\sqrt3&e^{i\pi/6}&3/4\\
31&\sqrt5&(\sqrt5-1)/2&5/4
\end{array}
\label{eq:numericsCharges}
\end{equation}
We use the spectral labels
\begin{equation}
 \nu(h)=
 \begin{cases}
 \sqrt{h-Q^2/4},& h\geq Q^2/4,\\
 i\sqrt{Q^2/4-h},&0\leq h<Q^2/4.
 \end{cases}
\end{equation}
Table~\ref{tab:freeBosonNumerics} lists the six external-weight configurations.

We normalize the Virasoro blocks by
$\cG_{h_1,h_2,h_3,h_4;h}(\chi)=\chi^{h-h_3-h_4}(1+O(\chi))$.
The exact correlators determine aggregate OPE products
\begin{equation}
 p_s(h)=\sum_{m:h_m=h}c_{ijm}c_{k\ell m},\qquad
 p_t(h)=\sum_{m:h_m=h}c_{\ell im}c_{jkm}.
 \label{eq:numericsOPEProducts}
\end{equation}
Resolving individual primaries at a degenerate weight is unnecessary,
since both the Virasoro blocks and the quantum-group forms depend only on
the representation labels. The $t$-channel blocks have external order
$(h_1,h_4,h_3,h_2)$ and argument $1-\chi$.

We determine $p_s$ and $p_t$ separately by recursive subtraction of the
Virasoro block series from the exact correlators. For identical fields the
possible exchanged weights $h_m$ lie in $\mathbb Z_{\geq0}$ or
$4h+\mathbb Z_{\geq0}$; for the orthogonal-charge examples they lie in
$\mathbb Z_{\geq0}$ in the $s$ channel and
$h_A+h_B+\mathbb Z_{\geq0}$ in the $t$ channel.

\subsection{Direct evaluation of the quantum-group forms}

For all six configurations we use the same test tensor
\begin{equation}
 F=f_1\otimes f_2\otimes f_3\otimes f_4,\quad
 f_j(x)=e^{-(x-a_j)^2},\quad
 (a_1,a_2,a_3,a_4)=(-0.3,0.1,0.2,-0.1).
 \label{eq:numericsGaussian}
\end{equation}
Note that $f_j$ belongs to the space $\cE_{\nu}$ of analytic test vectors for any nontrivial $\nu$. With $\boldsymbol\nu=(\nu(h_1),\ldots,\nu(h_4))$, define the partial sums of the two sides of~\eqref{eq:BAssoc2}, truncated at $h = H$:
\begin{equation}
 S_H(F)=\sum_{h\leq H}p_s(h)\cB^s_{\boldsymbol\nu;\nu(h)}(F),\qquad
 T_H(F)=\sum_{h\leq H}p_t(h)\cB^t_{\boldsymbol\nu;\nu(h)}(F).
 \label{eq:numericsChannelSums}
\end{equation}

We evaluate the $s$-channel form using the formula
\begin{equation}
 \cB^s_{\boldsymbol\nu;\nu_s}(F)
 =\cD_{\nu_s,\nu_s}\left(
 \cT^{\nu_s}_{\nu_1,\nu_2}(f_1,f_2),
 \cT^{\nu_s}_{\nu_3,\nu_4}(f_3,f_4)\right),
 \label{eq:numericsContraction}
\end{equation}
with precisely the Clebsch--Gordan kernel and normalization $\cN_b$ used
in the text. The bilinear
pairing is evaluated in momentum space:
\begin{equation}
 \cD_{\nu,\nu}(g_1,g_2)
 =\int_{\rr}e^{\pi Qp}\frac{s_b(p+\nu)}{s_b(p-\nu)}
 \widehat g_1(p)\widehat g_2(-p)\,dp.
 \label{eq:numericsDpairing}
\end{equation}
The $t$-channel form is computed by the KMS cyclic identity
\begin{equation}
 \cB^t_{\nu_1,\nu_2,\nu_3,\nu_4;\nu}(f_1,f_2,f_3,f_4)
 =\cB^s_{\nu_4,\nu_1,\nu_2,\nu_3;\nu}
 (\nabla^{-1}_{\nu_4}f_4,f_1,f_2,f_3),
\end{equation}
where $\nabla^{-1}_{\nu_4}f_4(x)=f_4(x+iQ)$. In the Gaussian
implementation this is the shift $a_4\mapsto a_4-iQ$. This uses cyclicity
of an individual form, not the associativity relation being tested.

Translation invariance of the Clebsch--Gordan kernel allows one Gaussian
integral to be performed analytically. The remaining integrals use composite Gauss--Legendre quadrature. For $c=19$, the special functions are evaluated using convergent product
representations. For real $b$ at $c=31$, we instead use the subtracted
integral
\begin{equation}
 \log s_b(z)=2i\int_0^\infty\left[
 \frac{z}{t^2}-\frac{\sin(zt)}{4t\sinh(bt/2)\sinh(t/(2b))}
 \right]dt,\qquad |\operatorname{Im}z|<Q/2,
\end{equation}
together with the shift identities.

\subsection{Results}

\begin{figure}[htbp]
\centering
\includegraphics[width=\linewidth]{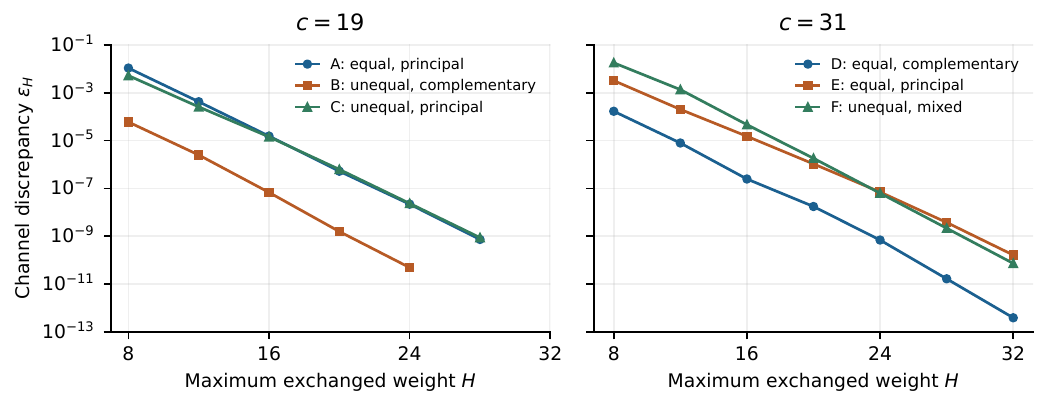}
\caption{Relative channel discrepancy as the spectral cutoff increases.
Each curve uses the values truncated at the indicated
$H$.}
\label{fig:freeBosonConvergence}
\end{figure}

We quantify channel agreement by
\begin{equation}
 \epsilon_H=\frac{|S_H(F)-T_H(F)|}
 {\max\{|S_H(F)|,|T_H(F)|\}}.
 \label{eq:numericsResidual}
\end{equation}
Table~\ref{tab:freeBosonNumerics} gives the final residuals, and
Figure~\ref{fig:freeBosonConvergence} shows their dependence on the maximum
exchanged weight. The results strongly indicate that the free-boson spectral data do satisfy the quantum group associativity equations.

\begin{table}[htbp]
\centering\small
\begin{tabular}{ccllrr}
\toprule
Case&$c$&$(h_1,h_2,h_3,h_4)$&External series&$H$&$\epsilon_H$\\
\midrule
A&19&$(1.1,1.1,1.1,1.1)$&Principal&28&$7.4\times10^{-10}$\\
B&19&$(0.1,0.1,0.18,0.18)$&Complementary&24&$4.9\times10^{-11}$\\
C&19&$(0.8,0.8,1.1,1.1)$&Principal&28&$8.8\times10^{-10}$\\
D&31&$(0.4,0.4,0.4,0.4)$&Complementary&32&$3.9\times10^{-13}$\\
E&31&$(1.6,1.6,1.6,1.6)$&Principal&32&$1.7\times10^{-10}$\\
F&31&$(0.8,0.8,1.6,1.6)$&Mixed&32&$7.4\times10^{-11}$\\
\bottomrule
\end{tabular}
\caption{Direct tests of quantum-group associativity with the Gaussian
tensor~\eqref{eq:numericsGaussian}. Equal-weight cases use
\eqref{eq:numericsEqualBoson}; unequal-weight cases use
\eqref{eq:numericsUnequalBoson}. ``Mixed'' means two complementary and
two principal external representations.}
\label{tab:freeBosonNumerics}
\end{table}

The identity contribution is included explicitly using
\begin{equation}
 \cB^s_{\nu_A,\nu_A,\nu_B,\nu_B;i\frac{Q}{2}}(F)
 =\cD_{\nu_A,\nu_A}(f_1,f_2)\cD_{\nu_B,\nu_B}(f_3,f_4),
\end{equation}
and its $t$-channel counterpart when the external labels permit it.
For example, in case D the identity contributions are
\begin{equation}
 S_0=1.138576236+0.070306346i,\qquad
 T_0=1.190502653+0.225371110i,
\end{equation}
whereas the full sums at $H=32$ both approach
$1.274725108+0.096157150i$. Thus the observed agreement involves a nontrivial
balance between the vacuum and non-vacuum contributions.

\section{A functional test of the quantum group--Virasoro correspondence}
\label{app:functionalNumerics}

This appendix tests the correspondence~\eqref{eq:BfromG}. For a fixed test tensor $F$, the question is whether a single functional on Virasoro blocks reproduces both $\cB^s(F)$ and $\cB^t(F)$.

The numerical calculations were carried out using Mathematica and commercially-available AI tools.

\subsection{Setup and fitting}

We summarize two tests, using the central charges and values of $b$ in
\eqref{eq:numericsCharges}. The external weights are
\begin{equation}
 (h_1,h_2,h_3,h_4)=
 \begin{cases}
 (0.5,0.5,0.5,0.5),&c=19,\\
 (1.6,1.6,1.6,1.6),&c=31.
 \end{cases}
\end{equation}
Thus all four external representations are complementary in the first
case and principal in the second. Set $h_*:=Q^2/4=(c-1)/24$ and use the
Gaussian tensor $F=\bigotimes_{j=1}^4 f_j$ with
\begin{equation}
f_j(x)=e^{-4(x-z_j)^2},\qquad
 (z_1,z_2,z_3,z_4)=(-0.3,0.1,0.2,-0.1)+\tfrac{iQ}{8}(-3,-1,1,3).
 \label{eq:functionalGaussian}
\end{equation}
Note that $f_j \in \cE_{\nu}$ for any nontrivial $\nu$. The quantum group forms are evaluated from the same integral kernels as in Appendix~\ref{app:freeBosonNumerics}.

Using the disk coordinate $w$ of~\eqref{eq:wVar}, write the finite derivative
functional in the Taylor-coefficient basis as
\begin{equation}
 \omega_N(g)=\sum_{n=0}^N d_n\frac{g^{(n)}(0)}{n!}.
 \label{eq:functionalNumericsFit}
\end{equation}
The coefficients $d_n$ are determined by weighted least-squares fitting of
$\omega_N(\cG^s_h)$ to $\cB^s_{\nu(h)}(F)$ at 73 principal-series weights:
\begin{equation}
 \begin{split}
 \,&h=h_*+j/16,\qquad j=0,\ldots,64,\quad \text{and}\\
&h-h_*\in\{5,6.75,8.75,10.75,12.75,14.75,16.75,18.75\}.
 \end{split}
\end{equation}
The weights account for the magnitudes of the target values and their
estimated quadrature errors.

Without changing the fitted coefficients, we evaluate $\omega_N$ on both
channels at the following weights, all excluded from the fit:
\begin{align}
 \mathcal W_{\mathrm{low}}
 &=\{h_*+(3.5+8j)/16:j=0,\ldots,7\},\\
 \mathcal W_{\mathrm{high}}
 &=\{h_*+5.5,h_*+9.75,h_*+15.75\},\\
 \mathcal W_{\mathrm{comp}}
 &=\{0.02h_*,0.2h_*,0.8h_*\},\qquad
 \mathcal W_{\mathrm{vac}}=\{0\}.
\end{align}
\subsection{Results}

For a set $\mathcal W$ of validation weights and a channel $\gamma=s,t$, define the relative error as
\begin{equation}
 \epsilon^\gamma_N(\mathcal W)=
 \left[
 \frac{\displaystyle\sum_{h\in\mathcal W}
 |\omega_N(\cG^\gamma_h)-\cB^\gamma_{\nu(h)}(F)|^2}
 {\displaystyle\sum_{h\in\mathcal W}|\cB^\gamma_{\nu(h)}(F)|^2}
 \right]^{1/2}.
 \label{eq:functionalResidual}
\end{equation}
Table~\ref{tab:functionalNumerics} reports the results at the common order $N=24$.

\begin{table}[htbp]
\centering\small
\begin{tabular}{llrr}
\toprule
Validation set&Channel&$c=19$, $h_j=0.5$&$c=31$, $h_j=1.6$\\
\midrule
Lower principal&$s$&$1.9\times10^{-12}$&$1.6\times10^{-12}$\\
Lower principal&$t$&$1.3\times10^{-6}$&$3.8\times10^{-9}$\\
Complementary&$s$&$3.7\times10^{-6}$&$3.0\times10^{-4}$\\
Complementary&$t$&$3.5\times10^{-10}$&$1.2\times10^{-6}$\\
Vacuum&$s$&$7.2\times10^{-6}$&$6.4\times10^{-4}$\\
Vacuum&$t$&$5.8\times10^{-10}$&$2.3\times10^{-6}$\\
\midrule
Higher principal&$s$&$3.8\times10^{-2}$&$9.2\times10^{-1}$\\
Higher principal&$t$&$8.0\times10^{3}$&$2.0\times10^{1}$\\
\bottomrule
\end{tabular}
\caption{Relative errors~\eqref{eq:functionalResidual} at $N=24$. The lower and higher principal sets are $\mathcal W_{\mathrm{low}}$ and $\mathcal W_{\mathrm{high}}$, respectively.}
\label{tab:functionalNumerics}
\end{table}

The agreement at lower principal weights extends to the complementary
series and the vacuum in both channels. This is particularly nontrivial
for the $t$-channel vacuum: its factorized value is predicted with relative
errors $5.8\times10^{-10}$ and $2.3\times10^{-6}$, despite the fit containing
neither vacuum nor $t$-channel input. More generally, the same functional
simultaneously reproduces all six lower-weight validation groups in the
table.

The finite derivative approximation has its limitations. In particular, agreement fails at the higher principal-series weights shown in the table. Improving accuracy with increasing $N$ requires adjustment of the fitting scheme. The present numerical experiment does not establish convergence as $N\to\infty$ or uniform control over the full spectrum. Its positive result is the simultaneous prediction of both channels, including complementary and vacuum exchange, from principal-series data in a single channel. Together with the direct free-boson associativity test, this provides substantial numerical evidence for the proposed equivalence.

\bibliographystyle{./ytphys}
\bibliography{./refs-v2}
\end{document}